\documentclass[traditabstract,longauth]{aa}
\usepackage{graphicx}
\usepackage{txfonts}
\usepackage{placeins}
\usepackage{todonotes}
\usepackage[normalem]{ulem}
\usepackage{amstext}
\usepackage[breaklinks=true]{hyperref}
\usepackage{xspace}
\usepackage{lscape}

\usepackage{listings}
\makeatletter
\renewcommand*\maketitle{%
  \thispagestyle{firstpage}
\begingroup
    \if@wideboxfn
    \setlength\bibindent{1.4\parindent}
    \else
    \setlength\bibindent{\parindent}
    \fi
    \renewcommand*\thefootnote{\@fnsymbol\c@footnote}%
    \renewcommand\@makefntext[1]{%
    \ifaa@longfn\hsize\textwidth\fi
    \noindent
    \hb@xt@\bibindent{\hss\@makefnmark\enspace}##1}
  \ifaa@twocolumn
  \begingroup
    \begin{aa@strip}
          \aa@maketitle
    \end{aa@strip}
    \@thanks	  	
  \endgroup
  \else
    \begingroup
      \let\thanks\footnote
      \aa@maketitle
    \endgroup
  \fi
\endgroup
  \setcounter{footnote}{0}%
}
\makeatother

\definecolor{dkgreen}{rgb}{0,0.6,0}
\definecolor{gray}{rgb}{0.5,0.5,0.5}
\definecolor{mauve}{rgb}{0.58,0,0.82}
\newcommand{\orcit}[1]{\protect\href{https://orcid.org/#1}{\protect\includegraphics[width=8pt]{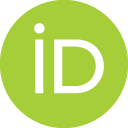}}}
\newcommand{\gdr}[1]{\gaia~DR#1\xspace}
\newcommand{\gedr}{\gaia~EDR3\xspace}

\newcommand{\hip}{{ Hipparcos}\xspace}
\newcommand{\tyc}{{ Thyco}\xspace}
\newcommand{\tyctwo}{{ Thyco~2}\xspace}
\newcommand{\cu}{CU}% We could change this to CalUnit if the referee complains about CU (as it happened for JMC's paper). This is why it is important to use the \cu command!!

\newcommand{\gaia}{\textit{Gaia}\xspace}
\newcommand{\gband}{$G$--band\xspace}
\newcommand{\gbp}{\ensuremath{G_{\rm BP}}\xspace}
\newcommand{\grp}{\ensuremath{G_{\rm RP}}\xspace}
\newcommand{\bprp}{\ensuremath{G_{\rm BP}-G_{\rm RP}}\xspace}
\newcommand{\ggrp}{\ensuremath{G-G_{\rm RP}}\xspace}

\newcommand{\xp}{BP and RP\xspace}

\newcommand{\wrt}{with respect to\xspace}

\newcommand{\webref}[1]{\href{#1}{#1}}
\newcommand{\secname}{Sect.}
\newcommand{\equref}[1]{Eq.~\ref{eq:#1}}
\newcommand{\secref}[1]{\secname~\ref{sec:#1}}

\newcommand{\appref}[1]{Appendix~\ref{sec:#1}}
\newcommand{\figref}[1]{Fig.~\ref{fig:#1}}
\newcommand{\figsref}[1]{Figs.~\ref{fig:#1}}
\newcommand{\afigref}[1]{Figure~\ref{fig:#1}}
\newcommand{\afigsref}[1]{Figures~\ref{fig:#1}}
\newcommand{\tabref}[1]{Table~\ref{tab:#1}}

\hypersetup{draft}
\begin{document}

\title{\gaia Data Release 3: The Galaxy in your preferred colours. Synthetic photometry from Gaia low-resolution spectra}
\authorrunning{Gaia Collaboration: Montegriffo, Bellazzini, De Angeli et al.}
\titlerunning{Gaia DR3: Synthetic photometry from low-resolution spectra}

%%%%%%%%%%%%%%%%%%%%%%%%%%%%%%%%%%%%%%% AUTHORS
%\input{authors_dv2}
\author{
{\it Gaia} Collaboration
\and         P.~                   Montegriffo\orcit{0000-0001-5013-5948}\inst{\ref{inst:0001}}
\and         M.~                    Bellazzini\orcit{0000-0001-8200-810X}\inst{\ref{inst:0001}}
\and         F.~                     De Angeli\orcit{0000-0003-1879-0488}\inst{\ref{inst:0003}}
\and         R.~                        Andrae\orcit{0000-0001-8006-6365}\inst{\ref{inst:0004}}
\and       M.A.~                       Barstow\orcit{0000-0002-7116-3259}\inst{\ref{inst:0005}}
\and         D.~                       Bossini\orcit{0000-0002-9480-8400}\inst{\ref{inst:0006}}
\and         A.~                     Bragaglia\orcit{0000-0002-0338-7883}\inst{\ref{inst:0001}}
\and       P.W.~                       Burgess\inst{\ref{inst:0003}}
\and         C.~                      Cacciari\orcit{0000-0001-5174-3179}\inst{\ref{inst:0001}}
\and       J.M.~                      Carrasco\orcit{0000-0002-3029-5853}\inst{\ref{inst:0010}}
\and         N.~                       Chornay\orcit{0000-0002-8767-3907}\inst{\ref{inst:0003}}
\and         L.~                    Delchambre\orcit{0000-0003-2559-408X}\inst{\ref{inst:0012}}
\and       D.W.~                         Evans\orcit{0000-0002-6685-5998}\inst{\ref{inst:0003}}
\and         M.~                     Fouesneau\orcit{0000-0001-9256-5516}\inst{\ref{inst:0004}}
\and         Y.~                    Fr\'{e}mat\orcit{0000-0002-4645-6017}\inst{\ref{inst:0015}}
\and         D.~                      Garabato\orcit{0000-0002-7133-6623}\inst{\ref{inst:0016}}
\and         C.~                         Jordi\orcit{0000-0001-5495-9602}\inst{\ref{inst:0010}}
\and         M.~                      Manteiga\orcit{0000-0002-7711-5581}\inst{\ref{inst:0018}}
\and         D.~                       Massari\orcit{0000-0001-8892-4301}\inst{\ref{inst:0001}}
\and         L.~                     Palaversa\orcit{0000-0003-3710-0331}\inst{\ref{inst:0020},\ref{inst:0003}}
\and         E.~                       Pancino\orcit{0000-0003-0788-5879}\inst{\ref{inst:0022},\ref{inst:0023}}
\and         M.~                        Riello\orcit{0000-0002-3134-0935}\inst{\ref{inst:0003}}
\and         D.~                    Ruz Mieres\orcit{0000-0002-9455-157X}\inst{\ref{inst:0003}}
\and         N.~                         Sanna\orcit{0000-0001-9275-9492}\inst{\ref{inst:0022}}
\and         R.~                 Santove\~{n}a\orcit{0000-0002-9257-2131}\inst{\ref{inst:0016}}
\and         R.~                         Sordo\orcit{0000-0003-4979-0659}\inst{\ref{inst:0028}}
\and         A.~                     Vallenari\orcit{0000-0003-0014-519X}\inst{\ref{inst:0028}}
\and       N.A.~                        Walton\orcit{0000-0003-3983-8778}\inst{\ref{inst:0003}}
\and     A.G.A.~                         Brown\orcit{0000-0002-7419-9679}\inst{\ref{inst:0031}}
\and         T.~                        Prusti\orcit{0000-0003-3120-7867}\inst{\ref{inst:0032}}
\and     J.H.J.~                    de Bruijne\orcit{0000-0001-6459-8599}\inst{\ref{inst:0032}}
\and         F.~                        Arenou\orcit{0000-0003-2837-3899}\inst{\ref{inst:0034}}
\and         C.~                     Babusiaux\orcit{0000-0002-7631-348X}\inst{\ref{inst:0035},\ref{inst:0034}}
\and         M.~                      Biermann\inst{\ref{inst:0037}}
\and       O.L.~                       Creevey\orcit{0000-0003-1853-6631}\inst{\ref{inst:0038}}
\and         C.~                     Ducourant\orcit{0000-0003-4843-8979}\inst{\ref{inst:0039}}
\and         L.~                          Eyer\orcit{0000-0002-0182-8040}\inst{\ref{inst:0040}}
\and         R.~                        Guerra\orcit{0000-0002-9850-8982}\inst{\ref{inst:0041}}
\and         A.~                        Hutton\inst{\ref{inst:0042}}
\and       S.A.~                       Klioner\orcit{0000-0003-4682-7831}\inst{\ref{inst:0043}}
\and       U.L.~                       Lammers\orcit{0000-0001-8309-3801}\inst{\ref{inst:0041}}
\and         L.~                     Lindegren\orcit{0000-0002-5443-3026}\inst{\ref{inst:0045}}
\and         X.~                          Luri\orcit{0000-0001-5428-9397}\inst{\ref{inst:0010}}
\and         F.~                       Mignard\inst{\ref{inst:0038}}
\and         C.~                         Panem\inst{\ref{inst:0048}}
\and         D.~            Pourbaix$^\dagger$\orcit{0000-0002-3020-1837}\inst{\ref{inst:0049},\ref{inst:0050}}
\and         S.~                       Randich\orcit{0000-0003-2438-0899}\inst{\ref{inst:0022}}
\and         P.~                    Sartoretti\inst{\ref{inst:0034}}
\and         C.~                      Soubiran\orcit{0000-0003-3304-8134}\inst{\ref{inst:0039}}
\and         P.~                         Tanga\orcit{0000-0002-2718-997X}\inst{\ref{inst:0038}}
\and     C.A.L.~                  Bailer-Jones\inst{\ref{inst:0004}}
\and         U.~                       Bastian\orcit{0000-0002-8667-1715}\inst{\ref{inst:0037}}
\and         R.~                       Drimmel\orcit{0000-0002-1777-5502}\inst{\ref{inst:0057}}
\and         F.~                        Jansen\inst{\ref{inst:0058}}
\and         D.~                          Katz\orcit{0000-0001-7986-3164}\inst{\ref{inst:0034}}
\and       M.G.~                      Lattanzi\orcit{0000-0003-0429-7748}\inst{\ref{inst:0057},\ref{inst:0061}}
\and         F.~                   van Leeuwen\inst{\ref{inst:0003}}
\and         J.~                        Bakker\inst{\ref{inst:0041}}
\and         J.~                 Casta\~{n}eda\orcit{0000-0001-7820-946X}\inst{\ref{inst:0064}}
\and         C.~                     Fabricius\orcit{0000-0003-2639-1372}\inst{\ref{inst:0010}}
\and         L.~                     Galluccio\orcit{0000-0002-8541-0476}\inst{\ref{inst:0038}}
\and         A.~                      Guerrier\inst{\ref{inst:0048}}
\and         U.~                        Heiter\orcit{0000-0001-6825-1066}\inst{\ref{inst:0068}}
\and         E.~                        Masana\orcit{0000-0002-4819-329X}\inst{\ref{inst:0010}}
\and         R.~                      Messineo\inst{\ref{inst:0070}}
\and         N.~                       Mowlavi\orcit{0000-0003-1578-6993}\inst{\ref{inst:0040}}
\and         C.~                       Nicolas\inst{\ref{inst:0048}}
\and         K.~                  Nienartowicz\orcit{0000-0001-5415-0547}\inst{\ref{inst:0073},\ref{inst:0074}}
\and         F.~                       Pailler\orcit{0000-0002-4834-481X}\inst{\ref{inst:0048}}
\and         P.~                       Panuzzo\orcit{0000-0002-0016-8271}\inst{\ref{inst:0034}}
\and         F.~                        Riclet\inst{\ref{inst:0048}}
\and         W.~                          Roux\orcit{0000-0002-7816-1950}\inst{\ref{inst:0048}}
\and       G.M.~                      Seabroke\orcit{0000-0003-4072-9536}\inst{\ref{inst:0079}}
\and         F.~                  Th\'{e}venin\inst{\ref{inst:0038}}
\and         G.~                  Gracia-Abril\inst{\ref{inst:0081},\ref{inst:0037}}
\and         J.~                       Portell\orcit{0000-0002-8886-8925}\inst{\ref{inst:0010}}
\and         D.~                      Teyssier\orcit{0000-0002-6261-5292}\inst{\ref{inst:0084}}
\and         M.~                       Altmann\orcit{0000-0002-0530-0913}\inst{\ref{inst:0037},\ref{inst:0086}}
\and         M.~                        Audard\orcit{0000-0003-4721-034X}\inst{\ref{inst:0040},\ref{inst:0074}}
\and         I.~                Bellas-Velidis\inst{\ref{inst:0089}}
\and         K.~                        Benson\inst{\ref{inst:0079}}
\and         J.~                      Berthier\orcit{0000-0003-1846-6485}\inst{\ref{inst:0091}}
\and         R.~                        Blomme\orcit{0000-0002-2526-346X}\inst{\ref{inst:0015}}
\and         D.~                      Busonero\orcit{0000-0002-3903-7076}\inst{\ref{inst:0057}}
\and         G.~                         Busso\orcit{0000-0003-0937-9849}\inst{\ref{inst:0003}}
\and         H.~                   C\'{a}novas\orcit{0000-0001-7668-8022}\inst{\ref{inst:0084}}
\and         B.~                         Carry\orcit{0000-0001-5242-3089}\inst{\ref{inst:0038}}
\and         A.~                       Cellino\orcit{0000-0002-6645-334X}\inst{\ref{inst:0057}}
\and         N.~                         Cheek\inst{\ref{inst:0098}}
\and         G.~                    Clementini\orcit{0000-0001-9206-9723}\inst{\ref{inst:0001}}
\and         Y.~                      Damerdji\orcit{0000-0002-3107-4024}\inst{\ref{inst:0012},\ref{inst:0101}}
\and         M.~                      Davidson\inst{\ref{inst:0102}}
\and         P.~                    de Teodoro\inst{\ref{inst:0041}}
\and         M.~              Nu\~{n}ez Campos\inst{\ref{inst:0042}}
\and         A.~                      Dell'Oro\orcit{0000-0003-1561-9685}\inst{\ref{inst:0022}}
\and         P.~                        Esquej\orcit{0000-0001-8195-628X}\inst{\ref{inst:0106}}
\and         J.~   Fern\'{a}ndez-Hern\'{a}ndez\inst{\ref{inst:0107}}
\and         E.~                        Fraile\inst{\ref{inst:0106}}
\and         P.~              Garc\'{i}a-Lario\orcit{0000-0003-4039-8212}\inst{\ref{inst:0041}}
\and         E.~                        Gosset\inst{\ref{inst:0012},\ref{inst:0050}}
\and         R.~                       Haigron\inst{\ref{inst:0034}}
\and      J.-L.~                     Halbwachs\orcit{0000-0003-2968-6395}\inst{\ref{inst:0113}}
\and       N.C.~                        Hambly\orcit{0000-0002-9901-9064}\inst{\ref{inst:0102}}
\and       D.L.~                      Harrison\orcit{0000-0001-8687-6588}\inst{\ref{inst:0003},\ref{inst:0116}}
\and         J.~                 Hern\'{a}ndez\orcit{0000-0002-0361-4994}\inst{\ref{inst:0041}}
\and         D.~                    Hestroffer\orcit{0000-0003-0472-9459}\inst{\ref{inst:0091}}
\and       S.T.~                       Hodgkin\orcit{0000-0002-5470-3962}\inst{\ref{inst:0003}}
\and         B.~                          Holl\orcit{0000-0001-6220-3266}\inst{\ref{inst:0040},\ref{inst:0074}}
\and         K.~                    Jan{\ss}en\orcit{0000-0002-8163-2493}\inst{\ref{inst:0122}}
\and         G.~          Jevardat de Fombelle\inst{\ref{inst:0040}}
\and         S.~                        Jordan\orcit{0000-0001-6316-6831}\inst{\ref{inst:0037}}
\and         A.~                 Krone-Martins\orcit{0000-0002-2308-6623}\inst{\ref{inst:0125},\ref{inst:0126}}
\and       A.C.~                     Lanzafame\orcit{0000-0002-2697-3607}\inst{\ref{inst:0127},\ref{inst:0128}}
\and         W.~                  L\"{ o}ffler\inst{\ref{inst:0037}}
\and         O.~                       Marchal\orcit{ 0000-0001-7461-892}\inst{\ref{inst:0113}}
\and       P.M.~                       Marrese\orcit{0000-0002-8162-3810}\inst{\ref{inst:0131},\ref{inst:0023}}
\and         A.~                      Moitinho\orcit{0000-0003-0822-5995}\inst{\ref{inst:0125}}
\and         K.~                      Muinonen\orcit{0000-0001-8058-2642}\inst{\ref{inst:0134},\ref{inst:0135}}
\and         P.~                       Osborne\inst{\ref{inst:0003}}
\and         T.~                       Pauwels\inst{\ref{inst:0015}}
\and         A.~                  Recio-Blanco\orcit{0000-0002-6550-7377}\inst{\ref{inst:0038}}
\and         C.~                     Reyl\'{e}\orcit{0000-0003-2258-2403}\inst{\ref{inst:0139}}
\and         L.~                     Rimoldini\orcit{0000-0002-0306-585X}\inst{\ref{inst:0074}}
\and         T.~                      Roegiers\orcit{0000-0002-1231-4440}\inst{\ref{inst:0141}}
\and         J.~                       Rybizki\orcit{0000-0002-0993-6089}\inst{\ref{inst:0004}}
\and       L.M.~                         Sarro\orcit{0000-0002-5622-5191}\inst{\ref{inst:0143}}
\and         C.~                        Siopis\orcit{0000-0002-6267-2924}\inst{\ref{inst:0049}}
\and         M.~                         Smith\inst{\ref{inst:0079}}
\and         A.~                      Sozzetti\orcit{0000-0002-7504-365X}\inst{\ref{inst:0057}}
\and         E.~                       Utrilla\inst{\ref{inst:0042}}
\and         M.~                   van Leeuwen\orcit{0000-0001-9698-2392}\inst{\ref{inst:0003}}
\and         U.~                         Abbas\orcit{0000-0002-5076-766X}\inst{\ref{inst:0057}}
\and         P.~               \'{A}brah\'{a}m\orcit{0000-0001-6015-646X}\inst{\ref{inst:0150},\ref{inst:0151}}
\and         A.~                Abreu Aramburu\inst{\ref{inst:0107}}
\and         C.~                         Aerts\orcit{0000-0003-1822-7126}\inst{\ref{inst:0153},\ref{inst:0154},\ref{inst:0004}}
\and       J.J.~                        Aguado\inst{\ref{inst:0143}}
\and         M.~                          Ajaj\inst{\ref{inst:0034}}
\and         F.~                 Aldea-Montero\inst{\ref{inst:0041}}
\and         G.~                     Altavilla\orcit{0000-0002-9934-1352}\inst{\ref{inst:0131},\ref{inst:0023}}
\and       M.A.~                   \'{A}lvarez\orcit{0000-0002-6786-2620}\inst{\ref{inst:0016}}
\and         J.~                         Alves\orcit{0000-0002-4355-0921}\inst{\ref{inst:0162}}
\and       R.I.~                      Anderson\orcit{0000-0001-8089-4419}\inst{\ref{inst:0163}}
\and         E.~                Anglada Varela\orcit{0000-0001-7563-0689}\inst{\ref{inst:0107}}
\and         T.~                        Antoja\orcit{0000-0003-2595-5148}\inst{\ref{inst:0010}}
\and         D.~                        Baines\orcit{0000-0002-6923-3756}\inst{\ref{inst:0084}}
\and       S.G.~                         Baker\orcit{0000-0002-6436-1257}\inst{\ref{inst:0079}}
\and         L.~        Balaguer-N\'{u}\~{n}ez\orcit{0000-0001-9789-7069}\inst{\ref{inst:0010}}
\and         E.~                      Balbinot\orcit{0000-0002-1322-3153}\inst{\ref{inst:0169}}
\and         Z.~                         Balog\orcit{0000-0003-1748-2926}\inst{\ref{inst:0037},\ref{inst:0004}}
\and         C.~                       Barache\inst{\ref{inst:0086}}
\and         D.~                       Barbato\inst{\ref{inst:0040},\ref{inst:0057}}
\and         M.~                        Barros\orcit{0000-0002-9728-9618}\inst{\ref{inst:0125}}
\and         S.~                 Bartolom\'{e}\orcit{0000-0002-6290-6030}\inst{\ref{inst:0010}}
\and      J.-L.~                     Bassilana\inst{\ref{inst:0177}}
\and         N.~                       Bauchet\inst{\ref{inst:0034}}
\and         U.~                      Becciani\orcit{0000-0002-4389-8688}\inst{\ref{inst:0127}}
\and         A.~                     Berihuete\orcit{0000-0002-8589-4423}\inst{\ref{inst:0180}}
\and         M.~                        Bernet\orcit{0000-0001-7503-1010}\inst{\ref{inst:0010}}
\and         S.~                       Bertone\orcit{0000-0001-9885-8440}\inst{\ref{inst:0182},\ref{inst:0183},\ref{inst:0057}}
\and         L.~                       Bianchi\orcit{0000-0002-7999-4372}\inst{\ref{inst:0185}}
\and         A.~                    Binnenfeld\orcit{0000-0002-9319-3838}\inst{\ref{inst:0186}}
\and         S.~               Blanco-Cuaresma\orcit{0000-0002-1584-0171}\inst{\ref{inst:0187}}
\and         T.~                          Boch\orcit{0000-0001-5818-2781}\inst{\ref{inst:0113}}
\and         A.~                       Bombrun\inst{\ref{inst:0189}}
\and         S.~                    Bouquillon\inst{\ref{inst:0086},\ref{inst:0191}}
\and         L.~                      Bramante\inst{\ref{inst:0070}}
\and         E.~                        Breedt\orcit{0000-0001-6180-3438}\inst{\ref{inst:0003}}
\and         A.~                       Bressan\orcit{0000-0002-7922-8440}\inst{\ref{inst:0194}}
\and         N.~                     Brouillet\orcit{0000-0002-3274-7024}\inst{\ref{inst:0039}}
\and         E.~                    Brugaletta\orcit{0000-0003-2598-6737}\inst{\ref{inst:0127}}
\and         B.~                   Bucciarelli\orcit{0000-0002-5303-0268}\inst{\ref{inst:0057},\ref{inst:0061}}
\and         A.~                       Burlacu\inst{\ref{inst:0199}}
\and       A.G.~                     Butkevich\orcit{0000-0002-4098-3588}\inst{\ref{inst:0057}}
\and         R.~                         Buzzi\orcit{0000-0001-9389-5701}\inst{\ref{inst:0057}}
\and         E.~                        Caffau\orcit{0000-0001-6011-6134}\inst{\ref{inst:0034}}
\and         R.~                   Cancelliere\orcit{0000-0002-9120-3799}\inst{\ref{inst:0203}}
\and         T.~                 Cantat-Gaudin\orcit{0000-0001-8726-2588}\inst{\ref{inst:0010},\ref{inst:0004}}
\and         R.~                      Carballo\orcit{0000-0001-7412-2498}\inst{\ref{inst:0206}}
\and         T.~                      Carlucci\inst{\ref{inst:0086}}
\and       M.I.~                     Carnerero\orcit{0000-0001-5843-5515}\inst{\ref{inst:0057}}
\and         L.~                   Casamiquela\orcit{0000-0001-5238-8674}\inst{\ref{inst:0039},\ref{inst:0034}}
\and         M.~                    Castellani\orcit{0000-0002-7650-7428}\inst{\ref{inst:0131}}
\and         A.~                 Castro-Ginard\orcit{0000-0002-9419-3725}\inst{\ref{inst:0031}}
\and         L.~                        Chaoul\inst{\ref{inst:0048}}
\and         P.~                       Charlot\orcit{0000-0002-9142-716X}\inst{\ref{inst:0039}}
\and         L.~                        Chemin\orcit{0000-0002-3834-7937}\inst{\ref{inst:0215}}
\and         V.~                    Chiaramida\inst{\ref{inst:0070}}
\and         A.~                     Chiavassa\orcit{0000-0003-3891-7554}\inst{\ref{inst:0038}}
\and         G.~                     Comoretto\inst{\ref{inst:0084},\ref{inst:0219}}
\and         G.~                      Contursi\orcit{0000-0001-5370-1511}\inst{\ref{inst:0038}}
\and       W.J.~                        Cooper\orcit{0000-0003-3501-8967}\inst{\ref{inst:0221},\ref{inst:0057}}
\and         T.~                        Cornez\inst{\ref{inst:0177}}
\and         S.~                        Cowell\inst{\ref{inst:0003}}
\and         F.~                         Crifo\inst{\ref{inst:0034}}
\and         M.~                       Cropper\orcit{0000-0003-4571-9468}\inst{\ref{inst:0079}}
\and         M.~                        Crosta\orcit{0000-0003-4369-3786}\inst{\ref{inst:0057},\ref{inst:0228}}
\and         C.~                       Crowley\inst{\ref{inst:0189}}
\and         C.~                       Dafonte\orcit{0000-0003-4693-7555}\inst{\ref{inst:0016}}
\and         A.~                    Dapergolas\inst{\ref{inst:0089}}
\and         P.~                         David\inst{\ref{inst:0091}}
\and         P.~                    de Laverny\orcit{0000-0002-2817-4104}\inst{\ref{inst:0038}}
\and         F.~                      De Luise\orcit{0000-0002-6570-8208}\inst{\ref{inst:0234}}
\and         R.~                      De March\orcit{0000-0003-0567-842X}\inst{\ref{inst:0070}}
\and         J.~                     De Ridder\orcit{0000-0001-6726-2863}\inst{\ref{inst:0153}}
\and         R.~                      de Souza\inst{\ref{inst:0237}}
\and         A.~                     de Torres\inst{\ref{inst:0189}}
\and       E.F.~                    del Peloso\inst{\ref{inst:0037}}
\and         E.~                      del Pozo\inst{\ref{inst:0042}}
\and         M.~                         Delbo\orcit{0000-0002-8963-2404}\inst{\ref{inst:0038}}
\and         A.~                       Delgado\inst{\ref{inst:0106}}
\and      J.-B.~                       Delisle\orcit{0000-0001-5844-9888}\inst{\ref{inst:0040}}
\and         C.~                      Demouchy\inst{\ref{inst:0244}}
\and       T.E.~                 Dharmawardena\orcit{0000-0002-9583-5216}\inst{\ref{inst:0004}}
\and         S.~                       Diakite\inst{\ref{inst:0246}}
\and         C.~                        Diener\inst{\ref{inst:0003}}
\and         E.~                     Distefano\orcit{0000-0002-2448-2513}\inst{\ref{inst:0127}}
\and         C.~                       Dolding\inst{\ref{inst:0079}}
\and         H.~                          Enke\orcit{0000-0002-2366-8316}\inst{\ref{inst:0122}}
\and         C.~                         Fabre\inst{\ref{inst:0251}}
\and         M.~                      Fabrizio\orcit{0000-0001-5829-111X}\inst{\ref{inst:0131},\ref{inst:0023}}
\and         S.~                       Faigler\orcit{0000-0002-8368-5724}\inst{\ref{inst:0254}}
\and         G.~                      Fedorets\orcit{0000-0002-8418-4809}\inst{\ref{inst:0134},\ref{inst:0256}}
\and         P.~                      Fernique\orcit{0000-0002-3304-2923}\inst{\ref{inst:0113},\ref{inst:0258}}
\and         F.~                      Figueras\orcit{0000-0002-3393-0007}\inst{\ref{inst:0010}}
\and         Y.~                      Fournier\orcit{0000-0002-6633-9088}\inst{\ref{inst:0122}}
\and         C.~                        Fouron\inst{\ref{inst:0199}}
\and         F.~                     Fragkoudi\orcit{0000-0002-0897-3013}\inst{\ref{inst:0262},\ref{inst:0263},\ref{inst:0264}}
\and         M.~                           Gai\orcit{0000-0001-9008-134X}\inst{\ref{inst:0057}}
\and         A.~              Garcia-Gutierrez\inst{\ref{inst:0010}}
\and         M.~              Garcia-Reinaldos\inst{\ref{inst:0041}}
\and         M.~             Garc\'{i}a-Torres\orcit{0000-0002-6867-7080}\inst{\ref{inst:0268}}
\and         A.~                      Garofalo\orcit{0000-0002-5907-0375}\inst{\ref{inst:0001}}
\and         A.~                         Gavel\orcit{0000-0002-2963-722X}\inst{\ref{inst:0068}}
\and         P.~                        Gavras\orcit{0000-0002-4383-4836}\inst{\ref{inst:0106}}
\and         E.~                       Gerlach\orcit{0000-0002-9533-2168}\inst{\ref{inst:0043}}
\and         R.~                         Geyer\orcit{0000-0001-6967-8707}\inst{\ref{inst:0043}}
\and         P.~                      Giacobbe\orcit{0000-0001-7034-7024}\inst{\ref{inst:0057}}
\and         G.~                       Gilmore\orcit{0000-0003-4632-0213}\inst{\ref{inst:0003}}
\and         S.~                        Girona\orcit{0000-0002-1975-1918}\inst{\ref{inst:0276}}
\and         G.~                     Giuffrida\inst{\ref{inst:0131}}
\and         R.~                         Gomel\inst{\ref{inst:0254}}
\and         A.~                         Gomez\orcit{0000-0002-3796-3690}\inst{\ref{inst:0016}}
\and         J.~    Gonz\'{a}lez-N\'{u}\~{n}ez\orcit{0000-0001-5311-5555}\inst{\ref{inst:0098},\ref{inst:0281}}
\and         I.~   Gonz\'{a}lez-Santamar\'{i}a\orcit{0000-0002-8537-9384}\inst{\ref{inst:0016}}
\and       J.J.~            Gonz\'{a}lez-Vidal\inst{\ref{inst:0010}}
\and         M.~                       Granvik\orcit{0000-0002-5624-1888}\inst{\ref{inst:0134},\ref{inst:0285}}
\and         P.~                      Guillout\inst{\ref{inst:0113}}
\and         J.~                       Guiraud\inst{\ref{inst:0048}}
\and         R.~     Guti\'{e}rrez-S\'{a}nchez\inst{\ref{inst:0084}}
\and       L.P.~                           Guy\orcit{0000-0003-0800-8755}\inst{\ref{inst:0074},\ref{inst:0290}}
\and         D.~                Hatzidimitriou\orcit{0000-0002-5415-0464}\inst{\ref{inst:0291},\ref{inst:0089}}
\and         M.~                        Hauser\inst{\ref{inst:0004},\ref{inst:0294}}
\and         M.~                       Haywood\orcit{0000-0003-0434-0400}\inst{\ref{inst:0034}}
\and         A.~                        Helmer\inst{\ref{inst:0177}}
\and         A.~                         Helmi\orcit{0000-0003-3937-7641}\inst{\ref{inst:0169}}
\and       M.H.~                     Sarmiento\orcit{0000-0003-4252-5115}\inst{\ref{inst:0042}}
\and       S.L.~                       Hidalgo\orcit{0000-0002-0002-9298}\inst{\ref{inst:0299},\ref{inst:0300}}
\and         N.~                   H\l{}adczuk\orcit{0000-0001-9163-4209}\inst{\ref{inst:0041},\ref{inst:0302}}
\and         D.~                         Hobbs\orcit{0000-0002-2696-1366}\inst{\ref{inst:0045}}
\and         G.~                       Holland\inst{\ref{inst:0003}}
\and       H.E.~                        Huckle\inst{\ref{inst:0079}}
\and         K.~                       Jardine\inst{\ref{inst:0306}}
\and         G.~                    Jasniewicz\inst{\ref{inst:0307}}
\and         A.~          Jean-Antoine Piccolo\orcit{0000-0001-8622-212X}\inst{\ref{inst:0048}}
\and     \'{O}.~            Jim\'{e}nez-Arranz\orcit{0000-0001-7434-5165}\inst{\ref{inst:0010}}
\and         J.~             Juaristi Campillo\inst{\ref{inst:0037}}
\and         F.~                         Julbe\inst{\ref{inst:0010}}
\and         L.~                     Karbevska\inst{\ref{inst:0074},\ref{inst:0313}}
\and         P.~                      Kervella\orcit{0000-0003-0626-1749}\inst{\ref{inst:0314}}
\and         S.~                        Khanna\orcit{0000-0002-2604-4277}\inst{\ref{inst:0169},\ref{inst:0057}}
\and         G.~                    Kordopatis\orcit{0000-0002-9035-3920}\inst{\ref{inst:0038}}
\and       A.J.~                          Korn\orcit{0000-0002-3881-6756}\inst{\ref{inst:0068}}
\and      \'{A}~                K\'{o}sp\'{a}l\orcit{\'{u}t 15-17, 1121 }\inst{\ref{inst:0150},\ref{inst:0004},\ref{inst:0151}}
\and         Z.~           Kostrzewa-Rutkowska\inst{\ref{inst:0031},\ref{inst:0323}}
\and         K.~                Kruszy\'{n}ska\orcit{0000-0002-2729-5369}\inst{\ref{inst:0324}}
\and         M.~                           Kun\orcit{0000-0002-7538-5166}\inst{\ref{inst:0150}}
\and         P.~                       Laizeau\inst{\ref{inst:0326}}
\and         S.~                       Lambert\orcit{0000-0001-6759-5502}\inst{\ref{inst:0086}}
\and       A.F.~                         Lanza\orcit{0000-0001-5928-7251}\inst{\ref{inst:0127}}
\and         Y.~                         Lasne\inst{\ref{inst:0177}}
\and      J.-F.~                    Le Campion\inst{\ref{inst:0039}}
\and         Y.~                      Lebreton\orcit{0000-0002-4834-2144}\inst{\ref{inst:0314},\ref{inst:0332}}
\and         T.~                     Lebzelter\orcit{0000-0002-0702-7551}\inst{\ref{inst:0162}}
\and         S.~                        Leccia\orcit{0000-0001-5685-6930}\inst{\ref{inst:0334}}
\and         N.~                       Leclerc\inst{\ref{inst:0034}}
\and         I.~                 Lecoeur-Taibi\orcit{0000-0003-0029-8575}\inst{\ref{inst:0074}}
\and         S.~                          Liao\orcit{0000-0002-9346-0211}\inst{\ref{inst:0337},\ref{inst:0057},\ref{inst:0339}}
\and       E.L.~                        Licata\orcit{0000-0002-5203-0135}\inst{\ref{inst:0057}}
\and     H.E.P.~                  Lindstr{\o}m\inst{\ref{inst:0057},\ref{inst:0342},\ref{inst:0343}}
\and       T.A.~                        Lister\orcit{0000-0002-3818-7769}\inst{\ref{inst:0344}}
\and         E.~                       Livanou\orcit{0000-0003-0628-2347}\inst{\ref{inst:0291}}
\and         A.~                         Lobel\orcit{0000-0001-5030-019X}\inst{\ref{inst:0015}}
\and         A.~                         Lorca\inst{\ref{inst:0042}}
\and         C.~                          Loup\inst{\ref{inst:0113}}
\and         P.~                 Madrero Pardo\inst{\ref{inst:0010}}
\and         A.~               Magdaleno Romeo\inst{\ref{inst:0199}}
\and         S.~                       Managau\inst{\ref{inst:0177}}
\and       R.G.~                          Mann\orcit{0000-0002-0194-325X}\inst{\ref{inst:0102}}
\and       J.M.~                      Marchant\orcit{0000-0002-3678-3145}\inst{\ref{inst:0353}}
\and         M.~                       Marconi\orcit{0000-0002-1330-2927}\inst{\ref{inst:0334}}
\and         J.~                        Marcos\inst{\ref{inst:0084}}
\and     M.M.S.~                 Marcos Santos\inst{\ref{inst:0098}}
\and         D.~                Mar\'{i}n Pina\orcit{0000-0001-6482-1842}\inst{\ref{inst:0010}}
\and         S.~                      Marinoni\orcit{0000-0001-7990-6849}\inst{\ref{inst:0131},\ref{inst:0023}}
\and         F.~                       Marocco\orcit{0000-0001-7519-1700}\inst{\ref{inst:0360}}
\and       D.J.~                      Marshall\orcit{0000-0003-3956-3524}\inst{\ref{inst:0361}}
\and         L.~                   Martin Polo\inst{\ref{inst:0098}}
\and       J.M.~            Mart\'{i}n-Fleitas\orcit{0000-0002-8594-569X}\inst{\ref{inst:0042}}
\and         G.~                        Marton\orcit{0000-0002-1326-1686}\inst{\ref{inst:0150}}
\and         N.~                          Mary\inst{\ref{inst:0177}}
\and         A.~                         Masip\orcit{0000-0003-1419-0020}\inst{\ref{inst:0010}}
\and         A.~          Mastrobuono-Battisti\orcit{0000-0002-2386-9142}\inst{\ref{inst:0034}}
\and         T.~                         Mazeh\orcit{0000-0002-3569-3391}\inst{\ref{inst:0254}}
\and       P.J.~                      McMillan\orcit{0000-0002-8861-2620}\inst{\ref{inst:0045}}
\and         S.~                       Messina\orcit{0000-0002-2851-2468}\inst{\ref{inst:0127}}
\and         D.~                      Michalik\orcit{0000-0002-7618-6556}\inst{\ref{inst:0032}}
\and       N.R.~                        Millar\inst{\ref{inst:0003}}
\and         A.~                         Mints\orcit{0000-0002-8440-1455}\inst{\ref{inst:0122}}
\and         D.~                        Molina\orcit{0000-0003-4814-0275}\inst{\ref{inst:0010}}
\and         R.~                      Molinaro\orcit{0000-0003-3055-6002}\inst{\ref{inst:0334}}
\and         L.~                    Moln\'{a}r\orcit{0000-0002-8159-1599}\inst{\ref{inst:0150},\ref{inst:0151},\ref{inst:0377}}
\and         G.~                        Monari\orcit{0000-0002-6863-0661}\inst{\ref{inst:0113}}
\and         M.~                   Mongui\'{o}\orcit{0000-0002-4519-6700}\inst{\ref{inst:0010}}
\and         A.~                       Montero\inst{\ref{inst:0042}}
\and         R.~                           Mor\orcit{0000-0002-8179-6527}\inst{\ref{inst:0010}}
\and         A.~                          Mora\inst{\ref{inst:0042}}
\and         R.~                    Morbidelli\orcit{0000-0001-7627-4946}\inst{\ref{inst:0057}}
\and         T.~                         Morel\orcit{0000-0002-8176-4816}\inst{\ref{inst:0012}}
\and         D.~                        Morris\inst{\ref{inst:0102}}
\and         T.~                      Muraveva\orcit{0000-0002-0969-1915}\inst{\ref{inst:0001}}
\and       C.P.~                        Murphy\inst{\ref{inst:0041}}
\and         I.~                       Musella\orcit{0000-0001-5909-6615}\inst{\ref{inst:0334}}
\and         Z.~                          Nagy\orcit{0000-0002-3632-1194}\inst{\ref{inst:0150}}
\and         L.~                         Noval\inst{\ref{inst:0177}}
\and         F.~                     Oca\~{n}a\inst{\ref{inst:0084},\ref{inst:0392}}
\and         A.~                         Ogden\inst{\ref{inst:0003}}
\and         C.~                     Ordenovic\inst{\ref{inst:0038}}
\and       J.O.~                        Osinde\inst{\ref{inst:0106}}
\and         C.~                        Pagani\orcit{0000-0001-5477-4720}\inst{\ref{inst:0005}}
\and         I.~                        Pagano\orcit{0000-0001-9573-4928}\inst{\ref{inst:0127}}
\and       P.A.~                       Palicio\orcit{0000-0002-7432-8709}\inst{\ref{inst:0038}}
\and         L.~               Pallas-Quintela\orcit{0000-0001-9296-3100}\inst{\ref{inst:0016}}
\and         A.~                        Panahi\orcit{0000-0001-5850-4373}\inst{\ref{inst:0254}}
\and         S.~               Payne-Wardenaar\inst{\ref{inst:0037}}
\and         X.~         Pe\~{n}alosa Esteller\inst{\ref{inst:0010}}
\and         A.~                 Penttil\"{ a}\orcit{0000-0001-7403-1721}\inst{\ref{inst:0134}}
\and         B.~                        Pichon\orcit{0000 0000 0062 1449}\inst{\ref{inst:0038}}
\and       A.M.~                    Piersimoni\orcit{0000-0002-8019-3708}\inst{\ref{inst:0234}}
\and      F.-X.~                        Pineau\orcit{0000-0002-2335-4499}\inst{\ref{inst:0113}}
\and         E.~                        Plachy\orcit{0000-0002-5481-3352}\inst{\ref{inst:0150},\ref{inst:0151},\ref{inst:0377}}
\and         G.~                          Plum\inst{\ref{inst:0034}}
\and         E.~                        Poggio\orcit{0000-0003-3793-8505}\inst{\ref{inst:0038},\ref{inst:0057}}
\and         A.~                      Pr\v{s}a\orcit{0000-0002-1913-0281}\inst{\ref{inst:0412}}
\and         L.~                        Pulone\orcit{0000-0002-5285-998X}\inst{\ref{inst:0131}}
\and         E.~                        Racero\orcit{0000-0002-6101-9050}\inst{\ref{inst:0098},\ref{inst:0392}}
\and         S.~                       Ragaini\inst{\ref{inst:0001}}
\and         M.~                        Rainer\orcit{0000-0002-8786-2572}\inst{\ref{inst:0022},\ref{inst:0418}}
\and       C.M.~                       Raiteri\orcit{0000-0003-1784-2784}\inst{\ref{inst:0057}}
\and         P.~                         Ramos\orcit{0000-0002-5080-7027}\inst{\ref{inst:0010},\ref{inst:0113}}
\and         M.~                  Ramos-Lerate\inst{\ref{inst:0084}}
\and         P.~                  Re Fiorentin\orcit{0000-0002-4995-0475}\inst{\ref{inst:0057}}
\and         S.~                        Regibo\inst{\ref{inst:0153}}
\and       P.J.~                      Richards\inst{\ref{inst:0425}}
\and         C.~                     Rios Diaz\inst{\ref{inst:0106}}
\and         V.~                        Ripepi\orcit{0000-0003-1801-426X}\inst{\ref{inst:0334}}
\and         A.~                          Riva\orcit{0000-0002-6928-8589}\inst{\ref{inst:0057}}
\and      H.-W.~                           Rix\orcit{0000-0003-4996-9069}\inst{\ref{inst:0004}}
\and         G.~                         Rixon\orcit{0000-0003-4399-6568}\inst{\ref{inst:0003}}
\and         N.~                      Robichon\orcit{0000-0003-4545-7517}\inst{\ref{inst:0034}}
\and       A.C.~                         Robin\orcit{0000-0001-8654-9499}\inst{\ref{inst:0139}}
\and         C.~                         Robin\inst{\ref{inst:0177}}
\and         M.~                       Roelens\orcit{0000-0003-0876-4673}\inst{\ref{inst:0040}}
\and     H.R.O.~                        Rogues\inst{\ref{inst:0244}}
\and         L.~                    Rohrbasser\inst{\ref{inst:0074}}
\and         M.~              Romero-G\'{o}mez\orcit{0000-0003-3936-1025}\inst{\ref{inst:0010}}
\and         N.~                        Rowell\orcit{0000-0003-3809-1895}\inst{\ref{inst:0102}}
\and         F.~                         Royer\orcit{0000-0002-9374-8645}\inst{\ref{inst:0034}}
\and       K.A.~                       Rybicki\orcit{0000-0002-9326-9329}\inst{\ref{inst:0324}}
\and         G.~                      Sadowski\orcit{0000-0002-3411-1003}\inst{\ref{inst:0049}}
\and         A.~        S\'{a}ez N\'{u}\~{n}ez\inst{\ref{inst:0010}}
\and         A.~       Sagrist\`{a} Sell\'{e}s\orcit{0000-0001-6191-2028}\inst{\ref{inst:0037}}
\and         J.~                      Sahlmann\orcit{0000-0001-9525-3673}\inst{\ref{inst:0106}}
\and         E.~                      Salguero\inst{\ref{inst:0107}}
\and         N.~                       Samaras\orcit{0000-0001-8375-6652}\inst{\ref{inst:0015},\ref{inst:0447}}
\and         V.~               Sanchez Gimenez\orcit{0000-0003-1797-3557}\inst{\ref{inst:0010}}
\and         M.~                       Sarasso\orcit{0000-0001-5121-0727}\inst{\ref{inst:0057}}
\and       M.S.~                    Schultheis\orcit{0000-0002-6590-1657}\inst{\ref{inst:0038}}
\and         E.~                       Sciacca\orcit{0000-0002-5574-2787}\inst{\ref{inst:0127}}
\and         M.~                         Segol\inst{\ref{inst:0244}}
\and       J.C.~                       Segovia\inst{\ref{inst:0098}}
\and         D.~                 S\'{e}gransan\orcit{0000-0003-2355-8034}\inst{\ref{inst:0040}}
\and         D.~                        Semeux\inst{\ref{inst:0251}}
\and         S.~                        Shahaf\orcit{0000-0001-9298-8068}\inst{\ref{inst:0456}}
\and       H.I.~                      Siddiqui\orcit{0000-0003-1853-6033}\inst{\ref{inst:0457}}
\and         A.~                       Siebert\orcit{0000-0001-8059-2840}\inst{\ref{inst:0113},\ref{inst:0258}}
\and         L.~                       Siltala\orcit{0000-0002-6938-794X}\inst{\ref{inst:0134}}
\and         A.~                       Silvelo\orcit{0000-0002-5126-6365}\inst{\ref{inst:0016}}
\and         E.~                        Slezak\inst{\ref{inst:0038}}
\and         I.~                        Slezak\inst{\ref{inst:0038}}
\and       R.L.~                         Smart\orcit{0000-0002-4424-4766}\inst{\ref{inst:0057}}
\and       O.N.~                        Snaith\inst{\ref{inst:0034}}
\and         E.~                        Solano\inst{\ref{inst:0466}}
\and         F.~                       Solitro\inst{\ref{inst:0070}}
\and         D.~                        Souami\orcit{0000-0003-4058-0815}\inst{\ref{inst:0314},\ref{inst:0469}}
\and         J.~                       Souchay\inst{\ref{inst:0086}}
\and         A.~                        Spagna\orcit{0000-0003-1732-2412}\inst{\ref{inst:0057}}
\and         L.~                         Spina\orcit{0000-0002-9760-6249}\inst{\ref{inst:0028}}
\and         F.~                         Spoto\orcit{0000-0001-7319-5847}\inst{\ref{inst:0187}}
\and       I.A.~                        Steele\orcit{0000-0001-8397-5759}\inst{\ref{inst:0353}}
\and         H.~            Steidelm\"{ u}ller\inst{\ref{inst:0043}}
\and       C.A.~                    Stephenson\inst{\ref{inst:0084},\ref{inst:0477}}
\and         M.~                  S\"{ u}veges\orcit{0000-0003-3017-5322}\inst{\ref{inst:0478}}
\and         J.~                        Surdej\orcit{0000-0002-7005-1976}\inst{\ref{inst:0012},\ref{inst:0480}}
\and         L.~                      Szabados\orcit{0000-0002-2046-4131}\inst{\ref{inst:0150}}
\and         E.~                  Szegedi-Elek\orcit{0000-0001-7807-6644}\inst{\ref{inst:0150}}
\and         F.~                         Taris\inst{\ref{inst:0086}}
\and       M.B.~                        Taylor\orcit{0000-0002-4209-1479}\inst{\ref{inst:0484}}
\and         R.~                      Teixeira\orcit{0000-0002-6806-6626}\inst{\ref{inst:0237}}
\and         L.~                       Tolomei\orcit{0000-0002-3541-3230}\inst{\ref{inst:0070}}
\and         N.~                       Tonello\orcit{0000-0003-0550-1667}\inst{\ref{inst:0276}}
\and         F.~                         Torra\orcit{0000-0002-8429-299X}\inst{\ref{inst:0064}}
\and         J.~               Torra$^\dagger$\inst{\ref{inst:0010}}
\and         G.~                Torralba Elipe\orcit{0000-0001-8738-194X}\inst{\ref{inst:0016}}
\and         M.~                     Trabucchi\orcit{0000-0002-1429-2388}\inst{\ref{inst:0491},\ref{inst:0040}}
\and       A.T.~                       Tsounis\inst{\ref{inst:0493}}
\and         C.~                         Turon\orcit{0000-0003-1236-5157}\inst{\ref{inst:0034}}
\and         A.~                          Ulla\orcit{0000-0001-6424-5005}\inst{\ref{inst:0495}}
\and         N.~                         Unger\orcit{0000-0003-3993-7127}\inst{\ref{inst:0040}}
\and       M.V.~                      Vaillant\inst{\ref{inst:0177}}
\and         E.~                    van Dillen\inst{\ref{inst:0244}}
\and         W.~                    van Reeven\inst{\ref{inst:0499}}
\and         O.~                         Vanel\orcit{0000-0002-7898-0454}\inst{\ref{inst:0034}}
\and         A.~                     Vecchiato\orcit{0000-0003-1399-5556}\inst{\ref{inst:0057}}
\and         Y.~                         Viala\inst{\ref{inst:0034}}
\and         D.~                       Vicente\orcit{0000-0002-1584-1182}\inst{\ref{inst:0276}}
\and         S.~                     
Voutsinas\inst{\ref{inst:0102}}
%%%%\and         M.~                        
%%%%% Weiler\inst{\ref{inst:0010}}
\and         T.~                        Wevers\orcit{0000-0002-4043-9400}\inst{\ref{inst:0003},\ref{inst:0507}}
\and      \L{}.~                   Wyrzykowski\orcit{0000-0002-9658-6151}\inst{\ref{inst:0324}}
\and         A.~                        Yoldas\inst{\ref{inst:0003}}
\and         P.~                         Yvard\inst{\ref{inst:0244}}
\and         H.~                          Zhao\orcit{0000-0003-2645-6869}\inst{\ref{inst:0038}}
\and         J.~                         Zorec\inst{\ref{inst:0512}}
\and         S.~                        Zucker\orcit{0000-0003-3173-3138}\inst{\ref{inst:0186}}
\and         T.~                       Zwitter\orcit{0000-0002-2325-8763}\inst{\ref{inst:0514}}
}
\institute{
     INAF - Osservatorio di Astrofisica e Scienza dello Spazio di Bologna, via Piero Gobetti 93/3, 40129 Bologna, Italy\relax                                                                                                                                                                                                                                      \label{inst:0001}
\and Institute of Astronomy, University of Cambridge, Madingley Road, Cambridge CB3 0HA, United Kingdom\relax                                                                                                                                                                                                                                                      \label{inst:0003}\vfill
\and Max Planck Institute for Astronomy, K\"{ o}nigstuhl 17, 69117 Heidelberg, Germany\relax                                                                                                                                                                                                                                                                       \label{inst:0004}\vfill
\and School of Physics and Astronomy / Space Park Leicester, University of Leicester, University Road, Leicester LE1 7RH, United Kingdom\relax                                                                                                                                                                                                                     \label{inst:0005}\vfill
\and Instituto de Astrof\'{i}sica e Ci\^{e}ncias do Espa\c{c}o, Universidade do Porto, CAUP, Rua das Estrelas, PT4150-762 Porto, Portugal\relax                                                                                                                                                                                                                    \label{inst:0006}\vfill
\and Institut de Ci\`{e}ncies del Cosmos (ICCUB), Universitat  de  Barcelona  (IEEC-UB), Mart\'{i} i  Franqu\`{e}s  1, 08028 Barcelona, Spain\relax                                                                                                                                                                                                                \label{inst:0010}\vfill
\and Institut d'Astrophysique et de G\'{e}ophysique, Universit\'{e} de Li\`{e}ge, 19c, All\'{e}e du 6 Ao\^{u}t, B-4000 Li\`{e}ge, Belgium\relax                                                                                                                                                                                                                    \label{inst:0012}\vfill
\and Royal Observatory of Belgium, Ringlaan 3, 1180 Brussels, Belgium\relax                                                                                                                                                                                                                                                                                        \label{inst:0015}\vfill
\and CIGUS CITIC - Department of Computer Science and Information Technologies, University of A Coru\~{n}a, Campus de Elvi\~{n}a s/n, A Coru\~{n}a, 15071, Spain\relax                                                                                                                                                                                             \label{inst:0016}\vfill
\and CIGUS CITIC, Department of Nautical Sciences and Marine Engineering, University of A Coru\~{n}a, Paseo de Ronda 51, 15071, A Coru\~{n}a, Spain\relax                                                                                                                                                                                                          \label{inst:0018}\vfill
\and Ru{\dj}er Bo\v{s}kovi\'{c} Institute, Bijeni\v{c}ka cesta 54, 10000 Zagreb, Croatia\relax                                                                                                                                                                                                                                                                     \label{inst:0020}\vfill
\and INAF - Osservatorio Astrofisico di Arcetri, Largo Enrico Fermi 5, 50125 Firenze, Italy\relax                                                                                                                                                                                                                                                                  \label{inst:0022}\vfill
\and Space Science Data Center - ASI, Via del Politecnico SNC, 00133 Roma, Italy\relax                                                                                                                                                                                                                                                                             \label{inst:0023}\vfill
\and INAF - Osservatorio astronomico di Padova, Vicolo Osservatorio 5, 35122 Padova, Italy\relax                                                                                                                                                                                                                                                                   \label{inst:0028}\vfill
\and Leiden Observatory, Leiden University, Niels Bohrweg 2, 2333 CA Leiden, The Netherlands\relax                                                                                                                                                                                                                                                                 \label{inst:0031}\vfill
\and European Space Agency (ESA), European Space Research and Technology Centre (ESTEC), Keplerlaan 1, 2201AZ, Noordwijk, The Netherlands\relax                                                                                                                                                                                                                    \label{inst:0032}\vfill
\and GEPI, Observatoire de Paris, Universit\'{e} PSL, CNRS, 5 Place Jules Janssen, 92190 Meudon, France\relax                                                                                                                                                                                                                                                      \label{inst:0034}\vfill
\and Univ. Grenoble Alpes, CNRS, IPAG, 38000 Grenoble, France\relax                                                                                                                                                                                                                                                                                                \label{inst:0035}\vfill
\and Astronomisches Rechen-Institut, Zentrum f\"{ u}r Astronomie der Universit\"{ a}t Heidelberg, M\"{ o}nchhofstr. 12-14, 69120 Heidelberg, Germany\relax                                                                                                                                                                                                         \label{inst:0037}\vfill
\and Universit\'{e} C\^{o}te d'Azur, Observatoire de la C\^{o}te d'Azur, CNRS, Laboratoire Lagrange, Bd de l'Observatoire, CS 34229, 06304 Nice Cedex 4, France\relax                                                                                                                                                                                              \label{inst:0038}\vfill
\and Laboratoire d'astrophysique de Bordeaux, Univ. Bordeaux, CNRS, B18N, all{\'e}e Geoffroy Saint-Hilaire, 33615 Pessac, France\relax                                                                                                                                                                                                                             \label{inst:0039}\vfill
\and Department of Astronomy, University of Geneva, Chemin Pegasi 51, 1290 Versoix, Switzerland\relax                                                                                                                                                                                                                                                              \label{inst:0040}\vfill
\and European Space Agency (ESA), European Space Astronomy Centre (ESAC), Camino bajo del Castillo, s/n, Urbanizacion Villafranca del Castillo, Villanueva de la Ca\~{n}ada, 28692 Madrid, Spain\relax                                                                                                                                                             \label{inst:0041}\vfill
\and Aurora Technology for European Space Agency (ESA), Camino bajo del Castillo, s/n, Urbanizacion Villafranca del Castillo, Villanueva de la Ca\~{n}ada, 28692 Madrid, Spain\relax                                                                                                                                                                               \label{inst:0042}\vfill
\and Lohrmann Observatory, Technische Universit\"{ a}t Dresden, Mommsenstra{\ss}e 13, 01062 Dresden, Germany\relax                                                                                                                                                                                                                                                 \label{inst:0043}\vfill
\and Lund Observatory, Department of Astronomy and Theoretical Physics, Lund University, Box 43, 22100 Lund, Sweden\relax                                                                                                                                                                                                                                          \label{inst:0045}\vfill
\and CNES Centre Spatial de Toulouse, 18 avenue Edouard Belin, 31401 Toulouse Cedex 9, France\relax                                                                                                                                                                                                                                                                \label{inst:0048}\vfill
\and Institut d'Astronomie et d'Astrophysique, Universit\'{e} Libre de Bruxelles CP 226, Boulevard du Triomphe, 1050 Brussels, Belgium\relax                                                                                                                                                                                                                       \label{inst:0049}\vfill
\and F.R.S.-FNRS, Rue d'Egmont 5, 1000 Brussels, Belgium\relax                                                                                                                                                                                                                                                                                                     \label{inst:0050}\vfill
\and INAF - Osservatorio Astrofisico di Torino, via Osservatorio 20, 10025 Pino Torinese (TO), Italy\relax                                                                                                                                                                                                                                                         \label{inst:0057}\vfill
\and European Space Agency (ESA, retired)\relax                                                                                                                                                                                                                                                                                                                    \label{inst:0058}\vfill
\and University of Turin, Department of Physics, Via Pietro Giuria 1, 10125 Torino, Italy\relax                                                                                                                                                                                                                                                                    \label{inst:0061}\vfill
\and DAPCOM for Institut de Ci\`{e}ncies del Cosmos (ICCUB), Universitat  de  Barcelona  (IEEC-UB), Mart\'{i} i  Franqu\`{e}s  1, 08028 Barcelona, Spain\relax                                                                                                                                                                                                     \label{inst:0064}\vfill
\and Observational Astrophysics, Division of Astronomy and Space Physics, Department of Physics and Astronomy, Uppsala University, Box 516, 751 20 Uppsala, Sweden\relax                                                                                                                                                                                           \label{inst:0068}\vfill
\and ALTEC S.p.a, Corso Marche, 79,10146 Torino, Italy\relax                                                                                                                                                                                                                                                                                                       \label{inst:0070}\vfill
\and S\`{a}rl, Geneva, Switzerland\relax                                                                                                                                                                                                                                                                                                                           \label{inst:0073}\vfill
\and Department of Astronomy, University of Geneva, Chemin d'Ecogia 16, 1290 Versoix, Switzerland\relax                                                                                                                                                                                                                                                            \label{inst:0074}\vfill
\and Mullard Space Science Laboratory, University College London, Holmbury St Mary, Dorking, Surrey RH5 6NT, United Kingdom\relax                                                                                                                                                                                                                                  \label{inst:0079}\vfill
\and Gaia DPAC Project Office, ESAC, Camino bajo del Castillo, s/n, Urbanizacion Villafranca del Castillo, Villanueva de la Ca\~{n}ada, 28692 Madrid, Spain\relax                                                                                                                                                                                                  \label{inst:0081}\vfill
\and Telespazio UK S.L. for European Space Agency (ESA), Camino bajo del Castillo, s/n, Urbanizacion Villafranca del Castillo, Villanueva de la Ca\~{n}ada, 28692 Madrid, Spain\relax                                                                                                                                                                              \label{inst:0084}\vfill
\and SYRTE, Observatoire de Paris, Universit\'{e} PSL, CNRS,  Sorbonne Universit\'{e}, LNE, 61 avenue de l'Observatoire 75014 Paris, France\relax                                                                                                                                                                                                                  \label{inst:0086}\vfill
\and National Observatory of Athens, I. Metaxa and Vas. Pavlou, Palaia Penteli, 15236 Athens, Greece\relax                                                                                                                                                                                                                                                         \label{inst:0089}\vfill
\and IMCCE, Observatoire de Paris, Universit\'{e} PSL, CNRS, Sorbonne Universit{\'e}, Univ. Lille, 77 av. Denfert-Rochereau, 75014 Paris, France\relax                                                                                                                                                                                                             \label{inst:0091}\vfill
\and Serco Gesti\'{o}n de Negocios for European Space Agency (ESA), Camino bajo del Castillo, s/n, Urbanizacion Villafranca del Castillo, Villanueva de la Ca\~{n}ada, 28692 Madrid, Spain\relax                                                                                                                                                                   \label{inst:0098}\vfill
\and CRAAG - Centre de Recherche en Astronomie, Astrophysique et G\'{e}ophysique, Route de l'Observatoire Bp 63 Bouzareah 16340 Algiers, Algeria\relax                                                                                                                                                                                                             \label{inst:0101}\vfill
\and Institute for Astronomy, University of Edinburgh, Royal Observatory, Blackford Hill, Edinburgh EH9 3HJ, United Kingdom\relax                                                                                                                                                                                                                                  \label{inst:0102}\vfill
\and RHEA for European Space Agency (ESA), Camino bajo del Castillo, s/n, Urbanizacion Villafranca del Castillo, Villanueva de la Ca\~{n}ada, 28692 Madrid, Spain\relax                                                                                                                                                                                            \label{inst:0106}\vfill
\and ATG Europe for European Space Agency (ESA), Camino bajo del Castillo, s/n, Urbanizacion Villafranca del Castillo, Villanueva de la Ca\~{n}ada, 28692 Madrid, Spain\relax                                                                                                                                                                                      \label{inst:0107}\vfill
\and Universit\'{e} de Strasbourg, CNRS, Observatoire astronomique de Strasbourg, UMR 7550,  11 rue de l'Universit\'{e}, 67000 Strasbourg, France\relax                                                                                                                                                                                                            \label{inst:0113}\vfill
\and Kavli Institute for Cosmology Cambridge, Institute of Astronomy, Madingley Road, Cambridge, CB3 0HA\relax                                                                                                                                                                                                                                                     \label{inst:0116}\vfill
\and Leibniz Institute for Astrophysics Potsdam (AIP), An der Sternwarte 16, 14482 Potsdam, Germany\relax                                                                                                                                                                                                                                                          \label{inst:0122}\vfill
\and CENTRA, Faculdade de Ci\^{e}ncias, Universidade de Lisboa, Edif. C8, Campo Grande, 1749-016 Lisboa, Portugal\relax                                                                                                                                                                                                                                            \label{inst:0125}\vfill
\and Department of Informatics, Donald Bren School of Information and Computer Sciences, University of California, Irvine, 5226 Donald Bren Hall, 92697-3440 CA Irvine, United States\relax                                                                                                                                                                        \label{inst:0126}\vfill
\and INAF - Osservatorio Astrofisico di Catania, via S. Sofia 78, 95123 Catania, Italy\relax                                                                                                                                                                                                                                                                       \label{inst:0127}\vfill
\and Dipartimento di Fisica e Astronomia ""Ettore Majorana"", Universit\`{a} di Catania, Via S. Sofia 64, 95123 Catania, Italy\relax                                                                                                                                                                                                                               \label{inst:0128}\vfill
\and INAF - Osservatorio Astronomico di Roma, Via Frascati 33, 00078 Monte Porzio Catone (Roma), Italy\relax                                                                                                                                                                                                                                                       \label{inst:0131}\vfill
\and Department of Physics, University of Helsinki, P.O. Box 64, 00014 Helsinki, Finland\relax                                                                                                                                                                                                                                                                     \label{inst:0134}\vfill
\and Finnish Geospatial Research Institute FGI, Geodeetinrinne 2, 02430 Masala, Finland\relax                                                                                                                                                                                                                                                                      \label{inst:0135}\vfill
\and Institut UTINAM CNRS UMR6213, Universit\'{e} Bourgogne Franche-Comt\'{e}, OSU THETA Franche-Comt\'{e} Bourgogne, Observatoire de Besan\c{c}on, BP1615, 25010 Besan\c{c}on Cedex, France\relax                                                                                                                                                                 \label{inst:0139}\vfill
\and HE Space Operations BV for European Space Agency (ESA), Keplerlaan 1, 2201AZ, Noordwijk, The Netherlands\relax                                                                                                                                                                                                                                                \label{inst:0141}\vfill
\and Dpto. de Inteligencia Artificial, UNED, c/ Juan del Rosal 16, 28040 Madrid, Spain\relax                                                                                                                                                                                                                                                                       \label{inst:0143}\vfill
\and Konkoly Observatory, Research Centre for Astronomy and Earth Sciences, E\"{ o}tv\"{ o}s Lor{\'a}nd Research Network (ELKH), MTA Centre of Excellence, Konkoly Thege Mikl\'{o}s \'{u}t 15-17, 1121 Budapest, Hungary\relax                                                                                                                                     \label{inst:0150}\vfill
\and ELTE E\"{ o}tv\"{ o}s Lor\'{a}nd University, Institute of Physics, 1117, P\'{a}zm\'{a}ny P\'{e}ter s\'{e}t\'{a}ny 1A, Budapest, Hungary\relax                                                                                                                                                                                                                 \label{inst:0151}\vfill
\and Instituut voor Sterrenkunde, KU Leuven, Celestijnenlaan 200D, 3001 Leuven, Belgium\relax                                                                                                                                                                                                                                                                      \label{inst:0153}\vfill
\and Department of Astrophysics/IMAPP, Radboud University, P.O.Box 9010, 6500 GL Nijmegen, The Netherlands\relax                                                                                                                                                                                                                                                   \label{inst:0154}\vfill
\and University of Vienna, Department of Astrophysics, T\"{ u}rkenschanzstra{\ss}e 17, A1180 Vienna, Austria\relax                                                                                                                                                                                                                                                 \label{inst:0162}\vfill
\and Institute of Physics, Laboratory of Astrophysics, Ecole Polytechnique F\'ed\'erale de Lausanne (EPFL), Observatoire de Sauverny, 1290 Versoix, Switzerland\relax                                                                                                                                                                                              \label{inst:0163}\vfill
\and Kapteyn Astronomical Institute, University of Groningen, Landleven 12, 9747 AD Groningen, The Netherlands\relax                                                                                                                                                                                                                                               \label{inst:0169}\vfill
\and Thales Services for CNES Centre Spatial de Toulouse, 18 avenue Edouard Belin, 31401 Toulouse Cedex 9, France\relax                                                                                                                                                                                                                                            \label{inst:0177}\vfill
\and Depto. Estad\'istica e Investigaci\'on Operativa. Universidad de C\'adiz, Avda. Rep\'ublica Saharaui s/n, 11510 Puerto Real, C\'adiz, Spain\relax                                                                                                                                                                                                             \label{inst:0180}\vfill
\and Center for Research and Exploration in Space Science and Technology, University of Maryland Baltimore County, 1000 Hilltop Circle, Baltimore MD, USA\relax                                                                                                                                                                                                    \label{inst:0182}\vfill
\and GSFC - Goddard Space Flight Center, Code 698, 8800 Greenbelt Rd, 20771 MD Greenbelt, United States\relax                                                                                                                                                                                                                                                      \label{inst:0183}\vfill
\and EURIX S.r.l., Corso Vittorio Emanuele II 61, 10128, Torino, Italy\relax                                                                                                                                                                                                                                                                                       \label{inst:0185}\vfill
\and Porter School of the Environment and Earth Sciences, Tel Aviv University, Tel Aviv 6997801, Israel\relax                                                                                                                                                                                                                                                      \label{inst:0186}\vfill
\and Harvard-Smithsonian Center for Astrophysics, 60 Garden St., MS 15, Cambridge, MA 02138, USA\relax                                                                                                                                                                                                                                                             \label{inst:0187}\vfill
\and HE Space Operations BV for European Space Agency (ESA), Camino bajo del Castillo, s/n, Urbanizacion Villafranca del Castillo, Villanueva de la Ca\~{n}ada, 28692 Madrid, Spain\relax                                                                                                                                                                          \label{inst:0189}\vfill
\and LFCA/DAS,Universidad de Chile,CNRS,Casilla 36-D, Santiago, Chile\relax                                                                                                                                                                                                                                                                                        \label{inst:0191}\vfill
\and SISSA - Scuola Internazionale Superiore di Studi Avanzati, via Bonomea 265, 34136 Trieste, Italy\relax                                                                                                                                                                                                                                                        \label{inst:0194}\vfill
\and Telespazio for CNES Centre Spatial de Toulouse, 18 avenue Edouard Belin, 31401 Toulouse Cedex 9, France\relax                                                                                                                                                                                                                                                 \label{inst:0199}\vfill
\and University of Turin, Department of Computer Sciences, Corso Svizzera 185, 10149 Torino, Italy\relax                                                                                                                                                                                                                                                           \label{inst:0203}\vfill
\and Dpto. de Matem\'{a}tica Aplicada y Ciencias de la Computaci\'{o}n, Univ. de Cantabria, ETS Ingenieros de Caminos, Canales y Puertos, Avda. de los Castros s/n, 39005 Santander, Spain\relax                                                                                                                                                                   \label{inst:0206}\vfill
\and Centro de Astronom\'{i}a - CITEVA, Universidad de Antofagasta, Avenida Angamos 601, Antofagasta 1270300, Chile\relax                                                                                                                                                                                                                                          \label{inst:0215}\vfill
\and DLR Gesellschaft f\"{ u}r Raumfahrtanwendungen (GfR) mbH M\"{ u}nchener Stra{\ss}e 20 , 82234 We{\ss}ling\relax                                                                                                                                                                                                                                               \label{inst:0219}\vfill
\and Centre for Astrophysics Research, University of Hertfordshire, College Lane, AL10 9AB, Hatfield, United Kingdom\relax                                                                                                                                                                                                                                         \label{inst:0221}\vfill
\and University of Turin, Mathematical Department ""G.Peano"", Via Carlo Alberto 10, 10123 Torino, Italy\relax                                                                                                                                                                                                                                                     \label{inst:0228}\vfill
\and INAF - Osservatorio Astronomico d'Abruzzo, Via Mentore Maggini, 64100 Teramo, Italy\relax                                                                                                                                                                                                                                                                     \label{inst:0234}\vfill
\and Instituto de Astronomia, Geof\`{i}sica e Ci\^{e}ncias Atmosf\'{e}ricas, Universidade de S\~{a}o Paulo, Rua do Mat\~{a}o, 1226, Cidade Universitaria, 05508-900 S\~{a}o Paulo, SP, Brazil\relax                                                                                                                                                                \label{inst:0237}\vfill
\and APAVE SUDEUROPE SAS for CNES Centre Spatial de Toulouse, 18 avenue Edouard Belin, 31401 Toulouse Cedex 9, France\relax                                                                                                                                                                                                                                        \label{inst:0244}\vfill
\and M\'{e}socentre de calcul de Franche-Comt\'{e}, Universit\'{e} de Franche-Comt\'{e}, 16 route de Gray, 25030 Besan\c{c}on Cedex, France\relax                                                                                                                                                                                                                  \label{inst:0246}\vfill
\and ATOS for CNES Centre Spatial de Toulouse, 18 avenue Edouard Belin, 31401 Toulouse Cedex 9, France\relax                                                                                                                                                                                                                                                       \label{inst:0251}\vfill
\and School of Physics and Astronomy, Tel Aviv University, Tel Aviv 6997801, Israel\relax                                                                                                                                                                                                                                                                          \label{inst:0254}\vfill
\and Astrophysics Research Centre, School of Mathematics and Physics, Queen's University Belfast, Belfast BT7 1NN, UK\relax                                                                                                                                                                                                                                        \label{inst:0256}\vfill
\and Centre de Donn\'{e}es Astronomique de Strasbourg, Strasbourg, France\relax                                                                                                                                                                                                                                                                                    \label{inst:0258}\vfill
\and Institute for Computational Cosmology, Department of Physics, Durham University, Durham DH1 3LE, UK\relax                                                                                                                                                                                                                                                     \label{inst:0262}\vfill
\and European Southern Observatory, Karl-Schwarzschild-Str. 2, 85748 Garching, Germany\relax                                                                                                                                                                                                                                                                       \label{inst:0263}\vfill
\and Max-Planck-Institut f\"{ u}r Astrophysik, Karl-Schwarzschild-Stra{\ss}e 1, 85748 Garching, Germany\relax                                                                                                                                                                                                                                                      \label{inst:0264}\vfill
\and Data Science and Big Data Lab, Pablo de Olavide University, 41013, Seville, Spain\relax                                                                                                                                                                                                                                                                       \label{inst:0268}\vfill
\and Barcelona Supercomputing Center (BSC), Pla\c{c}a Eusebi G\"{ u}ell 1-3, 08034-Barcelona, Spain\relax                                                                                                                                                                                                                                                          \label{inst:0276}\vfill
\and ETSE Telecomunicaci\'{o}n, Universidade de Vigo, Campus Lagoas-Marcosende, 36310 Vigo, Galicia, Spain\relax                                                                                                                                                                                                                                                   \label{inst:0281}\vfill
\and Asteroid Engineering Laboratory, Space Systems, Lule\aa{} University of Technology, Box 848, S-981 28 Kiruna, Sweden\relax                                                                                                                                                                                                                                    \label{inst:0285}\vfill
\and Vera C Rubin Observatory,  950 N. Cherry Avenue, Tucson, AZ 85719, USA\relax                                                                                                                                                                                                                                                                                  \label{inst:0290}\vfill
\and Department of Astrophysics, Astronomy and Mechanics, National and Kapodistrian University of Athens, Panepistimiopolis, Zografos, 15783 Athens, Greece\relax                                                                                                                                                                                                  \label{inst:0291}\vfill
\and TRUMPF Photonic Components GmbH, Lise-Meitner-Stra{\ss}e 13,  89081 Ulm, Germany\relax                                                                                                                                                                                                                                                                        \label{inst:0294}\vfill
\and IAC - Instituto de Astrofisica de Canarias, Via L\'{a}ctea s/n, 38200 La Laguna S.C., Tenerife, Spain\relax                                                                                                                                                                                                                                                   \label{inst:0299}\vfill
\and Department of Astrophysics, University of La Laguna, Via L\'{a}ctea s/n, 38200 La Laguna S.C., Tenerife, Spain\relax                                                                                                                                                                                                                                          \label{inst:0300}\vfill
\and Faculty of Aerospace Engineering, Delft University of Technology, Kluyverweg 1, 2629 HS Delft, The Netherlands\relax                                                                                                                                                                                                                                          \label{inst:0302}\vfill
\and Radagast Solutions\relax                                                                                                                                                                                                                                                                                                                                      \label{inst:0306}\vfill
\and Laboratoire Univers et Particules de Montpellier, CNRS Universit\'{e} Montpellier, Place Eug\`{e}ne Bataillon, CC72, 34095 Montpellier Cedex 05, France\relax                                                                                                                                                                                                 \label{inst:0307}\vfill
\and Universit\'{e} de Caen Normandie, C\^{o}te de Nacre Boulevard Mar\'{e}chal Juin, 14032 Caen, France\relax                                                                                                                                                                                                                                                     \label{inst:0313}\vfill
\and LESIA, Observatoire de Paris, Universit\'{e} PSL, CNRS, Sorbonne Universit\'{e}, Universit\'{e} de Paris, 5 Place Jules Janssen, 92190 Meudon, France\relax                                                                                                                                                                                                   \label{inst:0314}\vfill
\and SRON Netherlands Institute for Space Research, Niels Bohrweg 4, 2333 CA Leiden, The Netherlands\relax                                                                                                                                                                                                                                                         \label{inst:0323}\vfill
\and Astronomical Observatory, University of Warsaw,  Al. Ujazdowskie 4, 00-478 Warszawa, Poland\relax                                                                                                                                                                                                                                                             \label{inst:0324}\vfill
\and Scalian for CNES Centre Spatial de Toulouse, 18 avenue Edouard Belin, 31401 Toulouse Cedex 9, France\relax                                                                                                                                                                                                                                                    \label{inst:0326}\vfill
\and Universit\'{e} Rennes, CNRS, IPR (Institut de Physique de Rennes) - UMR 6251, 35000 Rennes, France\relax                                                                                                                                                                                                                                                      \label{inst:0332}\vfill
\and INAF - Osservatorio Astronomico di Capodimonte, Via Moiariello 16, 80131, Napoli, Italy\relax                                                                                                                                                                                                                                                                 \label{inst:0334}\vfill
\and Shanghai Astronomical Observatory, Chinese Academy of Sciences, 80 Nandan Road, Shanghai 200030, People's Republic of China\relax                                                                                                                                                                                                                             \label{inst:0337}\vfill
\and University of Chinese Academy of Sciences, No.19(A) Yuquan Road, Shijingshan District, Beijing 100049, People's Republic of China\relax                                                                                                                                                                                                                       \label{inst:0339}\vfill
\and Niels Bohr Institute, University of Copenhagen, Juliane Maries Vej 30, 2100 Copenhagen {\O}, Denmark\relax                                                                                                                                                                                                                                                    \label{inst:0342}\vfill
\and DXC Technology, Retortvej 8, 2500 Valby, Denmark\relax                                                                                                                                                                                                                                                                                                        \label{inst:0343}\vfill
\and Las Cumbres Observatory, 6740 Cortona Drive Suite 102, Goleta, CA 93117, USA\relax                                                                                                                                                                                                                                                                            \label{inst:0344}\vfill
\and Astrophysics Research Institute, Liverpool John Moores University, 146 Brownlow Hill, Liverpool L3 5RF, United Kingdom\relax                                                                                                                                                                                                                                  \label{inst:0353}\vfill
\and IPAC, Mail Code 100-22, California Institute of Technology, 1200 E. California Blvd., Pasadena, CA 91125, USA\relax                                                                                                                                                                                                                                           \label{inst:0360}\vfill
\and IRAP, Universit\'{e} de Toulouse, CNRS, UPS, CNES, 9 Av. colonel Roche, BP 44346, 31028 Toulouse Cedex 4, France\relax                                                                                                                                                                                                                                        \label{inst:0361}\vfill
\and MTA CSFK Lend\"{ u}let Near-Field Cosmology Research Group, Konkoly Observatory, MTA Research Centre for Astronomy and Earth Sciences, Konkoly Thege Mikl\'{o}s \'{u}t 15-17, 1121 Budapest, Hungary\relax                                                                                                                                                    \label{inst:0377}\vfill
\and Departmento de F\'{i}sica de la Tierra y Astrof\'{i}sica, Universidad Complutense de Madrid, 28040 Madrid, Spain\relax                                                                                                                                                                                                                                        \label{inst:0392}\vfill
\and Villanova University, Department of Astrophysics and Planetary Science, 800 E Lancaster Avenue, Villanova PA 19085, USA\relax                                                                                                                                                                                                                                 \label{inst:0412}\vfill
\and INAF - Osservatorio Astronomico di Brera, via E. Bianchi, 46, 23807 Merate (LC), Italy\relax                                                                                                                                                                                                                                                                  \label{inst:0418}\vfill
\and STFC, Rutherford Appleton Laboratory, Harwell, Didcot, OX11 0QX, United Kingdom\relax                                                                                                                                                                                                                                                                         \label{inst:0425}\vfill
\and Charles University, Faculty of Mathematics and Physics, Astronomical Institute of Charles University, V Holesovickach 2, 18000 Prague, Czech Republic\relax                                                                                                                                                                                                   \label{inst:0447}\vfill
\and Department of Particle Physics and Astrophysics, Weizmann Institute of Science, Rehovot 7610001, Israel\relax                                                                                                                                                                                                                                                 \label{inst:0456}\vfill
\and Department of Astrophysical Sciences, 4 Ivy Lane, Princeton University, Princeton NJ 08544, USA\relax                                                                                                                                                                                                                                                         \label{inst:0457}\vfill
\and Departamento de Astrof\'{i}sica, Centro de Astrobiolog\'{i}a (CSIC-INTA), ESA-ESAC. Camino Bajo del Castillo s/n. 28692 Villanueva de la Ca\~{n}ada, Madrid, Spain\relax                                                                                                                                                                                      \label{inst:0466}\vfill
\and naXys, University of Namur, Rempart de la Vierge, 5000 Namur, Belgium\relax                                                                                                                                                                                                                                                                                   \label{inst:0469}\vfill
\and CGI Deutschland B.V. \& Co. KG, Mornewegstr. 30, 64293 Darmstadt, Germany\relax                                                                                                                                                                                                                                                                               \label{inst:0477}\vfill
\and Institute of Global Health, University of Geneva\relax                                                                                                                                                                                                                                                                                                        \label{inst:0478}\vfill
\and Astronomical Observatory Institute, Faculty of Physics, Adam Mickiewicz University, Pozna\'{n}, Poland\relax                                                                                                                                                                                                                                                  \label{inst:0480}\vfill
\and H H Wills Physics Laboratory, University of Bristol, Tyndall Avenue, Bristol BS8 1TL, United Kingdom\relax                                                                                                                                                                                                                                                    \label{inst:0484}\vfill
\and Department of Physics and Astronomy G. Galilei, University of Padova, Vicolo dell'Osservatorio 3, 35122, Padova, Italy\relax                                                                                                                                                                                                                                  \label{inst:0491}\vfill
\and CERN, Geneva, Switzerland\relax                                                                                                                                                                                                                                                                                                                               \label{inst:0493}\vfill
\and Applied Physics Department, Universidade de Vigo, 36310 Vigo, Spain\relax                                                                                                                                                                                                                                                                                     \label{inst:0495}\vfill
\and Association of Universities for Research in Astronomy, 1331 Pennsylvania Ave. NW, Washington, DC 20004, USA\relax                                                                                                                                                                                                                                             \label{inst:0499}\vfill
\and European Southern Observatory, Alonso de C\'ordova 3107, Casilla 19, Santiago, Chile\relax                                                                                                                                                                                                                                                                    \label{inst:0507}\vfill
\and Sorbonne Universit\'{e}, CNRS, UMR7095, Institut d'Astrophysique de Paris, 98bis bd. Arago, 75014 Paris, France\relax                                                                                                                                                                                                                                         \label{inst:0512}\vfill
\and Faculty of Mathematics and Physics, University of Ljubljana, Jadranska ulica 19, 1000 Ljubljana, Slovenia\relax                                                                                                                                                                                                                                               \label{inst:0514}\vfill
}
%%%%%%%%%%%%%%%%%%%%%%%%%%%%%%%%%%%%%%%%%%%%%%%%%%%%%%%%%%%%%%%%%%%%%%%%%%%%%%%%%

\date{Received April 4, 2022; accepted May 19, 2022}

\abstract{\gaia Data Release 3 provides novel flux-calibrated low-resolution spectrophotometry for $\simeq 220$ million sources in the wavelength range 330~nm$\ \le \lambda\le\ $1050~nm (XP spectra). Synthetic photometry directly tied to a flux in physical units can be obtained from these spectra for any passband fully enclosed in this wavelength range. We describe how synthetic photometry can be obtained from XP spectra, illustrating the performance that can be achieved under a range of different conditions ---for example passband width and wavelength range--- as well as the limits and the problems affecting it. Existing 
top-quality photometry can be reproduced within a few per cent over a wide range of magnitudes and colour, for wide and medium bands, and with up to millimag accuracy when synthetic photometry is standardised with respect to these external sources. Some examples of potential scientific application are presented, including the detection of multiple populations in globular clusters, the estimation of metallicity extended to the very metal-poor regime, and the classification of white dwarfs. A catalogue providing standardised photometry for $\simeq 2.2\times 10^8$ sources in several wide bands of widely used photometric systems is provided (Gaia Synthetic Photometry Catalogue; GSPC) as well as a catalogue of $\simeq 10^5$ white dwarfs with 
DA/non-DA classification obtained with a Random Forest algorithm (Gaia Synthetic Photometry Catalogue for White Dwarfs; GSPC-WD).}


\maketitle

%%%%%%%%%%%%%%%%%%%%%%%%%%%%%%%%%%%%%%%%%%%%%%%% INTRO
%\input{sections/introduction.tex}

\section{Introduction}
\label{sec:introduction}

Photometry, together with astrometry,  are probably the most ancient and fundamental techniques upon which our knowledge of the Universe and of astrophysical phenomena relies, and can be traced  back to the ancient Greeks \citep[see e.g.][for a short historical introduction and references; SMY11 hereafter]{SMY2011}. Photometry consists in sampling the spectra of astronomical sources by measuring their incoming flux passing through a transmission curve (TC) that allows the user to detect only the light within a defined
wavelength range (spectral window). For a pure black-body spectrum, photometric measurements in two different spectral windows are sufficient to estimate the derivative of the black-body curve, unequivocally establishing its temperature. As the spectra of stars and other kinds of celestial sources deviate from black body, more than two  spectral windows must be sampled to properly infer the most relevant astrophysical properties and/or to obtain an adequate classification \citep{Young1992ASPC}.

The earliest TC used for photometry was the sensitivity curve of the human eye, the details of which depend on the physiology of the observer.  Today, actual TCs, which in the following we also refer to as {passbands}, are defined by the combination of the TC of an optical filter ---which is designed to select the desired spectral window---, the sensitivity curve of a photon-counting detector (typically a CCD for observations in the optical spectral range), and the TC of the optical elements that collect the light from a source and properly convey it to the detector (telescope and camera), plus a contribution from the terrestrial atmosphere if observations are performed on the Earth's surface. 

A photometric system is defined by a set of passbands and a set of standard stars observed in these passbands with an instrumental setup and a data-reduction procedure that is as controlled and homogeneous as possible
\citep{Bessel05,SMY2011}. The magnitude and colour differences between the standard stars define a {relative} photometric scale. 

Following SMY11, a {`closed'} photometric system is  established by taking all the relevant measurements with the same observing site and instrumental setup (or the best possible approximation of this condition; these latter authors report the Walraven \citep{walra07} and the Geneva \citep{geneva71} systems as examples of closed systems). 
This approach should maximise the homogeneity and consequently the {precision} of the measurements gathered.
On the other hand, we refer to {`open'} systems as those with a sufficient number of standard stars distributed over the sky to allow broad accessibility, such that any observer can attempt to obtain photometry in that scale using their own instrumental setup, chosen to match the original one as closely as possible. This, in principle, allows a general use of the system and fruitful comparison between observations obtained at different places and in different epochs\footnote{We can also refer to an open system as a {standard} system, as it offers the opportunity to standardise a given magnitude scale, transforming instrumental magnitudes into magnitudes in the desired open system.}. However, as the exact reproduction of the original observing conditions ---in particular of the actual TCs--- is virtually impossible to achieve, colour-dependent transformations are required to convert instrumental magnitudes into the desired scale.  Transformations require repeated observations of standard stars during an observing run, and, in general, they are  prone to subtle but sizable systematic errors \citep[][SMY11]{Young1992ASPC}. In general, a transformation should imply non-linear colour terms that may be hard to constrain and that are often neglected \citep[see][and references therein]{Young1992ASPC,Young1992AA,Young1994}. In any case, ground-based photometric measures must be corrected for time- and (slightly)colour-dependent atmospheric extinction, a complex process in itself that may significantly contribute to the systematic error budget. 

Finally, to convert magnitudes into physical fluxes, spectrophotometry is required, which comes in the form of synthetic photometry through the system TCs on the flux-calibrated spectrum of (at least) one standard star \citep[see, e.g.][]{Fukugita1996}. According to \citet{Landolt2011}, spectrophotometry is the only kind of photometry that can be considered {absolute}, as it is directly linked to fluxes in physical units\footnote{Transformation of instrumental magnitudes into a standard system is often referred to as {absolute photometric calibration}. This is not completely unjustified as, in principle, it is a process transforming magnitudes in an arbitrary scale into magnitudes in a standard system that in turn can be (and in most cases is) tied to physical fluxes by spectrophotometry of some of its standard stars.}. 

The synthetic description, the set of definitions, and the nomenclature above, as well as many general concepts that are used throughout this paper, are largely based on the reviews collected in \citet{MS2011}, in particular SMY11, \citet{Landolt2011}, but also in \citet{Bessel05}, \citet{Sirianni2005}, \citet{Sterken2007a,Sterken2007b}, \citet{Young1992ASPC,Young1992AA,Young1994}, and \citet{Manfroid1992}. We refer the interested reader to these papers and the references therein, as well as to \citet{Magnier2020} and \citet{Thanjavur2021}, for two examples of very recent, 
state-of-the-art applications to wide-area surveys. 

Here we limit our discussion to photometry in the {optical} wavelength range, that is approximately between 300~nm and 1100~nm. 
In this context, it is worth reiterating the definition of photometric precision and accuracy provided by \citet[][as reported by SMY11]{Young1994ASIC}, as a reference: {\em precision} refers to the repeatability of a measurement, while {\em accuracy} means the absence of error, as measured against some external reference, such as a set of standard stars.
The typical precision and accuracy of ground-based photometry in the past century is of the order of $\ga 1\%$ \citep{stubbs2006}. 
Such a limit is sufficient for many applications but is somewhat lacking when compared with other physical quantities that are known with an accuracy of better than one part in a million \citep{Young1992ASPC}.

A significant step forward in the precision of ground-based photometry was obtained by modern digital panoramic surveys, such as the Sloan Digital Sky Survey \citep[SDSS,][]{York2000} or Pan-STARRS1 \citep[PS1,][]{Chambers2016}. The acquisition of multi-colour photometry for many millions of stars over huge areas of the sky, with strictly the same setup and innovative techniques of photometric calibration, has allowed for the first time to achieve precision of $<0.01$ mag on an industrial scale. This achievement converted the de facto closed systems associated to these surveys into open systems, providing abundant standard stars with which to transform suitable observations taken outside the survey into the standard system that they define 
\citep[see][for a synthetic review and references on modern surveys and calibration techniques]{Huang21}.  

However, it is widely recognised \citep[see e.g.][]{Huang2021sm,Magnier2020} that the all-sky, space-based, three-band photometry provided by the ESA space mission \gaia \citep{Prusti2016} 
presents high-quality photometric measurements with photometric precision rivaling the best available, especially for wide sky coverage.
 In its Early Data Release 3 (EDR3) realisation \citep{riello2021}, it effectively reaches submillimag precision in the range $10.0 \le G\le 17.0$~mag. Indeed, this exquisite degree of internal homogeneity has been used to significantly reduce residual systematic errors in the best set of SDSS standard stars \citep[see e.g.][]{Thanjavur2021,Huang21}.

The new {\gaia} Data Release 3 \citep[\gdr3;][]{DR3-DPACP-185} provides ---for the first time--- internally \citep{Carrasco2021,EDR3-DPACP-118} and externally calibrated \citep[i.e. flux and wavelength calibrated;][]{EDR3-DPACP-120} very low resolution ($\lambda/\Delta \lambda \simeq 25-100$) spectra from the BP and RP spectrophotometers for about 220 million sources, mostly with $G<17.65$~mag \citep[see][for a complete list of sources with released BP/RP spectra]{DR3-DPACP-160,EDR3-DPACP-118}.
These spectra were used to infer astrophysical parameters, which are also released as part of {\gdr3} \citep{DR3-DPACP-157,DR3-DPACP-160,DR3-DPACP-156}. 

Another interesting product that can be obtained from externally calibrated\footnote{Meaning, in this context, ''flux-calibrated using spectrophotometric data external to the direct data product of the {\gaia} satellite'', namely the Spectro Photometric Standard Stars by \citet{pancino2021}, see \citet{EDR3-DPACP-120}} (EC) {\xp} 
(hereafter XP, for brevity) 
spectra is {synthetic photometry}.  In principle, synthetic photometry can be obtained from EC XP spectra in any photometric system and for any passband enclosed in the spectral range covered by XP spectra ($330$--$1050$~nm) and whose characteristic width is larger than the line spread function (LSF) of XP spectra at the relevant wavelength\footnote{Parametrised here as the ratio $Rf$ between the full width at half maximum (FWHM) of the passband  and of the XP Line Spread Function. It is useful to anticipate here the (conservative) criterion found in App.~\ref{sec:app_Rf}: flux-conserving SP from XP spectra in a given photometric band can be achieved (also in  presence of a strong spectral feature) if $Rf\ge 1.4$. See App.~\ref{sec:app_Rf} for additional details and discussion.}. In principle and in perspective, this may constitute a true revolution in optical photometry. 

For the passbands of a given photometric system fulfilling the above conditions, we can get all-sky space-based photometry for all the sources for which XP spectra are available, in terms of magnitudes and flux in physical units. This is limited to $\simeq 220$~million sources in {\gdr3} but will amount to the entire {\gaia} data set in future releases ($\sim 2$ billion sources down to $G\simeq 20.5$~mag).
The relative flux scale relies on the precision of the EC XP spectra, while
the absolute flux scale is based on the {\gaia} grid of SpectroPhotometric Standard Stars 
\citep[SPSS][]{pancino2021}, and references therein).

Therefore, in principle, synthetic photometry from XP spectra (XP Synthetic Photometry, XPSP, hereafter) can supply absolute optical photometry for hundreds of millions of stars in any suitable system over the entire sky, thus for example transforming any {closed system} into an {open system} (albeit limited by the exact knowledge of the TCs). This in turn can provide, among the various possibilities: (a) the basis for the validation and/or re-calibration of existing photometric surveys; (b) the basis for validation and/or calibration of {\em future} photometric surveys from the ground or from space; and (c) the opportunity to experiment with the performance of a photometric system on a huge data set of real data on real sources before its actual realisation.  
 As we show in the following, the potentiality of the method and of the product have not yet been fully realised because of systematic errors depending on spectral type that still affect EC XP spectra. The present contribution should be considered as one step in a process that is designed to maximise our exploitation of \gaia spectrophotometric data and will be continued in future data releases.
 
This paper is intended to illustrate how to get synthetic photometry from {\gdr3} data. We showcase
the performance of the synthetic photometry that can be currently obtained from XP spectra and outline its limitations. We also show a few examples of possible applications, and provide a few general-use photometric catalogues from XPSP, which are publicly accessible through the {\gaia} mission archive or other public repositories.

An obvious {internal} application made possible by XPSP is to provide additional means for validation of the EC XP spectra by comparison with huge external sets of high-quality photometry (SDSS, PS1, etc., see below and Montegriffo et al. 2022a).
For example, \citet{EDR3-DPACP-120} demonstrated that $H_p$, $B_T$, and $V_T$ Hipparcos photometry \citep{Hippaphot1997}, which is considered a benchmark of precision \citep{Bessel05}, is reproduced by XPSP with typical accuracy of better than 2.5~millimag over the entire sky (see their Fig.~44).
Similarly, we can provide some cross-validation with the results of DPAC/CU8\footnote{The Data Processing and Analysis Consortium (DPAC) is the consortium responsible of the processing of data from the \gaia mission \citep{Prusti2016}. It is structured in Coordination Units (CUs), each dealing with a specific subsystem of the processing system. The core mission of CU8 is to provide astrophysical parameters (AP) of the sources in the \gaia catalogue. } \citep{DR3-DPACP-157,DR3-DPACP-160}, by treating the same observational material in a completely different way.

The paper is organised as follows. In Sect.~\ref{sec:methods} we illustrate our formalism, starting from the representation of XP spectra in the \gaia context \citep{EDR3-DPACP-118}. We also introduce the concept of {standardisation} within the XPSP context.
In Sect.~\ref{sec:standa} we show the performance of XPSP for widely used wide-band photometric systems, including the effects of standardisation. We deal separately with TCs including the spectral range $\lambda\le 400$~nm, as this is particularly critical for XP spectra and requires special treatment. In Sect.~\ref{sec:narrow} we show some examples of XPSP using medium-width and narrow  passbands, including emission line photometry. We also illustrate the case of a photometric system brought into life for the first time by means of XPSP, the \gaia C1 system \citep{Jordi2006}. In Sect.~\ref{sec:perfver} we present some example of performance verification in a scientific context and in Sect.~\ref{sec:products} we illustrate the XPSP products offered to the general user in {\gdr3}, namely tools to get XPSP in the preferred system of the user and value-added catalogues. In Sect.~\ref{sec:recommend} some caveats and recommendations for best use are reported. Finally, in Sect.~\ref{sec:conclu} we summarise our results and discuss perspectives and developments of XPSP for future {\gaia} data releases. For increased readability, we collect some figures, tables, and discussions relevant to the quantitative understanding of the performance of XPSP  and for its actual use and provide these in a series of Appendices at the end of the paper. A list of the principal \gaia-related acronyms used throughout the paper is presented in Table~\ref{tab:acronyms}. Finally, as synthetic photometry can only be obtained from externally calibrated spectra, in the following we often drop the EC label for brevity, referring to the EC XP spectra used to get synthetic magnitudes simply as XP spectra.

%%%%%%%%%%%%%%%%%%%%%%%%%%%%%%%%%%%%%%%%%%%%%%%% END INTRO

%%%%%%%%%%%%%%%%%%%%%%%%%%%%%%%%%%%%%%%%%%%%%%%% METHODS
%\input{sections/methods.tex}

\section{Methods}\label{sec:methods}

Synthetic photometry is based on the computation of a properly normalised mean flux \citep[as defined in][]{BM12} obtained by integrating  the product of a transmission curve $S(\lambda)$ and a spectral energy distribution (SED) over a given
wavelength or frequency interval (depending on the photometric system definition). Following \citet{BM12}, for the photometric systems considered in this work, the mean flux can be expressed as

\begin{equation}
    <\!f_\lambda\!>\, = \frac{\int\, f_\lambda(\lambda)\, S(\lambda)\,\lambda \, {\rm d}\lambda }{\int\, S(\lambda)\,\lambda \, {\rm d}\lambda }
    \label{eq:vegaSynthFlux}
\end{equation}

in VEGAMAG and Johnson-Kron-Cousins systems, and

\begin{equation}
    <\!f_\nu\!>\, = \frac{\int\, f_\lambda(\lambda)\, S(\lambda)\,\lambda \, {\rm d}\lambda }{\int\, S(\lambda)\,\left(c/\lambda\right) \, {\rm d}\lambda } 
    \label{eq:abSynthFlux}
\end{equation}

in the AB system \citep[see][and references therein]{Fukugita1996,Bessel05,Sirianni2005}.

In this work we express wavelengths $\lambda$ in units of nanometres (nm), energy flux per wavelength units $f_\lambda$ in units of ${\rm W\,m^{-2}\,nm^{-1}}$, 
and energy flux per frequency units $f_\nu$ in units of ${\rm W\,m^{-2}\,Hz^{-1}}$. $S(\lambda)$ designs a photonic response curve (i.e. it includes the quantum efficiency curve of the CCD).

The synthetic flux can be converted into a magnitude by
\begin{equation}
   \mathrm{mag} = -2.5\,\mathrm{log} <\!f_{\lambda|\nu}\!>\! + ZP
    \label{eq:synthMag}
,\end{equation}
where the zero point (ZP) in VEGAMAG is computed \wrt a reference SED:
\begin{equation}
    ZP = + 2.5\,\mathrm{log}<\!{f_{\lambda}}^{ref}\!>\! + V
,\end{equation}

and in the AB case is
\begin{equation}
    ZP = -56.10
.\end{equation}

VEGAMAG and Johnson-Kron-Cousins only differ in the choice of the reference SED: in the first case, we adopt the same reference as \citet{riello2021}, that is, an unreddened A0V star with $V=0.0$, while in the other case we assume the Alpha Lyrae SED provided by \citet{Bohlin2014} and  $V=0.03$~mag as reference.

In this context, \gaia EC XP spectra are no exception, and synthetic fluxes and magnitudes can be derived as described above. However, in the \gaia Archive, the XP spectra are stored as the projection on a set of basis functions, that is, as coefficients and corresponding covariance matrix. The SEDs (BP and RP separately)  can then be reconstructed by linear combination of the bases, given the coefficients, as described in \citet{EDR3-DPACP-118} and \citet{EDR3-DPACP-120}:

\begin{equation}
     f_\lambda^{XP}(\lambda) = \sum_{i=1}^N b_i^{XP} \phi_i^{XP}(\lambda)
     \label{eq:ecsXp}
.\end{equation}
The two partially overlapping SEDs  can be combined into a single distribution by computing a weighted mean
with the weight for BP and RP given by:
\begin{equation}
    w_{BP}(\lambda) = \left\{
\begin{array}{l@{\quad}l}
1                & {\rm if}  ~\lambda < \lambda_{lo}\\~\\
1 - \frac{\lambda-\lambda_{lo}}{\lambda_{hi}-\lambda_{lo}}                   & {\rm if} ~\lambda_{lo}<\lambda<\lambda_{hi}\\~\\
0                & {\rm if} ~\lambda>\lambda_{hi},
\end{array}\right.
\end{equation}
\begin{equation}
    w_{RP}(\lambda) = 1 - w_{BP}(\lambda)
,\end{equation}
where $[\lambda_{lo}, \lambda_{hi}]$ is the overlapping region \citep[see][for further details]{EDR3-DPACP-120}.
Combining \equref{vegaSynthFlux} with \equref{ecsXp}, we obtain a very efficient algorithm to compute synthetic fluxes in a VEGAMAG system  by means of the quantities:
\begin{equation}
    s_i^{XP}\, = \frac{\int\, w_{XP}(\lambda) \, \phi_i^{XP}(\lambda)\, S(\lambda)\,\lambda \, {\rm d}\lambda }{\int\, S(\lambda)\,\lambda \, {\rm d}\lambda }
    \label{eq:filterCoeffs}
,\end{equation}
so that the mean synthetic flux of the source is simply given by
\begin{equation}
     <\!f_\lambda\!>\, = \sum_{i=1}^N b_i^{BP} s_i^{BP}
     + \sum_{i=1}^N b_i^{RP} s_i^{RP}
     \label{eq:synthFlux}
.\end{equation}
The computation of \equref{filterCoeffs} coefficients for an AB system is straightforward.
In practice, a photometric system containing $K$ passbands is reduced to a set of \emph{TC bases} composed of two $K\times N$ matrices $\mathbb{S}^{BP}$ and $\mathbb{S}^{RP}$ ; 
given a \gaia source with spectral coefficients $\mathbf{b}^{BP}$ and $\mathbf{b}^{RP}$, 
the $K$ synthetic fluxes $\mathbf{f}$ in the photometric system are readily given by
\begin{equation}
    \mathbf{f} = \mathbb{S}^{BP} \cdot \mathbf{b}^{BP}
    + \mathbb{S}^{RP} \cdot \mathbf{b}^{RP}
     \label{eq:synthFluxVec}
.\end{equation}
A covariance matrix can be computed for fluxes $\mathbf{f}$ as:
\begin{equation}
    \mathrm{K_{ff}} = \mathbb{S}^{BP}\cdot \mathrm{K_{bb}}^{BP} \cdot {\mathbb{S}^{BP}}^T + \mathbb{S}^{RP} \cdot \mathrm{K_{bb}}^{RP} \cdot {\mathbb{S}^{RP}}^T
.\end{equation}

The nominal
uncertainties on $\mathbf{f}$ fluxes are given by the square root of diagonal elements of $\mathrm{K_{ff}}$. In practice, for issues related to the uncertainties in the XP spectra \citep{EDR3-DPACP-118,EDR3-DPACP-120}, these can be significantly underestimated. In Sect.~\ref{sec:errorcorrections} we derive empirical corrections to properly trace the uncertainty in synthetic fluxes and magnitudes.
We computed the filter bases for a number of commonly used photometric systems, many of them discussed and validated below.
XPSP in these and other systems can be obtained from the \gaia Archive as described in Sect.~\ref{sec:gaiaXPy}.

\subsection{Empirical estimate of errors}
\label{sec:errorcorrections}

In order to validate the uncertainty estimates for the passband fluxes, as derived from XPSP, we took a sample of 43\,653 randomly selected sources covering a suitable range of colour and magnitude, and for each source we randomly split its epoch observations into two groups \citep[hereafter `BP/RP split-epoch validation dataset'; for further details see][]{EDR3-DPACP-118}. We then compute two separate mean {\xp} spectra and their resulting synthetic fluxes for every pair. This procedure results in two statistically independent measurements for each source, which should be consistent within their respective uncertainty estimates. We emphasise that the randomised grouping of epoch observations is essential because it prevents any potential intrinsic time variability of a source from compromising the uncertainty validation.

\begin{figure}
\center{
\includegraphics[width=\columnwidth]{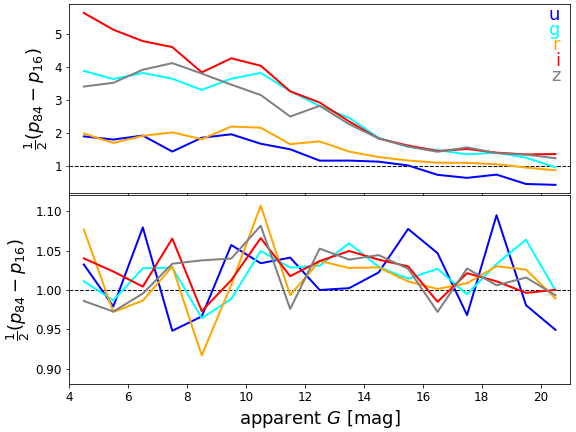}
}
\caption{Illustration of underestimated uncertainties for the standardised SDSS system. We summarise the underestimation as half the difference between the 84th and the 16th percentiles of randomly split sources falling into this apparent $G$ magnitude bin. If uncertainties are correctly estimated, this quantity should be 1, as indicated by the horizontal dashed line.
Top panel: Nominal uncertainties. Bottom panel: Calibrated uncertainties (we highlight the very different y-axis range).} 
\label{fig:corrections_factors_SdssDoiStd}
\end{figure}

\begin{figure}
\center{
\includegraphics[width=\columnwidth]{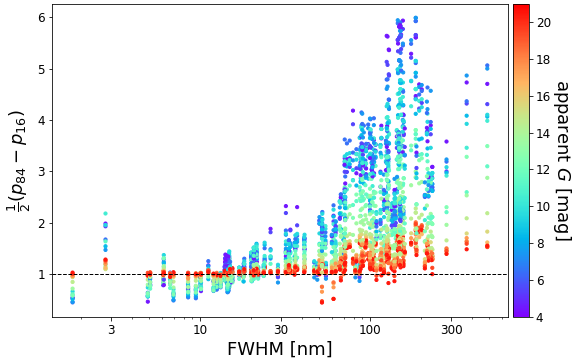}
}
\caption{Systematic underestimation of nominal uncertainties for synthetic fluxes as function of FWHM of each band in all photometric systems considered in this paper.} 
\label{fig:corrections_factors_vs_FWHM}
\end{figure}

As expected, this test revealed that the nominal uncertainty estimates of the synthetic fluxes are systematically underestimated for most photometric systems \citep[see][for a discussion on the underestimation of errors in the underlying XP spectra]{EDR3-DPACP-118}. In such cases, the distributions of flux differences within a pair of randomly split sources normalised by their combined uncertainties would be substantially broader than a unit Gaussian. In particular, we notice that this underestimation of uncertainties appears to depend on the apparent $G$ magnitude of a source. We illustrate this for the example of the standardised SDSS system (see \secref{standa_sdss}) in the top panel of \figref{corrections_factors_SdssDoiStd}. Here, we clearly see that the distribution of normalised flux differences in the randomly split sources is broader than a unit Gaussian, because half the difference between the 84th and 16th percentiles is larger than 1. We also see that this underestimation of uncertainties has a different effect from one passband to another; the underestimation appears to be stronger for broader synthetic bands, as is evident from \figref{corrections_factors_vs_FWHM}, yet we did not observe any dependence on the wavelength of the band.

In order to calibrate the uncertainty estimates for the synthetic fluxes,  for each band in every photometric system considered in this paper we tabulate the factors by which the distributions of normalised differences are too high, as a function of apparent $G$ magnitude (see top panel of \figref{corrections_factors_SdssDoiStd}). The calibrated uncertainties are then obtained by inflating the nominal uncertainties for every source according to the tabulated factors by which they are found to be too small. Again, this is illustrated in the bottom panel of \figref{corrections_factors_SdssDoiStd} for the example of the standardised SDSS system. Evidently, the calibrated uncertainties now fully account for the flux differences in the pairs of randomly split sources.

We note that the Python software tool to deal with XP spectra, GaiaXPy (Sect.~\ref{sec:gaiaXPy} and \citealt{EDR3-DPACP-118}), provides by default the nominal uncertainties for the synthetic fluxes, which are underestimated. 
However, it can optionally compute the calibrated uncertainties instead (by setting \texttt{error\_correction=True}), for all the sets of passbands currently included in the GaiaXPy repository. Please refer to the GaiaXPy documentation (link included in \secref{gaiaXPy}) for instructions and for a full list of the systems for which this is available.

\subsection{Standardisation}\label{sec:stand}

Externally calibrated XP spectra are known to suffer from systematic errors attributable to various factors \citep[see][]{EDR3-DPACP-120}. These issues manifest as systematic differences between XPSP magnitudes and the corresponding magnitudes of top-quality external sources that are taken as reference for a given photometric system (e.g. sets of primary and/or secondary standard stars). In general, for wide-band XPSP, these effects amount ZP differences within a few hundredths of a magnitude and/or to trends as a function of colour with a maximum amplitude of a few hundredths of a magnitude over wide colour ranges (with the exception of ultraviolet (UV) bands, $\lambda<400$~nm, which are discussed separately in Sect.~\ref{sec:ustand}; see Appendix~\ref{sec:app_stand}; see also \citealt{EDR3-DPACP-120}).

In future data releases, once we are able to keep these systematic errors on EC XP spectra to a minimum, XPSP will directly serve to re-define  optical photometric systems based on exquisitely homogeneous space-based spectrophotometry. However, in the meantime, users might be interested in reproducing the existing photometric systems at best, with currently available XPSP. This can be achieved by a process that we call {standardisation}, following \citet{Bessel05}. 
In our context, standardisation consists in (a) adopting an external photometric dataset as the reference set of standards for a given photometric system, such as SDSS Stripe 82 photometric standard stars \citep{I07,Thanjavur2021}, (b) comparing the XPSP magnitudes for these standard stars (mag$_{synth}$) with those from the reference source (mag$_{phot}$) as a function of magnitude and colour, and (c) finding a correction  that, when applied to XPSP magnitudes, minimises the differences ($\Delta {\rm mag}={\rm mag}_{phot}-{\rm mag}_{synth}$), thus providing the best reproduction of the external system. 

When dealing with {pure} magnitudes, that is, when the product of the measuring apparatus is an estimate of the integrated flux of the source through the considered TC, this kind of standardisation is typically achieved by means of polynomial transformation as a function of colour. As mentioned above, these may suffer from strong systematic effects, for example because a polynomial may not be adequate to model all the subtleties of the relations between the two systems.
This kind of problem can be mitigated if the set of magnitudes to be transformed is from synthetic photometry. In this case, the safest and most widely adopted way to standardise magnitudes is to tweak the profile of the TC adopted for synthetic photometry in order to minimise $\Delta {\rm mag}$ and its trends with colour \citep{Bessel05}. This process is designed to remove the small differences between the TC of the reference system and the one to be transformed, possibly taking into account the effects that would require high-order terms in a polynomial transformation\footnote{The underlying hypothesis is that  a TC should exist that removes all the systematic differences between the two sets of magnitudes, assuming that both accurately trace the original SED of the observed sources.}. 

In our specific case, the tweaking is mainly used to minimise the effects of the residual systematic errors of EC XP spectra on synthetic photometry using the external standards as a kind of second-level calibrator. As we see in \secref{standa}, for wide passbands in the range $\lambda\ga 400$~nm, standardisation allows us to reproduce existing systems with typical accuracy from a few millimag to submillimag, depending on the specific passband, over broad ranges in colour and for the large majority of well-measured stars with published XP spectra in {\gdr3}. 
In Sect.~\ref{sec:nonustand}, we describe the way in which we get standardisation by TC tweaking, and how we deal with passbands in the range $\lambda\la 400$~nm  (Sect.~\ref{sec:ustand}).

\subsubsection{Standardisation: general method}\label{sec:nonustand} 
\begin{figure}[t]
\center{
\includegraphics[width=\columnwidth]{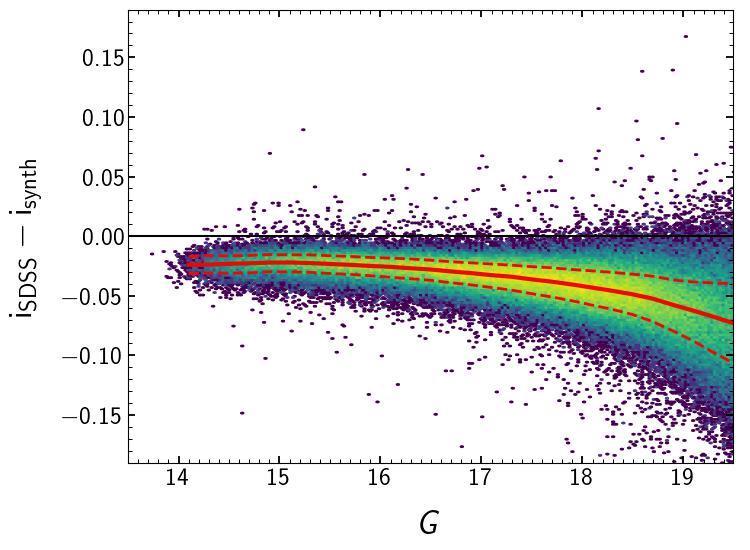}}
\center{
\includegraphics[width=\columnwidth]{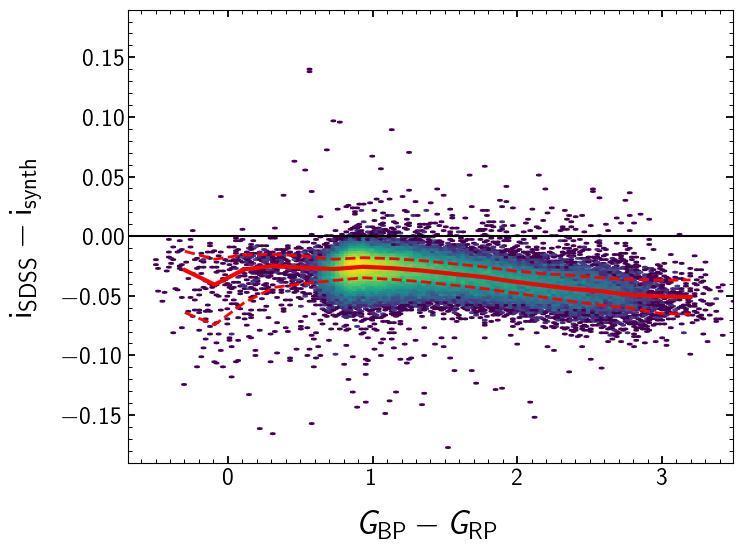}}
\caption{Residuals between reference and synthetic magnitudes computed through a nominal filter transmission curve \citep{Doi2010} for a set of standard stars plotted as a function of $G$ magnitudes (\emph{top}) and {\bprp} colour (\emph{bottom}). The red curves represent a smoothed median line of the data.}
\label{fig:standExample}
\end{figure}

\figref{standExample} shows residuals between standard and synthetic $i_{\rm SDSS}$ magnitudes obtained with the  \citet{Doi2010} TCs for the SDSS reference dataset presented in \secref{standa_sdss}. The figure provides an example illustrating all the effects that need to be corrected within the standardisation process. 
Residuals are plotted as a function of G magnitude (upper panel) and $G_{BP}-G_{RP}$ colour (lower panel). In both cases, the continuous red curve traces the median ($P_{50}$) of the residual distribution computed in bins of 0.4~mag in width,  while the dashed curves are the loci of the 15.87\% ($P_{16}$) and 84.13\% ($P_{84}$) percentiles. There is a clear trend as a function of magnitude that is common to all photometric systems. This is interpreted as being (mainly) due to systematic overestimation of the background, which produces a negative offset in measured XP fluxes \citep[see][for a detailed discussion]{EDR3-DPACP-118}. In the following, we refer to this general magnitude-dependent trend as the {`hockey-stick' effect}, described and discussed in \citet{EDR3-DPACP-120}\footnote{In the context of \gaia photometry, the hockey-stick effect is mentioned for the first time in \citet{Dafydd2018}. A realisation of the same effect we are dealing with here is shown in the top panel of Fig.~23 of \citet{riello2021}, and is briefly discussed there.}. Independently of the actual nature of this effect, which will be further investigated in preparation for future \gaia data releases, we find that it can be effectively mitigated by applying a background-like correction and, consequently, we adopted this approach in the standardisation process.

The presence of  additional offsets in the magnitude scale cannot be excluded, but the median of the residuals in the range where the hockey-stick effect is minimised ($G\la 15.5$~mag) constrains their amplitude to $< 0.01$~mag. A selection in magnitude ($G<17.65$~mag)\footnote{This is the general magnitude limit for XP spectra in \gaia DR3, see Sect.~\ref{sec:standa_sdss}.} has been applied to 
data plotted as a function of colour in order to minimise the disturbance due to the hockey-stick effect and to better appreciate the small colour term present in the data (linear trend with {\bprp} colour).

The standardisation process is composed of two phases that can be iterated a few times. For each passband: (1) the flux offset $f_{bg}$ to be added to synthetic fluxes for the removal of the hockey-stick is evaluated; and (2) the TC shape is tweaked to remove the colour term.

To minimise the entanglement 
of the two effects, we  perform process (1)\ on a subsample of available data by selecting a restricted {\bprp} colour range (to minimise disturbance due to the colour term) while process (2) is performed on a subsample selected in magnitude, avoiding fainter stars which are more affected by the background issue.
Finally, we evaluate a correction factor for the zero point $ZP_{std}$  in order to mitigate any residual {\em grey} offset.\\
A {standardised photometric system} thus consists in a new set of basis functions $(\mathbb{S}^{BP}, \mathbb{S}^{RP})_{STD}$ computed with  
the tweaked TC, an array of flux offsets $ \mathbf{f}_{bg}$ to be added to synthetic fluxes of \equref{synthFluxVec},
\begin{equation}
     \mathbf{f}_{STD} =  \mathbf{f} +  \mathbf{f}_{bg}
,\end{equation}
and the array of ZP correction factors to be included in \equref{synthMag},
\begin{equation}
    \mathrm{mag}_{Std} = -2.5\,\mathrm{log} <\!f_{Std}\!>\! + ZP + ZP_{STD}
    \label{eq:synthMagStd}
,\end{equation}

where all the involved vectors have one component for each passband of the considered system.
The evaluation of the background offset can be achieved only if available standards span a sufficiently wide range in magnitude (it must roughly cover from $G\simeq13$~mag to $G\ga18$~mag).  
We typically select standards with colours within $\simeq 0.5$~mag of  $\bprp\simeq 1.0$~mag; data are then partitioned in magnitude bins of $\simeq 0.5$~mag. For each bin, we evaluate the median of the differences,
\begin{equation}
    \mathrm{mag}_{phot}-\mathrm{mag}_{STD}
,\end{equation}
which are arranged in the array $\mathbf{P_{50}}$.
The background correction $f_{bg}$ is found as the value that minimises the cost function, 
\begin{equation}
    \rho = \sum_i \left({P_{50}}_{i} - <\mathbf{P_{50}}>\right)^2
.\end{equation}
To implement the filter-tweaking algorithm, we model the shape of the standardised filter response by multiplying the nominal transmission $S(\lambda)$ with a linear combination of a (low) number of basis functions $\mathcal{S}_k$:
\begin{equation}
    S^\dagger(\lambda) = S(\lambda) \cdot \sum_k \alpha_k \mathcal{S}_k(\lambda)
.\end{equation}

\begin{figure}[!htbp]
\center{
\includegraphics[width=\columnwidth]{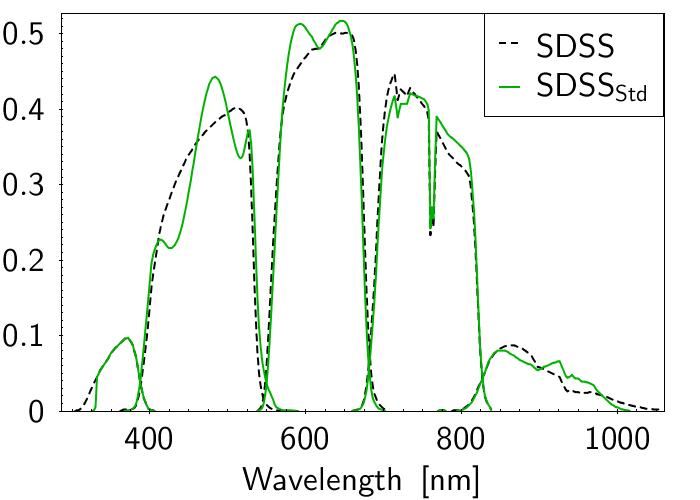}
}
\caption{Original SDSS transmission curves from \citet[][black dashed line]{Doi2010} are compared to their tweaked version obtained with the standardisation process (green continuous lines). It is important to remember that the shape of standardised TCs is designed to correct for the systematic errors that still affect EC XP spectra. No tweaking is applied to the $u$ passband, as the standardisation of $u$ magnitudes and fluxes is performed by means of polynomial transformation. The cut at 330~nm follows the TC of the BP spectrometer.} 
\label{fig:stdbands}
\end{figure}

The basis functions used for the present work include mainly Legendre polynomials and Hermite functions. 
An important issue to keep in mind is that this method has an intrinsically low sensitivity: large variations in the shape of the filter may result in very small changes in the residuals, meaning that there is no unique solution to the problem. When several models give comparable results, we arbitrarily select TCs with shapes closer to the nominal one.
The procedure for the optimisation of the model $S^\dagger(\lambda)$ is similar to that described for the correction of the hockey-stick effect: (1) We select calibrators with a $G$ magnitude brighter than a given value (depending on the specific data set); (2) we partition data in {\bprp} colour bins of $\sim0.2$ mag; (3) for each bin, we compute the median $P_{50}$ and the width $\sigma=0.5(P_{84}-P_{16})$ of the distribution of the difference $(\mathrm{mag}_{phot}-\mathrm{mag}_{Std})$; and (4) the model is optimised by minimising the cost function
    
    \begin{equation}
        \rho = \sum_i \left({P_{50}}_{i}^2 + \sigma_i^2\right),
    \end{equation}

where the $\sigma_i$ terms have been included as they were found to be effective in preventing odd solutions of the standardisation process that were sometimes found to arise. In all the cases considered here, the changes of the TC shapes induced by the standardisation are small; a typical example is shown in Fig.~\ref{fig:stdbands}.

As a final remark, while the $f_{bg}$ values we derive are representative of
the conditions of the adopted reference samples, which are
typical uncrowded field stars, we cannot guarantee their universal validity, because we have not been able to test their possible variation
as a function of position in the sky, local stellar density,  and so on. However, in
Sect.~\ref{sec:standa_hugs} we use a reference sample where the crowding
conditions are significantly poorer than in the typical reference sample (as e.g. in those described in Sect.~\ref{sec:standa_sdss} and
Sect.~\ref{sec:standa_jkc}), and we verified that the $f_{bg}$ values estimated
for broadly similar passbands in the different cases are similar, with typical
differences of $\la$20\%.

\subsubsection{The case of UV bands}\label{sec:ustand}

\begin{figure}[!htbp]
\center{
\includegraphics[width=\columnwidth]{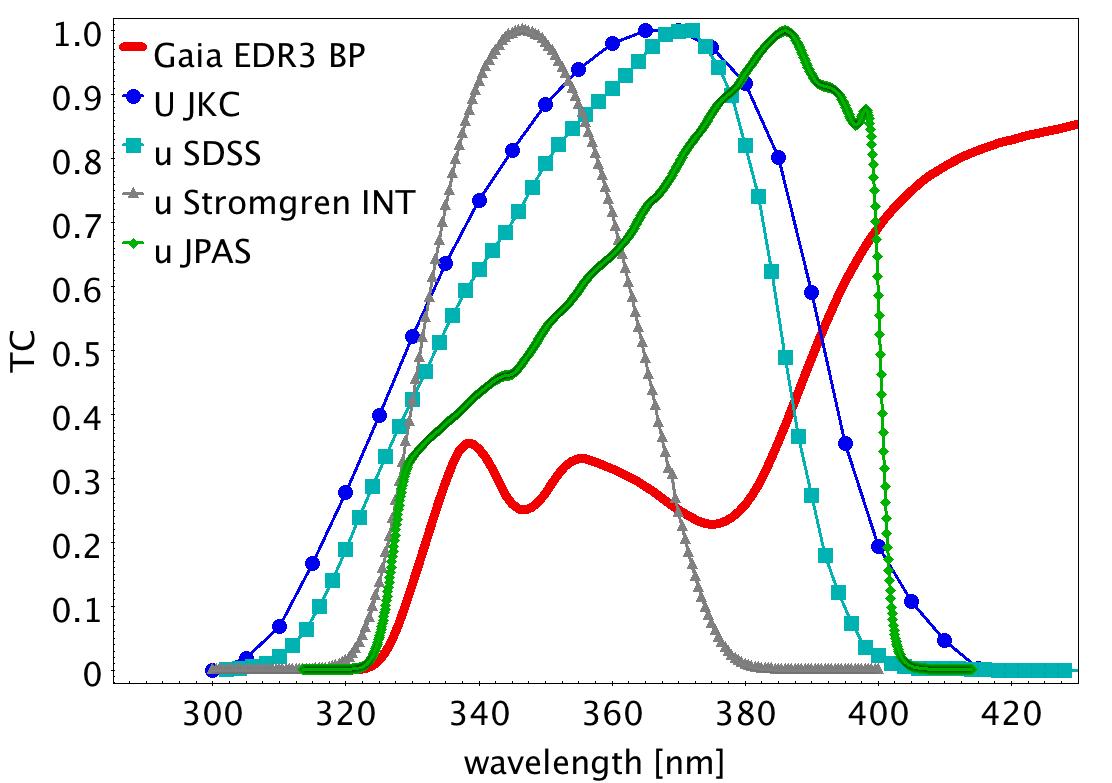}
}
\caption{Transmission curves of all the UV bands considered in this paper are compared with the transmission curve of the BP spectrometer. All the curves are normalised to their maximum.
} 
\label{fig:ubands}
\end{figure}

As anticipated above and discussed in detail in \citet{EDR3-DPACP-120}, the strongest colour-dependent systematic errors affecting EC XP spectra occur in the spectral range $\lambda\la 400$~nm, where the TC of the BP spectrophotometer is low and highly structured, with two very steep branches found at around 390~nm and at the blue cut-off at $\simeq 330$~nm, and two local maxima at $\lambda \simeq 338$~nm and $\simeq 355$~nm (see Fig.~\ref{fig:ubands}). In the following, for brevity, we refer to passbands whose predominant part of the spectral range is below 400~nm (and typically $\ga 300$~nm) as UV bands.

The most widely used UV bands (a) span this spectral window, with most of the throughput in the region bluer than $\simeq 375$~nm, which is especially critical for XP spectra, and (b) have a blue cut-off exceeding the blue limit of BP (Fig.~\ref{fig:ubands}). Therefore, reproducing the photometry in these passbands with XPSP is quite challenging, with factor (b) effectively preventing the possibility of a full standardisation\footnote{The information in the spectral range $\lambda\la 330$~nm is not present in the XP spectra and no correction can help to recover it. Hence, in cases where significant star-to-star differences in that wavelength range occur, the standardised $UV_{synth}$ magnitudes cannot adequately reproduce $UV_{phot}$ ones.}. 

However, as that region of the stellar spectra is especially important and informative, we attempt a standardisation of SDSS u and Johnson-Kron-Cousin's \citep[JKC, hereafter, as defined by][standard stars]{Landolt1992} U bands. 
In these cases we were not able to obtain satisfactory standardisations by tweaking the TCs and we used high-degree colour-dependent polynomial transformations instead.
Moreover, as the adopted solution does not provide satisfactory results over the whole {\gdr3} sample of XP spectra, the use of standardised u/U magnitudes is recommended only for a subset limited in signal-to-noise ratio (S/N; see Sect.~\ref{sec:standa_u} and Sect.~\ref{sec:gspc}). 

As a first step we produced new passbands identical to the original ones (from \citet{Doi2010} for u$_{SDSS}$ and from \citet{BM12} for U$_{JKC}$) but valued 0.0 for  $\lambda <330$~nm. We then proceeded in a similar way as for the non-UV passbands.
The hockey-stick correction was obtained, taking special care to minimise the effect of the large colour terms at work in this case.
The median and $\sigma$ of the resulting residuals as a function of colour computed over bins were then fitted  with high-order polynomials. We find that adopting colours from the same system as the considered UV bands provides simpler and more robust solutions, and therefore the polynomials are a function of (synthetic and non-standardised) $g-i$ and $B-V$ for $u_{\rm SDSS}$ and $U_{\rm JKC}$, respectively. The public tool to manage XP spectra (GaiaXPy, see Sect.~\ref{sec:gaiaXPy}) will allow the user to produce both raw and standardised XPSP (for the standardised systems), independently of the method adopted, that is, polynomial transformation for UV bands and TC tweaking for all the other cases.

The actual performance of the standardised version of the two UV bands considered in this section is discussed in Sect.~\ref{sec:standa_u} and Appendix~\ref{sec:app_stand}, while the recommendations for safe use are shown and discussed in Sect.~\ref{sec:gspc} and Sect.~\ref{sec:recommend}.  

%%%%%%%%%%%%%%%%%%%%%%%%%%%%%%%%%%%%%%%%%%%%%%%% END METHODS

%%%%%%%%%%%%%%%%%%%%%%%%%%%%%%%%%%%%%%%%%%%%%%%% WIDEBAND
%\input{sections/wideband.tex}
\section{Wide band synthetic photometry}
\label{sec:standa}

In this section we illustrate the performance of XPSP in reproducing the photometry of existing and widely used wide-band photometric systems. We also show how residual inaccuracies are reduced below the 1\% level by the process of standardisation (described in Sect.~\ref{sec:methods}) with respect to selected sets of reliable photometric standard stars. To illustrate the process, we treat the cases of the SDSS and JKC systems  more extensively, 
while for the other standardised systems, some of the relevant plots and tables are collected in Appendix~\ref{sec:app_stand}. Some experiments of validation using stellar models are also reported in Appendix~\ref{sec:app_isoc}.

%%%%%%%%%%%%%%
\subsection{SDSS system and its standardisation}
\label{sec:standa_sdss}

The Sloan Digital Sky Survey \citep[SDSS;][]{York2000} was the first modern digital survey producing precise photometry over a large portion of the Northern Sky. Its photometric system, defined in \citet{Fukugita1996}, established a new standard, now widely used in Galactic and extra-galactic astronomy \citep[see][and references therein]{I07,Thanjavur2021}.

As a reference set for the SDSS system, we used a selected subsample of the Stripe 82 standard stars recently presented and discussed by \citet[][T21 hereafter]{Thanjavur2021}. Compared to the previous realisation of the same set \citep{I07}, T21 has two to three times more epochs per source used in photometric averaging; systematic photometric zero-point errors as functions of RA and Dec are estimated and corrected for using {\gedr} photometry\footnote{It is important to note that, as a consequence, any spatial trend of the photometric zero-points in \gedr should have been transferred to the T21 photometry. However, we also note that (a) when comparing {standardised} XPSP photometry with T21 we find residual trends of amplitude $\la 10.0$~mmag as a function of position, and (b) the comparison of XPSP photometry with Hipparcos photometry presented in \citet{EDR3-DPACP-120} suggests that XPSP photometry should be spatially homogeneous to the level of a few mmag over most of the sky. This may suggest that spatial trends were not completely removed from T21 photometry.}; and the same is used to correct \textit{ugiz} magnitudes relative to the \textit{r}-band. This approach results in random photometric errors approximately 30\% smaller than in the I07 catalogue and below $\approx$ 0.01 mag for stars brighter than 20.0, 21.0, 21.0, 20.5, and 19.0 mag in \textit{u}, \textit{g}, \textit{r}, \textit{i,} and \textit{z}-bands, respectively.

To obtain our reference set to be used for comparison and standardisation of XP photometry, we cross-matched the EDR3 sources with XP spectra to the T21 sample and applied the following quality filters on Gaia data:

\begin{itemize}
    \item \texttt{XP\_num\_of\_transits} $>=$ 15,
    \item \texttt{XP\_num\_of\_contaminated\_transits / XP\_num\_of\_transits} $<$ 0.1,
    \item \texttt{XP\_num\_of\_blended\_transits / XP\_num\_of\_transits} $<$ 0.1,
    \item \texttt{XP\_number\_of\_neighbours}$<$ 2,
    \item \texttt{XP\_number\_of\_mates}$<$ 2,
    \item \texttt{XP\_number\_of\_visibility\_periods\_used}$>$ 10,
\end{itemize}

where \texttt{XP} stands for Gaia BP and RP. A set of broad quality filters was applied on parameters from the T21 sample as well:
\begin{itemize}
    \item \texttt{\{u,g,r,i,z\}Nobs} > 4,
    \item \texttt{\{g, r, i\}msig $\cdot \sqrt{\{g,r,i\}Nobs}$} < 0.03.
\end{itemize}

Detailed explanations of the used columns can be found in the \href{https:\\abc.def.ghi}{Gaia DR3 documentation} and the \href{https://www.sdss.org/dr17/}{SDSS Data Model}.
The final reference sample includes approximately 280 879 sources.
For the synthetic photometry, we adopt the official SDSS TCs from \citet{Doi2010}.

\begin{figure*}[!htbp]
    \centerline{
    \includegraphics[width=(\columnwidth)]{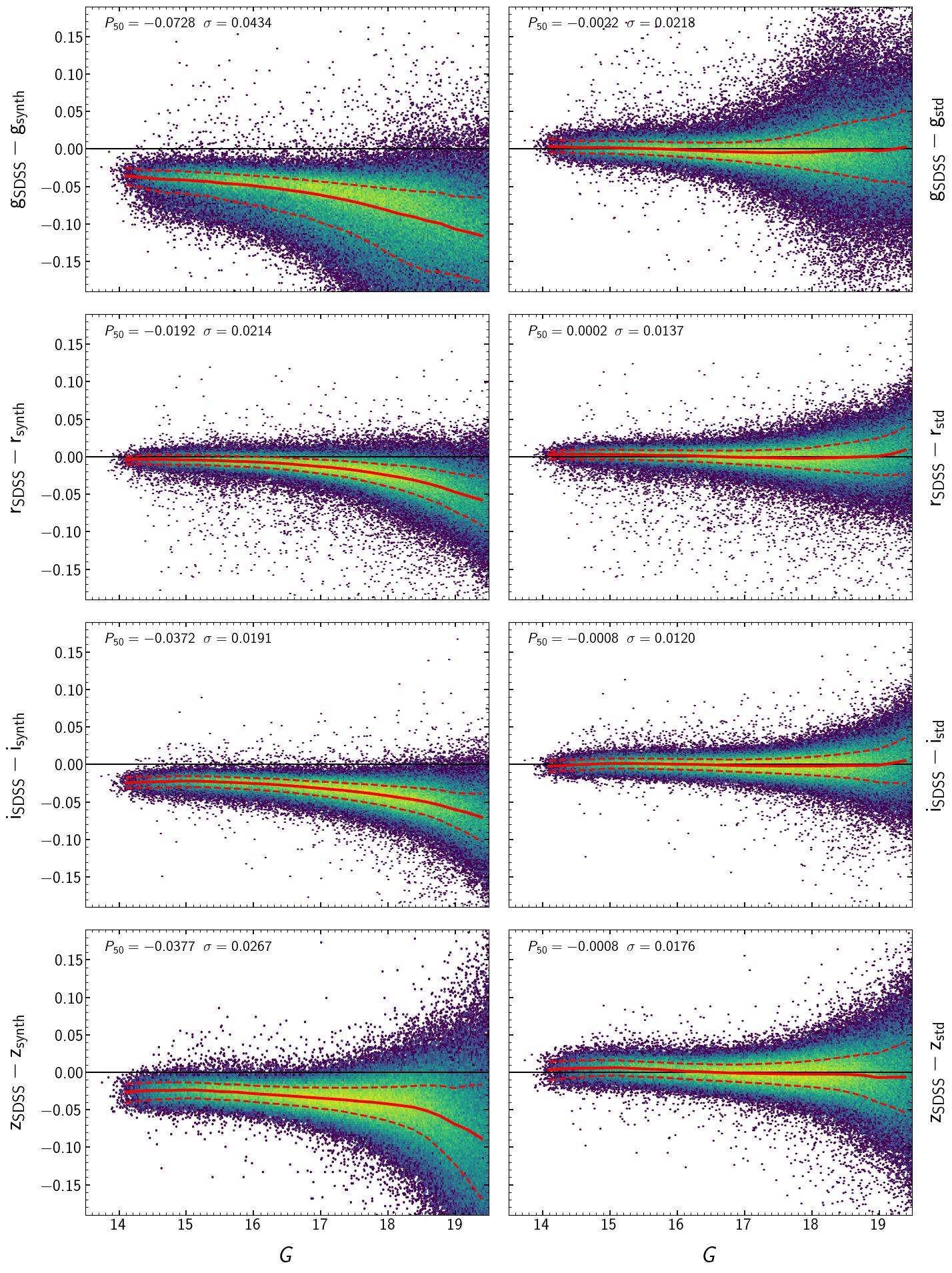}
    \includegraphics[width=(\columnwidth)]{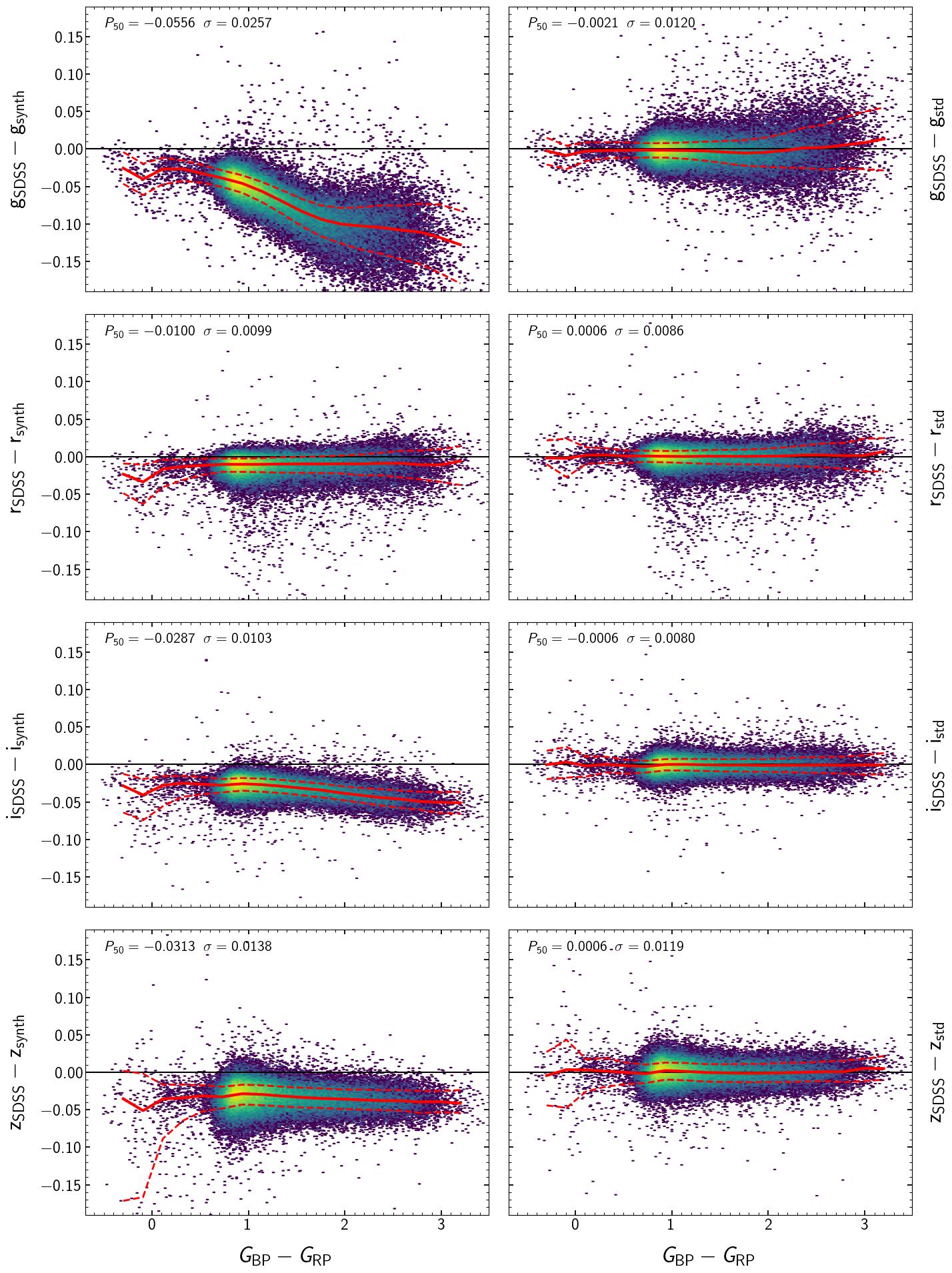}}
    \caption{Performance and standardisation of SDSS $ugriz$ XP synthetic magnitudes using the selected subset of the \citet{Thanjavur2021} sample, which we adopted as reference. Left set of panels: $\Delta$mag as a function of $G$ magnitude for the entire sample using nominal XP synthetic magnitudes (left panels) and standardised XP synthetic magnitudes (right panels). In each panel, the continuous red line connects the median $\Delta$mag computed in 0.2~mag wide bins, and the dashed red lines connect the loci of the 15.87\% ($P_{16}$) and the 84.13\% ($P_{84}$) percentile computed in the same bins. The median ($P_{50}$) and the difference between $P_{84}$ and $P_{16}$ ---here used as a proxy for the standard deviation $\sigma$--- for the entire sample are reported in the upper left corner of each panel. Right set of panels: Same for $\Delta$mag as a function of {\bprp} colour, limited to the subsample of reference stars with XP spectra released in {\gdr3}.}
    \label{fig:sdss_delta_app}
\end{figure*}

\begin{figure}[!htbp]
    \centering
    \includegraphics[width=(\columnwidth)]{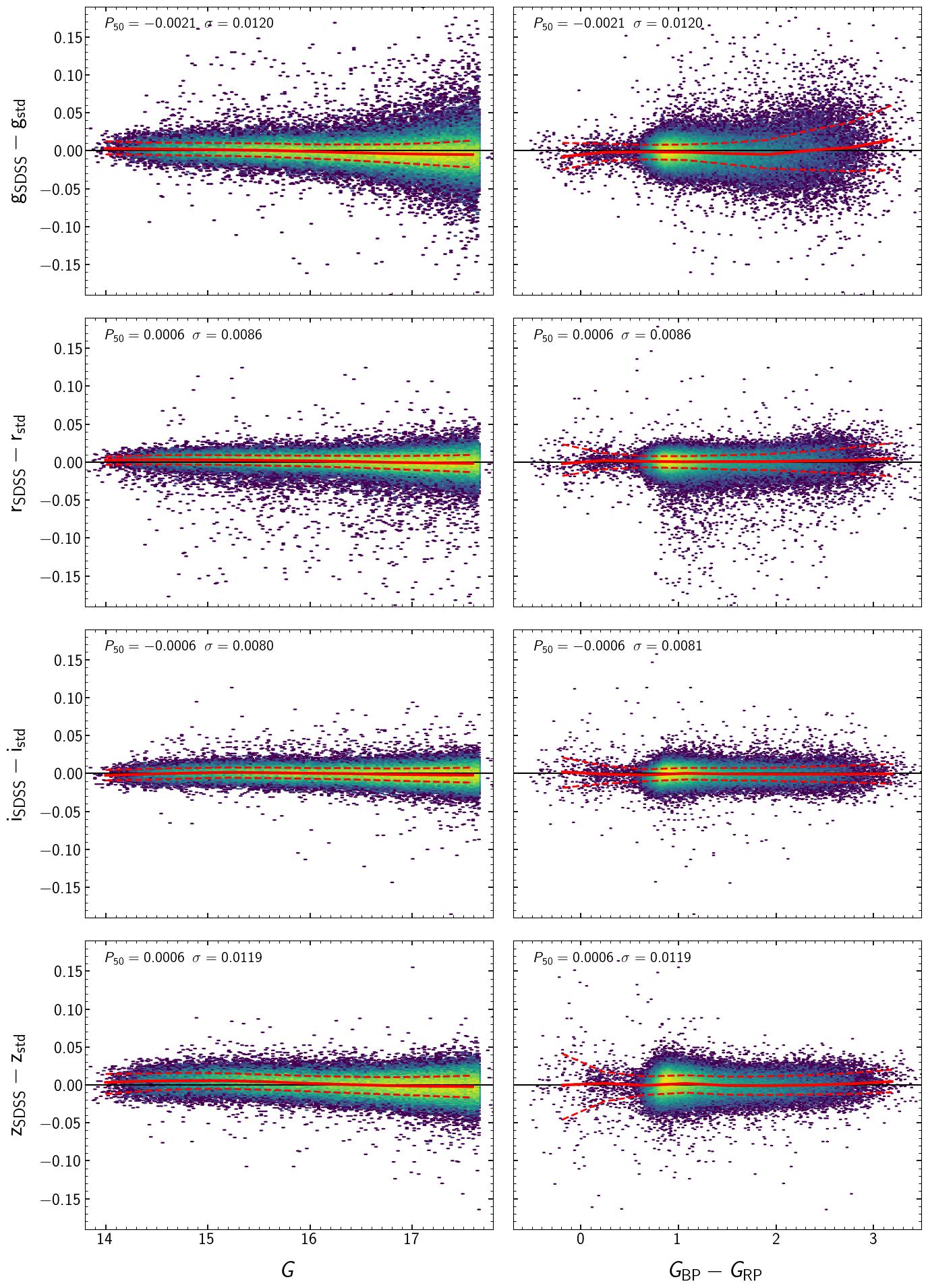}
    \caption{Performances of standardised XPSP in the SDSS system ($griz$). We show $\Delta$mag as a function of $G$ magnitude (left panels) and {\bprp} colour (right panels) for the subsample of T21 stars whose XP spectra have been released in {\gdr3}. The arrangement and the meaning of the symbols is the same as in Fig.~\ref{fig:sdss_delta_app}.}
    \label{fig:sdss_deltaindr3}
\end{figure}

Figure~\ref{fig:sdss_delta_app} is a good example of the typical plot with which we illustrate the performance of XPSP in reproducing the photometry of the external set adopted as a reference, for the various photometric systems. Two multi-panel sets of diagrams are presented, the set on the left showing $\Delta$~mag as a function of $G$ magnitude, and the set on the right showing $\Delta$~mag as a function of {\bprp} colour. Within each of the two sets, the left column displays the comparison with raw XPSP magnitudes before standardisation, while the right columns show the comparison {after} standardisation. In each panel, the continuous red line is the median ($P_{50}$) of the $\Delta$~mag distribution computed over independent bins of 0.4~mag in width, while the dashed red lines trace the 15.87\% ($P_{16}$) and the 84.13\% ($P_{84}$) percentiles computed in the same bins. It is important to recall that, in this figure, as well as in all other analogous figures for other systems shown below, if not otherwise stated, the plots as a function of magnitude refer to the entire reference sample, including stars fainter than $G=17.65$~mag, which in general do not have their XP spectra released. This is required to adequately constrain the hockey-stick effect in order to correct for it in the process of standardisation. On the other hand, the plots as a function of colour refer only to the subsample with $G<17.65$~mag in order to better trace genuine colour terms, minimising the additional noise due to the hockey-stick effect.

The left rows of the two panels of Fig.~\ref{fig:sdss_delta_app} show the performances of raw XPSP in reproducing SDSS magnitudes. The deviation that is apparent for $G\ga 16.0$~mag in the diagrams as a function of $G$ magnitude is due to the hockey-stick effect. Taking this factor into account, we conclude that $riz$ photometry is reproduced remarkably well, with zero-point differences of $<0.02$ mag (as traced by $G\la 15.5$~mag stars) and colour terms with amplitudes of $\la 0.02$ mag over the whole colour range covered by the reference sample. On the other hand, $\Delta G$ displays a colour term with an amplitude of $\simeq 0.05$~mag, and also produces a larger and asymmetric scatter about the median in the plot as a function of magnitude with respect to the other passbands. This reflects the coverage by $g$ band of regions of the XP spectra that suffer from colour-dependent systematic errors, including the first sudden drop of the BP TC around 390~nm. A fully analogous behaviour is observed for PanSTARRS g (Appendix~\ref{sec:app_stand}; see also the case of JKC $B$ band discussed below), confirming that XP spectra are to be blamed for the mismatch. Standardisation significantly reduces all the discrepancies described above, as can be readily appreciated from the direct comparison between standardised and non-standardised $\Delta G$ distributions shown in Fig.~\ref{fig:sdss_delta_app}. A larger scatter about the median remains in the $g$ band than in the redder passbands, and the scatter in $z$ is slightly greater than in $r$ and $i$, while in the latter passbands the perfomance of standardised SP appears to be excellent. 

Fig.~\ref{fig:sdss_deltaindr3}  shows the final result of the standardisation process for the stars of the reference sample whose XP spectra are released in {\gdr3}, that is, those for which XPSP can be obtained. This figure shows the excellent quality of the final product\footnote{There is some redundancy between this figure and the right  columns of panels of Fig.~\ref{fig:sdss_delta_app}, as well as in analogous sets of figures produced for other photometric systems. Still we feel that it is worth showing both kinds of plots, as those as in Figure~\ref{fig:sdss_delta_app} illustrate the comparison with raw XPSP and the effect of standardisation, while those as in  Fig.~\ref{fig:sdss_deltaindr3} give a direct view of the XPSP perfomance for the material that is actually made available in \gaia DR3, from the \gaia archive.}. The median difference over the entire subsample is $< 2.5$~mmag for all the passbands considered here, and the standard deviation $\sigma$\footnote{Estimated as half of the difference between the 84.13\% and the 15.87\% percentiles of the distribution of $\Delta mag$. In the following, we  refer to this quantity as $\sigma$, if not otherwise stated, for brevity.} is $\le 12$~mmag.

%%%%%%%%%%%%%%%%%%%%%%%%%%%%%%%%%%%%%%%%%%%%%%%%%%%%%%%%%%%%%%%%%
\begin{table*}[!htbp]
    \centering
        \caption{\label{tab:SDSS_median} SDSS system: median ($P_{50)}$), 15.87\% ($P_{16}$), and 84.13\% ($P_{84}$) percentiles 
of the $\Delta {\rm mag}$ distributions of Fig.~\ref{fig:sdss_deltaindr3}. n$_{\star}$ is the number of sources in the considered bin.}
{\small
    \begin{tabular}{lccccccccccccr}
  G  &$P_{50}(\Delta g)$  & P$_{16}$ & P$_{84}$ & $P_{50}(\Delta r)$ & P$_{16}$ & P$_{84}$ & $P_{50}(\Delta i)$ & P$_{16}$ & P$_{84}$ & $P_{50}(\Delta z$) & P$_{16}$ & P$_{84}$ & n$_{\star}$\\
 mag &    mmag &  mmag &  mmag &  mmag &    mmag &  mmag & mmag &    mmag &  mmag & mmag &    mmag &  mmag & \\
\hline
14.0  &  2.7 &  -5.0 &  12.6 &   2.7 &  -3.6 &  7.1 &   -2.4 &  -9.2 &  4.7 &        3.2 & -12.0 &  14.3 &   198 \\
14.4  &  2.2 &  -5.4 &  10.6 &   3.3 &  -2.9 &  8.8 &   -0.8 &  -8.2 &  6.0 &        5.0 &  -7.8 &  14.9 &  2208 \\
14.8  &  1.6 &  -6.3 &  10.3 &   3.0 &  -3.5 &  9.3 &    0.5 &  -6.6 &  7.2 &        5.5 &  -5.6 &  15.9 &  4827 \\
15.2  &  1.1 &  -7.0 &  10.0 &   2.8 &  -3.6 &  9.0 &    1.5 &  -5.6 &  7.9 &        5.5 &  -5.2 &  15.2 &  7550 \\
15.6  &  0.3 &  -8.2 &   9.5 &   2.1 &  -4.7 &  8.6 &    0.9 &  -6.1 &  7.8 &        4.0 &  -6.6 &  13.4 & 10406 \\
16.0  & -1.0 & -10.3 &   8.3 &   1.6 &  -5.6 &  8.4 &    0.3 &  -7.0 &  7.1 &        1.9 &  -8.5 &  11.7 & 11724 \\
16.4  & -2.6 & -12.7 &   8.1 &   0.7 &  -7.1 &  8.2 &   -0.2 &  -7.8 &  7.1 &        0.4 & -10.7 &  10.7 & 14678 \\
16.8  & -3.9 & -15.5 &   8.2 &  -0.5 &  -9.1 &  7.9 &   -1.2 &  -9.3 &  6.5 &       -1.0 & -12.6 &  10.0 & 18968 \\
17.2  & -4.7 & -18.8 &  10.0 &  -1.1 & -11.2 &  8.5 &   -1.8 & -10.8 &  6.9 &       -1.8 & -14.7 &  10.8 & 22809 \\
17.6  & -5.1 & -22.1 &  13.4 &  -1.6 & -13.3 &  9.4 &   -2.1 & -12.0 &  7.7 &       -2.2 & -16.7 &  12.1 & 16586 \\
\hline
    \end{tabular}
}
\end{table*}

%%%%%%%%%%%%%%%%%%%%%%%%%%%%%%%%%%%%%%%%%%%%%%%%%%%%%%%%%%%%%%%%

$P_{50}$, $P_{16}$, and $P_{84}$ values from  Fig.~\ref{fig:sdss_deltaindr3} are listed as a function of $G$ magnitude in Table~\ref{tab:SDSS_median}.
The median $\Delta {\rm mag}$ are within $\pm 6.0$~mmag for all the considered passbands and for the entire magnitude range sampled, and are typically lower than $\pm 3.0$~mmag in wide ranges of magnitudes, especially in $r$ and $i$ bands.
The typical scatter, as parametrised by $\sigma=0.5(P_{84}-P_{16})$, amounts to $\la 10$~mmag down to $G=16.5$~mag, for $riz$.

The adopted reference sample is dominated by dwarf stars, and almost completely lacks giants redder than $G_{BP}-G_{RP}=1.5$. In Appendix~\ref{sec:app_stand} we test our standardised XPSP in the SDSS system against a selected sample of giant stars reaching $G_{BP}-G_{RP}=3.5$, showing that it is accurate within $\simeq \pm 10$~mmag for these stars as well.

\begin{figure}[!ht]
    \centering
    \includegraphics[width=(\columnwidth)]{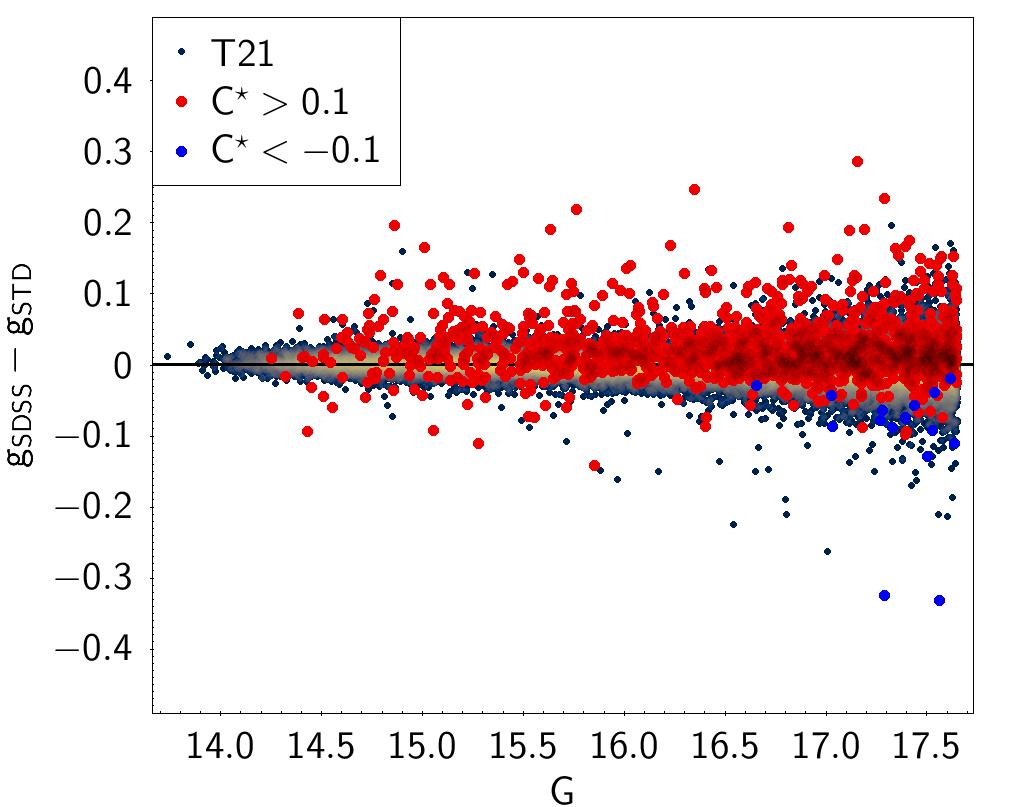}
    \caption{Tracing outliers in the $\Delta g$ vs. $G$ plot for the subset of the T21 reference sample having XP spectra in \gaia DR3 and $G<17.65$. Stars with a relatively large absolute value of $C^{\star}$ are highlighted in red, for $C^{\star}>0.1$, and in blue, for $C^{\star}<-0.1$).}
    \label{fig:dgcstar}
\end{figure}

We verified that the bulk of the $\Delta {\rm mag}$ distributions are very similar to Gaussian curves. However, a few outliers, for example those with $\Delta {\rm mag}> 50$~mmag at any $G$, can be noted in all the panels of Fig.~\ref{fig:sdss_deltaindr3}. We explored whether or not some quality parameter included in the \gaia Archive correlates with these outliers. Our far-from-exhaustive exploration led to the conclusion illustrated in Fig.~\ref{fig:dgcstar}, shown as an example. Many of the outliers have $|C^{\star}|>0.1$ \citep{riello2021}. In this sample, there was no source  with {\tt phot\_variable\_flag =VARIABLE} from the 
{\tt gaiaedr3.gaia\_source} table, but in other cases we verified that sources classified as VARIABLE according to this flag account for several outliers in $\Delta {\rm mag}$ (see e.g. Fig.~\ref{fig:ps1_cstar_app}).

This was found to be the case for all the passbands in all the photometric systems we tested in this way. We therefore conclude that the main reasons for anomalous individual inaccuracies in XPSP magnitudes are (a) contamination from nearby sources or, in any case, issues traced by
$C^{\star}$,
and (b) mean spectra obtained by combining epoch spectra of a variable source. It is worth noting that the majority of high-$C^{\star}$ outliers lie  on the same side of the $\Delta {\rm mag}$ distribution, either preferentially positive or negative, as in
Fig.~\ref{fig:dgcstar}, depending on the considered passband (see Appendix~\ref{sec:app_stand}, for an example).

%%%%%%%%%%%%%%%%%%%%%%%%%%%%
\subsection{Johnson-Kron-Cousins system and its standardisation}
\label{sec:standa_jkc}
Of the several photometric systems proposed since the advent of photoelectric and CCD (charge-coupled device) photometry, the Johnson-Kron-Cousins system (JKC hereafter) was ---and still is--- one of the most widely adopted. It was built starting from the Johnson $UBV$ \citep{johnson53, johnson63}, Kron RI \citep{kron53}, and Cousins VRI \citep{cousins73,cousins83,cousins84} 
systems. In 1992, Arlo U. Landolt published a catalogue of equatorial standard stars, which from then on became the fundamental defining set for the $UBVRI$ JKC system, and has  been used over the last three decades to calibrate the vast majority of all imaging observations in the $UBVRI$ passbands. The original 1992 photoelectric set was later extended with observations far from the celestial equator and also with a large amount of CCD observations \citep[hereafter Landolt collection,][]{Landolt1992,landolt07a,landolt07b,landolt09,landolt13,clem13,clem16}. 
Moreover, from 1983, P. B. Stetson collected observations for approximately 10$^5$ secondary $UBVRI$ standards using about half a million proprietary and archival CCD images (hereafter Stetson collection) of various fields of astrophysical interest, including star clusters, supernova remnants, and dwarfs galaxies. We used the Landolt and Stetson collections to respectively standardise and validate (see Appendix~\ref{sec:app_stand}) the UBVRI synthetic photometry obtained from {\gaia} XP spectra presented here. The Landolt and Stetson collections are described in detail by \citet{panci22}; here we briefly describe the quality selections that were applied for the purpose of the present work. 

First, we used {\gaia} and other literature catalogues to clean the collections from variables stars, binaries, blends, and stars with lower photometric quality. For the variables, we made use of the the {\gdr2} catalogue \citep{gdr2_var}, the ASAS-SN catalogue of variable stars \citep{shappee14,jaya18,jaya19a,jaya19b}, and the Zwicky Transient Facility catalogue of periodic variable stars \citep{chen20}. For binaries, we profited from the work done by the Survey of Surveys team \citep{tsantaki21}, who compiled all known spectroscopic binaries in large spectroscopic surveys and astroseismology missions \citep{price20,kounkel21,traven20,merle17,birko19,qian19,tian20,deleuil18,kirk16}. In addition, we used the following cuts on parameters from the main {\tt gaia\_source} table to further remove possible contaminated and blended sources: {\tt IPD\_Frac\_Odd\_Win} and {\tt IPD\_Frac\_Multi\_Peak} above 7\%,
Renormalised Unit Weight Error ({\tt RUWE}) above 1.4, and the recommended cut by \citet{riello2021} on the renormalised {\xp} flux excess, $|\rm{C^*}|>2\,\sigma_{\rm{C^*}}$, as well as a cut on the $\beta$\footnote{Defined as ({\tt phot\_bp\_n\_blended\_transits} + {\tt phot\_rp\_n\_blended\_transits}) / ({\tt phot\_bp\_n\_obs} + {\tt phot\_rp\_n\_obs})} parameter by \citet{riello2021} above 20\%.

The original Landolt and Stetson collections agree very well with each other, with zero-point offsets of below 1\% in all bands, and spreads of $\pm$1--2\%. However, some disagreement (3\%--5\%) was found for the reddest stars, which are less represented in the original \citet{Landolt1992} set, with only half a dozen stars \citep{panci22}. This is particularly evident for the $R$ and $I$ bands. For this reason, we decided to use only the Landolt collection for the standardisation and the Stetson one for the validation.  
This uncertainty for redder stars makes both the Landolt and Stetson collections less reliable for stars redder than $\bprp\simeq 2$~mag, although both collections are rigorously calibrated on the original \citet{Landolt1992} set. We used the Landolt sample to standardise the synthetic photometry in the $UBVRI$ system, which we obtained using the passbands by \citet{BM12}, as described in Section~\ref{sec:stand}. 

\begin{figure*}[!htbp]
    \centerline{
    \includegraphics[width=(\columnwidth)]{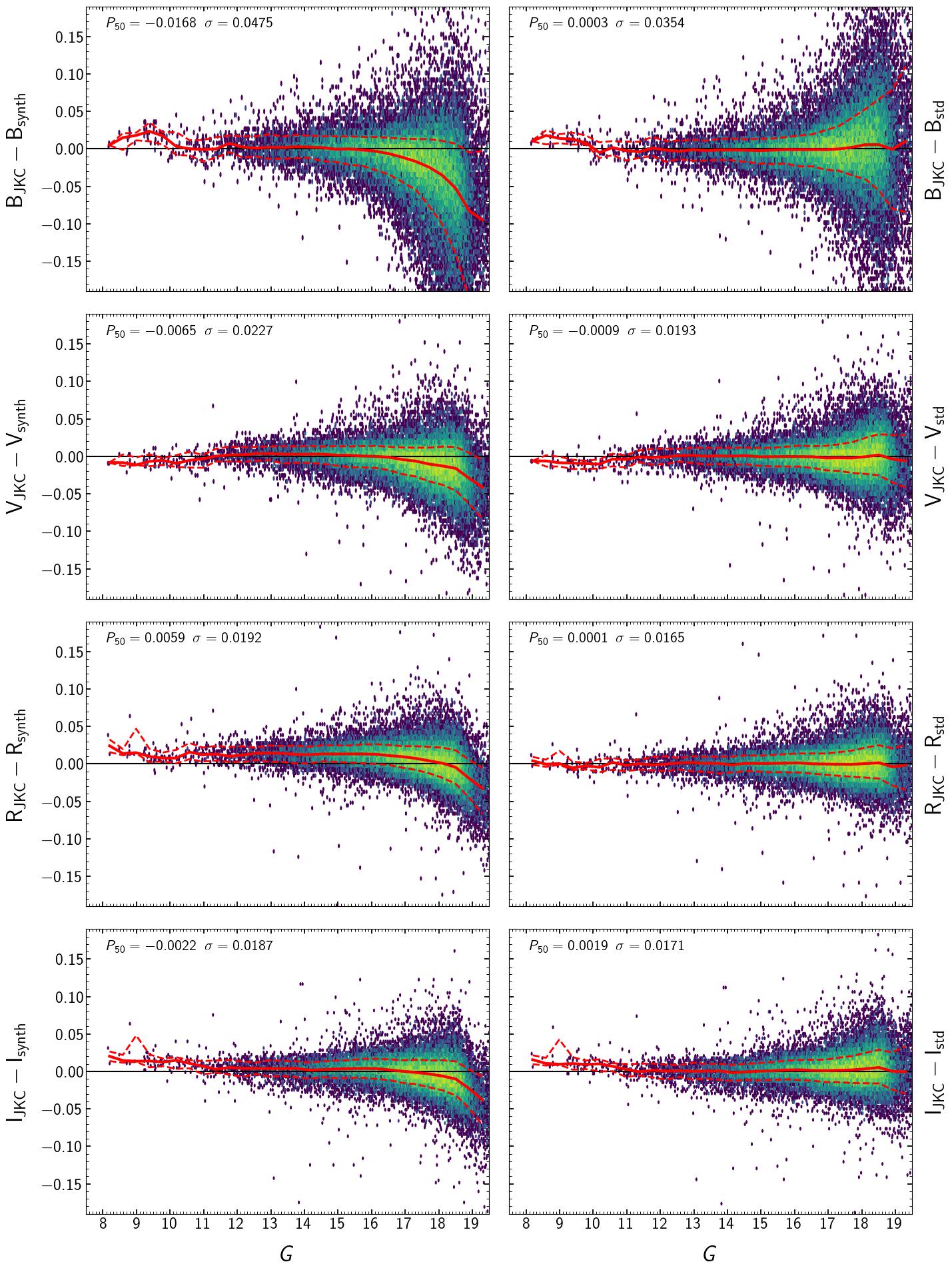}
    \includegraphics[width=(\columnwidth)]{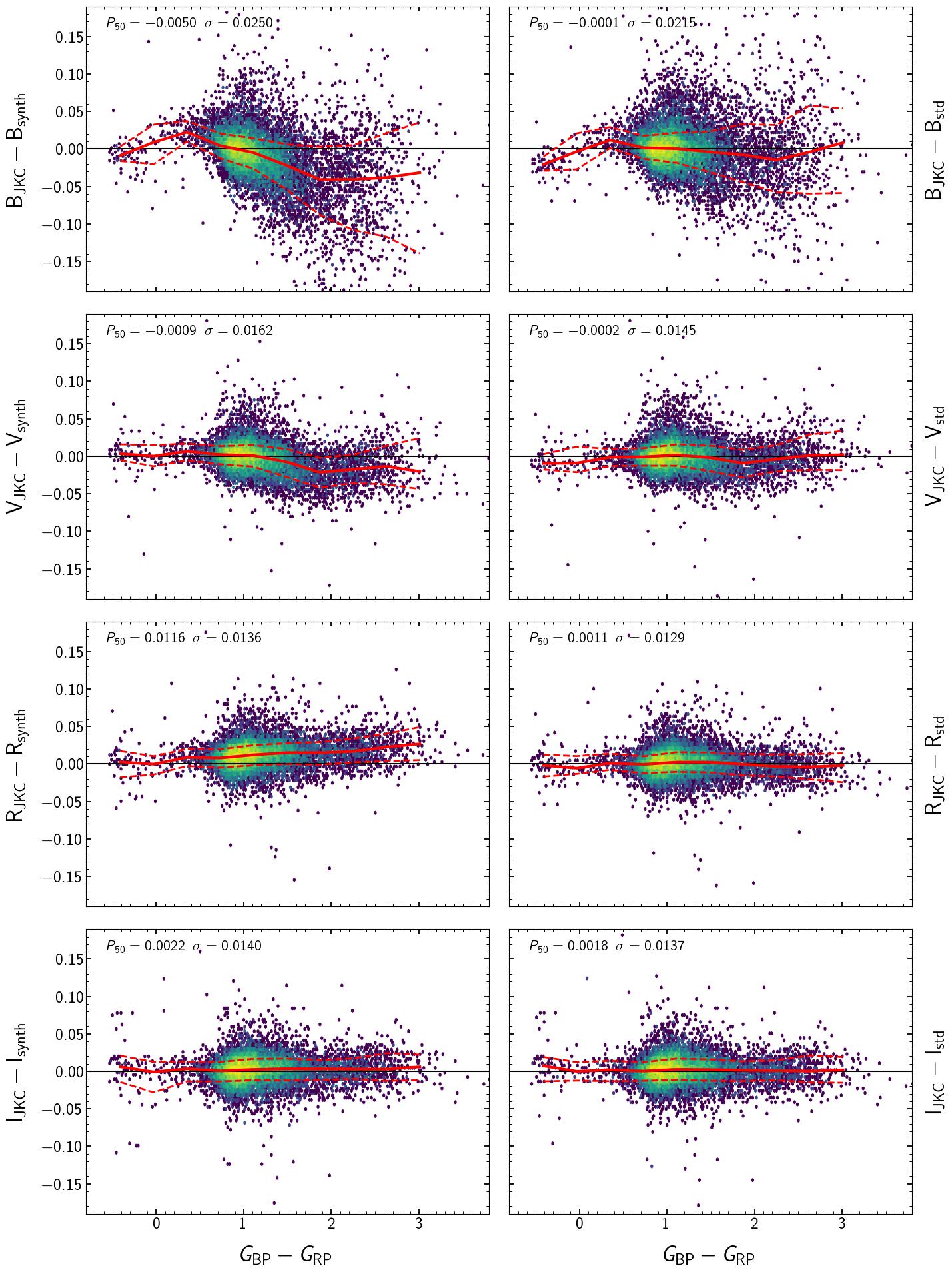}
    }
    \caption{Performance and standardisation of JKC $BVRI$ XP synthetic magnitudes using the reference set of standard stars described in the text. The arrangement and symbols are the same as in Fig.~\ref{fig:sdss_delta_app}.}
    \label{fig:jkc_delta_app}
\end{figure*}

Figure~\ref{fig:jkc_delta_app}, which is fully analogous to Fig.~\ref{fig:sdss_delta_app}, shows the comparison between XPSP and reference $BVRI$ magnitudes before and after standardisation. The overall ZP, for stars not seriously affected by the hockey-stick effect, are reproduced by raw XPSP to better than $\simeq 0.02$ mag in all the considered bands. A significant colour trend is observed for B that is very similar to the case of SDSS $g$. This analogy is not surprising because the two filters sample a similar range of the BP spectrum. Similarly, the performance of raw $V$ is significantly poorer than raw r, with larger scatter and colour terms, which is likely due to $V$ sampling more problematic regions of BP than $r$ ($\simeq 490-660$~nm versus $\simeq 540-699$~nm, respectively). We address the reader to \citet{EDR3-DPACP-120} for an additional discussion on this specific system. Here we note that standardisation significantly mitigates the amplitude of the residual systematic errors displayed by raw XPSP.

\begin{figure}[!htbp]
    \centering
    \includegraphics[width=(\columnwidth)]{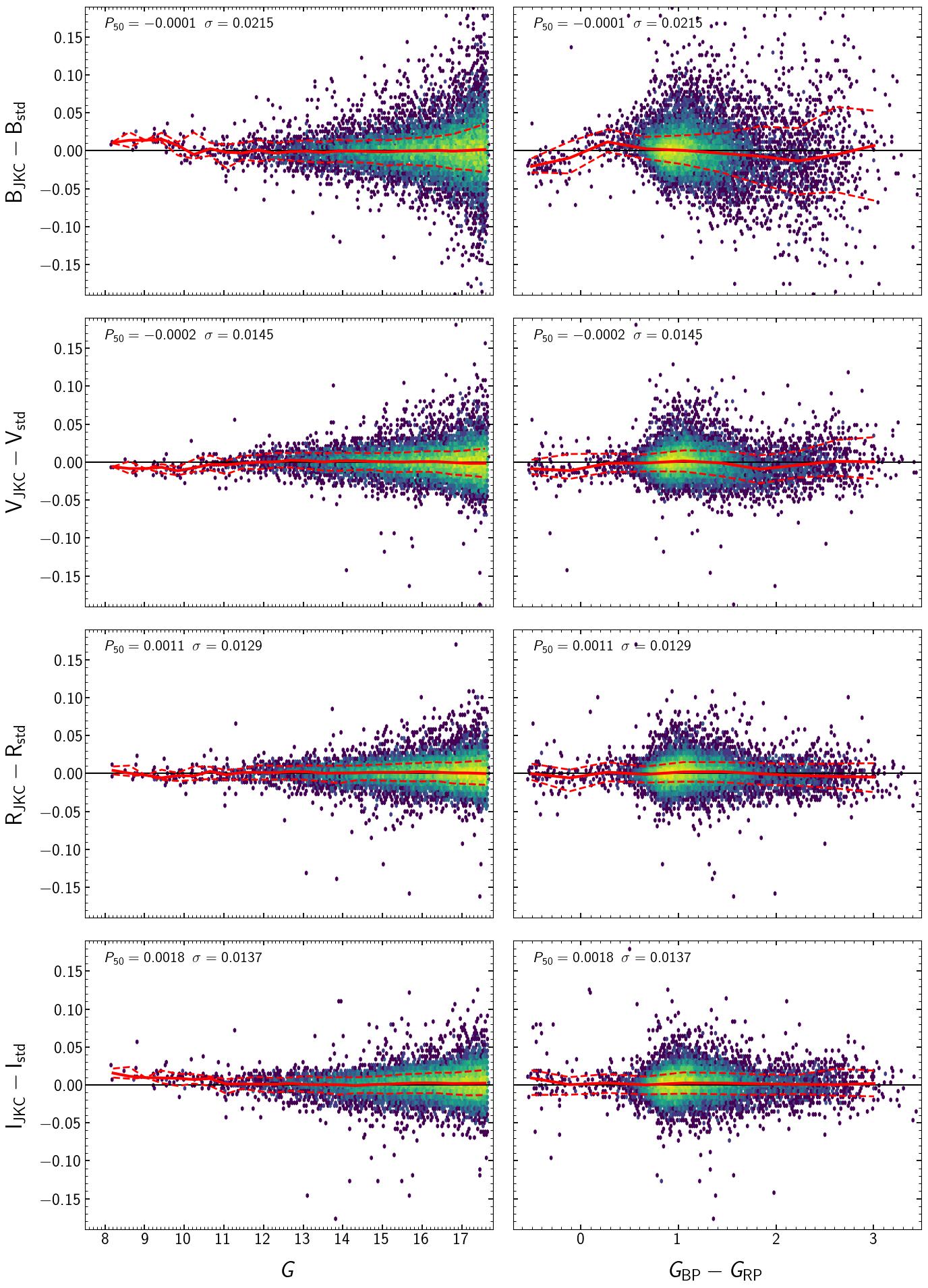}
    \caption{Performances of standardised XPSP in the JKC system ($BVRI$). We show $\Delta$mag as a function of $G$ magnitude (left panels) and {\bprp} colour (right panels) for the subsample of reference stars whose XP spectra has been released in {\gdr3}. The arrangement and the meaning of the symbols is the same as in
    \figref{sdss_delta_app}.}
    \label{fig:jkc_deltaindr3}
\end{figure}

The comparison between standardised XPSP magnitudes and those from the reference sample for $G<17.65$~mag stars is presented in Fig.~\ref{fig:jkc_deltaindr3}, as a function of magnitude and colour. Performances are very similar to the SDSS case described above. The remarkable differences are: (a) the loss of millimag accuracy for $G\la 11.5$~mag in correspondence with a transition to different setups of the BP and RP spectrometers which affects the internal calibration of
XP spectra in this bright magnitude range \citep[onset of {\em gates}, change of window class, etc.; see e.g.][]{EDR3-DPACP-118,EDR3-DPACP-120}, which is not sampled by SDSS stars\footnote{Because the saturation limit of SDSS occurs around $G=14.0$~mag.}; and (b) the residual colour terms of order $\simeq 10$~mmag remaining in some colour range for the $B$ and ---to a lesser extent--- $V$ passbands. The median, and the $P_{16}$ and $P_{84}$ percentiles of the $\Delta {\rm mag}$ distributions for JKC $BVRI$ magnitudes shown in Fig.~\ref{fig:jkc_deltaindr3} are listed in Table~\ref{tab:JKC_median}. The scatter about the median for $G<16.5$~mag is $\la 15$~mmag in $VRI$, and $\la 20$~mmag in $B$. 

Similarly to the case of the T21 sample, red giants are also relatively rare in the Landolt reference sample used here, with just a handful in the range $1.5<G_{BP}-G_{RP}<3.5$. We carefully verified that these red giants match the same locus of the bulk of the other stars in the sample in the $G_{BP}-G_{RP}$ versus $\Delta {\rm mag}$ diagrams, within $<10.0$~mmag.

%%%%%%%%%%%%%%%%%%%%%%%%%%%%%%%%%%%%%%%%%%%%%%%%%%%%%%%%%%%%
\begin{table*}[!htbp]
    \centering
        \caption{\label{tab:JKC_median} JKC system: median ($P_{50)}$) and  15.87\% ($P_{16}$) and 84.13\% ($P_{84}$) percentiles 
of the $\Delta {\rm mag}$ distributions of Fig.~\ref{fig:jkc_deltaindr3}. Here, n$_{\star}$ is the number of sources in the considered bin.}
{\small
    \begin{tabular}{lccccccccccccr}
  G  &$P_{50}(\Delta B)$  & P$_{16}$ & P$_{84}$ & $P_{50}(\Delta V)$ & P$_{16}$ & P$_{84}$ & $P_{50}(\Delta R)$ & P$_{16}$ & P$_{84}$ & $P_{50}(\Delta I$) & P$_{16}$ & P$_{84}$ & n$_{\star}$\\
 mag &    mmag &  mmag &  mmag &  mmag &    mmag &  mmag & mmag &    mmag &  mmag & mmag &    mmag &  mmag & \\
\hline
 9.4  & 15.1 &  11.9 &  23.6 &  -6.9 &  -14.4 & -2.7   & -5.7 &  -8.3 &  4.9   &   9.2 &    2.4 &  18.9 &   17 \\
 9.8  &  7.0 &  -6.1 &  11.3 & -11.6 &  -17.3 & -6.3   & -2.8 &  -9.8 &  0.9   &   8.7 &   -1.4 &  14.6 &   15 \\
10.2  & -5.0 & -10.6 &  23.9 &  -7.6 &  -13.1 &  3.4   & -3.2 &  -8.7 &  9.2   &   7.1 &    1.7 &  14.2 &   11 \\
10.7  &  2.5 &  -6.5 &  10.2 &  -3.1 &  -10.7 &  8.7   &  2.9 &  -7.2 &  8.8   &   8.4 &   -1.2 &  11.0 &   20 \\
11.1  & -2.0 & -23.1 &  8.1  &  -3.7 &  -15.8 &  1.9   & -1.8 &  -8.2 &  4.6   &   1.1 &  -11.5 &  7.7  &   31 \\
11.5  & -3.1 & -11.8 &  10.5 &  -1.5 &   -8.5 &  8.1   &  1.4 &  -8.0 &  6.6   &   1.2 &   -5.1 &  9.2   &   42 \\
11.9  &  0.8 & -14.2 &  14.0 &  -0.2 &   -9.7 &  9.6   &  1.5 &  -8.2 &  10.8  &   1.0 &   -8.4 &  10.4 &   70 \\
12.3  & -3.3 & -11.2 &  10.0 &   0.3 &   -7.3 &  12.0  &  1.1 &  -8.1 &  10.1  &   0.8 &  -10.1 &  9.1  &  105 \\
12.7  & -1.2 & -13.1 &  12.2 &   2.4 &   -7.1 &  11.9  &  2.5 &  -7.7 &  9.6   &   1.8 &   -7.7 &  9.7   &  148 \\
13.1  & -0.7 & -14.5 &  13.6 &   1.9 &   -9.0 &  12.4  &  2.2 &  -8.6 &  10.6  &   0.4 &   -9.5 &  11.6 &  210 \\
13.5  & -2.3 & -15.9 &  11.2 &   0.2 &  -11.0 &  11.6  &  0.4 & -10.4 &  10.2  &   0.3 &  -10.4 &  9.5   &  282 \\
13.9  & -0.4 & -14.6 &  14.5 &   1.7 &  -11.4 &  12.1  &  0.7 & -12.9 &  10.3  &   0.3 &  -12.8 &  10.1 &  381 \\
14.3  & -0.6 & -15.6 &  12.4 &   1.7 &  -10.3 &  12.0  &  0.8 &  -9.0 &  10.3  &  -1.0 &  -10.9 &  10.0 &  555 \\
14.7  & -1.1 & -16.2 &  14.1 &   0.8 &  -10.3 &  11.7  &  1.5 &  -8.8 &  11.4  &   0.4 &  -10.6 &  10.9 &  706 \\
15.1  & -1.5 & -19.5 &  13.6 &   0.3 &  -13.1 &  12.1  &  1.4 &  -9.6 &  12.1  &   0.9 &  -11.8 &  12.2 &  875 \\
15.5  & -0.5 & -17.7 &  16.0 &   0.1 &  -13.1 &  12.7  &  1.4 & -10.1 &  13.0  &   1.9 &  -10.8 &  14.1 & 1149 \\
16.0  & -0.3 & -19.6 &  17.1 &   0.6 &  -12.8 &  13.9  &  2.3 & -10.5 &  13.7  &   2.5 &  -11.0 &  14.8 & 1395 \\
16.4  &  0.5 & -20.1 &  18.9 &   0.5 &  -13.3 &  14.5  &  1.6 & -10.0 &  14.9  &   2.4 &  -10.8 &  16.1 & 1692 \\
16.8  & -0.6 & -24.0 &  21.7 &  -0.6 &  -16.4 &  13.8  &  0.8 & -13.0 &  14.1  &   1.8 &  -12.5 &  16.0 & 2270 \\
17.2  &  0.3 & -25.4 &  28.1 &  -1.6 &  -17.7 &  15.8  &  0.7 & -13.8 &  15.7  &   2.0 &  -13.7 &  17.7 & 2973 \\
17.6  &  1.5 & -28.8 &  35.6 &  -1.6 &  -20.0 &  17.6  & -0.0 & -14.9 &  16.7  &   2.3 &  -13.4 &  19.6 & 2295 \\
\hline
    \end{tabular}
}
\end{table*}

%%%%%%%%%%%%%%%%%%%%%%%%%%%%
%%%%%%%%%%%%%%%%%%%%%%%%%%%%%%%%%%%%%%%%%%%%%%%%%%%%%%%%%%%%

%%%%%%%%%%%%%%%%%%%%%%%%%%%%
\subsection{Standardised ultraviolet bands}
\label{sec:standa_u}
\begin{figure}[!htbp]
    \centering
     \includegraphics[width=(\columnwidth)]{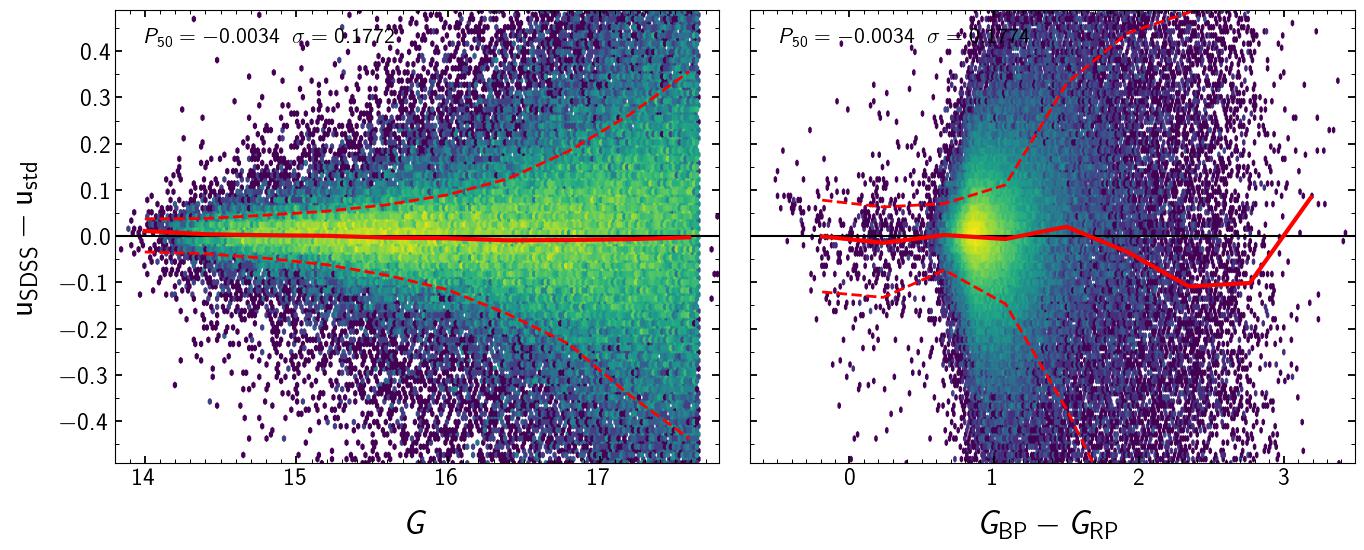}     
     \includegraphics[width=(\columnwidth)]{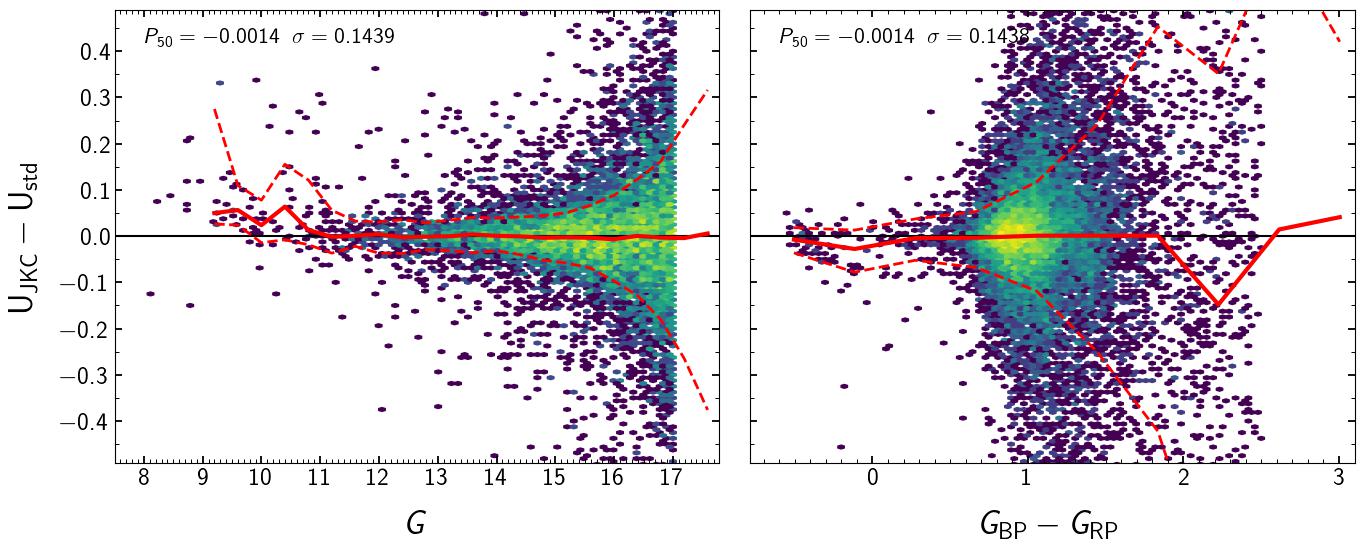}
    \caption{Upper panel: Performance of standardised XPSP in the SDSS $u$ band. The reference sample is the same as in Fig.~\ref{fig:sdss_deltaindr3}.
    Lower panel: Performance of standardised XPSP in the JKC $U$ band. The reference sample is the same as in Fig.~\ref{fig:jkc_deltaindr3}.
    Please note that the bright limit of the two reference samples is very different. }
    \label{fig:deltau}
\end{figure}

Figure~\ref{fig:deltau} shows the performances of the JKC $U$ band and SDSS $u$ band magnitudes, which are standardised as described in Sect.~\ref{sec:ustand}. These are shown in comparison with the respective reference samples (see Sect.~\ref{sec:standa_sdss} and Sect.~\ref{sec:standa_jkc}, respectively) for the subset of sources that will have XP spectra released in {\gdr3} (see Appendix~\ref{sec:app_stand} for comparison of raw and standardised magnitudes).
The median $\Delta {\rm mag}$ is a few millimag for $15.2\le G\le 17.6$~mag in $u_{\rm SDSS}$ and for $11.5\le G\le 17.6$~mag in $U_{\rm JKC}$, with the issue related to sources brighter than $G=11.5$~mag discussed above decreasing the accuracy in this range (see Table~\ref{tab:sdss_u_median} and Table~\ref{tab:jkc_u_median}). However, the scatter is significantly larger than in all redder wide-band magnitudes considered here, reaching $0.12$--$0.15$ mag at $G=16.4$~mag and $>0.3$ mag at $G=17.6$~mag. In particular, the accuracy is generally poor for red sources, missing sufficient signal in the wavelength range covered by UV bands for reliable magnitudes to be provided. 

\begin{figure}[!htbp]
    \centering
     \includegraphics[width=(\columnwidth)]{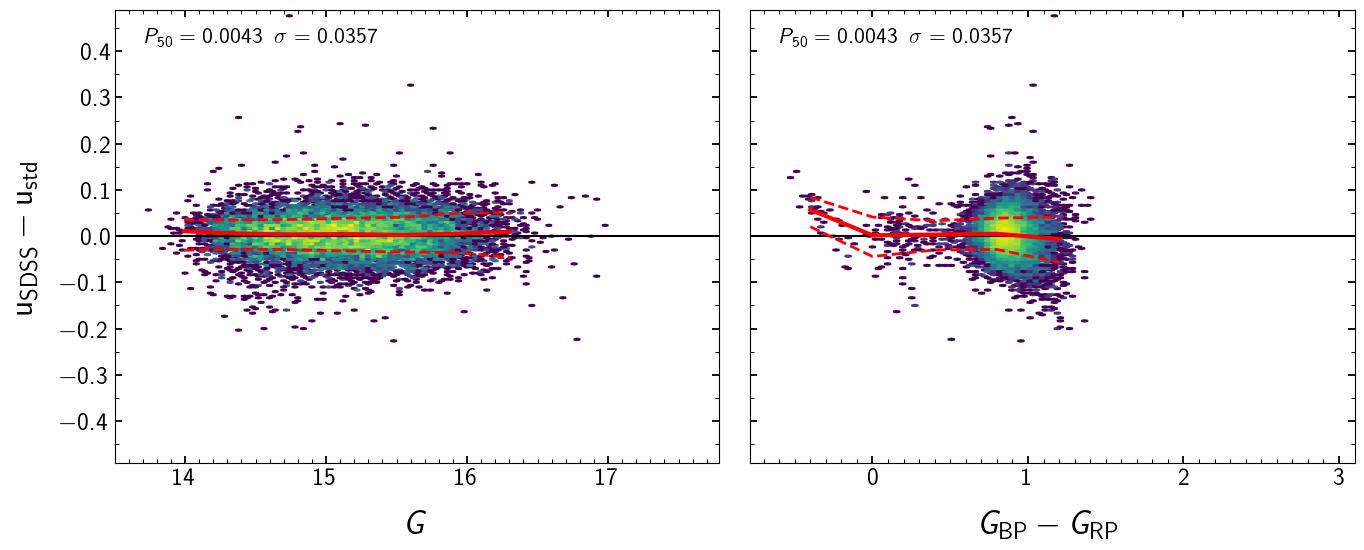}     
     \includegraphics[width=(\columnwidth)]{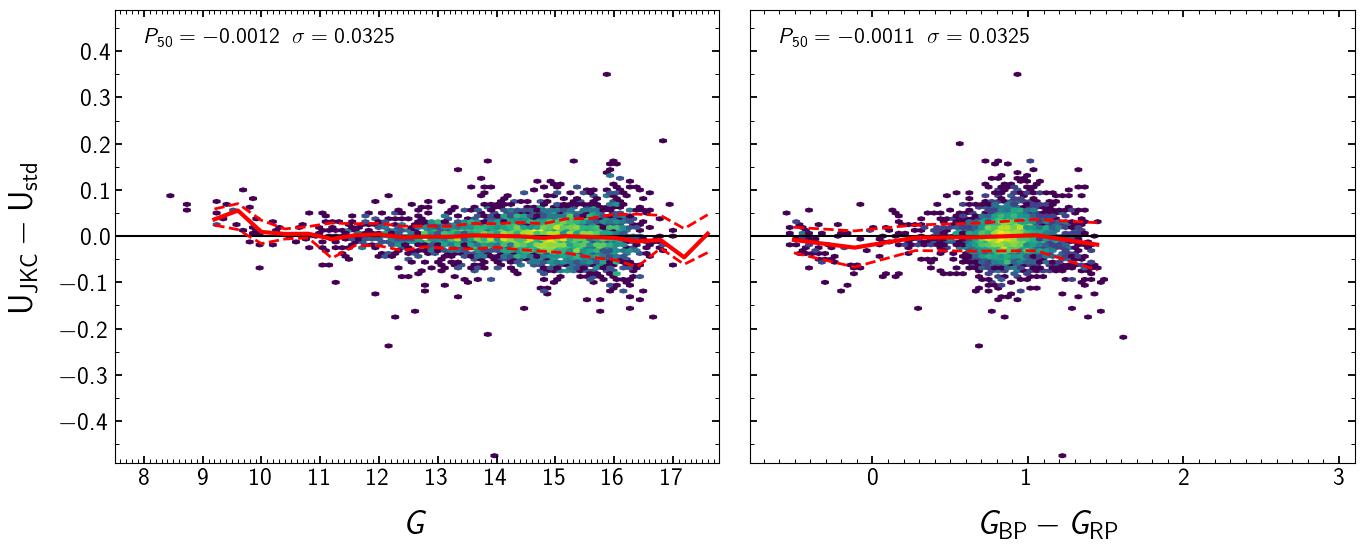}
    \caption{Same as Fig.~\ref{fig:deltau} but limited to stars with 
    $flux/flux_error>30$ in $u_{\rm SDSS}$ and $U_{\rm JKC}$, respectively. This is the S/N limit on individual magnitudes that we adopt for the GSPC (Sect.~\ref{sec:gspc}).}
    \label{fig:deltau30}
\end{figure}

Figure~\ref{fig:deltau30}  shows the effect of adopting the selection in S/N,  {\tt flux\_x/flux\_x\_error$>30$}, for $x=u_{\rm SDSS}$ and $U_{\rm JKC}$, respectively, on the $\Delta {\rm mag}$ distribution of Fig.~\ref{fig:deltau}. This selection greatly reduces the scatter about the median, thus providing much more reliable individual magnitudes, but it implies a strong selection in magnitude and in colour in these samples, in practice removing all stars with ${\bprp}\ga 1.3$~mag and $G\ga 16.5$. In Sect.~\ref{sec:gspc} we show that when applied to larger samples, reliable UV magnitudes can be obtained for stars as red as ${\bprp}\simeq 3.0$~mag, depending on their apparent magnitude, still maintaining a strong bias against red and faint stars.

In summary, as anticipated in Sect.~\ref{sec:ustand}, the performances for any band covering the XP range $\le 400$~nm are significantly poorer than in all the redder passbands. We do not discuss similar bands from other systems any further, as the results would be very similar to those shown in Fig.~\ref{fig:sdss_u_delta_app} and Fig.~\ref{fig:JKC_U_delta_app}  for example, namely. strong colour-dependent trends with respect to reference external photometry. Given the high astrophysical relevance of these UV bands and the lack of all-sky sources for them, we made a concerted effort to provide  standardised $u_{\rm SDSS}$ and $U_{\rm JKC}$ magnitudes and managed to obtain reasonably accurate and precise photometry for the subset of stars with sufficient signal in that region of the spectrum due to favourable combinations of magnitude and colour. While XPSP magnitudes in bands at $\le 400$~nm can be obtained for all the sources with XP spectra released in {\gdr3}, and sometimes a highly uncertain measurement can be better than no measure at all, we strongly recommend using these magnitudes only if {\tt flux\_x/flux\_x\_error$>30$}, and, in particular, using preferentially the standardised $u_{\rm SDSS}$ and $U_{\rm JKC}$ provided in the Gaia Synthetic Photometry Catalogue (GSPC; Sect.~\ref{sec:gspc}). In any case, even standardised UV XPSP must be used with caution (Sect.~\ref{sec:recommend} for further caveats). 

%############################################################
\begin{table}[!htbp]
    \centering
        \caption{\label{tab:sdss_u_median} SDSS system: median ($P_{50}$) and 15.87\% ($P_{16}$), and 84.13\% ($P_{84}$) percentiles 
of the $\Delta u$ distribution of Fig.~\ref{fig:deltau}. Here, n$_{\star}$ is the number of sources in the considered bin.}
{\small
    \begin{tabular}{lcccr}
  G  &$P_{50}(\Delta u)$  & P$_{16}$ & P$_{84}$ &n$_{star}$ \\ 
 mag  &   mmag &   mmag &  mmag  & \\
\hline
14.0  & 11.3 &  -34.3 &   37.2 &    198 \\
14.4  &  4.2 &  -38.2 &   38.1 &   2208 \\
14.8  &  2.0 &  -47.8 &   45.5 &   4827 \\
15.2  &  0.7 &  -61.7 &   53.8 &   7550 \\
15.6  & -3.3 &  -84.6 &   67.9 &  10406 \\
16.0  & -4.4 & -115.6 &   89.2 &  11724 \\
16.4  & -9.0 & -168.4 &  123.9 &  14678 \\
16.8  & -8.3 & -232.2 &  183.4 &  18968 \\
17.2  & -6.6 & -342.5 &  261.3 &  22809 \\
17.6  & -2.6 & -437.8 &  355.9 &  16586 \\
\hline
    \end{tabular}
}
\end{table}
%############################################################

%############################################################
\begin{table}[!htbp]
    \centering
        \caption{\label{tab:jkc_u_median} JKC system: median and 15.87\% ($P_{16}$) and 84.13\% ($P_{84}$) percentiles 
of the $\Delta {\rm U}$ distributions of Fig.~\ref{fig:deltau}. Here, n$_{\star}$ is the number of sources in the considered bin.}
{\small
    \begin{tabular}{lcccr}
  G  &$P_{50}(\Delta U)$  & P$_{16}$ & P$_{84}$ & n$_{\star}$\\ 
 mag  &   mmag &   mmag &  mmag  & \\
\hline
10.0  &  23.7 &  -14.1 &   77.8 &    14 \\
10.4  &  63.9 &   -7.9 &  155.5 &    13 \\
10.8  &  13.2 &  -18.5 &  121.3 &    22 \\
11.2  &  -1.6 &  -36.8 &   55.9 &    35 \\
11.6  &   1.0 &  -22.1 &   32.5 &    51 \\
12.0  &   4.0 &  -36.9 &   30.5 &    73 \\
12.4  &  -2.4 &  -37.6 &   35.4 &   109 \\
12.8  &  -1.2 &  -29.2 &   28.9 &   158 \\
13.2  &  -0.2 &  -31.4 &   34.3 &   236 \\
13.6  &   3.7 &  -35.8 &   40.4 &   287 \\
14.0  &   0.7 &  -31.5 &   39.4 &   402 \\
14.4  &  -0.6 &  -40.0 &   41.5 &   566 \\
14.8  &  -3.5 &  -51.9 &   44.2 &   719 \\
15.2  &  -2.7 &  -58.2 &   50.6 &   880 \\
15.6  &  -3.5 &  -69.0 &   66.6 &  1192 \\
16.0  &  -6.7 &  -96.8 &   88.0 &  1357 \\
16.4  &   0.5 & -128.0 &  123.0 &  1689 \\
16.8  &  -3.2 & -180.5 &  161.3 &  2264 \\
17.2  &  -4.2 & -264.5 &  240.6 &  2923 \\
17.6  &   5.7 & -376.0 &  316.2 &  2230 \\
\hline
    \end{tabular}
}
\end{table}

%########################################################

\subsection{PanSTARRS-1 system and its standardisation}
\label{sec:standa_ps1}

PanSTARRS-1 (hereafter PS1, for brevity) is an ambitious multi-task project, the main aim of which is to survey all the sky above dec=-30$\degr$ in five passbands, $grizy$ \citep[3$\pi$ survey, see][and references therein]{Chambers2016}. The wide sky coverage and the high photometric precision reached qualify PS1 as one of the most widely used sources of stellar photometry \citep{Magnier2020,xiao22}. 

As a reference sample of standard stars, here we adopt two $15\degr\times 15\degr$ patches located at the Galactic caps ($|b|>60.0\degr$). Bona fide point sources with multi-epoch observations were selected using the difference between PSF magnitudes and Kron magnitudes as a diagnostic, following a kind suggestion by E. Magnier (private 
communication)\footnote{In particular, in our catalogue, we only kept sources with {\tt x\_chp\_psf-x\_chp\_kron$>-0.3$} and {\tt x\_chp\_psf-x\_chp\_kron $<0.1$}, and {\tt x\_chp\_psf\_nphot$\ge10$} for {\tt x=grizy}.} and requiring the uncertainty on magnitudes to be $<0.02$~mag in all passbands. Once matched with the {\it Gaia} source catalogue, with a $1\arcsec$ cone search, a reference sample of 76\ 491 stars was finally adopted.
XPSP magnitudes are compared with PS1 PSF magnitudes based on the average of the chip measurements ({\tt x\_chp\_psf}, where {\tt x=grizy}), because these have the best corrections for systematic effects.

\begin{figure}[!htbp]
    \centering
    \includegraphics[width=(\columnwidth)]{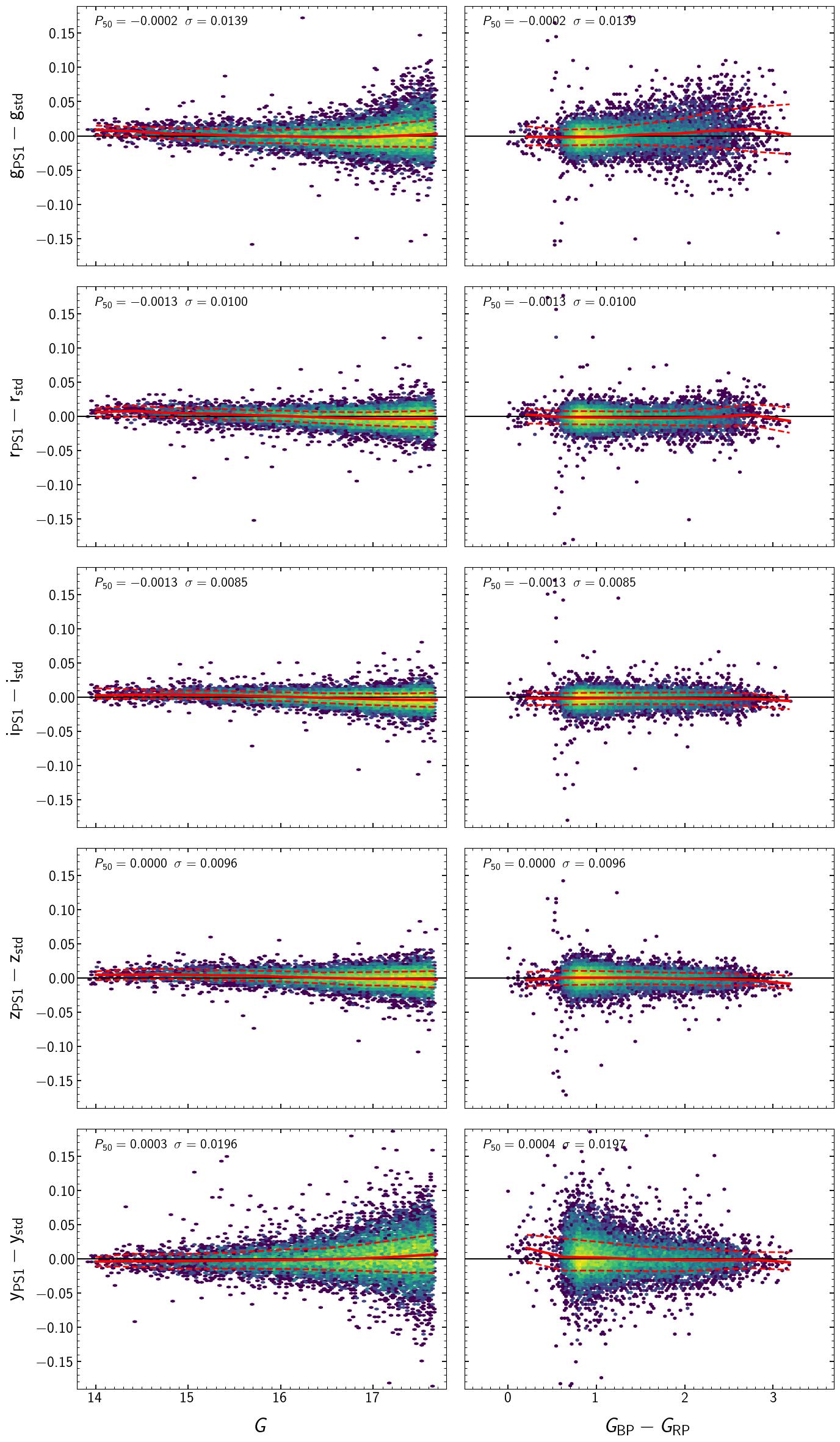}
    \caption{Performance of standardised XPSP in the PanSTARRS-1 system ($grizy$). We show $\Delta$mag as a function of $G$ magnitude (left panels) and {\bprp} colour (right panels) for the subsample of reference stars whose XP spectra has been released in {\gdr3}. The arrangement and the meaning of the symbols is the same as in Fig.~\ref{fig:sdss_delta_app}}
    \label{fig:ps1_deltaindr3}
\end{figure}

The performance before and after standardisation is similar to that obtained for SDSS magnitudes and is shown in Fig.~\ref{fig:ps1_delta_app}. 
Figure~\ref{fig:ps1_deltaindr3}  shows the performance for the subset that will be included in {\gdr3} for $G<17.65$~mag. The accuracy of standardised XPSP is good in all passbands; see also Table~\ref{tab:PS1_median}. The median $\Delta$mag amounts to a few millimag over the entire range of magnitudes considered, while the typical $\sigma$ ranges between 10 and 15 mmag for $G\le 16.5$~mag. Slightly larger deviations are observed at the extremes of the colour range spanned by the reference sample in the $g$ band (red side) and in $y$ band (blue side). It is worth noting that the reference sample adopted has a limited coverage of colour and spectral type compared to those we use for SDSS and JKC systems for example. We therefore recommend special caution in using PS1 XPSP magnitudes outside the  validated colour and magnitude ranges.

%%%%%%%%%%%%%%%%%%%%%%%%%%%%%%%%
\subsection{Standardised HST magnitudes}
\label{sec:standa_hugs}

The Hubble Space Telescope (HST) is one of the most successful space missions ever, with a long-standing and huge impact on virtually all branches of astrophysics (see e.g. Macchetto 2020, and references therein). 
Unprecedented photometric precision is one of the many excellent achievements of the optical-IR cameras on board this iconic space observatory \citep[see][for a recent example]{Bedin19}. 

HST cameras are equipped with large sets of narrow, medium, and wide filters, making for 
several very powerful and flexible photometric systems. 
XPSP for the subset of those that are enclosed within the XP spectral range may be very useful for many scientific applications. For instance, all-sky XPSP for the 220~M stars with XP spectra released in {\gdr3} will hugely extend the realm reachable by photometry in HST systems to the entire sky and in a bright range of magnitudes ($4.0\la G\le 17.65$~mag) not usually easily accessible to HST (and/or not convenient to be sampled with HST). It may be worth recalling an additional desirable feature, namely that these extensions come from space-based spectrophotometry. 

Unfortunately, this high degree of complementarity between HST and {\gaia}, makes it extremely difficult to find proper samples for validation and standardisation of XPSP in the HST systems.
Only a handful of well-measured {\gaia} sources can be found in the typical FoV of HST cameras ($\la 4$ arcmin$^2$) and HST is mainly used to measure very faint stars that would otherwise be unreachable for ground-based instruments, and therefore the typical overlap in magnitude between HST and {\gaia} is limited. Finally, even if there were samples with a significant number of stars common to both of them, HST observations would only be available in a limited number of passbands for any given camera.  

\begin{figure}[!htbp]
    \centering
    \includegraphics[width=(\columnwidth)]{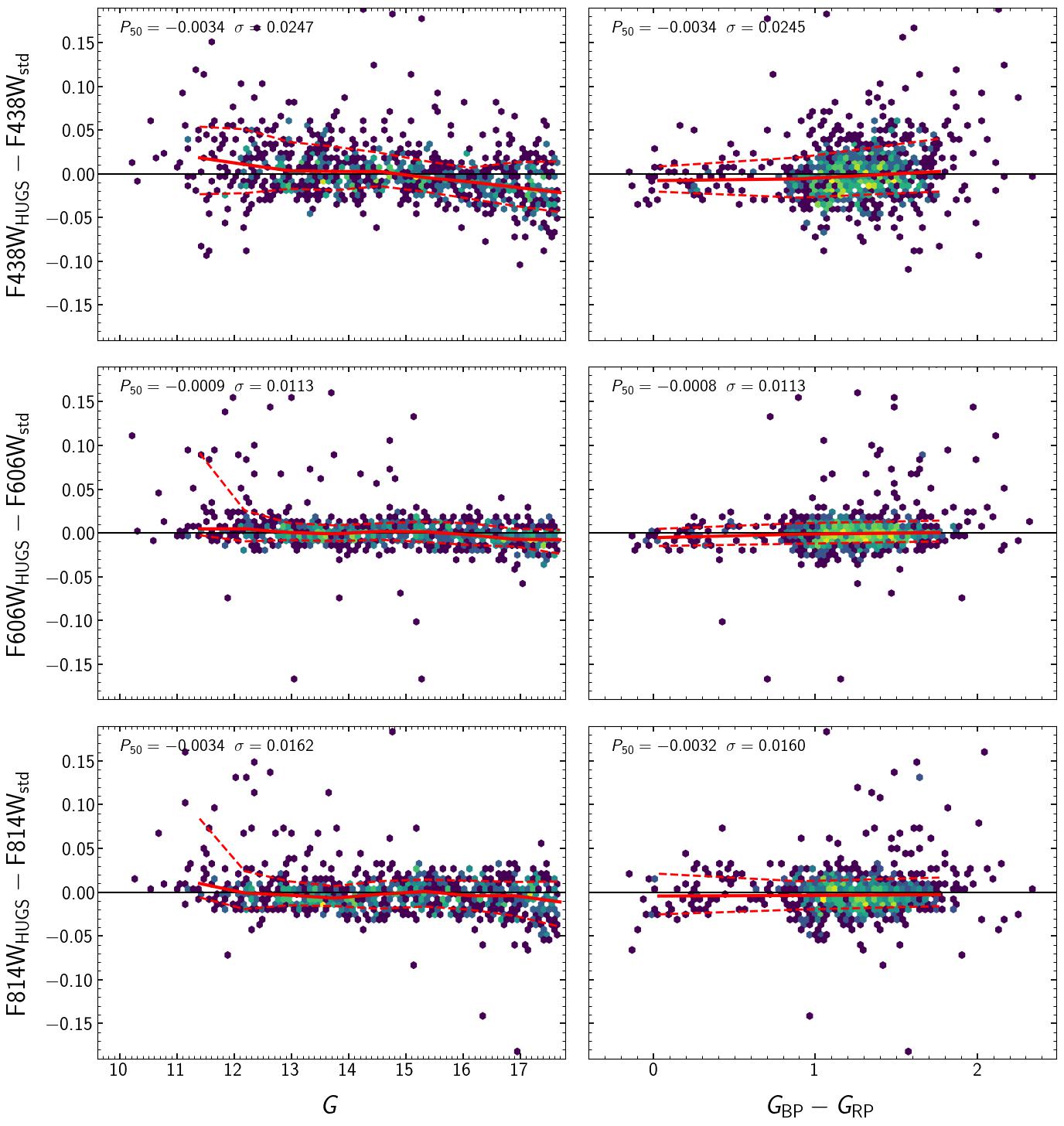}
    \caption{Performances of standardised XPSP in the HST WFC3/UVIS (F438W, VEGAMAG) and ACS/WFC systems (F606W, F814W, VEGAMAG). We show $\Delta$mag as a function of $G$ magnitude (left panels) and {\bprp} colour (right panels) for the subsample of reference stars whose XP spectra has been released in {\gdr3}. The arrangement and the meaning of the symbols is the same as in Fig.~\ref{fig:sdss_deltaindr3} but the percentiles are computed in bins of 0.8~mag in width.}
    \label{fig:hugs_deltaindr3}
\end{figure}

For these reasons, the sample we used for validation and standardisation of a few HST passbands, while absolutely excellent for the scientific application for which it was acquired, is clearly not ideal for our purpose. Still, it is adequate for testing the accuracy and precision of XPSP at the $\simeq$~1\% level over a limited range in colour and magnitude. 

We used the photometry in the WFC3/UVIS F438W and ACS/WFC F606W, F814W bands of a set of Galactic globular clusters (GCs) from the HUGS project  \citep{nardiello18} as a reference sample. These observations have the advantage of providing a large number of stars in the small FoV covered by the considered cameras, and remarkable overlap in magnitude with the {\gaia} source catalogue. Unfortunately, for obvious scientific reasons, HUGS fields target crowded areas in the central region of the clusters, where blending and contamination of the relatively wide BP and RP apertures by nearby sources and/or high background may severely affect most of the XP spectra. For this reason, we applied very strong selections on the original HUGS samples,  only keeping in the final reference sample used for standardisation stars (a) with $|RADXS|<0.1$,  $QFIT>0.9$, and photometric uncertainty on individual HUGS magnitudes $<0.1~{\rm mag}$\footnote{RADXS and QFIT are quality parameters from the HUGS catalogs, see \citet{nardiello18} for details and discussion. The original comparison with the HUGS photometry included also the F336W filter, that was later abandoned because its XPSP counterpart suffers from the strong systematics affecting U bands. However the selections of RADXS and QFIT were imposed an all the considered bands including F336W. Analogously, a star was accepted for the final sample only if it had a valid magnitude in all the four passbands.}; and (b) with a number of BP and RP epoch spectra sufficient to ensure high S/N in XP spectra, adopting the same criterion for the release of XP spectra, that is,  {\tt bp\_num\_of\_transits$>15$} and {\tt rp\_num\_of\_transits>$15$}, and $|C^{\star}|<0.05$ in order to minimise the impact of contamination of the spectra. After cross-correlation with the {\gedr} source catalogue, we end up with a sample of 1113 stars in the range $10.0\le G\le 19.0$~mag, 968 of which have $G<17.65$~mag. We used `method 1' magnitudes from the HUGS catalogue, as they are presented as the best choice for the bright magnitude range we are considering \citep{nardiello18}. Here we compare magnitudes in the VEGAMAG systems, but the comparison has general validity, as transforming into STMAG or ABMAG would imply a simple zero-point shift.

The comparison of HUGS magnitudes with raw XPSP is shown in the left rows of the two panels of Fig.~\ref{fig:hugs_delta_app}. The typical scatter about the median is larger than in the cases described above. We verified that, in spite of the severe selection in $C^{\star}$, residuals still correlate with this parameter, showing that this extra scatter is due to the fact that we are not dealing with a sample dominated by fully isolated stars, as is the case for the reference samples considered above, but with many sources whose spectrophotometry (and possibly also HUGS magnitudes, to a lesser extent) suffer from some degree of contamination, effectively limiting the achievable precision. However, at least for F606W and F814W, the original photometry is reproduced within $\simeq 1$\% over the entire range of colour covered by the reference sample, the only systematic deviations being attributable to the hockey-stick effect. For this reason, we decided to limit the process of standardisation to the correction of this effect, avoiding any modification of the original passbands. The final results for the subsample of stars with $G<17.65$~mag are shown in Fig.~\ref{fig:hugs_deltaindr3}. The performances in F606W and F814W, possibly the most widely used HST passbands, are satisfactory given the non-ideal conditions. The higher scatter and the lower accuracy of the F438W XPSP are attributable to the same kind of problems affecting passbands sampling the blue end of the BP spectra. 

While validation is limited to the passbands and the colour and magnitude ranges considered above, the results presented here and those of a limited set of additional tests we performed may suggest that HST photometry for $Rf>1.4$ passbands should be reasonably well reproduced by XPSP, while
the issues related to the blue and UV end of BP spectra remain valid also in this case. 

%%%%%%%%%%%%%%%%%%%%%%%%%%%%%%%%%%%%%%%%%%%%%%%% END WIDEBAND

%%%%%%%%%%%%%%%%%%%%%%%%%%%%%%%%%%%%%%%%%%%%%%%% NARROW BAND
%\input{sections/narrow.tex}
\section{Narrow-band photometry}
\label{sec:narrow}

In this section we explore the performance of XPSP in the realm of medium- and narrow-band photometry using a few widely used systems as test cases. Standardisation is attempted only for the version of the Str\"omgren system considered here. The J-PAS and J-Plus systems sample the performances of narrow-band XPSP over the entire range covered by XP spectra. General guidelines to use narrow band XPSP to calibrate surveys aimed at measuring emission line fluxes are also provided. Finally, we show an example of how XPSP can be used to take the design of a photometric system, and bring it into real life, measuring fluxes and magnitudes of real sources through its wide and medium-width passbands \citep[the Gaia C1 system,][]{Jordi2006}.

\subsection{Str\"{o}mgren photometry and its standardisation}
\label{sec:standa_strom}

According to \citet{SMY2011} the Str\"{o}mgren system \citep{strom56} was originally designed to investigate the astrophysical properties of low-reddening main sequence stars. However, colour indices obtained from its {\it uvby} bands have been widely used to estimate the stellar effective temperature and surface gravity, as well as other parameters such as reddening and metallicity, over a wide range of stellar types and classes. For instance, see the use of the $(b-y)$ temperature sensitive colour in the  \citet{Alonso99} relations, or of $m_1=(v-b)-(b-y)$ to derive metallicity
 \citep[e.g. in stellar clusters, as done by][among others]{frank15,piatti19}, or the correlation between $c_1=(u-v) - (v-b)$ and nitrogen abundance \citep{grundahl02}. 
 
In contrast to the cases discussed in Sect.~\ref{sec:standa}, the Str\"{o}mgren system lacks a generally accepted standard version, with its set of TCs and, especially, a large set of reliable standard stars.
Among the many available versions of Str\"{o}mgren TCs \citep[see e.g.][]{Bessel11}, here we adopt those provided by the Spanish Virtual Observatory\footnote{\url{http://svo2.cab.inta-csic.es/}} , which describe the filters mounted on the Wide Field Camera (WFC) of the Isaac Newton Telescope (INT) at El Roque de los Muchachos in the Canary Islands, as we have had some previous successful experience in using them \citep{massari16}.

Str\"{o}mgren bands are entirely located in the BP realm. We limit our analysis to $bvy$ bands here, because XPSP $u$ magnitudes suffer from the problems described in Sect.~\ref{sec:ustand}, affecting all the TCs in the range $\lambda <400$~nm; moreover, its blue edge exceeds the blue limit of BP (see Fig.~\ref{fig:ubands}). While $b$ and $v$ passbands have $Rf=2.0$ and $Rf=1.8$, respectively, $y$ is slightly below the nominal $Rf=1.4$ threshold for reproducible photometry established in Appendix~\ref{sec:app_Rf}, with $Rf=1.3$. This should not be considered as a serious issue, because the adopted limit is conservative and also because $Rf$ values below that threshold may only be troublesome in the presence of strong spectral features. 

 As a reference set for comparison and standardisation we chose the largest sample available in the literature of stars observed with the aformentioned TCs, which is the sample of stars located around the Galactic anticentre analysed by \citet{monguio13}.
 This sample contains 23 992 stars, covering many spectral types and a wide colour range, $0\le {\bprp} \le3$~mag, and, in turn, is calibrated on the set of standards defined in \citet{crawford1970}. 
 
 We performed a quality check in order to select only well-measured stars. We first  considered the photometric error in each of the three considered Str\"{o}mgren bands as a function of {\gband} magnitude and traced the median trend in steps of $\Delta G=0.5$~mag. Only the sources with an error smaller than the 95th percentile of the distribution for each bin were kept in the sample. To be conservative, we further rejected all the sources with a photometric error of $>0.1$ in any band, as their calibrating power would be poor in any case. This effectively limits the reference sample to $G<15.0$~mag stars. In turn, this implies that no correction for the hockey-stick effect is possible in the standardisation process.  After this first selection based on the Str\"{o}mgren photometry, we applied two other quality cuts based on {\gedr} parameters by requesting ${\tt ruwe<1.4}$ (from {\tt gaia\_source}) and $-0.03< C^* <0.03$. After all these selections, the final reference sample includes $6158$ stars.

\begin{figure}[!htbp]
    \centering
    \includegraphics[width=(\columnwidth)]{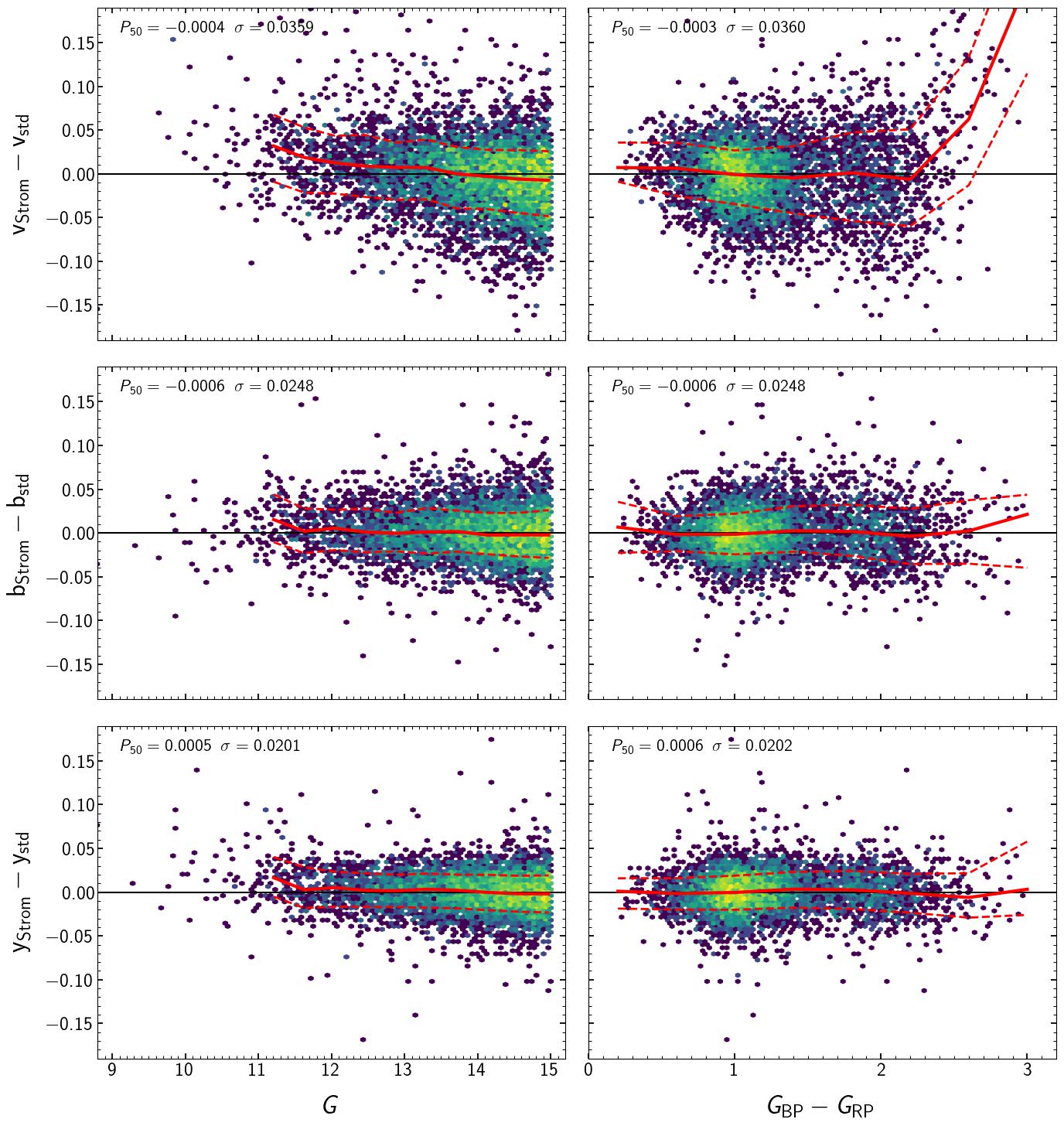}
    \caption{Performances of standardised XPSP in the Str\"{o}mgren system. The arrangement and the meaning of the symbols is the same as in Fig.~\ref{fig:sdss_deltaindr3}}
    \label{fig:strom_deltaindr3}
\end{figure}

The median difference between reference and XPSP raw magnitudes amounts to $\simeq 0.14$~mag in $v$, $\simeq 0.00$~mag in $b$, and $\simeq 0.03$~mag in $y$, with colour trends of amplitude $\la 0.05$~mag over the colour range covered by the reference sample (see Fig.~\ref{fig:strom_delta_app}). The performance is worse for the bluer TCs, in line with the already mentioned issues with the blue part of the BP spectra, which are probably exacerbated by the lower S/N unavoidably associated with passband widths smaller than those discussed in Sect.~\ref{sec:standa}. The median deviation of $v$ magnitudes is the largest among the non-UV
passbands considered here. Magnitudes in J-PAS passbands of similarly narrow
width and covering the same wavelength range (397-427~nm) have median deviations
$\simeq 0.08-0.10$~mag (see Sect.~\ref{sec:javalambre}). The relatively large
deviation of the raw synthetic $v$ magnitudes might be ascribed to a
combination of the systematic errors of EC XP spectra in the relevant wavelength range
and the intrinsic problems historically affecting photometric calibration of
the Str\"{o}mgren system \citep[see e.g.][SMY11]{Bessel05}.

In Fig.~\ref{fig:strom_deltaindr3} we show the usual $\Delta$~mag plots for the standardised magnitudes (see also Table~\ref{tab:strom_median}, for the corresponding $P_{50}$, $P_{16}$, and $P_{84}$ values). A small residual trend with magnitude remains in the bluest passband ($v$), as well as a strong colour term for $\bprp \ga 2.2$~mag, where the number of reference stars is low. On the other hand, the accuracy of the $b$ and $y$ magnitudes is very good, for $G>11.5$~mag. The typical scatter at $G\le 15.0$~mag is $\sigma\sim 40$~mmag in $v$, $\sigma\sim 30$~mmag in $b$, and $\sigma\sim 20$~mmag in $y$.

\begin{figure}
\center{
\includegraphics[width=\columnwidth]{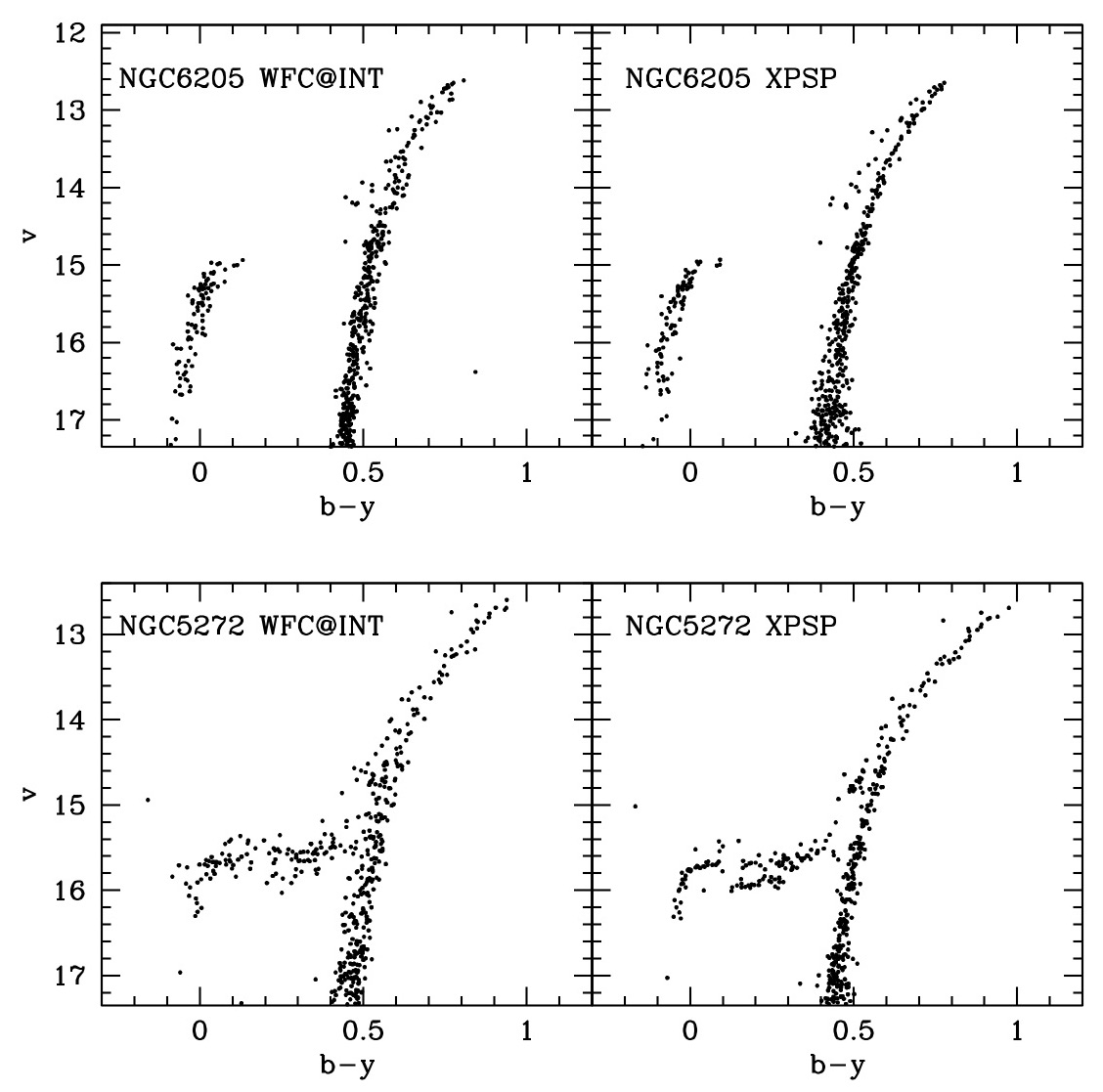}
}
\caption{Comparison between the observed (left columns) and synthetic (right columns) Str\"{o}mgren CMDs of the GC NGC~5272 (bottom panels) and NGC~6205 (upper panels). The same stars are plotted in the left and right panels.} 
\label{fig:stromgren_gcs}
\end{figure}

An immediate demonstration of the high photometric performance achieved by the standardised XPSP comes from a direct comparison between the Color Magnitude Diagrams (CMDs) of two Galactic GCs, namely NGC~5272 and NGC~6205 obtained from XPSP and from direct ground-based photometry taken from \citet{massari16} and \citet{savino18}, respectively. Figure \ref{fig:stromgren_gcs} shows that, once a strong selection to mitigate the effect of contamination of XP spectra is adopted ($-0.03<C^*<0.03$) for the stars in common between {\it Gaia} and the INT observations, the overall quality of the CMDs from synthetic photometry is clearly higher than their ground-based `observed' counterparts.
 All sequences are significantly tighter in general, especially towards the AGB and the tip of the red giant branch (RGB), where the ground-based photometry may suffer from saturation and non-linearity effects. Also, the horizontal branch (HB), particularly that of NGC~5272, appears cleaner and better defined in the XPSP diagrams.

Finally, we stress again that this is a particular realisation of the Str\"omgren system. For instance, our standardised XPSP $vby$ photometry fails to reproduce the colour indices provided by \citet{hauck98}. The size of the mismatch depends on the stellar type and on the colour index, and ranges from $\sim0.05$~mag in $b-y$ for red giant stars, up to $\sim0.1$~mag in $m_1$ for blue giants. On the other hand, an indirect validation of the adopted standardised magnitudes is provided in Sect.~\ref{sec:metalStromgren}, where metallicity estimates matching their spectroscopic counterparts within the uncertainties are obtained from standardised XPSP Str\"{o}mgren colour indices.

\subsection{Javalambre surveys}
\label{sec:javalambre}

In this section, we test XPSP performance against the medium- and narrow-band photometry from the two surveys obtained at the Javalambre Observatory in Teruel, Spain \citep[see][]{javalambre}. 
The first is the J-PAS survey \citep{JPAS}, which includes a photometric system of 54 narrow and contiguous passbands and 6 wider passbands (including SDSS filters), covering a similar wavelength range to {\gaia}. 
In preparation for the full J-PAS catalogue, recently a small region in the sky was observed and released (the mini-JPAS catalogue, \citealt{miniJPAS}), covering 1~deg$^2$ towards the Galactic halo (RA, Dec)$ = (+215^\circ,+53^\circ)$, up to magnitude 22-23 in the narrow bands and 24 in the broader passbands, with an absolute error of smaller than $\sim 0.04$ mag.

The cross-match with mini-JPAS yields 636 sources in common with the XP spectra in {\gdr3}. Although this is a small number of sources, they offer very detailed wavelength information when compared with XPSP results, which is very useful for estimating the level of detail that {\xp} spectra can provide, as XP spectra and J-PAS are of similar spectral resolution. 

The second Javalambre photometric catalogue used here is the J-PLUS survey \citep{JPLUS}. The 
J-PLUS set of passbands 
includes five broad (similar to SDSS) and seven medium passbands (similar to some of the C1 passbands originally designed for {\gaia} purposes; see \secref{Gaia2C1}). The J-PLUS project made its DR2 catalogue available in November 2020, including 31.5 million sources with $r < 21$~mag with absolute calibration errors of 8-19~mmag (depending on the passband; \citealt{JPLUScalibration}). Among all these sources, we used only sources in common with APOGEE DR16 \citep{apogee16}.
We also considered a set of white dwarfs (WDs) from  \cite{GF2019} based on {\gdr2}.

We applied extra quality filters to avoid the most obvious problems in the XP spectra (blending, contamination, multiple sources, gating, issues in the astrometric solution, and large photometric excess flux).
After all this filtering, the remaining set of sources for our analysis comprises 17\,465 APOGEE sources and 337 WDs in J-PLUS DR2, and 583 sources in mini-JPAS. No standardisation of any kind has been attempted with the derived synthetic photometry in either J-PAS or J-PLUS systems.

\begin{figure}[!htb]
\center{
\includegraphics[width=0.95\columnwidth]{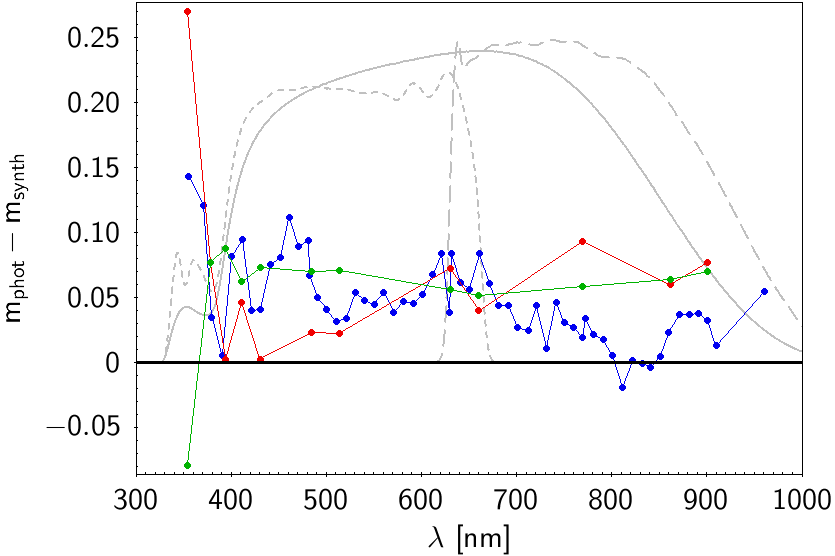}
}
\caption{Median residuals between the observed and synthetic magnitudes as a function of the mean wavelengths of the passbands for mini-JPAS (blue), J-PLUS in APOGEE DR16 (red), and J-PLUS WDs (green). 
In grey lines, we plot the {\gaia} passband transmissivity as in {\gedr} \citep{riello2021} divided by a factor three to fit in the same scale of the residuals. The solid grey line represents $G$ band, the short dashed line  BP band (at shorter wavelengths), and the long dashed line RP band (at longer wavelengths).
\label{fig:Resvswavelength_miniJPASJPLUS}
} 
\end{figure}

The median $\Delta {\rm mag}$ (reference minus synthetic magnitudes) for these samples as a function of mean wavelength of the passband can be seen in \figref{Resvswavelength_miniJPASJPLUS}.
In addition to a general ZP(of about $0.05$~mag) in the comparison, \figref{Resvswavelength_miniJPASJPLUS} indicates larger discrepancies at short wavelengths ($\lambda<400$~nm), which are due to the known issues of externally calibrated XP spectra in this range (Sect.~\ref{sec:methods} and Sect.~\ref{sec:standa}).  Moreover, the bluest passband, $u_{JAVA}$, has a blue edge located beyond the blue edge of XP spectra (see \secref{methods} and Fig.~\ref{fig:ubands}). The lower accuracy of XPSP for filters in this spectral region is therefore not surprising, and neither is the fact that passbands whose SP comes from BP spectrophotometry have larger median residuals than those from RP, on average.

\begin{figure}[!htb]
\center{
\includegraphics[width=0.9\columnwidth]{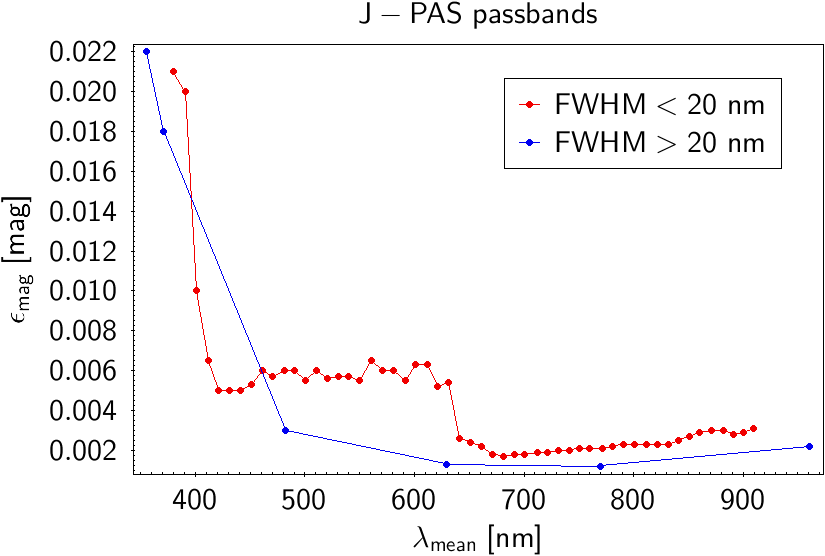}
}
\caption{Median magnitude uncertainties at the bright regime derived for J-PAS passbands as a function of wavelength. The red and blue lines show the behaviour for the narrow and wide passbands 
 ($u_{\rm JAVA}$ and SDSS passbands), respectively.
\label{fig:sigmacalvslambda}
} 
\end{figure}

The uncertainty on the XPSP does not decrease when the magnitude decreases for the bright regime ($G\lesssim 12$~mag). This is due to the fact that, in this range, the calibration errors dominate the estimated uncertainty. Therefore, we analyse here the J-PAS median uncertainty at the bright end as derived for the synthetic photometry. The reader should note that this does not depend in any way on J-PAS data, only on XP spectra. The resulting calibration errors as a function of the central wavelength of the filter are shown in \figref{sigmacalvslambda}. We can see that passbands with short central wavelengths suffer an increase in systematic effects and also that the minimum uncertainties from the BP and RP instrument wavelength range are at different levels (being systematically larger for BP). For the reddest passbands in RP, the calibration error also increases progressively.

\begin{figure}[!htb]
\center{
\includegraphics[width=0.9\columnwidth]{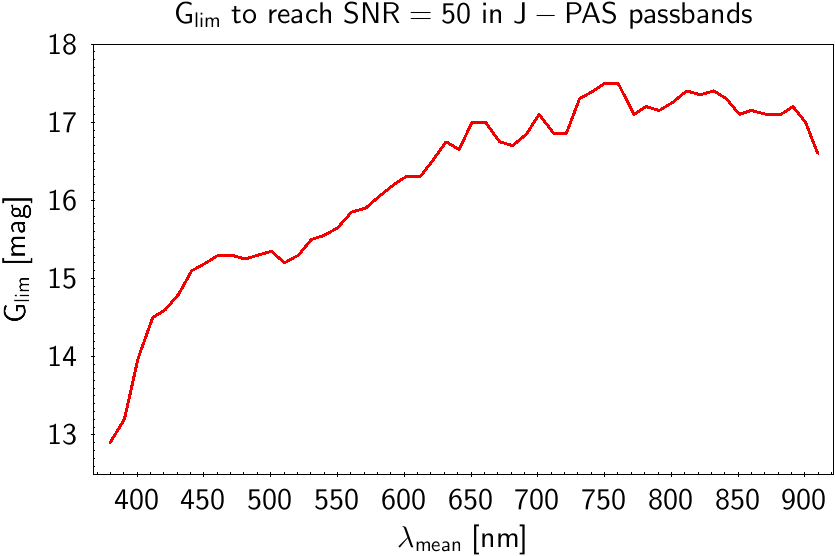}
}
\caption{{\gband} magnitude ($G_{\rm lim}$) needed to reach a median S/N$=50$ in every one of the narrow (FWHM$\sim 15$~nm) J-PAS passbands with mean wavelength equal to $\lambda_{\rm mean}$.
\label{fig:snr50JPAS}} 
\end{figure}

The homogeneous wavelength coverage in the J-PAS system allows us to evaluate the variation of the S/N obtained whilst performing synthetic photometry on the {\xp} spectra. \afigref{snr50JPAS} shows the $G$ magnitude needed in order to get a median S/N$=50$ for every one of the narrow J-PAS passbands (the width of these passbands is about 14 nm). 

\subsection{Emission line photometry: the IPHAS system}
\label{sec:narrow_iphas}

The INT Photometric H$\alpha$ Survey of the Northern Galactic Plane \citep[IPHAS,][]{drew2005} is designed to identify and characterise emission line stars and extended objects such as planetary nebulae. It uses passbands similar to SDSS $r$ and $i$ together with a narrow H$\alpha$ filter, leading to a H$\alpha$ TC with a full width at half maximum (FWHM) of 9.5 nm and $Rf \approx 1.13$. Broadband data are calibrated based on PanSTARRS, while the narrowband H$\alpha$ data rely on a fixed offset from the enclosing Sloan $r$ band, with further refinement based on overlapping fields anchored to fields with the best photometry (i.e. taken under stable photometric conditions). The final data release is presented in \citet{monguio2020} as part of the INT Galactic Plane Survey (IGAPS).

\begin{figure}
\center{
\includegraphics[width=\columnwidth]{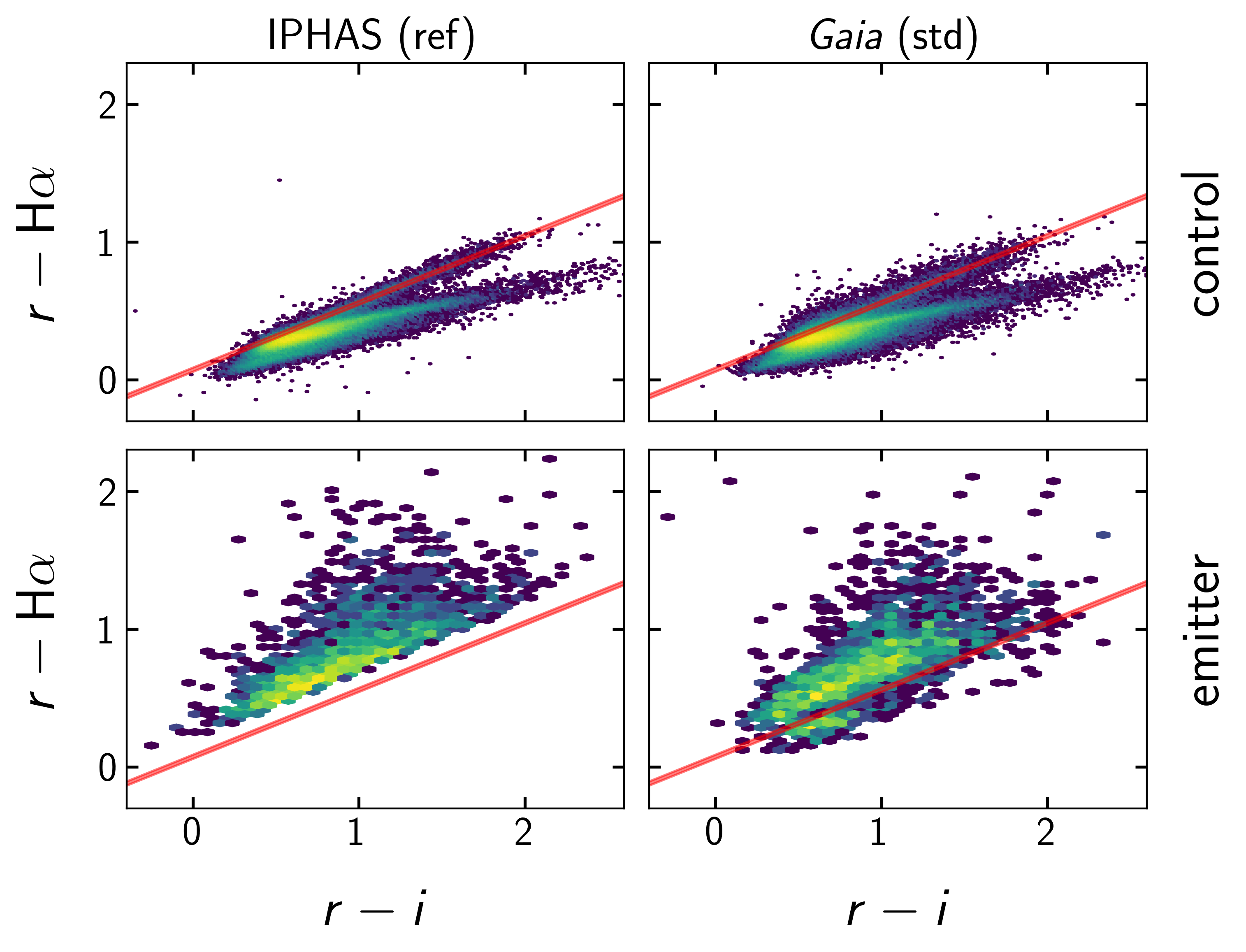}
}
\caption{IPHAS ($r - i$, $r-\textup{H}\alpha$) for control (top) and emitter (bottom) samples after applying the cuts discussed in the text. The original IPHAS magnitudes are on the left and the \gaia synthetic magnitudes are on the right. The diagonal red line is the cut used by \citet{monguio2020} to select emitting objects.}
\label{fig:iphas_colour_planes}
\end{figure}

We select two comparison sets of stars within the survey footprint ($-5$\degr\ $< b < +5$\degr,\ $30$\degr\ $< \ell < 215$\degr): a control sample consisting of a 1 in 40 sample of \gaia DR3 sources with relative parallax errors of better than 20\%, and the set of emission line objects identified by \citet{monguio2020} in the IGAPS catalogue through a linear cut in the ($r - i$, $r-\textup{H}\alpha$) colour plane. The former sample contains approximately 500,000 sources, and the latter just over 8,000. Sources are matched between the two catalogues using a 1\arcsec\ positional cross-match. We remove sources that are saturated, have error flags set in the IGAPS catalogue, or have broadband uncertainties > 0.02 mag or H$\alpha$ uncertainties > 0.05 mag. We also require $G \leq 17.65$~mag, \texttt{xp\_summary.bp\_n\_transits} $\geq$ 15 and \texttt{xp\_summary.rp\_n\_transits} $\geq$ 15, and 
\texttt{phot\_g\_flux\_over\_error$>50$}, and \texttt{phot\_x\_flux\_over\_error$>10$}, for \texttt{x=bp,rp}. We do not filter based on C$_{\star}$ as a larger value of this parameter can also reflect the presence of emission lines \citep{riello2021}. This does allow some extended sources to pass the selection, but such sources are rare and thus unlikely to affect the comparison (the vast majority of sources in both samples are classified in IGAPS as stellar). About one-third of the control sample and one quarter of the emitters sample pass these cuts, yielding 168,688 and 2165 sources respectively. These reduced samples are shown in \figref{iphas_colour_planes}.

\begin{figure}
\center{
\includegraphics[width=\columnwidth]{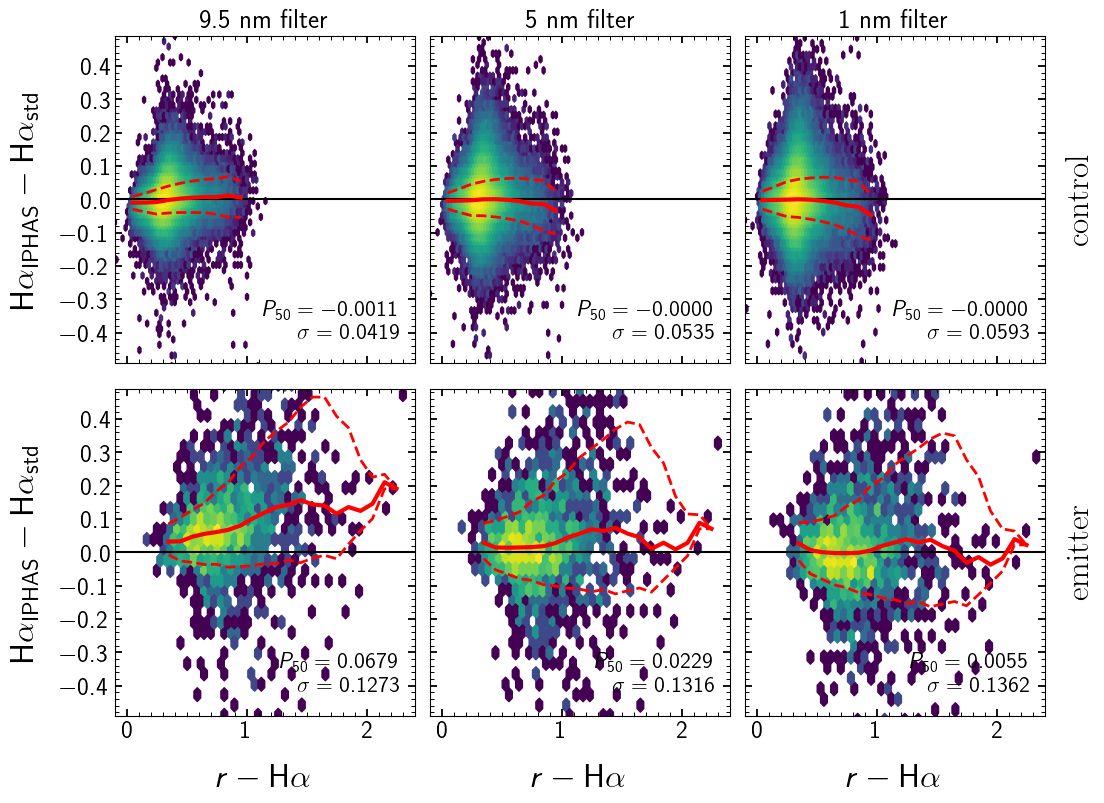}
}
\caption{Difference between IPHAS and synthetic H$\alpha$ magnitudes versus $r - \textup{H}\alpha$ colour for the control (top) and emitter (bottom) samples, using synthetic H$\alpha$ filters of different widths. Potentially variable sources have been removed.}
\label{fig:iphas_halpha_ref_synth}
\end{figure}

The broadband magnitudes of the control sample are well reproduced after standard corrections, which for simplicity we apply using polynomials rather than tweaking the shapes of the filters (we also note  that we do not provide means to obtain this photometry with GaiaXPy). Here we do not intend to provide standardised XPSP, but instead to show that \gaia XPSP can be used to calibrate narrow-band photometry aimed at tracing line emission, especially in view of future surveys. We add 0.02 and 0.01 to the $r$ and H$\alpha$ ZP; respectively, while for $i$ we also apply a linear correction, with $i_\textup{std}$ = $i_\textup{synth}$ + 0.01 + 0.11 $(r_\textup{synth} - i_\textup{synth})$. With these corrections, the IPHAS ($r - i$, $r-\textup{H}\alpha$) colour plane is qualitatively reproduced.

However; after calibration using the control sample, the H$\alpha$ magnitudes in the emitter sample show both a dependency on the $r-\textup{H}\alpha$ colour (which corresponds to the strength of the emission line) and a median offset of 0.07 mag (\figref{iphas_halpha_ref_synth}, left column). The H$\alpha$ fluxes in the synthetic photometry are generally lower than those from IPHAS, with the discrepancy being greater for sources with stronger H$\alpha$ emission. We believe that this is due to flux in the emission line being lost outside the edges of the IPHAS H$\alpha$ passband (see Appendix~\ref{sec:app_Rf}). For the vast majority of stars that do not have strong spectral features  at this wavelength, such as these mentioned above, the flux lost is not a problem as it is replaced by flux bleeding in from outside the nominal filter wavelength range.

We considered other explanations for this discrepancy, such as variability or selection effects, but these can be ruled out by checking for consistency with photometry from second detections in the IPHAS data. Much of the IPHAS footprint is observed multiple times due to field overlaps, offsets to fill CCD gaps, and repeated observations to improve upon data taken in poor conditions. The emitter selection in \citet{monguio2020} was based only on the primary detection. A small set of objects indeed show no emission in their second detections (largely corresponding to \gaia detections showing no emission in \figref{iphas_colour_planes}); aside from these, there is no systematic bias when comparing different IPHAS detections, ruling out selection effects or small-scale variability as the source of the discrepancy. In \figref{iphas_halpha_ref_synth} we only include sources that have second detections, with the $r-\textup{H}\alpha$ of the two detections consistent within 0.1 mag, which leaves 1682 sources in the emitter sample. This eliminates most of the sources lying below the selection line in the lower right plot of \figref{iphas_colour_planes}. As an additional check, we also compared a small set of emitters selected from the IPHAS sample with spectra published in \citet{rodriguezflores2014}. The high-resolution spectroscopy in that work is consistent with the IPHAS photometry for those sources, but the magnitude discrepancy was present for them in both simulated and actual \gaia photometry.

The sensitivity of the H$\alpha$ filter is related to its width, with narrower filters producing a greater magnitude difference between emitting and non-emitting sources. One way to better match the behaviour of the original filter is to use a narrower synthetic H$\alpha$ filter on the \gaia data. Doing so reduces the overall shift as well as the colour dependency, at the expense of greater scatter, particularly in the non-emitting sources (\figref{iphas_halpha_ref_synth}, centre and left columns). Nevertheless, we do not necessarily expect the narrowest filter to completely reproduce the range of colours from IPHAS (see Appendix~\ref{sec:app_Rf}).

Despite the limitations discussed here, the practical functionality of the narrow-band filter ---separating out emitters from non-emitting stars--- is well reproduced by both the original IPHAS passband and the narrower versions, and moreover, the consistent performance in the H$\alpha$ passband of stars without strong H$\alpha$ emission enables flux calibration of survey fields, which in turn allows the selection of emission line stars even fainter than the publishing limit of the {\gdr3} spectrophotometry. This is true despite the filter violating the $Rf$ limitations discussed in Appendix~\ref{sec:app_Rf}. Indeed, it should be possible to calibrate almost any narrow-band imagery taken with a well-characterised filter provided there are enough sufficiently bright, well-exposed stars in the field.

\subsection{The project of a photometric system brought to life: C1}
\label{sec:Gaia2C1}

The original design for {\gaia} included a set of photometric passbands \citep{Jordi2006}, called C1B and C1M systems 
for broad and medium band photometry, 
respectively. The C1 system was especially thought to maximise the scientific return in terms of stellar astrophysical parameters. The spectral resolution requirements on the alternative prisms finally flying with the mission were made based on those passbands. 

Although some of the passbands in the C1 photometric system were finally implemented in the J-PLUS survey (\secref{javalambre}), the synthetic photometry study in this paper provides the perfect opportunity to test the performance of the full C1 system. This illustrates the investigations that can be done even with future sets of passbands using EC XP spectra, which should be more accurate than only relying on simulated spectra from synthetic spectral libraries. Moreover, it serves as a good example of a photometric system that is realised in practice using only Gaia DR3 data. In principle, a general user of the Gaia Archive may conceive their own set of passbands designed for a specific science goal and get XPSP in that system for all the stars with XP spectra released in DR3. The example of applications shown here and in Sect.~\ref{sec:metalStromgrenC1} for C1 showcases the performance that can be achieved for a well-designed system.

We can use the C1 synthetic photometry to learn about the performance of XP spectra, checking if they are able to trace the astrophysical information (see also \secref{metalStromgrenC1}). The aim here is not to repeat the work done by the {\gaia} DPAC, deriving again the astrophysical parameters of the sources (already available in the {\gaia} catalogue; \citealt{DR3-DPACP-156,DR3-DPACP-160,DR3-DPACP-157}), but is simply to evaluate whether or not the synthetic photometry derived with the C1 system is able to keep this information. 

\begin{figure}[!htbp]
\center{
\includegraphics[width=0.8\columnwidth]{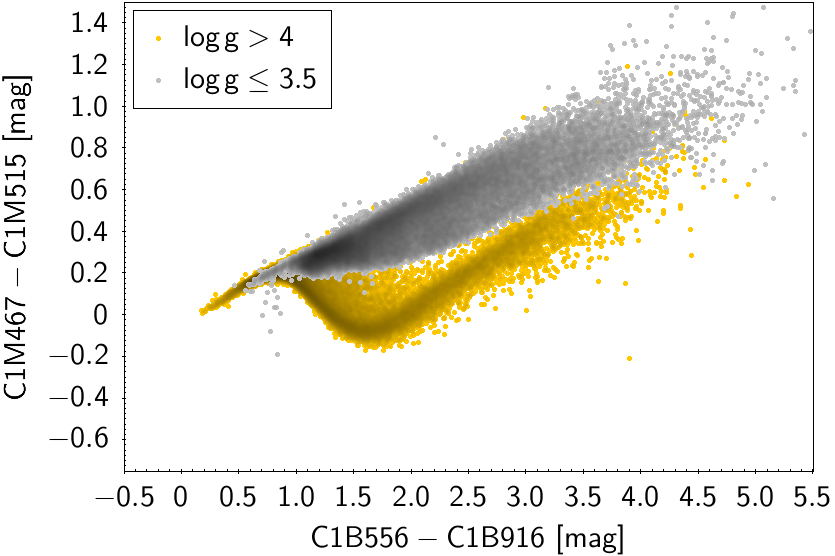}
}
\caption{Colour--colour diagram using the C1 photometric system,  able to separate giant (grey) and main sequence stars (orange). log~$g$ values are from GSP-Phot \citep{DR3-DPACP-156}.
\label{fig:loggcolours}
} 
\end{figure}

Using the C1 synthetic colour indices, we can perform for example a rough classification between giants and main sequence stars. For example, \figref{loggcolours} shows the C1M467$-$C1M515 colour (sensitive to surface gravity) plotted against C1B556$-$C1B996 (sensitive to the effective temperature).  Giants (in grey) and main sequence stars (in orange), which have different ranges of surface gravity ($\log g$) values, as derived by  DPAC/CU8 with the GSP-Phot module\footnote{GSP-phot is the  DPAC/CU8 module of the astrophysical parameters inference system \citep[Apsis;][]{DR3-DPACP-160} aimed at deriving the astrophysical parameters of the stars from XP spectra, parallaxes and $G$ apparent magnitudes, using a Bayesian full-forward modelling \citep{DR3-DPACP-156}.} \citep{DR3-DPACP-156}, are found at different positions in this diagram.

\begin{figure*}[!htbp]
\center{
\includegraphics[width=\textwidth]{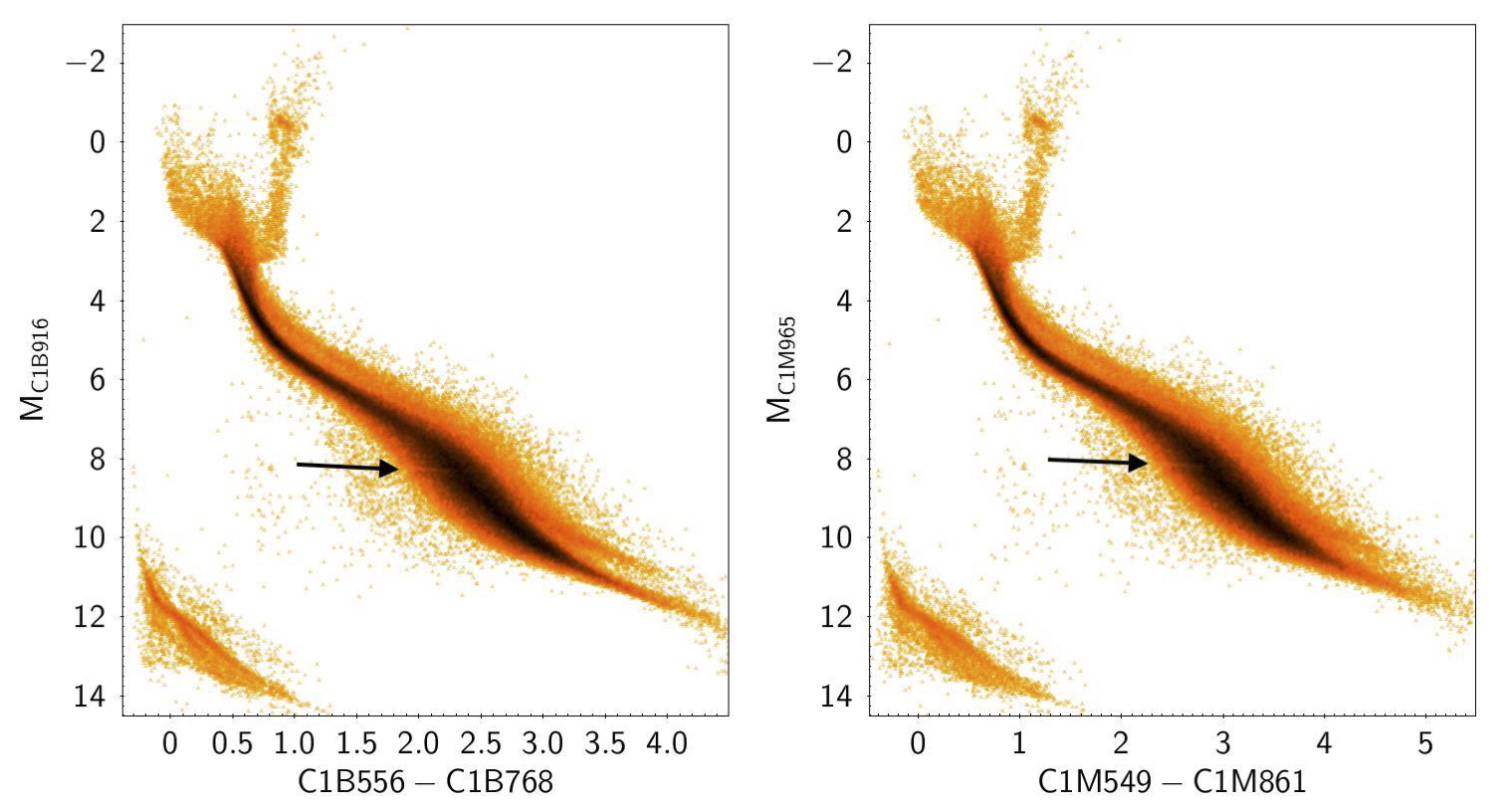}
}
\caption{Two examples of distance-corrected colour magnitude diagrams of the subset of the GCNS catalogue for which XPSP can be obtained, using combinations of broad (left panel) and medium width (right panel) passbands of the C1 system. Black arrows highlight the location of the Jao gap \citep{jao2018}, which is clearly visible in both diagrams, showing the high precision of the XPSP in these bands.} 
\label{fig:C1_HRD}
\end{figure*}

The C1 system can also be useful to estimate the metallicity of the studied sources, as we show in \secref{metalStromgrenC1}.
On the other hand, Figure~\ref{fig:C1_HRD} is intended to demonstrate the precision attainable with XPSP in this system. Two distance-corrected CMDs for the subset of stars from the Gaia Catalogue of Nearby Stars \citep[GCNS;][]{GCNS21} with $G<17.65$ and the selection criteria listed in Sect.~\ref{sec:gspc} are shown here. The CMD in 
the left panel is based on a combination of three C1B passbands, while the one in the right panel is based on a combination of three C1M passbands. In both diagrams all the sequences typical of CMDs of the solar neighbourhood are very well defined, including the various WD subsequences (see Sect.~\ref{sec:gwdc}, and references therein). 
The extremely subtle feature on the lower main sequence known as the Jao gap is clearly visible in both diagrams \citep{jao2018,jao2020,GCNS21}. This suggests that the precision of C1 XPSP is comparable to that achieved with the Gaia broadband system, as defined by \citet{riello2021}.

%%%%%%%%%%%%%%%%%%%%%%%%%%%%%%%%%%%%%%%%%%%%%%%% END NARROW BAND

%%%%%%%%%%%%%%%%%%%%%%%%%%%%%%%%%%%%%%%%%%%%%%%% PERFORMANCE VERIFICATION
%\input{sections/perfver.tex}
\section{Performances verification experiments}\label{sec:perfver}

In this section, we show a few examples of performance verification of XPSP against real science goals. In particular, we show that, in some cases, XPSP can be used to trace multiple populations (MPs) in GCs, or to estimate metallicity (also for extremely metal-poor stars) and even the abundance of $\alpha$ elements. It is also shown that XPSP can be used to identify emission line sources (ELS) with accuracy similar to that achieved from direct analysis of XP spectra.
We discuss a further example of application in Sect.~\ref{sec:gwdc}, namely classification of WDs.

\subsection{Multiple populations in globular clusters}
\label{sec:mps}

In the last four decades, the concept of GCs as a coeval and homogeneous simple stellar population has dramatically changed thanks to the discovery of star-to-star abundance variations in almost all GCs, which produce multiple photometric evolutionary sequences in the CMD \citep[and references therein]{bastian18,gratton19}. These MPs can therefore be studied not only with spectroscopy, but also with high-quality photometry \citep[see, e.g.][]{piotto07,milone12,milone13,milone15,Lee2019}. In particular, UV passbands are sensitive to the deep CN molecular bands at 388~nm \citep{pancino2010,sbordone2011}, and therefore stars with normal and enhanced N can be efficiently separated photometrically \citep[e.g.][and references therein]{Yong2008,lardo11,carretta11}. About 20\% of GCs also show multiple photometric sequences in CMDs not involving the U band, which are the result of different He and C+N+O abundances \citep{pancino00, sbordone2011, milone15, monelli13}.

\begin{figure}[t]
    \centering
    \resizebox{\hsize}{!}{\includegraphics[clip]{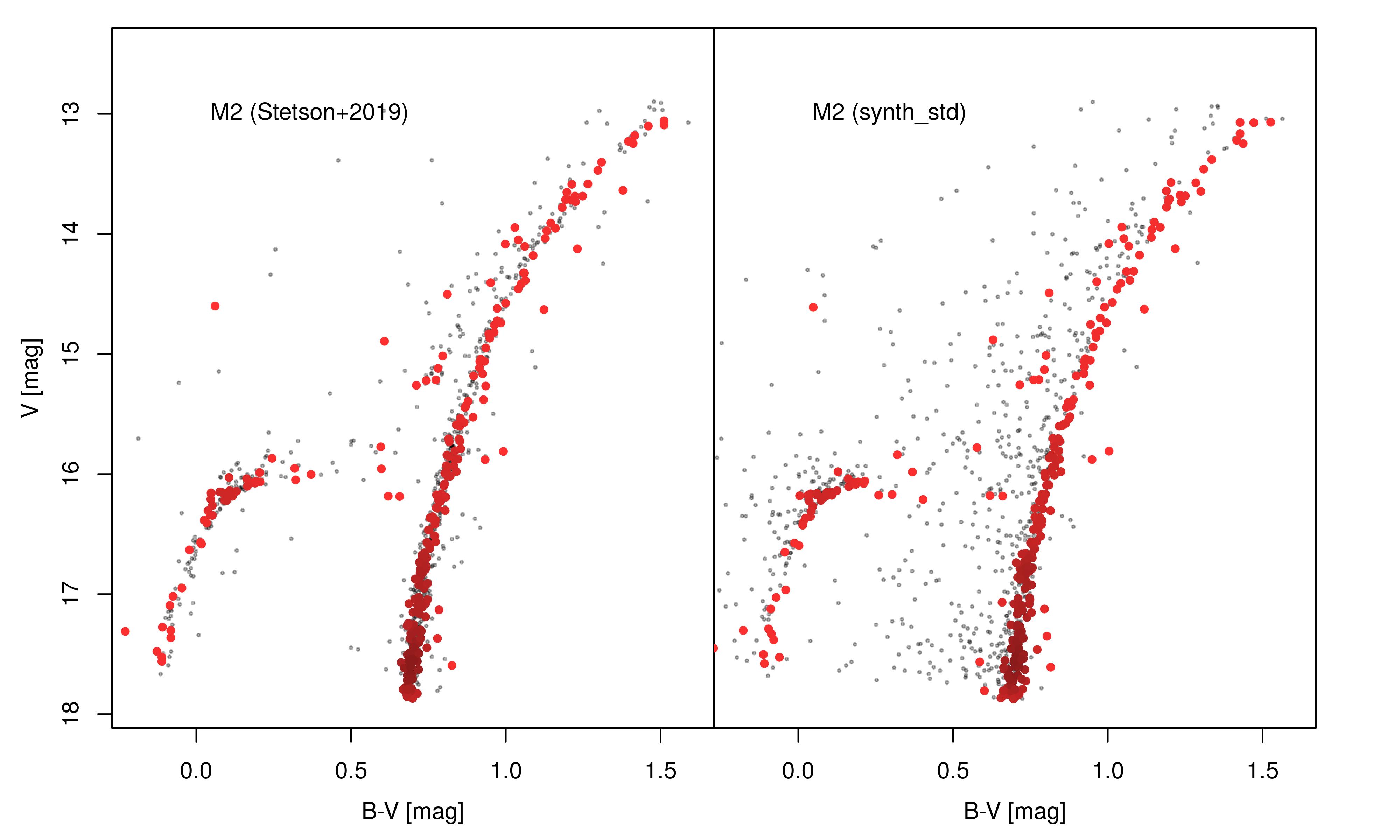}}    
    \resizebox{\hsize}{!}{\includegraphics[clip]{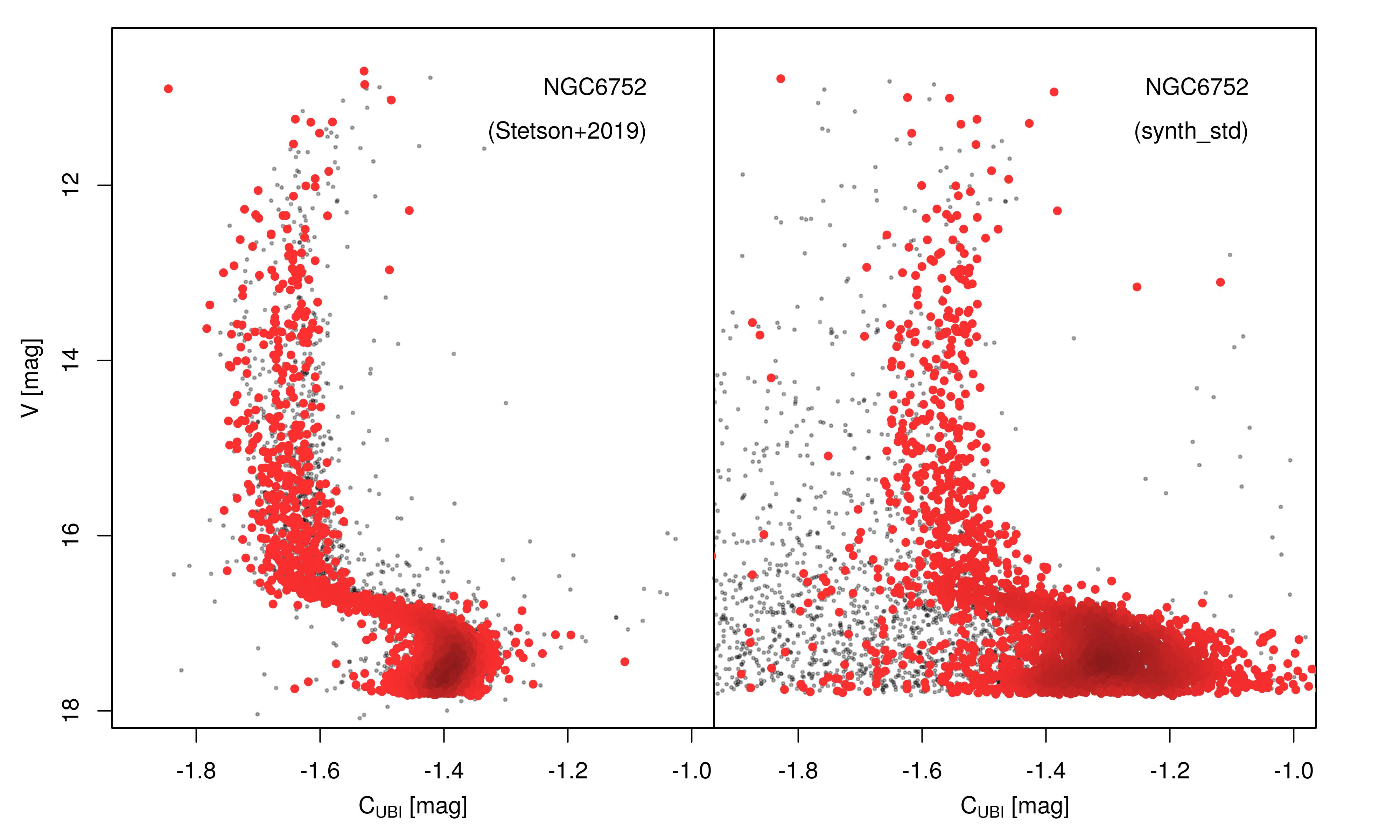}}    
    \caption{ 
    Top panels: Ground-based $V$, $B-V$ photometry of M2 \citep[left,][]{Stetson2019} and the corresponding synthetic {\gaia} photometry (right). The samples selected using the \citet{vasiliev21} criteria are plotted in grey, while the ones further selected with the criteria described in Section~\ref{sec:mps} are plotted in red. Bottom panels: Similar to the top panels, but for the case of the $V$, $C_{UBI}$ CMDs of NGC\,6752.}
    \label{fig:mpgcs}
\end{figure}

Here we use MPs in GCs to demonstrate the performance of the JKC synthetic standardised photometry presented in Section~\ref{sec:standa_jkc}. For this purpose, we tested all the GCs selected for other performance verification cases (Sect.~\ref{sec:metalStromgren}), complemented with a selection of half a dozen GCs hosting spectacular and well-studied MPs. In each GC, we selected the sample stars using the membership probability by \citet[][$>$0.9]{vasiliev21}. In doing so, we implicitly adopted their careful and complex selection on the quality of the {\em Gaia} astrometry based on {\tt RUWE}, the 
IPD parameters, and other indicators (see their Section~2). This selection appears in grey in Figure~\ref{fig:mpgcs}. We further selected stars according to the following criteria (see also Section~\ref{sec:standa_jkc}): (i) $|\rm{C*}|<\sigma_{\rm{C*}}$ \citep[Section~9.4]{riello2021}; (ii) {\tt RUWE}$<$1.4; (iii) 
{\tt IPD\_frac\_multi\_peak}$<$7; (iv) {\tt IPD\_frac\_odd\_win}$<$7; and (v) $\beta<$0.2 \citep[][Section~9.3]{riello2021}. This selection appears in red in Figure~\ref{fig:mpgcs}. We note that of all the applied selections, only the one on $\beta$ really makes a difference, because selections on the other parameters were already explicitly or implicitly applied by \citet{vasiliev21}.

In the top panels of Figure~\ref{fig:mpgcs}, we compared the $V$, $B-V$ synthetic standardised photometry with the ground-based photometry by \citet{Stetson2019} for M\,2 (NGC\,7089), a GC well-known for hosting an anomalous RGB, containing a few percent of the stars and being redder than the main RGB \citep{lardo12,lardo13}. As can be noted, the sample selected with the \citet{vasiliev21} membership and quality criteria (in grey in the figure) displays a `wind' of stars that are bluer than the red giant branch that is not present in the \citet{Stetson2019} photometry. This is caused by the fact that the typical seeing in the \citet{Stetson2019} data was of about $1.0\arcsec$, while the typical aperture of XP spectra is of $3.5\arcsec\times2.1\arcsec$. The {\gaia} XP synthetic photometry therefore suffers more from crowding and blending effects. Additionally, the \citet{Stetson2019} photometry is obtained by PSF fitting and with sophisticated deblending routines, while in DR3 no detailed deblending has been performed. Future {\em Gaia} releases will tackle blending and contamination with ad hoc processing pipelines, but in {\gdr3} we can use several indicators of crowding, such as the $\beta$ parameter defined by \citet{riello2021}. If we further select the sample as described above (red stars in Figure~\ref{fig:mpgcs}), a cleaner RGB is obtained, but at the expense of completeness. In any case, in both the ground-based and the {\em Gaia} XPSP CMDs the anomalous branch is clearly visible as a sparsely populated sequence $\simeq$0.2~mag redder than the main RGB, which demonstrates the very high performance of the synthetic photometry presented here. 

In order to investigate the case of the U band, we adopt the colour index C$_{UBI}$ \citep[defined as $(U-B)-(B-I)$,][]{monelli13} that combines and amplifies the effect of the variations in both N and He. The bottom panels of Figure~\ref{fig:mpgcs} show the case of NGC\,6752. In the ground-based photometry, the presence of MPs is indicated by the well-separated RGBs, while in the synthetic photometry the separation is not so clearly visible, but the MP presence is clear because of the large width of the RGB ($>$0.1~mag) compared to the typical photometric errors ($<$0.03~mag, see Sections~\ref{sec:standa_jkc} and \ref{sec:standa_u}). As in the case of M\,2, a further selection including $\beta$ is necessary (red points in Figure~\ref{fig:mpgcs}) in the synthetic photometry in order to clean the sample of untreated blends, also at the expense of completeness. We would like to highlight that NGC\,6752 is among the closest GCs (less than 4~kpc) and is one of the
cases in which MPs can be more clearly identified using XPSP. Another iconic GC, NGC\,1851, which is more distant, very compact, and displays a rather complex RGB substructure in the ground-based U-band CMD \citep[see Figure~10 by][]{Stetson2019}, does not clearly reveal any substructure in the synthetic {\em Gaia} U-band and C$_{UBI}$ CMDs. This is likely because (i)  the {\em Gaia} wavelength range does not fully include the $U$-band (as discussed in Sections~\ref{sec:ustand}, \ref{sec:standa_u}); (ii)   the cluster is more compact than NGC~6752, with a half-light radius of $r_h=0.51\arcmin$, to be compared with $r_h=1.91\arcmin$ for NGC~6752 \citep{harris1996}, thus crowding effects are necessarily expected to produce a stronger effect on NGC~1851\citep[see also][]{pancino17}; and (iii) the treatment of blending and contamination introduced in {\gedr} \citep{riello2021} is still not the complete treatment planned for {\gdr4}. We also note that there is a zero-point offset between the XPSP and the ground-based photometry in the U band that varies from GC to GC (0--0.2~mag). This is also likely caused by the above effects and is also due in part to the fact that the ground-based photometry 
\citet[][see in particular their Figure~4; see also App.~\ref{sec:stet}, below]{Stetson2019} is based on a collection of images taken with different facilities and filters, and that the U-band is notoriously difficult to standardise \citep{Alta2021}.

In conclusion, the $BVRI$ XPSP is of very high quality already, even for relatively distant GCs, in spite of the fact that the  treatment of blends is not yet fully implemented in {\gdr3,}  provided that one carefully selects the stars whilst considering parameters such as $\beta$ \citep{riello2021}. In the case of the $U$-band, the photometric performance is unavoidably lower than what is needed to study these fine details. Further improvements are eagerly awaited in the next {\gaia} releases.

\subsection{Metallicity from the Str\"omgren system}
\label{sec:metalStromgren}

\begin{figure}
\center{
\includegraphics[width=\columnwidth]{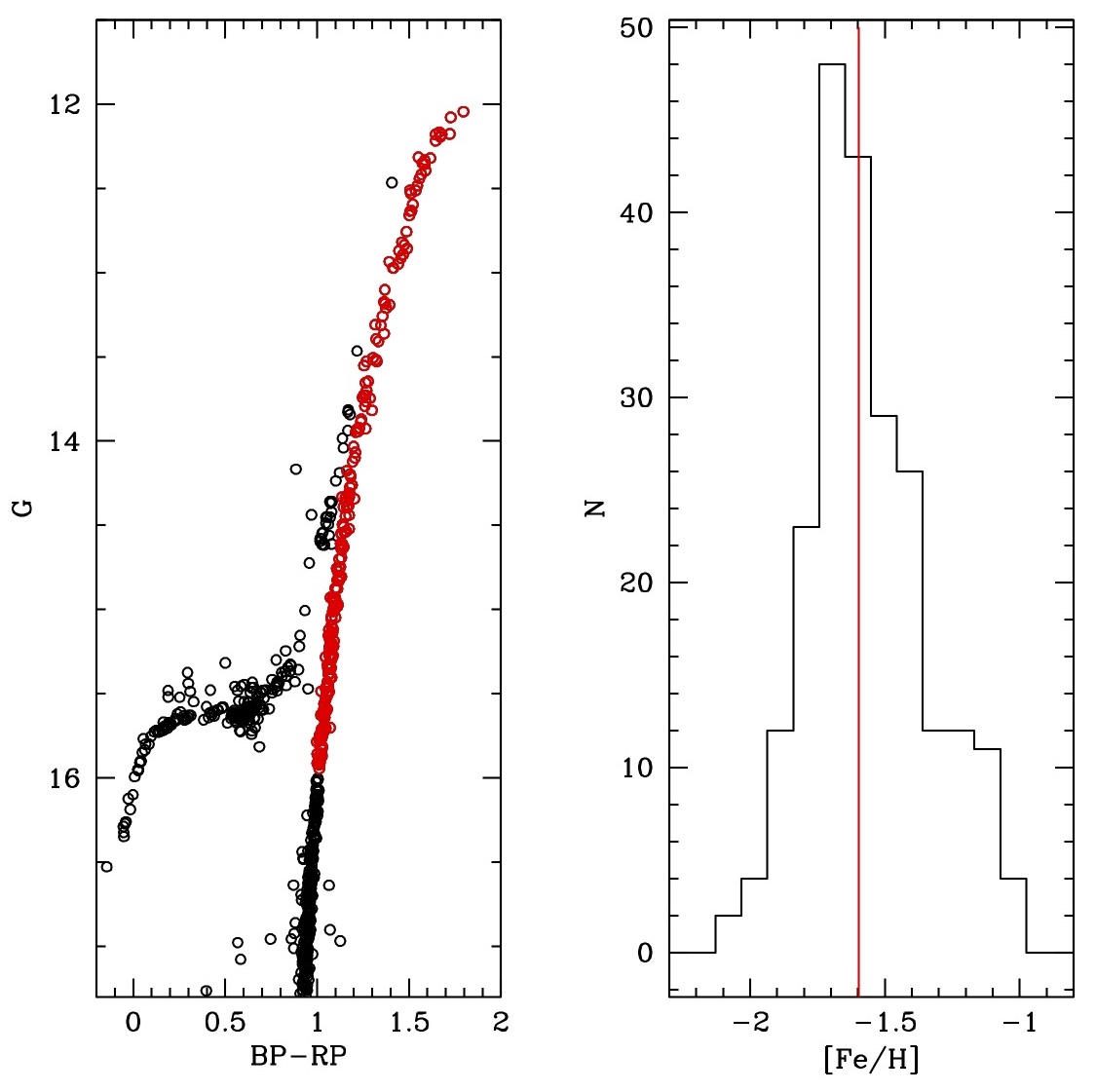}
}
\caption{{\it Left-hand} panel: {\it Gaia} CMD for the globular cluster NGC~5272. Red symbols indicate the giants for which a metallicity estimate was derived by means of the synthetic Str\"omgren photometry. {\it Right-hand} panel: Metallicity distribution inferred from the synthetic $m_0$ index for the selected stars. The vertical red line marks the 2.5$\sigma$-clipped mean value of [Fe/H]$=-1.59$ ($\sigma=0.14$). The cluster spectroscopic metallicity as quoted in \citet{carretta09} is [Fe/H]$=-1.50\pm0.05$.} 
\label{fig:m3_strom_met}
\end{figure}

The Str\"omgren index $m_1 = (v-b)-(b-y)$ has  been widely used as a tool to infer the metal abundance of giant stars \citep[see e.g.][]{richter99, att2000}. To explore the efficacy of our synthetic Str\"omgren photometry in recovering this parameter, we selected a sample of Galactic globular and open clusters (OCs), for which high-resolution spectroscopic [Fe/H] estimates exist. We determined their mean metallicity from \gaia XPSP by adopting the $m_0$-$(v-y)_0$-[Fe/H] relation provided by \citet[][based on GCs]{calamida07} for RGB stars, where $m_0$ and $(v-y)_0$ are the de-reddened version of the $m_1$ index and the ($v-y$) colour, respectively. It worth noting that such a relation is based on photometry that is calibrated on the same list of standards upon which our standardised XPSP ultimately relies, namely that provided by \citet[][see Sect.~\ref{sec:standa_strom}]{crawford1970}. The extinction law by \citet{cardelli89}, which provides A$_{v}$/A$_V=1.397$, A$_{b}$/A$_V=1.240$ and A$_{y}$/A$_V=1.005$ as extinction coefficients, has been adopted to correct the Str\"omgren magnitudes for reddening.
The 12 selected GCs span a metallicity range  from [Fe/H]$\sim-2.5$ to [Fe/H]$\sim-0.7$. To extend this range towards higher values, we included four metal-rich OCs with metallicities in the range $-0.2<$[Fe/H]$<+0.4$. 
In order to homogeneously select red giant stars in the analysed stellar clusters, we focused our analysis on all the stars from the red giant branch tip down to about $4$ mag fainter, and manually excluded obvious asymptotic giant branch (AGB) stars. Quality cuts were applied to the selection in order to select stars with $-0.03<C^*<0.03$. Finally, the stellar membership to the GCs was ensured by setting a minimum membership probability as determined by \citet{vasiliev21} of 90\%. This typically led to samples of several tens, or hundreds, of stars per cluster. Figure~\ref{fig:m3_strom_met} summarises an example of such a procedure, by showing the {\it Gaia} CMD of the globular cluster NGC~5272 in the left-hand panel, by highlighting the selected targets with red symbols, and by reconstructing their metallicity distribution derived from the synthetic Str\"omgren index $m_0$ in the right-hand panel.

\begin{figure}
\center{
\includegraphics[width=\columnwidth]{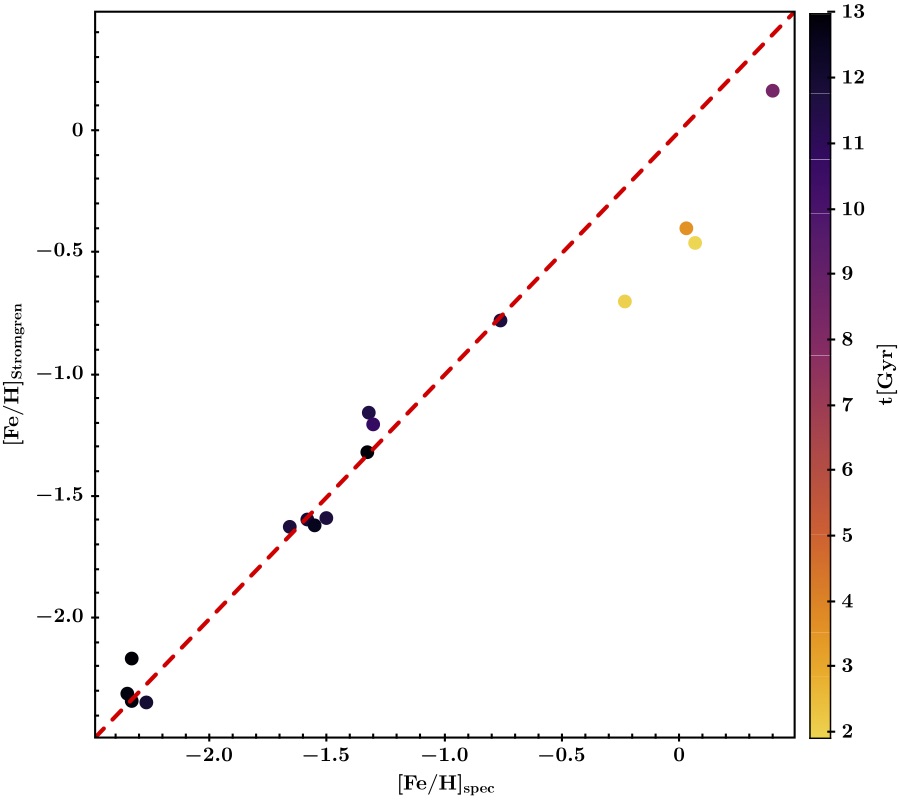}
}
\caption{Spectroscopic metallicity vs. photometric metallicity derived from the synthetic $m_0$ Str\"omgren index. The colour-coding indicates the age of each star cluster. Error bars are smaller than the symbol size.}. 
\label{fig:all_strom_met}
\end{figure}

The list of 12 GCs includes NGC~104, NGC~288, NGC~362, NGC~4590, NGC~5272, NGC~6205, NGC~6218, NGC~6341, NGC~6752, NGC~7078, and NGC~7099. Their spectroscopic metallicity is taken from the homogeneous scale provided by \citet{carretta09}, while we adopt the values provided in \citet[][with the 2010 update, available at https://physics.mcmaster.ca/$\sim$harris/mwgc.dat]{harris1996} for the extinction.  The clusters have in any case been purposely selected among the low-extinction ones. The additional four open stellar clusters are NGC~2506, NGC~6791, NGC~6819, and M67.  Their spectroscopic metallicity and reddening are taken from \citet{carretta04}, \citet{bragaglia14}, \citet{bragaglia01}, and \citet{zhang21}, respectively. The membership probability of the stars of these clusters is instead taken from \citet{cantat20}.
Figure \ref{fig:all_strom_met} shows the one-to-one comparison between the spectroscopic metallicity and the 2.5$\sigma$-clipped mean metallicity determined from our Str\"omgren photometry for the 16 star clusters, which are colour-coded according to their age. The error bars correspond to the error on the mean of each cluster metallicity distribution. Clusters ages are taken from \citet{vandenberg13} for the GCs, and from \citet{bossini19} for the OCs.

From Fig.\ref{fig:all_strom_met}, it is clear that while the comparison for the GCs is good, the OCs are systematically offset by about ~0.4 dex. Among these, the best behaved is NGC~6791, which is the oldest (t$=8.4$ Gyr). When interpreting these results, we should bear in mind that the colour of a giant star depends primarily on metallicity, but also on age. The sample of OCs has been chosen in such a way as to sample the high-metallicity part of the [Fe/H] distribution, and consists of systems that are much younger than the GCs used by \citet{calamida07} to calibrate their relation, that is, younger by between~4 and~10 Gyr. Therefore, the systematic offset of the OCs  is likely due to this intrinsic age difference rather than to a poor sensitivity of the \citet{calamida07} relation at this high [Fe/H] (which is an effect that could nevertheless still contribute). As supporting evidence, we note that the observed offset goes in the direction of our interpretation, in the sense that young OCs have intrinsically bluer RGBs, thus mimicking an old and more metal-poor population.

To assess the precision and accuracy of the metallicity estimates obtained via synthetic photometry, we therefore restricted our analysis to the GCs sample, finding rather good results. The mean difference between our photometric estimates and the spectroscopic values is $0.02$ dex, with a dispersion of $0.08$ dex. Such a dispersion closely matches the findings by \citet{calamida07}, who quote a precision   for their relations of $\sim0.1$ dex. As a last remark, the nominal error on the mean metallicity obtained from the synthetic Str\"omgren photometry is quite small (because of the large number of available stars), and ranges from $0.01$ dex in the case of NGC~104 up to $0.04$ dex for NGC~7099 (i.e. from the more metal-rich to the more metal-poor GCs). This in turn means that, in the considered case, the dominant contribution to the observed dispersion comes from the precision of the $m_0$-$(v-y)_0$--[Fe/H] relation itself.

Another way of testing the performance of our synthetic Str\"omgren photometry in determining the metallicty of giants is to directly compare with the spectroscopic measurements for nearby stars from large surveys such as GALAH \citep{galah} and APOGEE \citep{apogee16}.
To do so, after cross-matching {\it Gaia} DR3 sources with 
GALAH DR3 and APOGEE DR16, we select giants by requiring the spectroscopic log$~g$ measurements of our sample to be smaller than 2.5 (this is a strict selection, excluding lower RGB stars). After inspecting the {\it Gaia} HR diagram of these giants (which we obtained by correcting the observed magnitudes for reddening and distance using {\it Gaia} DR3 parameters), a further cut at G$_0<4.5$ was applied to exclude obvious dwarfs with likely uncertain log$~g$. To avoid the inclusion of highly reddened sources, we also imposed $0<E(G_{BP}-G_{RP})<0.1$. Finally, a quality cut at $-0.1<C^*<0.1$ was required to exclude low-quality {\it Gaia} measurements.
We are then left with a sample of $3,202$ giants in common with GALAH DR3, and $5,573$ giants in common with APOGEE DR16 (all of these are located at a mean distance of $\sim2$ kpc, with $\sigma\sim1$ kpc). 

\begin{figure}
\center{
\includegraphics[width=\columnwidth]{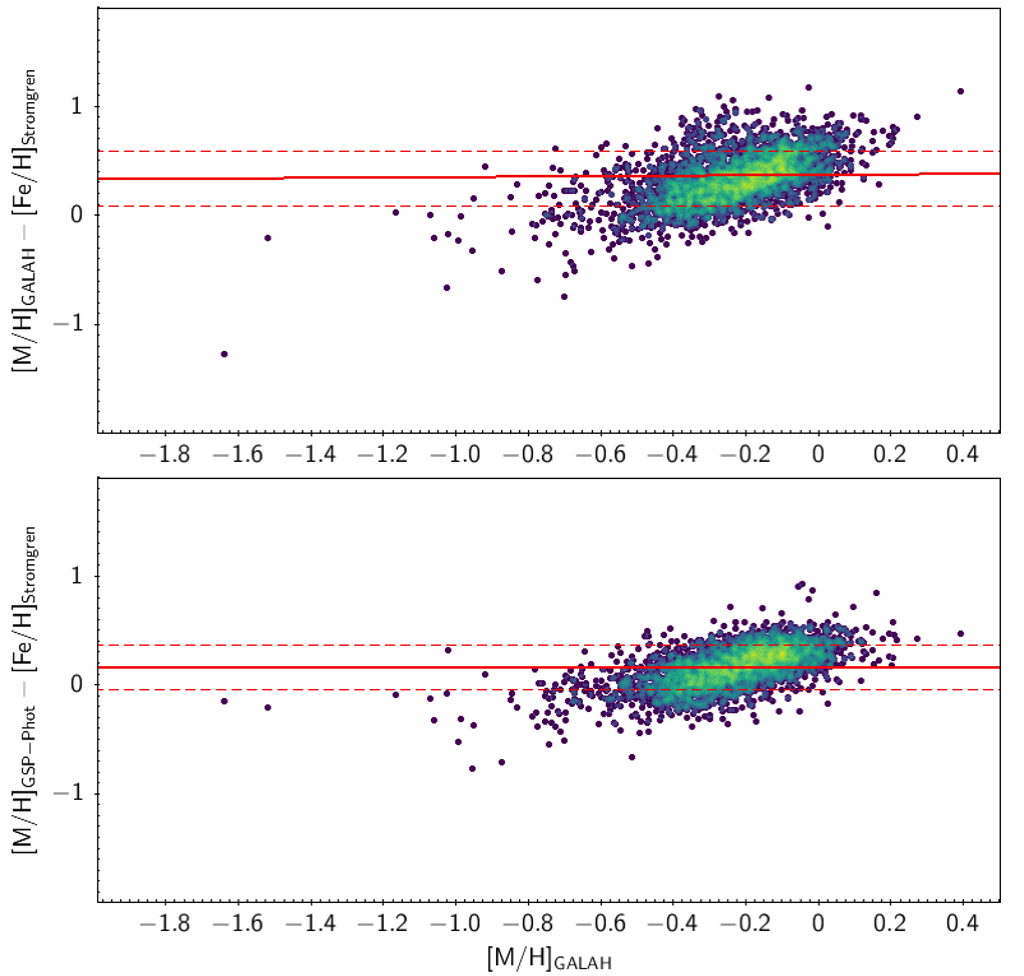}

\includegraphics[width=\columnwidth]{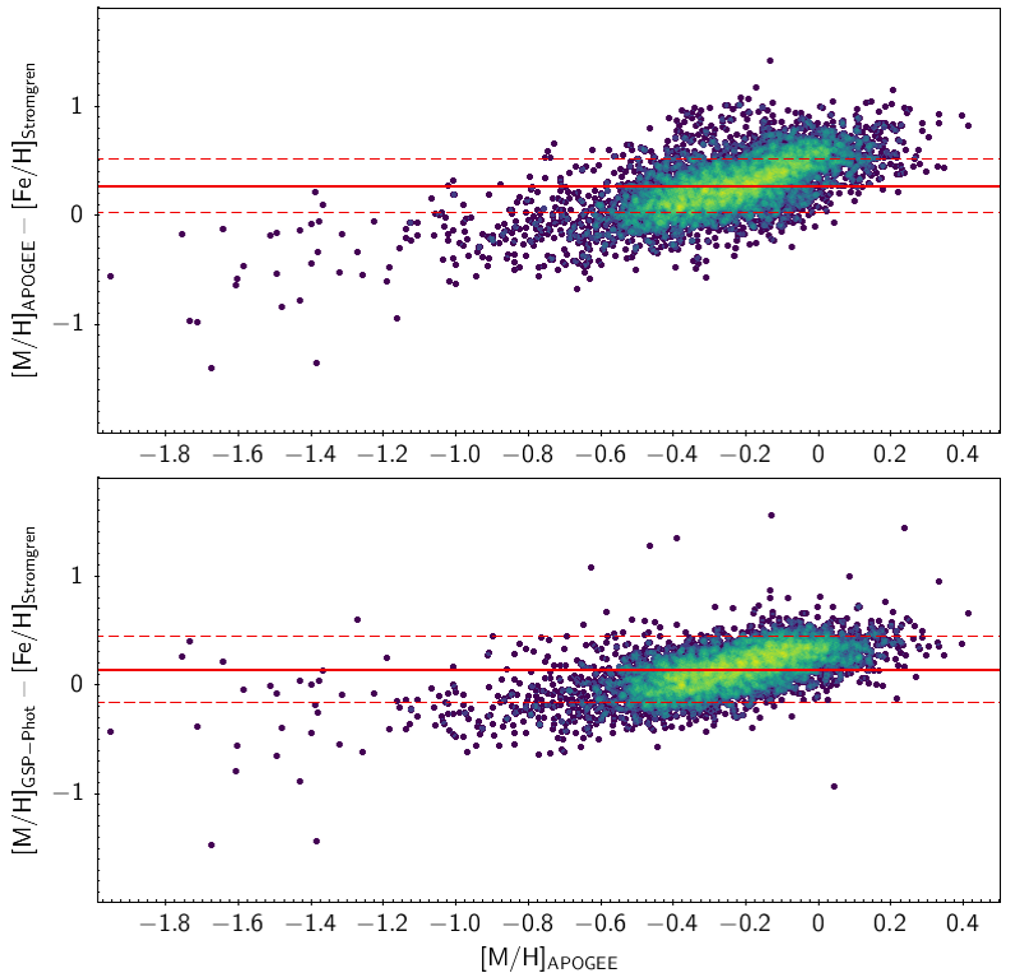}}
\caption{{\it Top panel}: Difference between the spectroscopic metallicity from GALAH DR3 and that coming from our synthetic Str\"omgren $m_0$ index. The red lines mark the mean value and the 1$\sigma$ dispersion. {\it Second panel}: Same but with {\it Gaia} GSP-Phot metallicty instead of that from GALAH.
{\it Third panel}: Difference between the spectroscopic metallicity from APOGEE DR16 and that coming from our synthetic Str\"omgren $m_0$ index. The red lines mark the mean value and the 1$\sigma$ dispersion. {\it Bottom panel}: Same but with {\it Gaia} GSP-Phot metallicity instead of that from APOGEE.} 
\label{fig:diff_galah_apo}
\end{figure}

The upper panel of Fig.~\ref{fig:diff_galah_apo} shows the difference between the GALAH spectroscopic metallicity and that derived from the Str\"omgren $m_0$ index, as a function of the former. Overall, the agreement looks reasonable. The mean value of the distribution is $\Delta[$Fe/H$]=0.33$, with a dispersion of $\sigma=0.25$. The distribution itself shows a positive trend for higher metallicity and flattens for [Fe/H]$_{\rm GALAH}<-0.5$.
Such a trend is consistent with the age effect that has already been observed and discussed for the star clusters (see Figure \ref{fig:all_strom_met}). More metal-rich stars are likely among the youngest of the sample, and the Calamida relation tends to underestimate their metallicity, while still performing reasonably well for the  stars that are older and more metal-poor. Unfortunately, the GALAH sample is intrinsically lacking in metal-deficient stars, meaning that we cannot robustly test the behaviour at lower metallicity. For the sake of cross-validation, the second-row panel of Fig.~\ref{fig:diff_galah_apo} shows the difference between the photometric metallicity estimated from XP spectra by {\it Gaia} GSP-Phot \citep{DR3-DPACP-156} and that coming from the synthetic Str\"omgren photometry, as a function of [Fe/H]$_{\rm GALAH}$. In this case, the agreement is even better, with a mean difference of $\Delta[$Fe/H$]=0.16$ and a somewhat tighter dispersion $\sigma=0.20$.

The lower panels of Figure~\ref{fig:diff_galah_apo} show the same kind of comparison,  with the metallicity measurements coming from APOGEE. The behaviour is very similar to that described for GALAH. The mean difference between the spectroscopic and the Str\"omgren metallicity is $\Delta[$Fe/H$]=0.27$, with a dispersion of $\sigma=0.25$ and a similar positive trend. As in the previous case, the comparison improves when using Gaia GSP-Phot measurements as a reference, as the mean difference drops to $\Delta[$Fe/H$]=0.14$ with a dispersion of $\sigma=0.30$.
Consistency within $\simeq \pm 0.2-0.3$~dex is also achieved in comparison to {\it Gaia} GSP-Phot metallicity.

Our analysis is particularly relevant in the case of distant sources, especially in the metal-poor regime. These are the cases where our primary source of metallicity from XP spectra encounters some limitations, while otherwise, GSP-Phot provides  astrophysical parameters with good accuracy for the large majority of stars \citep{DR3-DPACP-156,DR3-DPACP-160}.  For example, when estimating the mean metallicity of the GCs analysed above, GSP-Phot provides values that tend to significantly overestimate the metal content of these stellar systems. Hence, when robust estimates of the extinction exist,
metallicity-sensitive distance-independent colour indices obtained from synthetic photometry from XP spectra, like that presented here and in the following sections, can provide an useful alternative solution that is highly complementary to GSP-Phot.

\subsection{Metallicity from the C1 system}
\label{sec:metalStromgrenC1}

Metallicity and $\alpha$-element abundance information is more difficult to retrieve than temperature and surface gravity (see \secref{Gaia2C1}). Abundances leave an imprint in the spectra in narrow ranges of wavelength, and narrow passbands are more sensitive to fluctuations in the spectra.
Nevertheless, metallicity and even the $\alpha$-element abundance can be studied with the C1 system.

\begin{figure}[!htbp]
\center{
\includegraphics[width=0.9\columnwidth]{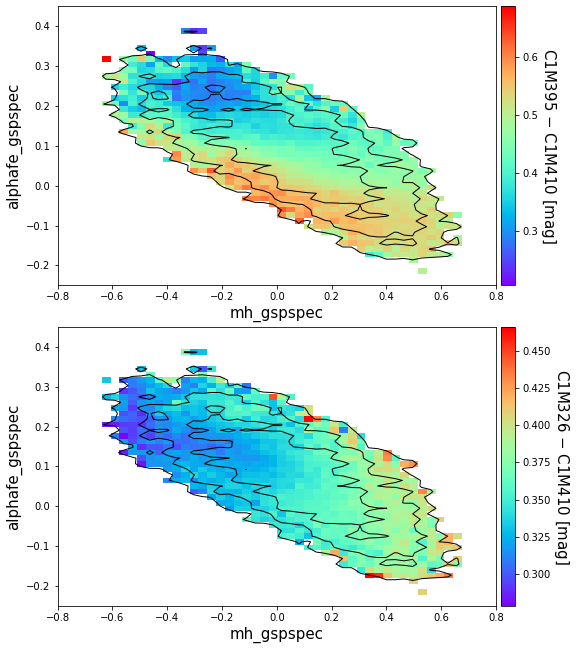}
}
\caption{
$\alpha$-element abundance ([$\alpha$/Fe]) as a function of global metallicity ([M/H]) of the main sequence stars ($\log g>4.0$~dex) and good quality flags (the first 13 digits in \texttt{flags\_gspspec} equal to zero). All parameters were derived by GSP-Spec \citep{DR3-DPACP-186}. Colour indices show the values of the colour C1M395$-$C1M410 (top panel), which is changing due to the $\alpha$ abundance and C1M326$-$C1M410 (bottom panel), which depends more on the global metallicity. Contours indicate density dropping by factors of 5.
\label{fig:C1-vs-GSPSpec-metallicity}
} 
\end{figure}

For example, 
\figref{C1-vs-GSPSpec-metallicity}, where spectroscopic metallicities and [$\alpha$/Fe] values are taken from GSP-spec\footnote{GSP-spec is the DPAC/CU8 Apsis module designed to derive chemical abundances from \gaia RVS spectra \citep{DR3-DPACP-186}. Here, to select well-measured abundances we considered only stars with the first 13 digits in \texttt{flags\_gspspec} equal to zero. We note that very similar results as those shown in \figref{C1-vs-GSPSpec-metallicity} are obtained if APOGEE DR16 abundances are used instead of GSP-spec ones.} \citep{DR3-DPACP-186}, shows
that when the pseudo-continuum at C1M410 is compared with C1M395 affected by CaII HK lines, we see a dependence with [$\alpha$/Fe] abundance. On the other hand, when the same C1M410 pseudo-continuum passband is compared with C1M326 (measuring the UV Balmer jump), this colour index has a stronger variation with metallicity ([M/H]) than with [$\alpha$/Fe] abundance.
It is interesting to note that these passbands are able to trace chemical composition in spite of the fact that they sample a critical region of the XP spectra.

\begin{figure}[!htbp]
\center{
\includegraphics[width=0.9\columnwidth]{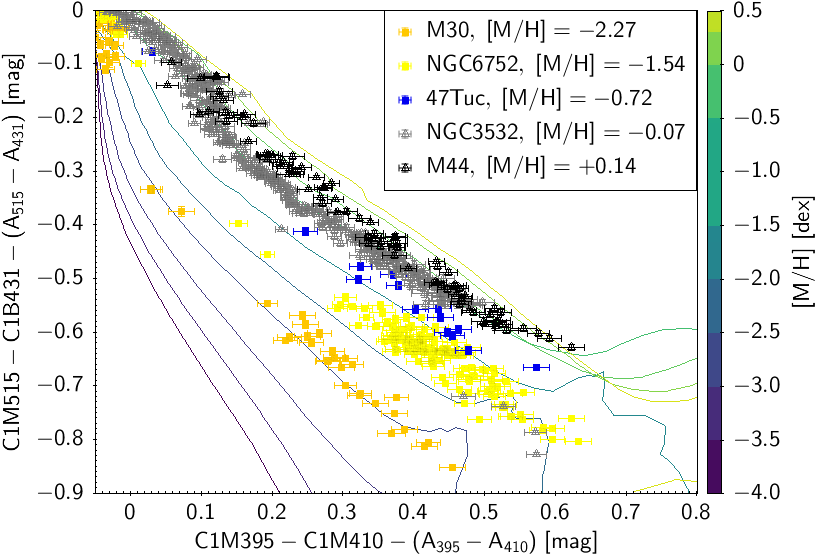}
}
\caption{C1 colour--colour diagram sensitive to global metallicity for a set of clusters corrected for reddening using $A_{X}$ absorption values in that passband derived as indicated in \appref{app_dereddening}.
Lines represent the simulations performed using the BTSettl library \citep{BTSettl} (with a line colour depending on the global metallicity, [M/H]). 
Solid squares represent the stars in GCs and empty triangles the stars in OCs, all of them with their error bars.
 BTSettl models with $\log g=2.0$ are plotted for GCs, and $\log g=3.0$ for OCs. 
\label{fig:clustersmetalC1}
} 
\end{figure}

Using well-studied open and globular clusters we can test the relationship between chemical abundances and C1M colour indices. \afigref{clustersmetalC1} shows the colour C1M515$-$C1B431 plotted against C1M395$-$C1M410, which is able to separate different metallicities. Only sources with total uncertainty $\sigma_C<0.02$~mag were plotted in the figure, where

\begin{equation}
\label{eq:sigmaC}
\sigma_C\equiv \sqrt{\sigma_{\rm C1M395}^2+\sigma_{\rm C1M410}^2+\sigma_{\rm C1M515}^2+\sigma_{\rm C1B431}^2}
.\end{equation}

The C1 synthetic photometry in \figref{clustersmetalC1} has been corrected from reddening effects using the relationships included in \appref{app_dereddening}. The absorption values used to perform this correction were obtained from literature estimates (\citealt{harris1996} for M30, NGC~6752, and NGC~104 GCs, \citealt{fritzewski2019} for NGC~3532, and \citealt{taylor2006} for M44 OCs).

The lines in \figref{clustersmetalC1} show the iso-metallicity lines derived from the BTSettl library  \citep{BTSettl}. In the low metallicity range, the BTSettl lines were derived using a surface gravity value equal to $\log g=2.0$, as only the giant stars in the GCs can be observed with enough accuracy in this colour--colour diagram. For higher metallicity, a value equal to $\log g=3.0$ was used. 
As can be seen in \figref{clustersmetalC1}, each cluster follows a  metallicity track. Therefore, we can conclude that, as we have also seen for temperature and surface gravity, 
the XP spectra (and therefore the synthetic photometry) allow us to discriminate between the abundance effects present in the spectra. 

\begin{figure}[!htbp]
\center{
\includegraphics[width=0.9\columnwidth]{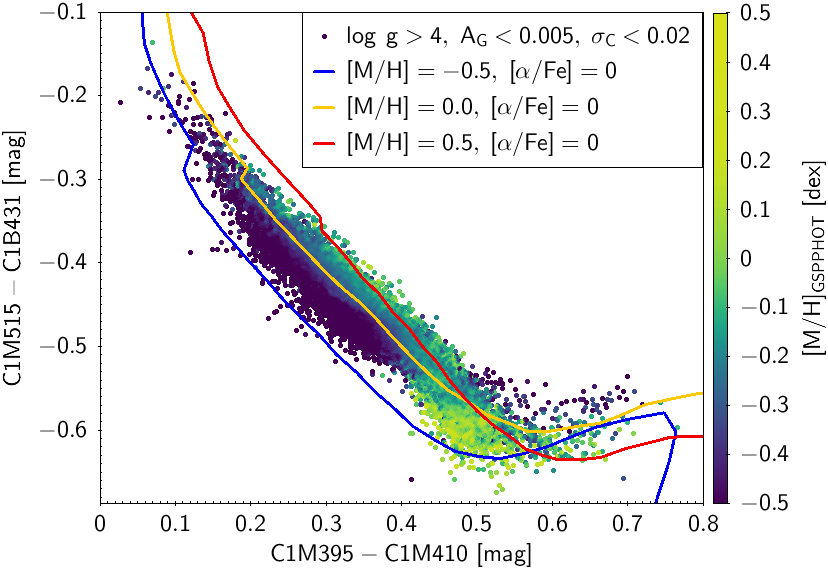}
}
\caption{C1 colour-colour diagram sensitive to global metallicity for a set of field stars. Lines in the figure represent the simulated photometry from BTSettl SEDs, plotting only solar alpha abundance ([$\alpha$/Fe]$=0$) for $\log\ g=4$. Dots represent the synthetic photometry derived from {\xp} spectra with a colour index showing the DPAC CU8 metallicity derived using the GSP-Phot algorithm (mh\_GSP-Phot).
\label{fig:fieldmetalC1}
} 
\end{figure}

The same diagram used for clusters in \figref{clustersmetalC1} can also be used for field stars. In \figref{fieldmetalC1} we show an example of this with a set of sources selected to compare with the results obtained with GSP-Phot \citep{DR3-DPACP-156}. We include sources with $\log g>4$~dex, $A_G<0.005$~mag, and $\sigma_C<0.02$~mag from \equref{sigmaC}. The results of this cross-validation test are satisfactory.

\subsection{Very metal-poor stars}
\label{sec:vmp}

In this section, we push the metallicity analysis one step further into the lower metallicity star regimes, that is, [Fe/H] $<$ -2\,dex. The more metal-poor the star, the more pristine it is. Ultra-metal poor (UMP) stars ([Fe/H] $<$ -4\,dex) belong to the earliest generations of stars formed in the Universe. Because of their very low abundance in metal elements, they are critical anchors to address questions on the formation of the first generation of stars, the (non-)universality of the initial mass function (IMF), the early formation stages of galaxies, and the first supernovae \citep[e.g.][]{2005ARA&A..43..531B}. However, the minimum metallicity at which low-mass stars can form is still an open question (see \citealt{2015ComAC...2....3G}, and references therein). Only 42  UMPs are known to date in our Galaxy despite simulations predicting multiple thousands of them \citep{2013RvMP...85..809K}. These stars are scarce objects, and are relatively faint sources because of their low masses. Finding them is therefore a challenge, and we are limited to mostly finding them in our Galaxy.

\begin{figure}
    \centering
    \includegraphics[width=0.475\textwidth]{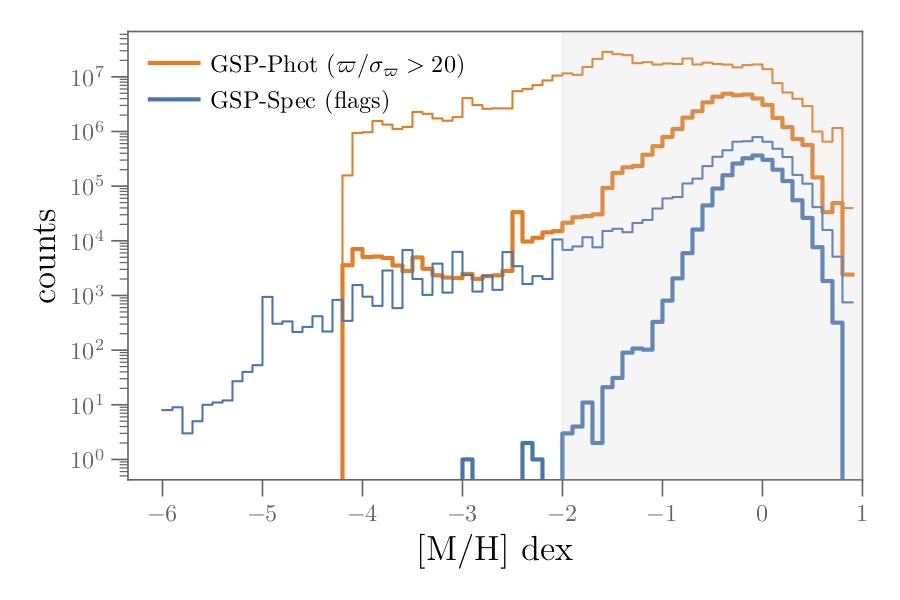}
    \caption{Distribution of [M/H] estimates in \gdr{3} from GSP-Phot (orange) and GSP-Spec (blue). As recommended in \citet{DR3-DPACP-186} and \citet{DR3-DPACP-156}, we selected the GSP-Spec estimates that have their 13 first bits flagged as equal to zero and GSP-Phot estimates with good parallax S/N. Nevertheless, we indicate the full distribution with the thin blue and orange lines, respectively. \citet{DR3-DPACP-156} suggest caution be exercised when using their estimates below [M/H] $\sim$ -2\,dex as their uncertainties are large. We shaded the region above the metal-poor regime. Overall the \gdr{3} astrophysical parameter estimates have limited coverage of metal-poor stars: there are only four stars with [M/H] < -2\,dex after using the flags from GSP-Spec and of unclear quality for GSP-Phot.}
    \label{fig:mh_cu8}
\end{figure}

One could imagine that \gdr{3} could be key to unlocking a systematic and efficient search for metal-poor candidates either from the metallicity estimates or from the spectra. However, the \gdr{3} astrophysical parameter estimates have limited power in finding these stars for the following reasons: (i) Metallicity is a weak signal in the XP spectra and significant limiting factors hamper the extraction of metallicity parameters from BP and RP \citep[e.g.][]{DR3-DPACP-157, DR3-DPACP-156}. (ii) The stellar atmosphere and evolution models have a limited calibration in these regimes making the absolute scale of [Fe/H] possibly biased. In particular, most metal-poor isochrones publicly available in this regime are solar-scaled $\alpha$-abundances. Indeed, Fig.~\ref{fig:mh_cu8} suggests that \gdr{3}  contains relatively few reliable estimates of chemical abundances for metal-poor stars. From the medium-resolution spectroscopy (radial velocity spectrometer (RVS) $[845-872]$\,nm, $\lambda/\Delta\lambda\sim ~ 11500$), GSP-Spec measured multiple iron lines to provide us with a [Fe/M] and a global model fit [M/H] parameter estimate for millions of stars.  However, \citet{DR3-DPACP-186} strongly advocated filtering on the flags (13 first bits \texttt{astrophysical\_parameters.flags\_gspspec} equal to zero), which leaves us with only a handful of metal-poor stars (thick blue line in Fig.~\ref{fig:mh_cu8}). We note that these authors also suggested that adjusting this filtering could provide a few thousand stars, but with large uncertainties. 
GSP-Phot also produced [M/H] estimates from the analysis of a combination of BP/RP spectra, parallax, and $G$ magnitude.  \citet{DR3-DPACP-156} on the other hand warned that caution should be exercised when using any values below [M/H]$\sim-2$~dex.

As the \gdr{3} APs are unreliable in this regime, we can take a step back and use XP spectra through XPSP. Indeed, survey programs dedicated to finding metal-poor stars ([Fe/H] < -2\,dex) use some special pre-selection through prism techniques (e.g. the HK and HES surveys; \citealt{1985AJ.....90.2089B}; \citealt{2002A&A...391..397C}) or narrow-band photometry (such as the SkyMapper and Pristine survey programs; \citealt{2018PASA...35...10W}, \citealt{2017MNRAS.471.2587S}). Other stars were discovered in blind but spectroscopic surveys such as SDSS/SEGUE/BOSS (\citealt{2000AJ....120.1579Y}; \citealt{2009AJ....137.4377Y}; \citealt{2011AJ....142...72E}) or LAMOST \citep{2012RAA....12.1197C}. Such endeavours are expensive in terms of telescope time (especially spectroscopic observations) and are not always fruitful.

As a first experiment, we selected from the Pristine survey public catalogue \citep{2019MNRAS.490.2241A} the stars with spectroscopic confirmation of [Fe/H] < -2\,dex and with \xp spectra in \gdr{3}. This represents $48$ stars out of $636$ ($\sim$7\%). The Pristine survey technique to find these stars consists in combining a custom-built CaII\,H and K narrow passband ($\sim$ 10\,nm wide) for the MegaCam wide-field imager on the 3.6-m  Canadian-France-Hawaii Telescope (CFHT) with existing broad-band photometry from SDSS \citep{2017MNRAS.471.2587S}.  This passband covers the wavelengths of the Ca H and K doublet lines (at 396.85 and 393.37nm), which are very sensitive to abundance variations, especially [Fe/H] but also carbon. Their success rate of uncovering stars with [Fe/H] < -2\,dex is 85\%, but is only 25\% for stars with [Fe/H] < -3\,dex. Facing this incredible efficiency, multiple surveys are now exploring the adoption of the same passband (e.g. the Dark Energy Camera; DECam\footnote{\url{https://noirlab.edu/science/news/announcements/sci22020}}
; J-PLUS \citealt{2021A&A...654A..61L}).

\begin{figure}
    \centering
    \includegraphics[width=0.475\textwidth]{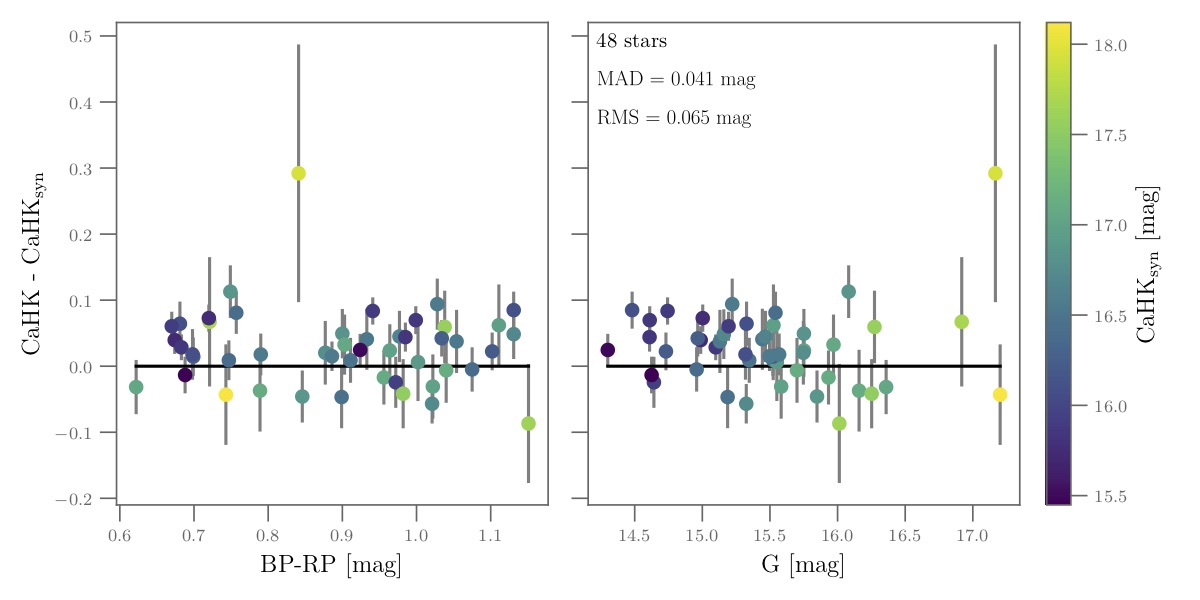}
    \caption{Residuals of the synthetic photometry of Ca\,H and K for a sample of $48$ stars from \citet{2019MNRAS.490.2241A} as a function of {\bprp} colour and $G$ magnitude, left and right panels, respectively. These stars have spectroscopic [Fe/H] < -2\,dex. The colours of the symbols reflect their CaHK synthetic magnitudes. We highlight the zero deviation line in grey. The residuals do not seem to correlate with {\bprp} colour.}
    \label{fig:cahk_residuals}
\end{figure}

 The width of the CaHK passband satisfies the criterion for flux conservation in the presence of strong spectral features of Eq.~\ref{eq:Rf_definition}, with 
$Rf=1.5$. Figure~\ref{fig:cahk_residuals} demonstrates the reliability of the XPSP on these objects. We find an overall MAD of $0.04$\,mag, but this is unsurprisingly dependent on the apparent magnitude of the stars; however, it seems independent of their $G_{\rm BP}-G_{\rm RP}$ colours.

Similarly to \citet{2017MNRAS.471.2587S}, we applied a regularised linear regression of the synthetic standardised SDSS ugri, the CaHK, to predict their photometric metallicities. We find a root mean square (RMS) of 0.3\,dex and a mean absolute error (MAE) of 0.2\,dex. However, the sample is small, which limits the comparison. 

\begin{figure}
    \centering
    \includegraphics[width=0.475\textwidth]{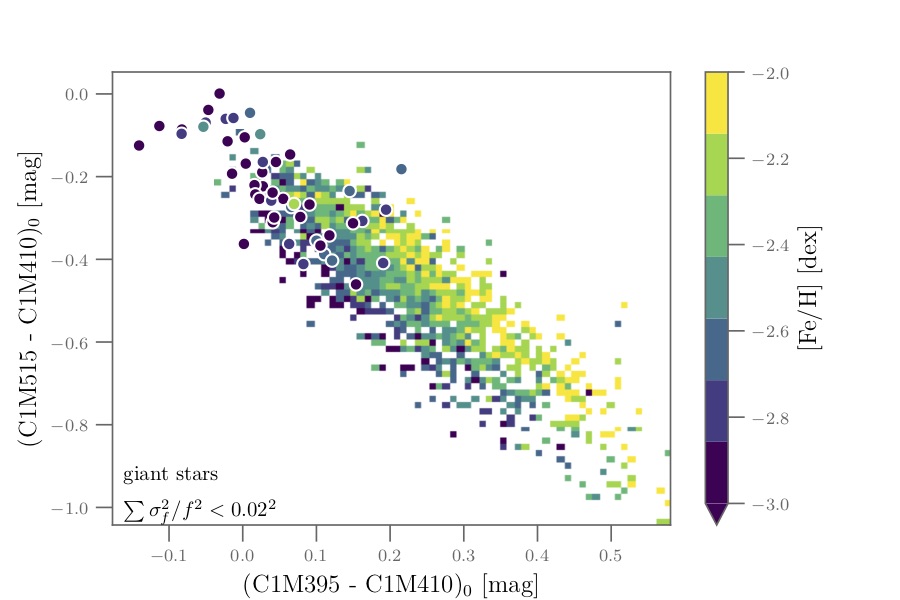}
    \caption{C1 colour-colour diagram sensitive to metallicity and corrected for extinction.
    The histogram shows the distribution of a random subset of $5\,112$ giant stars with [Fe/H]$< -2$\,dex from \citet{2022ApJ...925..164H} with a total fractional flux uncertainty in the C1 bands of below $0.02$. The colours refer to their photometric iron abundance estimates with the scale on the right-hand size. The round symbols indicate the $48$ stars (not only giants) from Pristine discussed in Sect.~\ref{sec:vmp} on the same scale using their spectroscopic estimates. The x and y axes are equivalent to CaII and MgH indices, respectively.
    }
    \label{fig:c1m_vmp}
\end{figure} 

In contrast with using a tailored passband, \citet{2022ApJ...925..164H} exploited the SkyMapper photometry (SMSS DR2) and the {\gedr} photometry and astrometry to estimate the metallicities (and other APs) of 20 million stars, and in particular half a million very metal-poor stars. Their method to estimate APs exploits multiple colour relations and often depends on distinguishing giants and dwarfs. 
Of those, we extracted a random sample of about $26\,000$ giant stars with XP spectra in \gdr{3}, which provided us with [Fe/H] and $A_0$ estimates. We used the {\gedr} extinction relations to obtain $A_G$ and further the relations from Appendix \ref{sec:app_dereddening} to obtain the coefficients in the C1 bands.

Figure~\ref{fig:c1m_vmp} shows in the (C1M396 -- C1M410) versus (C1M515 -- C1M410) colours the stars that have a total fractional uncertainty of 
\begin{equation}
\frac{\sigma_f}{f} \equiv \sqrt{\left(\frac{\sigma_{\rm C1M395}}{f_{\rm C1M395}}\right)^2
+ \left(\frac{\sigma_{\rm C1M410}}{f_{\rm C1M410}}\right)^2
+ \left(\frac{\sigma_{\rm C1M515}}{f_{\rm C1M515}}\right)^2} < 0.02.
\end{equation}
The metallicity gradient is strikingly visible, demonstrating the reliability of the the C1-based indices even in the very metal-poor regime. 

We remark that (C1M396 -- C1M410) colour is nearly equivalent to the (CaHK--C1M410) colour. 
Therefore, we indicated on Fig.~\ref{fig:c1m_vmp} the previously mentioned Pristine stars for comparison. The latter are not specifically giant stars, but are mostly turn-off stars, and therefore concentrate in the top left corner of the plot. However, they also agree with the scale from Pristine. 
This comparison allows us to draw the conclusion that XPSP can transfer knowledge from SkyMapper to Pristine (or the other way around if the latter sample is larger). In particular, this means a common metallicity scale, which is often an issue when comparing surveys. Finally, the XP spectra will offer a significantly large suite of passbands to explore metallicity estimates in a very new manner across the entire sky.

One major limitation of the XPSP is that very metal-poor stars are intrinsically faint in BP. As \gdr{3} limits the availability of the stellar XP spectra to $G < 17.65$\,mag, a large fraction of the sources in \citet{2019MNRAS.490.2241A} and \citet{2022ApJ...925..164H} remain beyond the reach of \gdr{3}. However, XPSP data offer a robust set of photometric calibrators across the entire sky and therefore a larger common ground to transfer knowledge between surveys.

\subsection{Classification of emission line sources}

\begin{figure}
    \centering
    \includegraphics[width=0.5\textwidth]{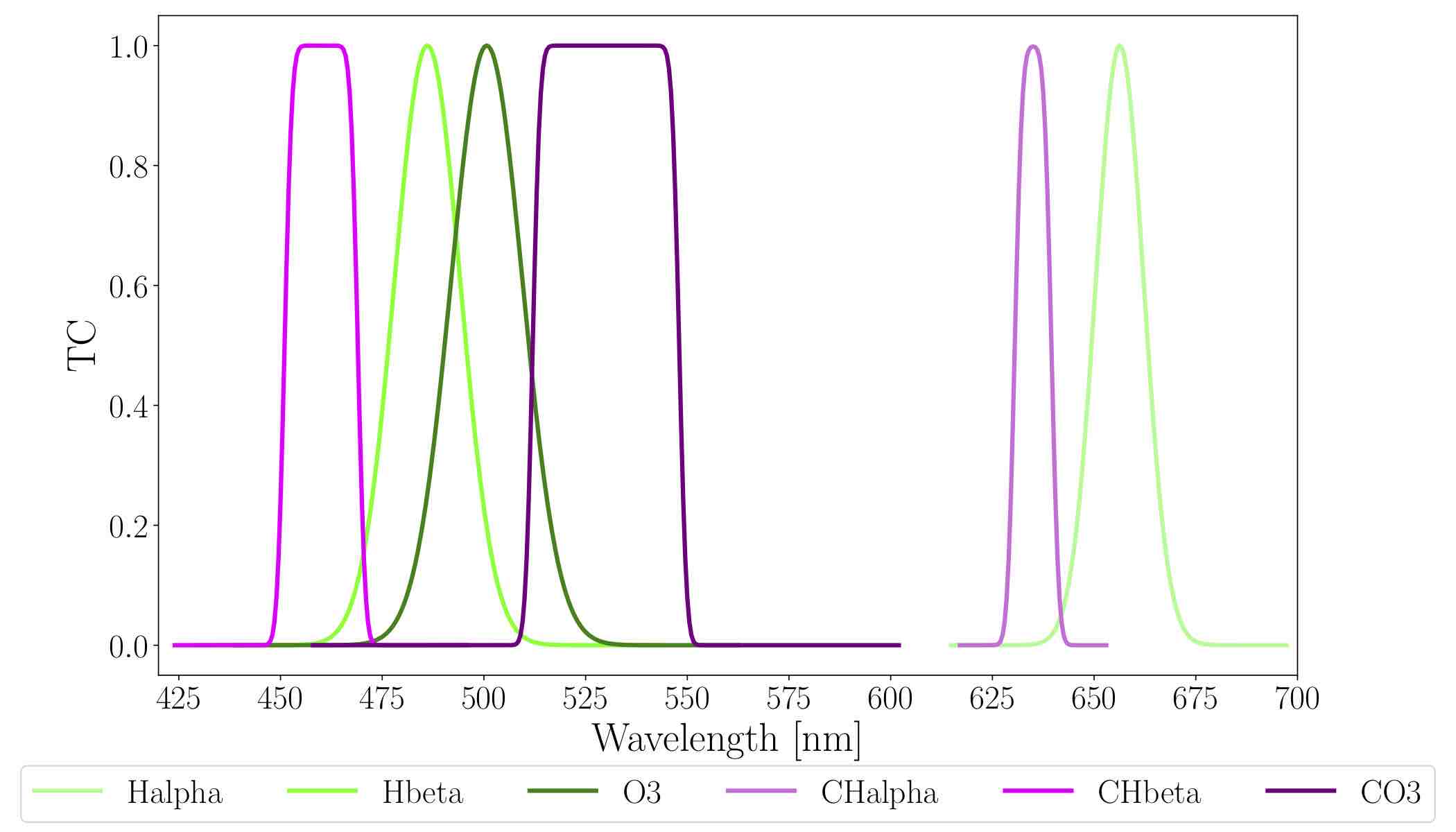}
    \caption{ELS\_custom\_w09\_s2 normalised transmission curves.}
    \label{fig:iphas_profiles}
\end{figure}

Among the algorithms devoted to the analysis of XP spectra for \gaia DR3, the ESP-ELS Apsis module is designed to identify six classes of ELS: Be stars, Herbig Ae-Be stars, T Tauri stars, active M dwarf stars, Wolf-Rayet (WR) stars, and planetary nebulae \citep[PNe;][]{DR3-DPACP-160}. The selection and classification is based on the use of two Random Forest classifiers trained on libraries of synthetic spectra as well as on observed BP/RP data obtained for a sample of reference ELSs (see detailed description of ESP-ELS in online documentation). To study the extent to which ESP-ELS results can be reproduced using XP-based synthetic photometry, we chose a custom system, {\tt ELS\_custom\_w09\_s2}\footnote{This newly defined system is included in the list that can be used to get XPSP by means of GaiaXPy (Sect.~\ref{sec:gaiaXPy}).}, which is illustrated in Fig.~\ref{fig:iphas_profiles}. This latter is composed of three narrow passbands with Gaussian shape and located at the rest frame wavelength of the H$_\beta$, [OIII] 5007, and H$_\alpha$ lines. It is complemented with three passbands aimed at sampling the
continuum in spectral regions adjacent to the lines of interest and with minimal contamination from other emission lines, and by the wide  SDSS $r$ and $i$ bands. All the TCs have $R_f\geq1.4$ (see Appendix~\ref{sec:app_Rf}).

\begin{figure}
    \centering
    \includegraphics[width=0.51\textwidth]{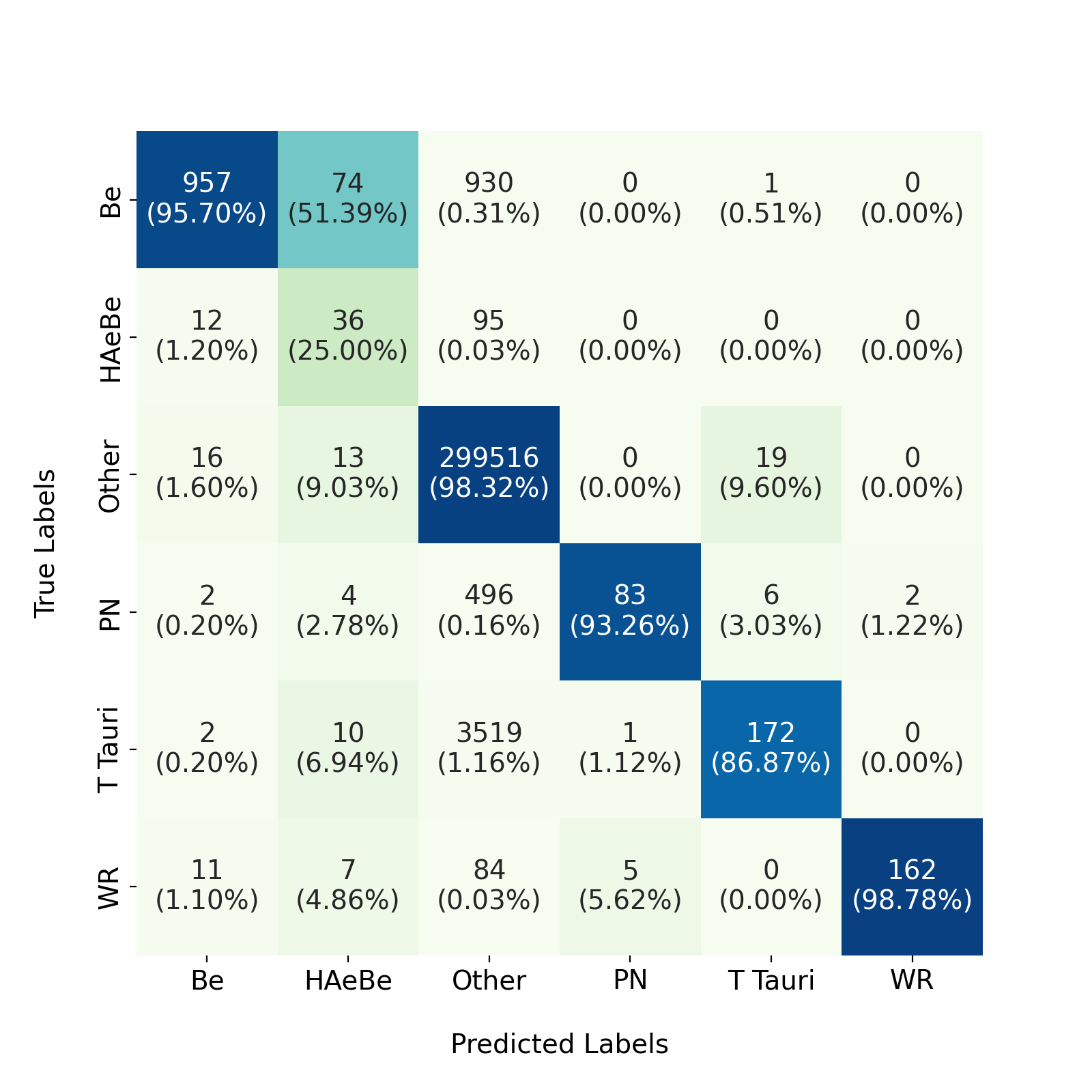}
    \caption{Confusion matrix between ESP-ELS classification (predicted label) and Simbad (true label). For each class, the percentages refer to the fraction of true positives with respect to the total number of objects predicted by ESP-ELS for such a class (precision).}
    \label{fig:simbad_cm}
\end{figure}

In order to have a reference sample, we extracted ELS from the SIMBAD\footnote{\url{http://simbad.u-strasbg.fr/simbad/}} database and obtained a total of 
1962 Be stars, 143 Herbig Ae/Be objects, 3704 T Tauri stars, 269 WR stars, and 593 PNe, while active M dwarfs are not considered in our experiment. We also selected 102763 targets with no ELS classification and 196801 targets randomly taken from the IPHAS catalogue, taken from \citet{scaringi2018}, with no overlap between them. All of these stars are labelled `Other' and represent normal non-emitting stars.
We then cross-matched the sources with eDR3 using the CDS x-matcher, and selected the closest target within $1.0\arcsec$. 
The comparison between the classification provided by Simbad and that by ESP-ELS is shown in Figure \ref{fig:simbad_cm}. The precision of ESP-ELS is excellent, between 87\% and 99\% in the different classes, except for the Herbig Ae/Be stars, for which ESP-ELS correctly classify only 25\% of the predicted sources. Despite the good results, the number of ELSs predicted as NO-ELSs demonstrates the quite conservative approach adopted by ESP-ELS; there are 5124 objects that are labelled as non-emitters, whereas  these objects are ELSs according
to SIMBAD.

\begin{figure}[hb!]
    \centering
    \includegraphics[width=0.5\columnwidth]{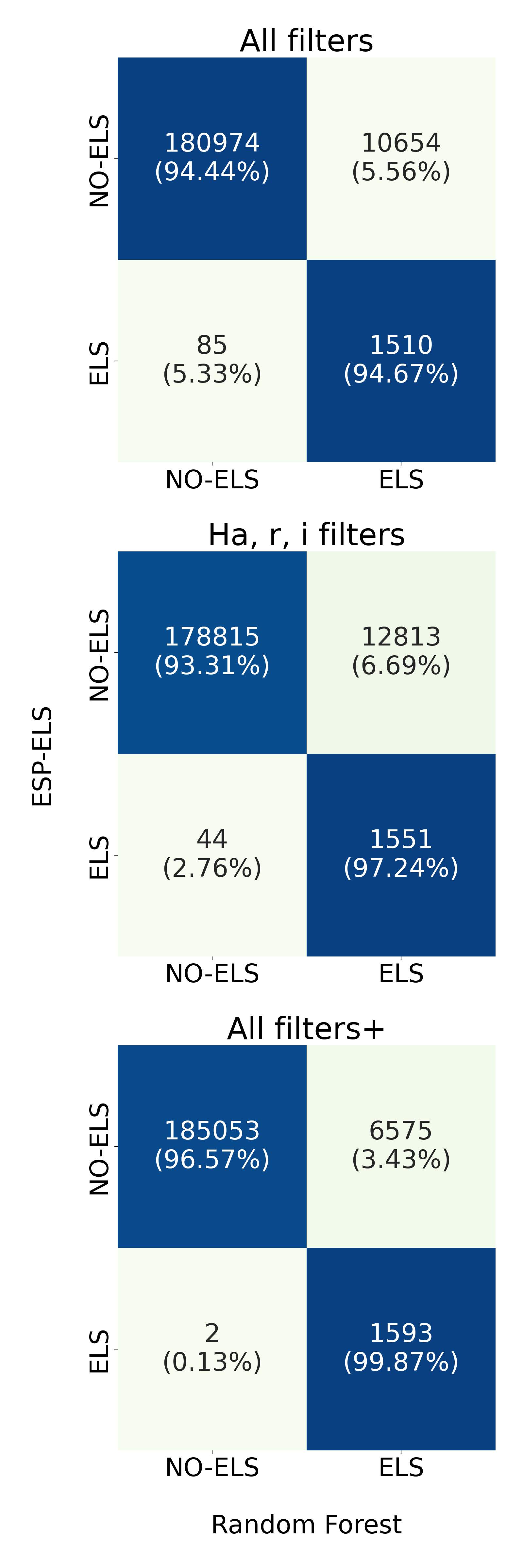}
    \caption{\label{fig:RF-CM1}Confusion matrices obtained using a Random Forest algorithm to separate between ESP-ELS emitting and non-emitting objects for experiments 1, 2, and 3 (see text for details). For each class, the percentages refer to the fraction of true positives with respect to the total number of objects predicted by ESP-ELS for such a class (recall).}
    \label{fig:rf_cm}
\end{figure}

\begin{figure*}[h!]
    \centering
    \includegraphics[width=\textwidth]{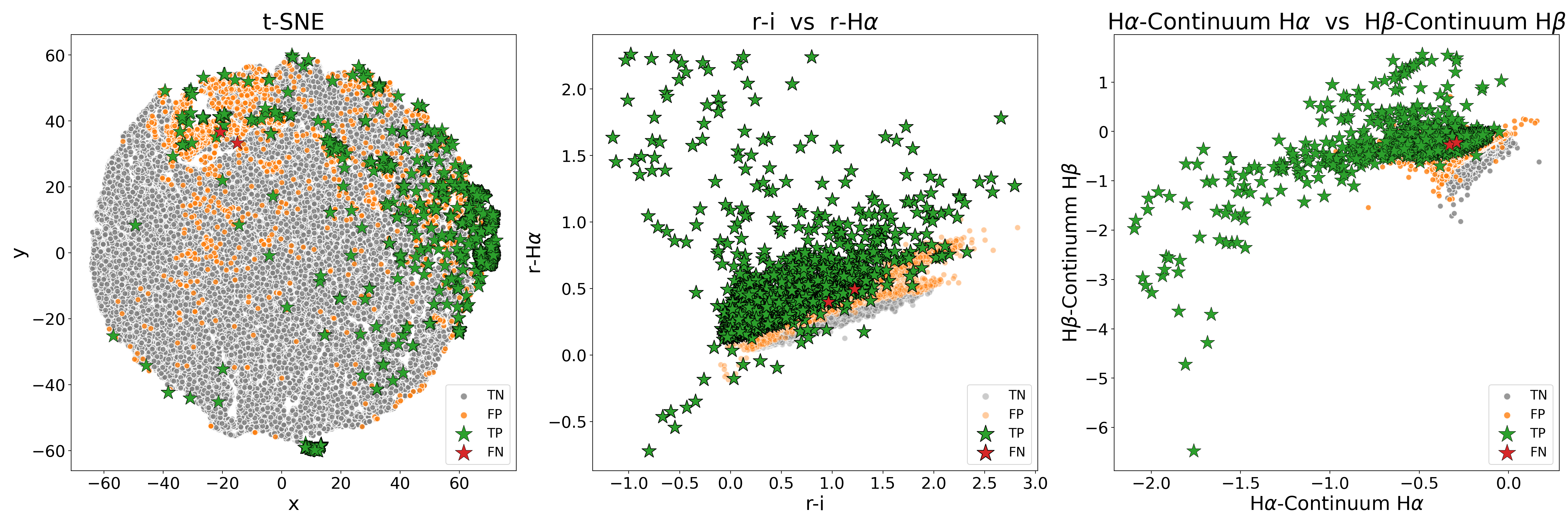}
    \caption{t-SNE and colour--colour diagrams representing the sample of emitting and non-emitting stars, classified     using a Random Forest algorithm on XP synthetic photometry (experiment 3). Legend as follows: TP: true positives; TN:     true negative; FP: false positive; FN: false negative.}
    \label{fig:tsne_rf_tags}
\end{figure*}

 The following validation datasets were selected in order to cross-validate ESP-ELS classification results and, at the same
time, demonstrate the classification capabilities of the chosen photometric
system: (i) the selection of ELS described above, which add up to a total of 6671 objects; and (ii) the 196801 IPHAS targets  also described above, representing non-emitting stars (IPHAS sample, hereafter). Specifically, we wish to decipher the degree to which a classification based on narrow-band XPSP reproduces the results by ESP-ELS. For simplicity, we limit to the classification between ELS and non-ELS.

We combine a supervised method  to classify the objects, namely Random Forest, with an unsupervised algorithm, t-SNE \citep{maaten2008}, to group and picture the classification results. We also show two colour--colour diagnostic plots to visualise them.
Three experiments were performed on different inputs:
 \begin{enumerate}
  \item {All filters in {\tt ELS\_custom\_w09\_s2} synthetic photometry.}
  \item {{\tt ELS\_custom\_w09\_s2} synthetic photometry except bands for H$_\beta$, O3, and their respective continua.}
  \item {{\tt ELS\_custom\_w09\_s2} synthetic photometry plus information from several band combinations:  r- H$_\alpha$, r-i, H$_\alpha$-H$_{\alpha cont}$, and H$_\beta$-H$_{\beta cont}$}.
\end{enumerate}
We divided the objects into a training set composed of 5124 ELSs (labelled as non-emitters by ESP-ELS, but found to be true emitters in Simbad) and the same number of non-ELSs randomly taken from our IPHAS sample. Once trained, Random Forest was tested on the remaining ESP-ELS emission line stars (1595 objects) plus the remaining IPHAS objects.
The confusion matrices obtained for experiments 1, 2, and 3 are shown in Figure \ref{fig:rf_cm}. In this case, the fractions reported in the confusion matrices are not the {\em precision} (the number of true positives divided by the number of objects predicted in that class) as in Fig.~\ref{fig:simbad_cm}, but the {\em recall}, that is, the fraction of predictions matching the classification taken as `truth'; in this case the ESP-ELS classification\footnote{Adopting the notation introduced below, $precision=TP/(TP+FP)$, and $recall=TP/(TP+FN)$.}

The confusion between classes is below 7.1\% in both emitting and non-emitting objects. This indicates that all the tested combinations of synthetic passbands are well suited for the classification of ELSs.
By including the H$_\beta$ and O3 passbands, the number of false positives (FPs) diminishes (from 6\% to 2.8\%) but the number of false negatives (FNs) increases (from 5.5\% to 7.1\%).
The best classification results are obtained in experiment 3, where only two ELS are predicted as non-ELS, while keeping the number of FNs to only 3.5\%.
   
Subsequently, we applied the t-SNE algorithm to the validation data, obtaining a 2D representation of the whole sample. t-SNE is an unsupervised algorithm, and does not use the labels to group the data, which are clustered according to the similarity of the XPSP fluxes. Once the clustering is done, it can be visualised using the labels (true negatives (TNs), FNs, FPs, and true positives (TPs), which are obtained with the Random Forest for experiment 3, the one with the highest score in the confusion matrix. The results are shown in Figure \ref{fig:tsne_rf_tags}, where we also show two different colour--colour diagrams. Figure \ref{fig:tsne_rf_tags} is clearly dominated by TNs, with high confusion with TPs in the same area. Colour--colour diagrams are better suited to distinguishing the regions corresponding to each of the object classes.

In conclusion, through simple machine learning experiments and using both supervised (Random Forest) and unsupervised (t-SNE) algorithms, we show that the synthetic photometry in the system {\tt ELS\_custom\_w09\_s2} obtained from the Gaia XPSP is adequate to separate stars with emission lines from those that do not emit with a reliability that reproduces that achievable by the XP spectra themselves by the ESP-ELS module with errors below 3.5\% for FNs and as low as 0.1\% for FPs. The ability to go further and separate different classes of ELS objects strongly depends on the possibility to train the algorithms with sufficiently representative sets of each class and to use additional passbands as a possible way to improve the performance.

%%%%%%%%%%%%%%%%%%%%%%%%%%%%%%%%%%%%%%%%%%%%%%%% END PERFORMANCE VERIFICATION

%%%%%%%%%%%%%%%%%%%%%%%%%%%%%%%%%%%%%%%%%%%%%%%% PRODUCTS
%\input{sections/products.tex}

\section{Products}
\label{sec:products}

\subsection{How to get synthetic photometry in your preferred system}
\label{sec:gaiaXPy}

Synthetic photometry in all the photometric systems used throughout this paper can be generated from the \gdr{3} XP spectra served by the archive\footnote{\url{https://gea.esac.esa.int/archive/}} via Datalink \citep[see Sect. 4 in ][for further instructions]{EDR3-DPACP-118}. The GaiaXPy\footnote{\url{ https://gaia-dpci.github.io/GaiaXPy-website/}} Python package offers several utilities to help users to maximise the potential of \xp spectra. The generation of synthetic photometry in a number of predefined photometric systems is one of the available functionalities. This is achieved by a simple matrix multiplication of the array of coefficients defining the mean spectra by a design matrix which is generated taking into account the specific photonic TC. Contributions from both \xp spectra in the case of filters spanning the range covered by both are taken into account \citep{EDR3-DPACP-120}. Colour corrections for the UV bands of some of the standardised systems (see Sec. \ref{sec:ustand}) and uncertainty correction factors (see Sec. \ref{sec:errorcorrections}) are also optionally available. To obtain standardised photometry, the properly tweaked passbands  must be used (see Sect.~\ref{sec:nonustand}), being denoted with {\tt \_STD} in their name.

GaiaXPy allows synthetic photometry to be generated in any of the available photometric systems or in a list of those in a single call. Users can either provide a list of {\tt source\_id} or input the XP spectra as downloaded from the \gaia archive in their continuous representation and in all file formats currently offered for their download. For updated and detailed instructions, readers are referred to the package documentation. 

New photometric systems can be added to those already available in the latest release of GaiaXPy (see the GaiaXPy web page). However, synthetic photometry in any system can also be obtained from EC XP spectra in the usual way (Equations~1 to 5; Sect.~\ref{sec:methods}) by any user of the \gaia archive, without the need for computing new basis functions.

\subsection{The Gaia Synthetic Photometry Catalogue}
\label{sec:gspc}

To make XPSP more readily available in the most widely used photometric systems, we produced the Gaia Synthetic Photometry Catalogue (GSPC), which includes the vast majority of the approximately $220$ million stars with XP spectra released in \gaia DR3. 
We limited the content of the GSPC to the sources brighter than $G=17.65$~mag, thus excluding  most of the sources in the special catalogue of WDs (which is treated separately below), and unresolved galaxies and quasars \citep[included into the unresolved galaxy catalogue (UGC) and quasi-stellar objects catalogue (QSOC) of \gaia DR3, respectively;][]{DR3-DPACP-101}. The catalogue will be accessible and queryable through the Gaia Archive  (table {\tt gaiadr3.synthetic\_photometry\_gspc}). Examples of queries are provided in Appendix~\ref{sec:ex_query}. GSPC is focused on wide-band photometry and is limited to standardised systems.

In particular, it includes standardised magnitudes, fluxes, and errors on fluxes for the following passbands:

\begin{itemize} 

\item{} $U_{JKC}$,$B_{JKC}$,$V_{JKC}$,$R_{JKC}$,$I_{JKC}$,

\item{} $u_{SDSS}$,$g_{SDSS}$,$r_{SDSS}$,$i_{SDSS}$,$z_{SDSS}$,

\item{} $y_{PS1}$,

\item{} $F606W_{ACS/WFC}$, and $F814W_{ACS/WFC}$.

\end{itemize}

We decided to include only y$_{PS1}$ from the PS1 system to avoid the redundancy implied by two different but very similar versions of the same set of magnitudes ($griz$). Moreover, the standardisation of SDSS performed here is more extensive and robust than what we achieved for PS1 magnitudes. 

\begin{figure*}[!htbp]
    \centerline{
    \includegraphics[width=(\columnwidth)]{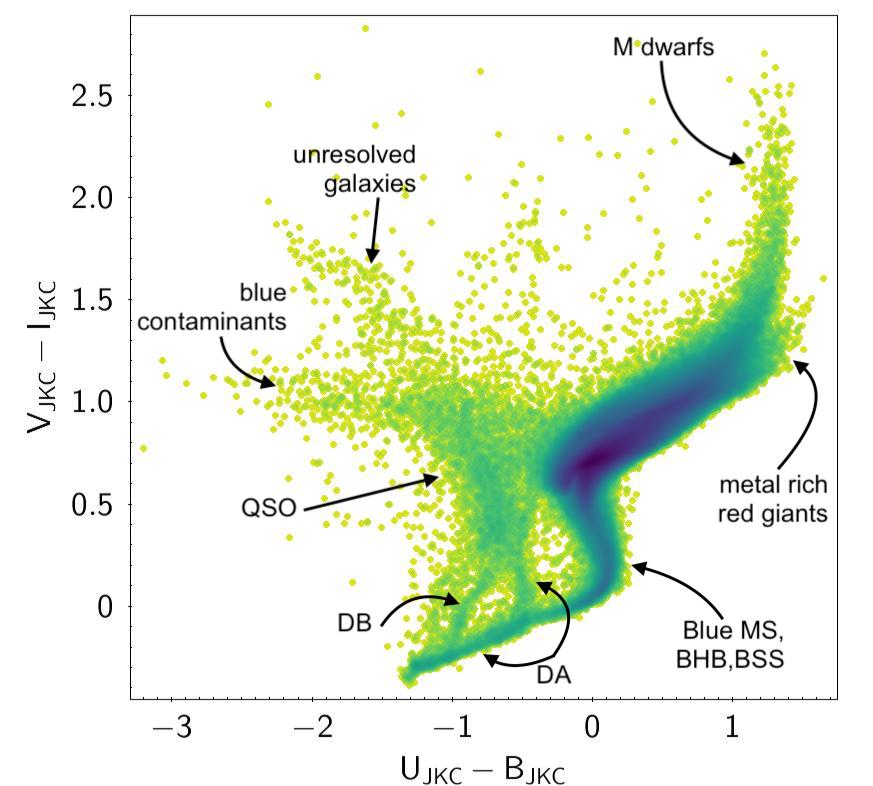}
    \includegraphics[width=(\columnwidth)]{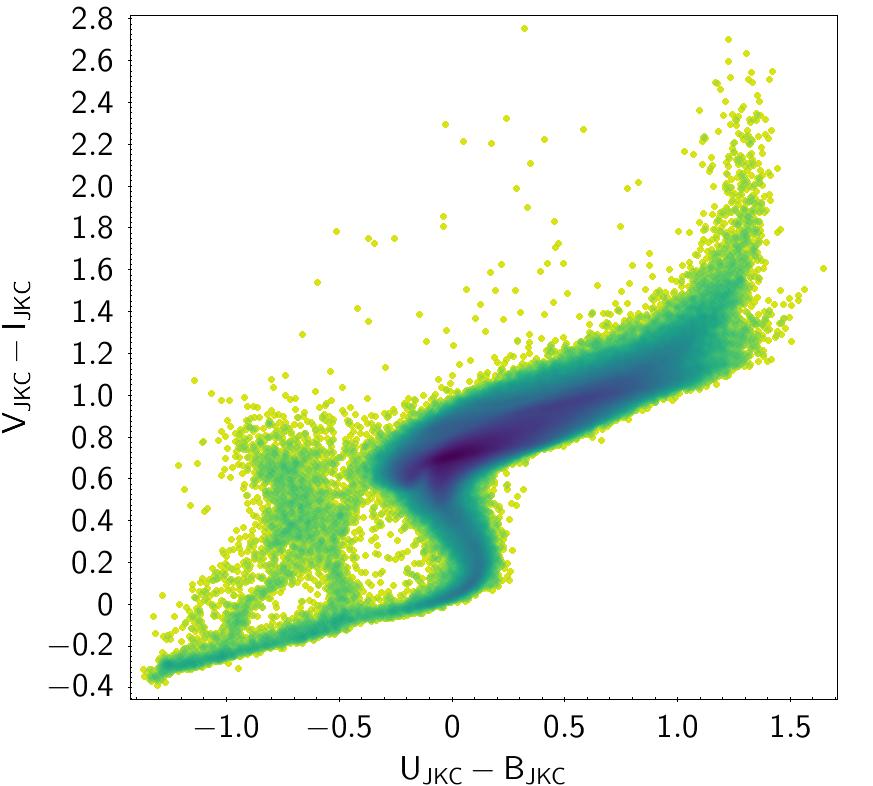}}
    \caption{$U-B$ vs. $V-I$ colour--colour diagram for the Galactic Caps subset of the GSPC ($|b|>50\degr$). Left panel: All sources, with labels for remarkable loci (2\ 025\ 048 sources with valid photometry in all the involved passbands). Right panel: Subset with $|C^{\star}|<0.05$ (1\ 985\ 565 sources).}
    \label{fig:gspc_UBVI}
\end{figure*}

In addition to the XP synthetic photometry listed above, the GSPC contains:
\gaia DR3 {$\tt source\_id$}, allowing a direct cross-match with other catalogues in the Gaia archive by means of {$\tt JOIN$} ADQL queries, the quality parameter C$^{\star}$ \citep{riello2021}, which can be used to select the sources with the most reliable photometry (but see also Sect.~\ref{sec:recommend}), a flag for each passband ($\tt Xflag$, where {\tt X=Ujkc,Bjkc,...}), which has a value of 1 if the \bprp colour and G magnitude of the considered star are within the ranges where standardisation and validation have been performed. In practice, the {\tt X} magnitude of a source with $\tt Xflag=0$ should be considered as an extrapolation of the adopted standardisation.

To keep only good-quality measurements, we adopted a unique criterion based on S/N for all the magnitudes in all the systems. A given source has valid photometry in the passband ${\tt X}$ only if

\begin{equation}
{\tt XFlux/XFluxError>30.0} 
,\end{equation}

that is, the S/N in that passband is higher that 30. As can be clearly appreciated from Table~\ref{tab:gspc_summary}, this constraint has a modest effect on the number of sources with valid photometry for all the considered passbands except for U$_{JKC}$ and u$_{SDSS}$. In these passbands, the sample with valid measures is reduced to $\la 17$\% of the entire content. As a reference, the next most affected passband is B$_{JKC}$, for which the same constraint leads to valid magnitudes for $\simeq 87$\% of GSPC sources.
There are 30220 sources with XP spectra in DR3 and $G<17.65$ but without a single GSPC magnitude satisfying the S/N>30 criterion; these are therefore not included in the final catalogue. It turns out that the overwhelming majority of them are very red AGB stars, possibly carbon stars. Most of them are classified as long-period variables ({\tt in\_vari\_long\_period\_variable=True} in {\tt dr3.vari\_summary}), and 352 of them are classified as carbon stars ({\tt spectraltype\_esphs==CSTAR} in {\tt dr3.astrophysical\_parameters}).

\begin{table}[!h]
    \centering
        \caption{Summary of the GSPC content for each passband: the parameters provided are the number of sources with synthetic photometry available in the catalogue, the magnitude range covered and the number of sources that are within the ranges in magnitude and colour that are fully validated (i.e. that have the corresponding flag set to \texttt{1}).}
    \begin{tabular}{l|rrr}
    \hline\hline
         Passband & Present & Mag range & Validated \\
         \hline
         U$_{JKC}$ & 32835800 & 2.22,18.97 & 32279743 \\
         B$_{JKC}$ & 191343258 & 2.96,20.31 & 160437248 \\
         V$_{JKC}$ & 217577173 & 3.10,20.58 & 206285205 \\
         R$_{JKC}$ & 218861537      &  2.59,19.99      & 206329396 \\
         I$_{JKC}$ & 218910521      &  1.98,19.14      & 206346825 \\
         u$_{SDSS}$         &  37990533      &   3.07,19.33   &    21965164 \\
         g$_{SDSS}$         &  210697330     &   2.89,20.55   &    191247211 \\
         r$_{SDSS}$         &  218262272     &   2.85,20.24   &    194747198 \\
         i$_{SDSS}$         &  218890040     &   2.46,19.78   &    194853547 \\ 
         z$_{SDSS}$        &  218840583     &   1.86,18.78   &    194788330 \\
         y$_{PS1}$ &  214043127     &   1.12,18.48   &    187461656 \\
         F606W$_{ACS/WFC}$  &  218549069     &   2.81,20.42   &    172587968 \\
         F814W$_{ACS/WFC}$  &  218919373     &   1.91,18.95    &   172588424 \\
         \hline
    \end{tabular}
    \label{tab:gspc_summary}
\end{table}
As already anticipated in Sect.~\ref{sec:standa_u}, the S/N$>30$ selection criterion imposes a strong colour bias on UV magnitudes. Considering the subsample of all GSPC sources with Galactic latitude $|b|>50\degr$ (Galactic Caps sample), while there are stars with valid U$_{JKC}$/u$_{SDSS}$ magnitudes as red as $\bprp\simeq 3.0$~mag, 95\% of those with valid U$_{JKC}$ have $\bprp\le 1.16$~mag and 95\% of those with valid u$_{SDSS}$ have $\bprp\le 1.18$~mag.

As a first glance at the quality of GSPC photometry, in Fig.~\ref{fig:gspc_UBVI} we show the JKC $U-B$ versus $V-I$ colour--colour diagram of the Galactic Caps sample introduced above for the entire sample (left panel) and for the best-quality subsample with $|C^{\star}|<0.05$, containing about $87$\% of the sources (right panel). The high-latitude selection is especially useful as it minimises the effect of blending and/or contamination and makes the effect of interstellar reddening negligible. In the left panel, some remarkable loci are labelled \citep[similarly to Fig.~22 in][]{I07}. We note that a significant residual population of unresolved galaxies and QSOs brighter than $G=17.65$ is included, some of them with U-B colours far exceeding those of the bluest bona fide stars. This is likely due to a combination of two main factors: first, the spectrum of some of these sources may have a significant non-thermal component (from active nuclei, nebular  emission, etc.), and, second, some of them may be partially resolved, thus making XPSP not fully reliable. The `blue contaminants' class is a mixture of source types including, among others, significantly blended stars, compact blue sources in relatively nearby galaxies (young stars clusters, stellar nuclei, HII regions), and distant compact blue galaxies.
It is interesting to note that most of these non-stellar sources are efficiently removed with a simple cut in $|C^{\star}|$, leaving in the left panel of Fig.~\ref{fig:gspc_UBVI} only well-defined stellar loci and a compact clump of truly point-source, bright QSOs. 

\begin{figure}[!htbp]
    \center{
    \includegraphics[width=(\columnwidth)]{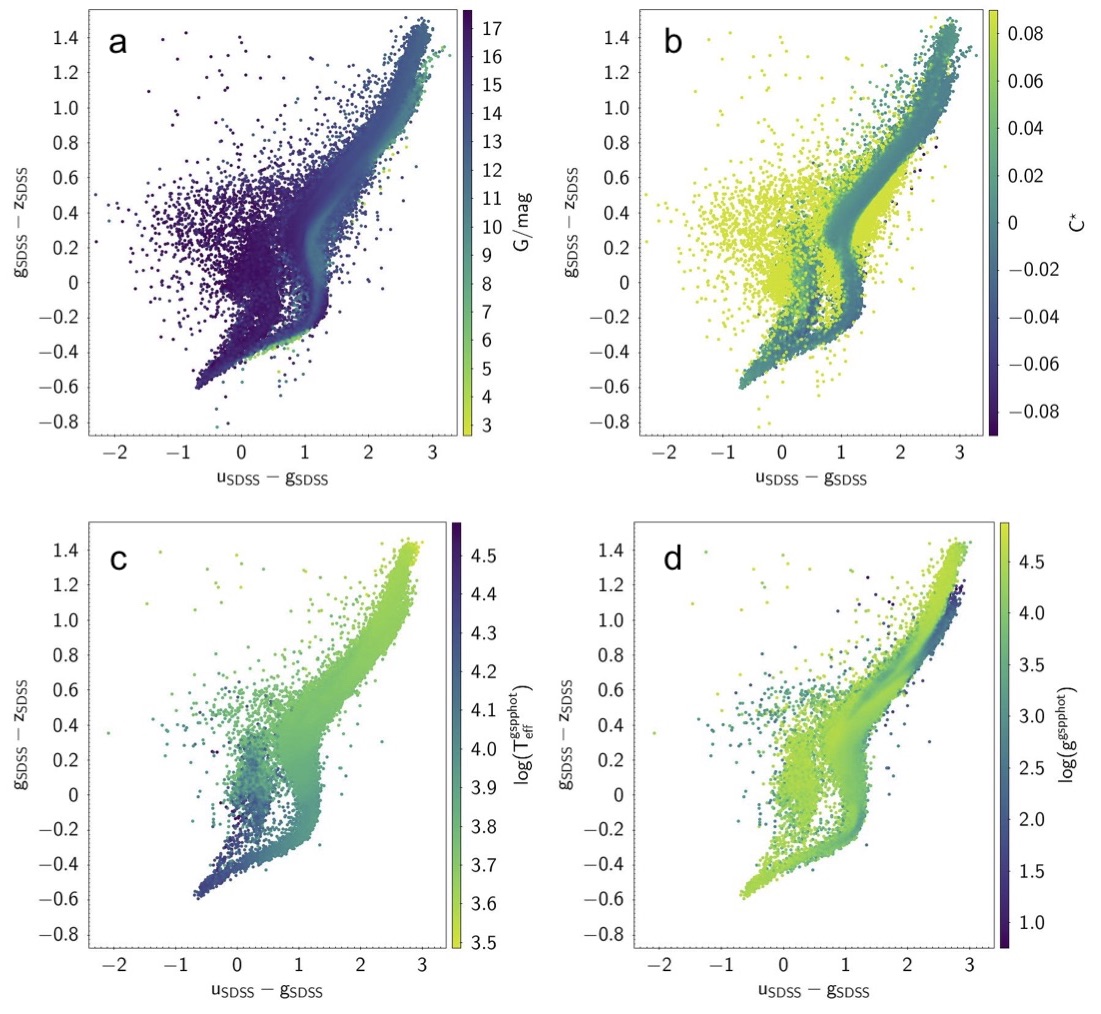} } 
    \caption{$u-g$ vs. $g-r$ colour--colour diagram for the Galactic Caps subset of the GSPC ($|b|>50\degr$; 2\ 003\ 727 sources having valid photometry in all the involved passbands), colour coded according to different parameters. Panel a: $G$ magnitude; Panel b: $C^{\star}$ ; Panel c: log $T_{eff}$ from GSP-Phot; Panel d: log $g$ from GSP-Phot. }
    \label{fig:gspc_colors}
\end{figure}

The GSPC is intended to provide accurate and precise all-sky photometry down to $G=17.65$~mag, with the limitations described above, in Sect. \ref{sec:stand}, and in Sect.~\ref{sec:recommend}, and is by no means a complete sample. Strong colour- and magnitude-dependent biases are unavoidably affecting the sample, induced by the selection criteria on the quality of the photometry. Moreover, the stellar populations sampled, the degree of crowding, and, consequently, the fraction of stars with excellent photometry, changes with position in the sky, depending on the mix of Galactic components encountered along the line of sight as well as on the amount of interstellar extinction\footnote{Please note that the cuts on magnitude and on the minimum number of BP and RP observations imposed for the release of XP spectra makes the footprints of the Gaia scanning law clearly visible in maps of GSPC sources.}.
Figure~\ref{fig:gspc_colors} gives overview examples of (i) the kind of selection bias at work (panel a), (ii) the effects of selection on the $C^{\star}$ parameter (panel (b)), as an example of a mean for additional cleaning of the sample, and (iii) the sensitivity of colour--colour diagrams to astrophysical parameters (panels (c) and (d); parameters from from GSP-Phot \citealt{DR3-DPACP-156}), 

\begin{figure}[!htbp]
    \center{
    \includegraphics[width=(\columnwidth)]{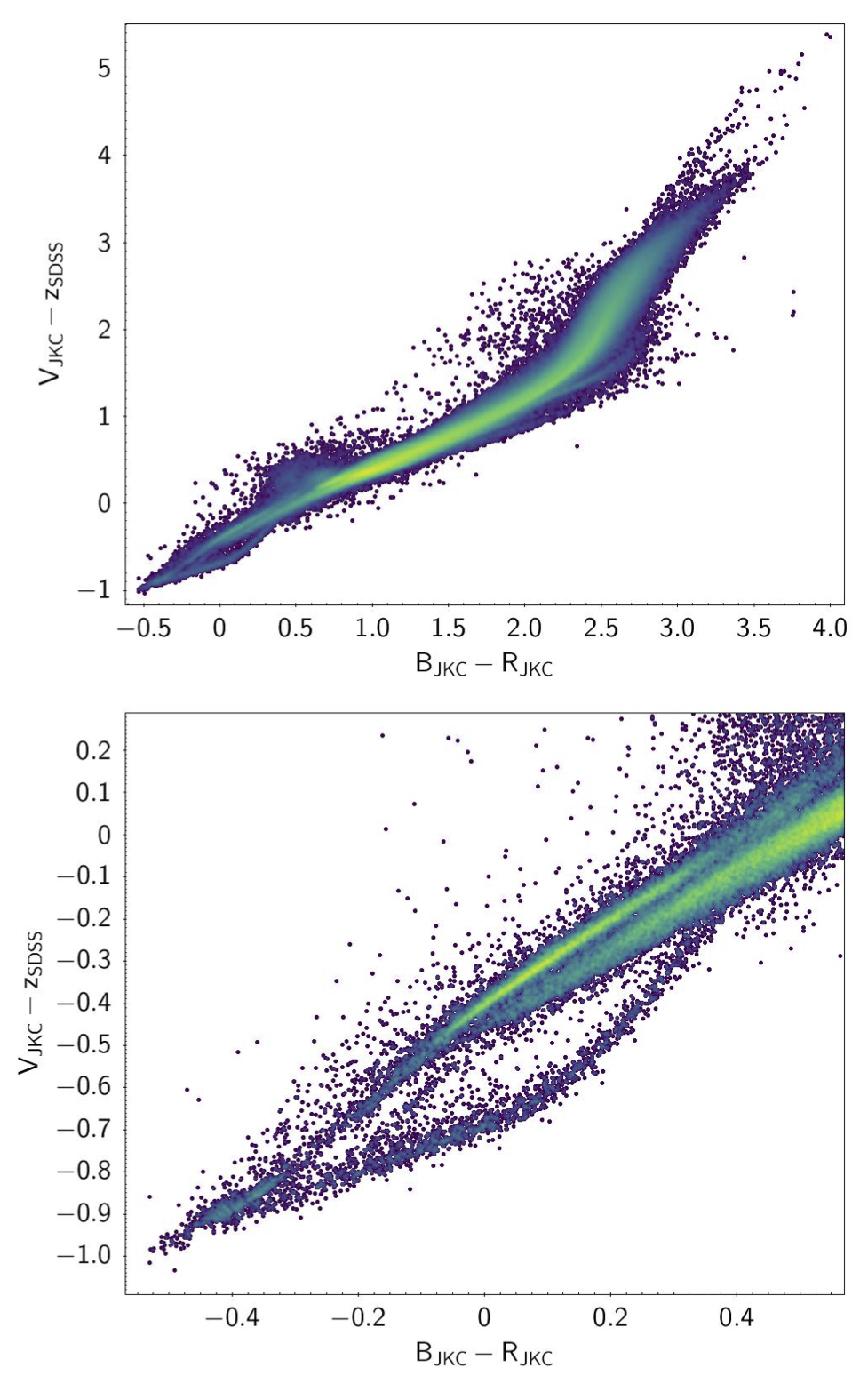}}
    \caption{Colour--colour diagrams for the Galactic Caps subset of the GSPC ($|b|>50\degr$), with colours obtained by mixing magnitudes from the JKC and SDSS photometric systems. To highlight the stellar loci more clearly, we included only stars with $|C^{\star}|<0.05$ (5\ 968\ 495 of the 6\ 266\ 882 GSPC sources in the Galactic Caps sample). Upper panel: Overall colour--colour diagram. Lower panel: Zoom into the highly structured blue region hosting sequences of DA and non-DA WDs, hot subdwarfs and extreme horizontal branch stars, blue horizontal branch stars, and so on, better demonstrating the level of detail emerging in these diagrams.  }
    \label{fig:gspc_manymag}
\end{figure}

Finally, to give an idea of the diagnostic power made available by GSPC (and by XPSP in general), in Fig.~\ref{fig:gspc_manymag} we present a colour--colour diagram of the Galactic Caps sample obtained by mixing magnitudes from two of the four photometric systems included in the GSPC. 

The diagrams display many well-defined features, suggesting a great potential to select various classes of sources. In spite of the $|C^{\star}|<0.05$ selection imposed to remove most non-stellar non-best-quality sources, the diagram includes more than 95\% of the sources of the original sample.

\subsection{The synthetic photometry catalogue for white dwarfs}
\label{sec:gwdc}

White dwarfs are important objects, and as well as meriting their own dedicated investigation, they can be used as tools to explore other areas of astrophysics.
For example, there is strong evidence that many WDs are accreting the remains of extrasolar planetary systems, which provides the only means of measuring their bulk composition. Furthermore, measured ages for the coolest known WDs can provide a limit to the age of the Galactic disc.
Many such studies require knowledge of the spectral type of the WDs in question, and whether or not they have H- or He-rich atmospheres. As the number of discoveries of  WDs grows, increasing numbers of spectroscopic observing campaigns are carried out to provide this information. The Sloan Digital Sky Survey (SDSS, \citealt{ahn2012}) has produced the largest spectroscopic catalogue of WDs so far (e.g. \citealt{Kleinman2013}), a data set that has allowed classification of approximately 10 000 WDs, the largest statistical sample of such stars prior to the publication of the \gdr2 and \gedr catalogues (\citealt{gaia2018} and \citealt{gaia2021}).
The quality- and distance-selected samples of \citet{Babusiaux2018} and \citet{GCNS21} each contain approximately 20 000 to 25 000 WDs and the works of \citet{GF2019} and \citet{GF2021} indicate that there might be as many as roughly 300 000 in the whole {\gaia} catalogue.
However, obtaining follow-up spectroscopy to classify all these candidates will be an enormous challenge and is not likely to be feasible using the currently available telescope resources.

The {\gaia} data release 2 (DR2) H-R diagram presented by \citet{Babusiaux2018} shows a clear WD cooling sequence and a degree of separation between the populations of H-rich (DA) and He-rich (DB) stars. However, while the WDs have a narrow range of masses, there is a significant overlap between the H- and He- groups over much of the parameter space.
The {\gaia} $G$, \gbp, and \grp integrated bands are broad in order to provide maximum sensitivity and the best possible photometric accuracy, and this limits the ability to distinguish between WDs of different spectral types. 
Wide- to narrow-band synthetic photometry generated from the {\gaia} XP spectra can be used to mimic the narrower band photometry available from surveys such as SDSS, allowing the diagnostic power to be applied to the larger number of WDs present in the {\gaia} catalogue. 

To test and illustrate this potential, we constructed a catalogue of about 100 000 WDs initially drawn from the {\gedr} data release, for which we have generated synthetic photometry in JKC, SDSS, J-PAS, and J-PLUS bands. This well-defined sample of WDs is designed to span the complete range of colours and magnitudes occupied by WDs. All the objects have a high probability of being a WD by virtue of their location in the H-R diagram. We followed the methodologies applied by \citet{GF2019} and the GCNS \citep{gcns}. The selection criteria are designed to remove contaminants whilst retaining as many high-probability WDs as possible. The sample extends to greater distance than the GCNS, and yields 100 786 WDs, a factor five increase compared to that catalogue.
Specifically, the following H-R diagram location and quality cuts were applied:

\begin{itemize}
    \item Equations 1-9 detailed in \citet{GF2019},
    \item $\rm astrometric\_ excess\_ noise \le 5,$
    \item $\rm phot\_ bp\_ mean\_ flux\_ over\_ error \ge 20,$
    \item $\rm phot\_ rp\_ mean\_ flux\_ over\_ error \ge 20,$
    \item $\rm parallax/parallax\_ error \ge 10,$
    \item $\rm phot\_ g\_ mean\_ flux\_ over\_ error \ge 20,$
    \item $\rm log(parallax/parallax\_ error)<-1.56(log(10^3/parallax) - 3.17) + 0.96.$
\end{itemize}

The majority of the stars in the sample have $G<19$~mag, but about 30\%\ are fainter. The effective \gband magnitude cut-off is $\approx $20~mag.

\begin{figure}[!htb]
\center{
\includegraphics[width=0.95\columnwidth]{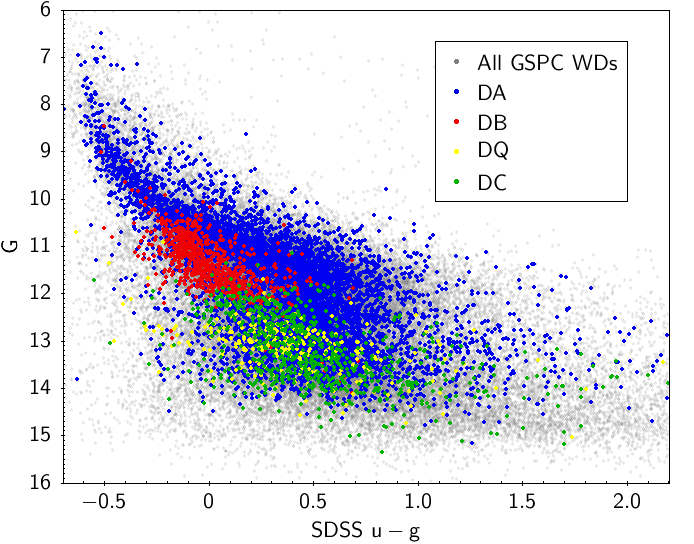}
}
\caption{Synthetic $u-g$ colour vs. absolute \gband magnitude diagram for white dwarfs in the GSPD-WD catalogue (grey). Data points are colour-coded according to SDSS spectral type where known:  blue = DA, red = DB, green = DC, yellow = DQ.
\label{fig:SDSSu_gsynth}
} 
\end{figure}

We derived synthetic photometry in SDSS, JKC, J-PLUS, and J-PAS systems for the 
full set of available WDs in the sample, which we designate the {\gaia} Synthetic Photometry Catalogue for WDs (GSPC-WD), which is published with this paper (see below, for the actual contents of the published table). Among these objects, 9758 have WD subtypes assigned by SDSS observations.
\figref{SDSSu_gsynth} shows absolute {\gband} magnitude plotted against the synthetic SDSS $u-g$ colour for these stars, indicating the main classifications.
The SDSS photometry gives a much better separation between the DA and DB WD spectral types than the {\gaia} photometry (see Figure 13 of \citealt{Babusiaux2018}). The DB stars also occupy a different region of the diagram compared to DC and DQ, but their ranges overlap substantially, and they also overlap with that of the cooler DAs.
Therefore, the synthetic magnitudes for bands that are narrower than \gband, \gbp, and \grp, provide a potential classification mechanism for all WDs in the {\gaia} catalogue.

\begin{figure}[!htb]
\center{
\includegraphics[width=0.95\columnwidth]{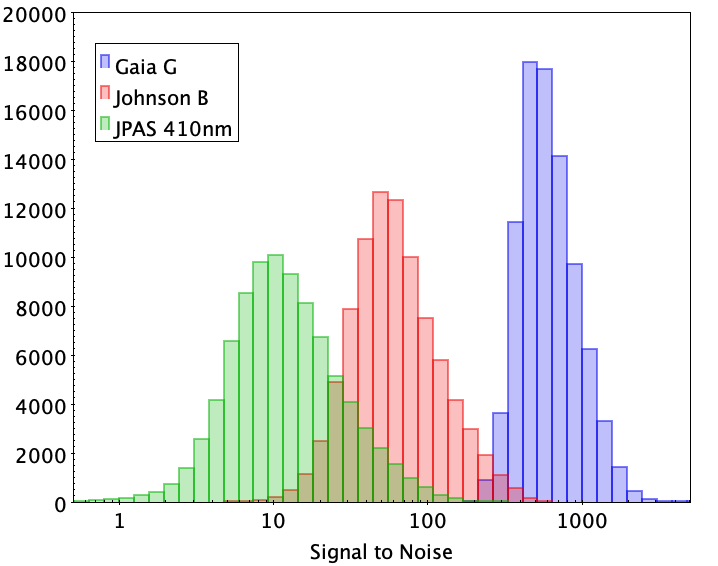}
}
\caption{S/N distributions for the stars in the GSPC-WD sample in three different filter bands: blue - \gband, red - synthetic Johnson $B$, and green - synthetic J-PAS 410~nm.
\label{fig:stonbands}
} 
\end{figure}

The choice of $u-g$ as an indicator of H-atmosphere DA spectral type compared to He-atmosphere DB WDs is related to the relative wavelengths of the bands compared to the Balmer jump at 364.5~nm, where the H Balmer series of lines converges. Compared to the DB WDs, the flux of DA WDs is suppressed shortward of this wavelength, making the stars appear redder, as seen in \figref{SDSSu_gsynth}.
In principle, there are many potential passband combinations available that may provide better or similar discrimination between DA and non-DA stars. For example, narrower passbands will allow better discrimination than wider ones, as discussed above. However, finer subdivision of the spectroscopic data reduces the S/N of individual bands. 
This is illustrated in \figref{stonbands}, which compares the S/N (flux over flux error) for the {\gband}, synthetic Johnson $B,$ and, as an example of the narrowest bands, J-PAS 410~nm. The Johnson $B$ band has approximately one-tenth of the S/N of {\gband} and the J-PAS 410~nm band one-fifth of Johnson $B$.

Whichever combination of filter bands is used to attempt the classification of the WDs, it is not straightforward to separate star types where their parameter space overlaps. Furthermore, it is a complex exercise to simultaneously use more than one colour--magnitude diagram to define the locus of particular spectral types. In trying to create `clean' samples of single WD classes for population studies, we need an objective method for carrying out the classification to distinguish among hydrogen type WDs (DA) and other types of WDs among these intrinsically faint sources.
A Random Forest algorithm allows us to make use of all the available photometry to carry out this task and determine a classification probability for DA WDs.
We used the SDSS dataset of 9758 WDs with known classifications to train the Random Forest algorithm to distinguish between DA and non-DA spectral types.
For the DA selection, we included all SDSS subtypes whose main type is `DA' in the classification scheme\footnote{This includes the following subtypes: "DA","DA(He)","DA(He)Z",\\
"DA+BD","DA+M","DA+M3","DA+M4","DA+M5","DA+M7",\\
"DA+M:","DA+Me","DA:","DA:DC","DAB","DAB+M","DABH",\\
"DAZ","DAE","DAH","DAH:","DAO","DAQ","DAQ:","DAZ",\\
"DAZ:","DAZB","DAZE:","DAZH:","DAe","DA+DB","DA:DC:"}. Among all the WDs with known subtypes, we were able to select 7567 DA and 2191 non-DA stars to be used to train or test the Random Forest algorithm. For training, we selected 1500 DA and 1500 non-DA, using the rest for testing purposes.
The input parameters used to perform the classification included all the SDSS, Johnson, J-PAS, and J-PLUS synthetic magnitudes, their uncertainties, and other information from {\gaia} (parallaxes, proper motion, integrated magnitudes and their uncertainties).

Using the Random Forest classification, we can use the obtained probabilities of being a DA to create a clean 
DA-type WD sample (\tabref{SourcesDAprobab}). For example, if we use only sources with probability larger than 0.7 of being a DA, derived using J-PAS filters, only 0.11\% non-DAs will contaminate our sample of selected sources.

\begin{table}[!htbp]
    \centering
        \caption{\label{tab:SourcesDAprobab} Percentage of non-DA sources contaminating our sample if selecting sources with probability of being a DA larger than $x$ when using different input passbands for classification.}
{\small
    \begin{tabular}{cccc
}
Input & $x=0.5$ & $x=0.6$ & $x=0.7$\\
\hline
SDSS         & 2.45 & 0.86 & 0.30\\
J-PLUS        & 1.43 & 0.55 & 0.27\\
J-PAS         & 0.77 & 0.26 & 0.11\\
Source coefficients & 0.50 & 0.15 & 0.03\\ 
    \end{tabular}
}
\end{table}

\begin{figure*}[!htbp]
\center{
\includegraphics[width=0.4\textwidth]{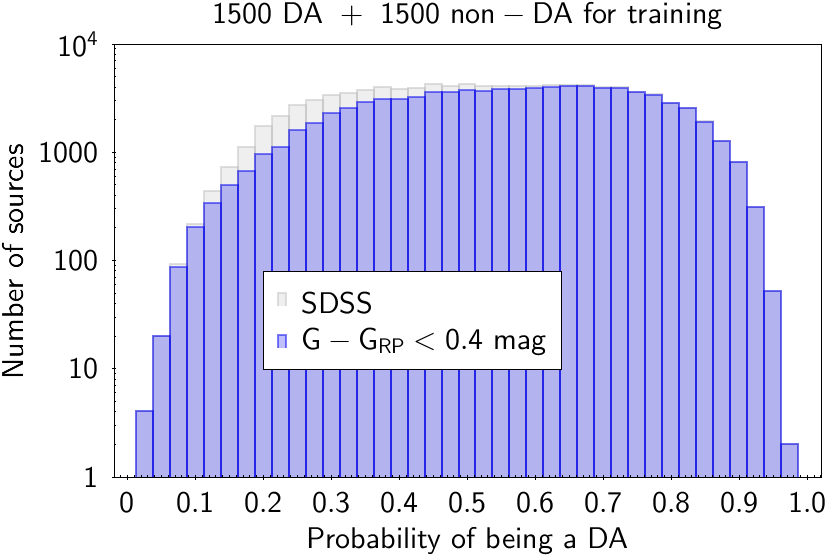}
\includegraphics[width=0.4\textwidth]{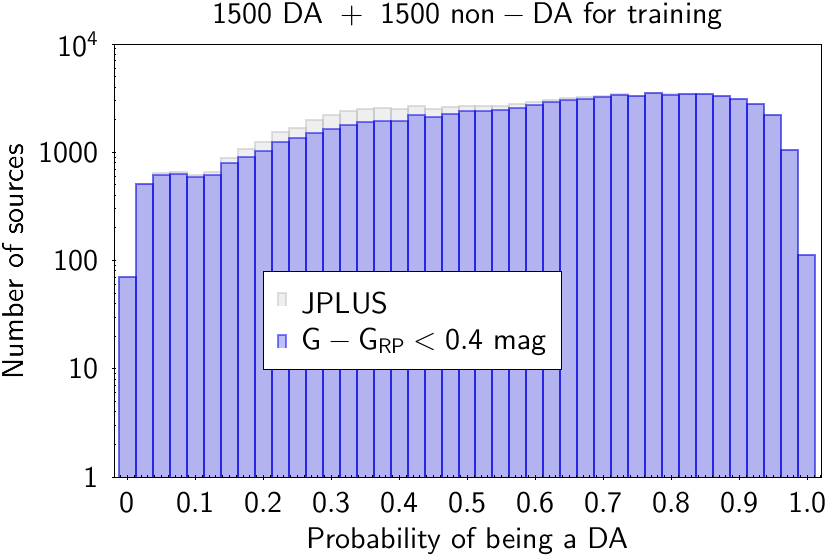}
\includegraphics[width=0.4\textwidth]{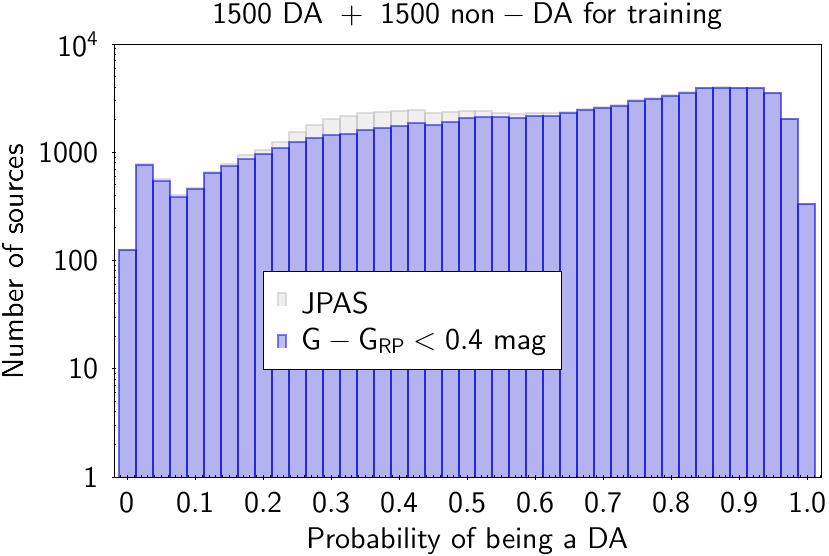}
\includegraphics[width=0.4\textwidth]{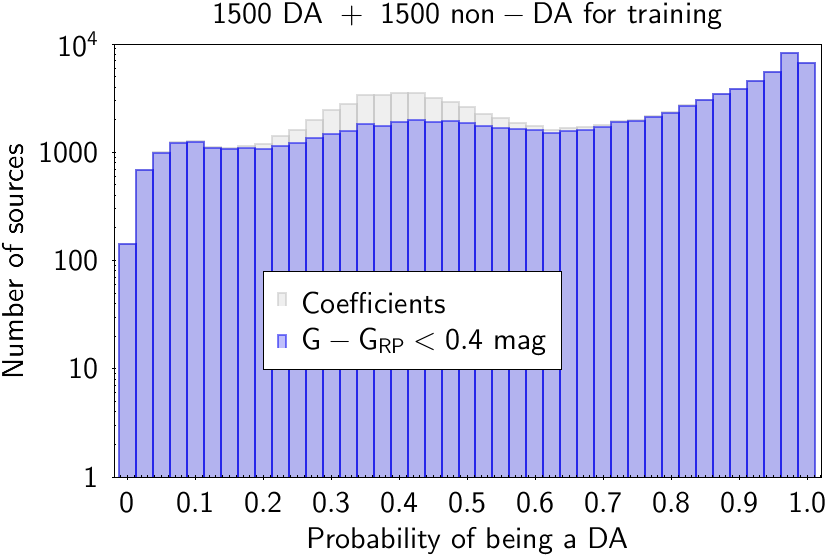}
}
\caption{Probabilities obtained for all classified WDs when using only SDSS (top-left), J-PLUS (top-right), and J-PAS (bottom-left) pass bands and the {\xp} source coefficients directly (bottom-right). In blue, the same histogram when we filter the reddest sources (keeping only those with $\ggrp<0.4$~mag), as they are not covered by the training dataset and lower output probabilities are expected.
\label{fig:DAprobability}
} 
\end{figure*}

Once the algorithm has been trained and validated, we can apply it to all the white dwarfs in the GSPC-WD catalogue, including those where an SDSS classification is not available. \afigref{DAprobability} shows the probability distribution for all four cases studied (SDSS, J-PLUS, J-PAS and source coefficients). 
Based on our results, the narrower the pass bands, the better the classification (increasing their probabilities and obtaining a less centred distribution), improving also when more pass bands are considered, covering the whole wavelength range. Nevertheless, it can also be seen that the best results are obtained when using the {\xp} coefficients representing the spectra, rather than the synthetic photometry.

When analysing the colour distribution of sources with SDSS types available to train our algorithm
we see that 96\% of the sources fall in the range $\ggrp<0.4$~mag.
For this reason it is expected that the algorithm is not working optimally for colours outside this range. In order to verify this, we show only those sources with $\ggrp<0.4$~mag in an overlapped distribution in \figref{DAprobability}. Indeed, sources with larger values for \ggrp are those located at intermediate probabilities, and this method is not able to properly classify them. 
Not all WDs in the sample have a complete set of JPAS magnitudes for classification, as some sources are too faint to generate significant magnitude measurements. Therefore, the total number of WDs classified by the Random Forest algorithm is 86783, to which we can add the 9758 WDs already classified by SDSS: a total of 96541 WDs.

\begin{figure*}[!htb]
\center{
\includegraphics[width=0.45\textwidth]{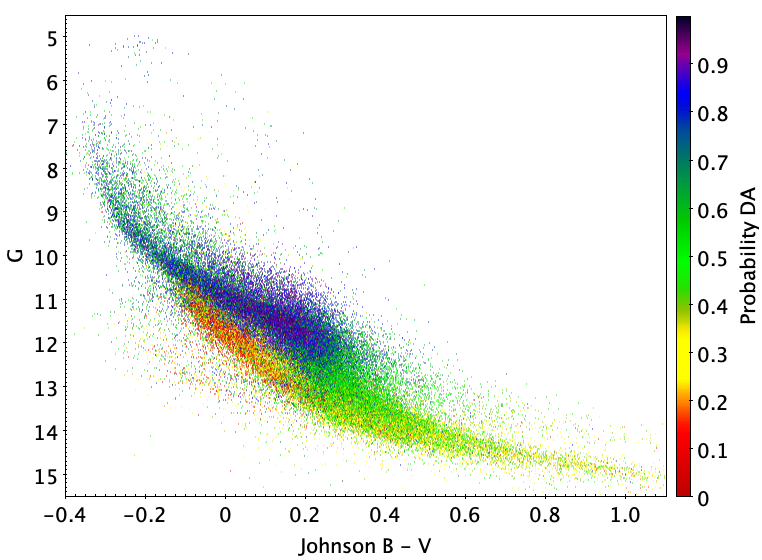}
\includegraphics[width=0.45\textwidth]{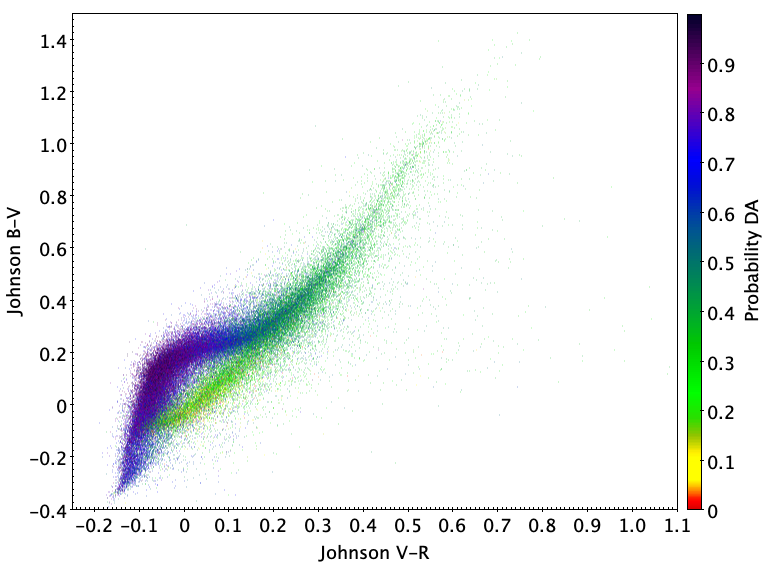}
}
\caption{(left) \gaia \ absolute G magnitude vs. B-V colour--magnitude diagram for the GSPC-WD sample colour-coded with the probability of a WD being a DA. (right) {\gaia} $B-V$ vs. $V-R$ colour--colour diagram for the GSPC-WD sample colour-coded according to the probability of a WD being a DA.
\label{fig:B-V}
} 
\end{figure*}

\begin{figure}[!htb]
\center{
\includegraphics[width=0.95\columnwidth]{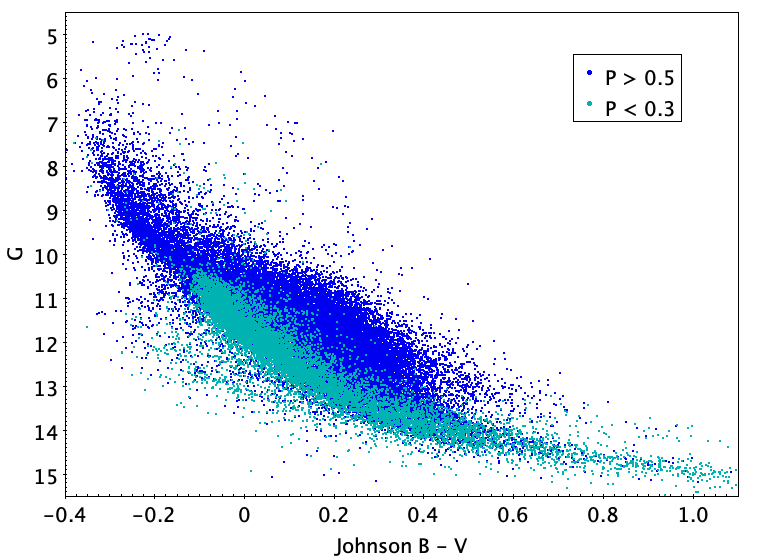}

}
\caption{(blue) \gaia absolute $G$ magnitude vs. $B-V$ colour--magnitude diagram for the GSPC-WD sample with probability of being a DA > 0.5. (cyan)) Gaia absolute G magnitude vs. $B-V$ colour--magnitude diagram for the GSPC-WD sample with the probability of being a DA $<0.3$.
\label{fig:DA-nonDA2}
} 
\end{figure}

The usefulness of the DA/non-DA classification scheme can be illustrated by considering the \gband versus Johnson $B-V$ colour--magnitude diagram (left hand panel of \figref{B-V}). 
The distribution is colour coded by the probability of a WD being a DA. The DA and non-DA cooling tracks appear to be clearly separated by the $B-V$ colour. Similarly,  the $B-V$ versus $V-R$ colour--colour diagram (right hand panel of \figref{B-V}) shows very good isolation of the DA and non-DA components. However, when we examine the distributions of DA and non-DA classifications separately, we see that there is considerable overlap of these in the parameter space of the colour--magnitude diagram (\figref{DA-nonDA2}). The figure shows all WDs with probability of being a DA above 0.5 (blue), with a non-DA contamination fraction of 0.77\% (\tabref{SourcesDAprobab}). Overlaying this distribution are those WDs with probability of being a DA of less than 0.3 (cyan). This shows the great difficulty in separating out DAs and non-DAs on the basis of a cut in any colour--magnitude diagram and underlines the importance of the Random Forest classification method.

In conclusion, we demonstrate that synthetic photometry for a range of standard systems can be used to classify white dwarfs in the {\gaia} catalogue into DA and non-DA types. However, we note that better results are achieved using the coefficients of the {\xp } spectra, without the need to compute the synthetic photometry. Nevertheless, the differences are small and, although our classification is not perfect, a catalogue of the synthetic photometry for our sample of \gaia \ white dwarfs provides a useful resource that can be applied to generating samples of white dwarf types without the need for further computation. As we use J-PAS synthetic photometry to classify white dwarfs in \gdr3, there is a strong prospect for the real J-PAS survey \citep{JPAS} to be used in a similar way. Indeed, the J-PLUS data have been used to classify and parameterise approximately $6000$ WDs \citep{JPLUSWD}. Clearly, this is a much smaller number than the WDs included in this work, but the S/N of the observations is potentially greater than we were able to achieve with the \gaia DR3 data. Therefore, the full J-PAS and J-PLUS surveys could provide valuable complementary data for analysis of the GSPC-WD catalogue.

We have made the GSPC-WD synthetic photometry available as a stand-alone catalogue\footnote{\url{https://zenodo.org/record/6637717\#.YqcREC8RpAY}}, including  SDSS, JKC, and J-PLUS XPSP and the DA classification probability. The photometry of the individual J-PAS bands used in the Random Forest analysis is not included because of their low S/N.
For WDs classified in SDSS, a subset of which were used in the training and validation of the Random Forest algorithm, we also include the full SDSS classifications as a separate column in the GSPC-WD catalogue table.
As can be seen from the example in \figref{stonbands}, when the synthetic spectral bands are very narrow, a significant number of sources will have low S/N. Furthermore, at the edges of the \gaia \ spectral range, away from the peak of the effective area, this is also true for some stars in the wider bands included in the catalogue. 
In some extreme cases, there is no significant detection of the object. The Random Forest algorithm is only able to classify a WD when valid flux measurements are available for every photometric band we include in the analysis. Therefore, no classification is recorded in the catalogue when data for one or more bands is `missing'. In total, 15003 WDs from the total sample of 101 783 are not classified.
For completeness, we have made all the flux measurements and corresponding magnitudes available for all objects in the GSPC-WD.
Therefore, magnitudes and fluxes with very large errors up to several times the flux itself are included.
However, where fluxes are negative, the magnitudes are not defined. 
When using the catalogue, appropriate S/N cuts are advisable for specific scientific objectives in order to ensure data quality.

%%%%%%%%%%%%%%%%%%%%%%%%%%%%%%%%%%%%%%%%%%%%%%%% END PRODUCTS

%%%%%%%%%%%%%%%%%%%%%%%%%%%%%%%%%%%%%%%%%%%%%%%% RECOMMENDATIONS
%\input{sections/recommendations}
\section{Recommendations and caveats}
\label{sec:recommend}

In this section we provide some caveats and recommendations that can serve as guidelines for best use of the products described and provided here.
It is important to be aware that, in spite of the huge effort made to check and verify XPSP, to results of which are only partially shown and discussed here because of obvious constraints on publishing space, the validation we provide is in any case partial, being unavoidably limited to high-quality reference samples that may not be perfect nor fully representative. In the discussion of the comparisons presented here, we focus almost exclusively on the known problems affecting XP spectra\footnote{It is not necessarily easy to disentangle problems due to the process of external calibration \citep{EDR3-DPACP-120} and to the internal calibration of XP spectra \citep{EDR3-DPACP-118}. Here we generally consider colour trends as due to imperfections in the instrument model, and therefore associated to EC XPs. On the other hand, external calibration cannot be responsible for the trends with magnitudes, such as e.g., the hockey-stick effects or the {\em blue dip}, which is due to imperfection in the internal calibration process. Hopefully, both sides of the process should significantly improve in future data releases.}, as described in \citet{EDR3-DPACP-120} and \citet{EDR3-DPACP-118}, but in fact, some of the observed anomalies may be due to issues in the reference samples. 

In any case, the users are invited to further validate the XPSP data they use, depending on their science goals and applications. The performances illustrated here should be considered in a statistical sense, and the individual magnitudes may still suffer from problems not traced by the available quality parameters. 

We did not make extensive tests to verify whether $\Delta {\rm mag}$ distributions for a given system and/or reference set depend on the luminosity type of the considered stars (e.g. giants, dwarfs, WDs, etc.) or on the interstellar extinction. In general, the sets we adopted for validation and/or standardisation, while being predominantly composed of dwarfs, includes all types of stars. For example, we verified that, in the Landolt's sample, WDs do not show a different $\Delta {\rm mag}$ distribution as a function of colour with respect to other kinds of stars in the same colour and magnitude range, within the uncertainties. In Sect.~\ref{sec:javalambre} it is shown that, in the J-PAS and J-PLUS systems, XPSP has very similar performances for WD and normal stars, except for the UV passbands. In our experience the most problematic range in this respect is that of cool stars, especially M-type stars, where giants and dwarfs may also behave differently in response to tiny differences in the TCs. In the cases of the SDSS  system, we explicitly checked that red ($1.0<G_{BP}-G_{RP}<3.5$) giants and dwarfs have compatible $\Delta {\rm mag}$ distributions, within $\simeq 10$~mmag (however, limited to K stars; see Appendix~\ref{sec:tha21} for further discussion). The same is true for the JKC, albeit tested with a much smaller sample of red giants (Sect.~\ref{sec:standa_jkc}).

In general, the accuracy of the standardised photometry presented here has not been tested against large variations in the interstellar extinction. Hence, in cases of highly reddened stars, XPSP should be used with caution. However, the analysis presented in Appendix~\ref{sec:tha21} suggests that, at least in the considered case (red giants in the SDSS system), stars with extinction as large as $A_0\la 5.0$~mag have $\Delta$~mag virtually indistinguishable  distributions from their 
low-extinction counterparts, where $A_0$ is the monochromatic extinction at $\lambda=547.7$~nm as estimated by GSP-phot \citep{DR3-DPACP-156}.

We also note that the XPSP performance has not been tested, or only partially (see Sect.~\ref{sec:standa_hugs}), in the presence of a significant degree of crowding or of a strong astrophysical background (see also the cautionary note at the end of Sect.~\ref{sec:nonustand}).
We are not aware of spatial variation of the systematic errors affecting XP ECS but we cannot exclude their existence. However, their amplitude should be very small, owing to the careful process of internal calibration of BP and RP spectra \citep{Carrasco2021, EDR3-DPACP-118}.

The main goal of this paper is to show the potential of XPSP, a new product available for the first time in {\gdr3}. We are confident that the astrophysical community will explore this potential much more extensively, seeking and extracting the greatest scientific return.
Within the limits of our resources, we will be happy to support extensions of the available photometric systems (see Sect.~\ref{sec:gaiaXPy}). We stress again that standardised $U_{\rm JKC}$ and $u_{\rm SDSS}$ XPSP {cannot} provide an exact reproduction of the corresponding reference magnitudes because they lack the bluest part of the wavelength coverage. Moreover, in general, standardisation of any magnitude is strictly valid only in the colour and magnitude range and in the range of astrophysical parameters where the processes have been performed, the range covered by the adopted reference sample (see Sect.~\ref{sec:gspc}).

It is important to recall that the calibration of XP ECS and the instrument model used to get XPSP are best suited to dealing only with point sources. Synthetic magnitudes of extended or even marginally resolved sources may (and, in fact, should) be affected by systematic errors depending of their extension, their spectrum, and the width and wavelength range of the considered passbands. Also, the entire chain of {\gaia} data processing leading to XPSP is designed for single stars and calibrated on single non-variable stars: magnitudes of sources with relevant non-stellar components in their spectra, of variable sources, and of unresolved multiple stars are not expected to have accurate XPSP. However, this does not imply that their synthetic magnitudes, fluxes, and colours do not carry useful information on these sources.

Saturation of portions of XP spectra may occur in a variety of circumstances, depending on the magnitude, colour, and detailed spectral shape of the sources (i.e. presence of emission lines), with obvious effects on the accuracy of the derived XPSP. As a general rule of thumb, derived from the analyses by \citet{riello2021}, \citet{EDR3-DPACP-118}, and especially  \citet{EDR3-DPACP-120}, we can assume that XPSP should be free from saturation effects for $G\ga 5.0$. XPSP from the BP spectral ranges are more easily affected, while reasonable photometry in the RP range should be possible up to $G\simeq 3.0$ in most cases. 
Finally, the
performance of the internal calibration of BP and RP spectra for $G\la 11.5$ is not as reliable as for fainter sources. The onset of different window classes and gate setups to extend the linear regime of the detectors up to $G\simeq 5.0$ and beyond makes the Gaia spectrophotometric system in this bright regime not perfectly matched with that established for $G\ga 11.5$.
\citep[see][and referenced therein]{EDR3-DPACP-118,EDR3-DPACP-120}. For these reasons, the accuracy and precision of XPSP should be poorer at very bright magnitudes and, in general,  above the $G\simeq 11.5$ limit, than for high-S/N measures below it.

A large number of parameters are available from the \gdr{3} archive for all sources with XP spectra. Here, we provide a few suggestions for how to select the best data. Users will have to consider which ones are appropriate and at which level, depending on their science case.
\begin{itemize}
    \item The renormalised unit weight error \texttt{ruwe} (available in \texttt{gaia\_source}) can be used to clean a sample from cases showing photocentric motions due to unresolved objects, such as astrometric binaries. Some guidance on filtering based on this parameter is provided in \cite{ruwe}.
    The criterion \texttt{ruwe}$<1.4$ retains about $93$\% of the sources with XP spectra in \gdr{3}.
    \item The corrected G$_{BP}$ and G$_{RP}$ flux excess factor $C^{*}$ defined in \cite{riello2021} and available from the GSPC table as \texttt{c\_star} is useful to clean the dataset from objects affected by inconsistencies in the photometry in the various bands (\gband, \gbp, \grp). These inconsistencies can be due to different source properties (e.g. in the case of extended sources) or systematic errors in the calibration procedures (e.g. in the case of residual background due to nearby bright sources). See \cite{riello2021} for more details. The same paper (Sect. 9.4) provides a function reproducing the $1\sigma$ scatter for a sample of well-behaved isolated stellar sources with good-quality photometry. The criterion $C^{\star}<1\sigma$ retains $79$\% of the sources, while a more generous $C^{\star}<3\sigma$ retains $90$\% of the sources.
    \item The photometric errors can be used to define a variability proxy as $\sqrt{n}\, \sigma_f/f$, where $n$ is the number of observations and $f$ and $\sigma_f$ are the flux and its uncertainty in the \gband \citep{varproxy}. All required parameters are available from the \texttt{gaia\_source} table. This can be used effectively to remove objects that vary in flux. A possible criterion could be defined selecting sources that have a variability proxy value within $K$ sigma from the average value at a given magnitude: this would retain $95$\% of the sources for $K=1$ and $99$\% of the sources for $K=3$.
    \item Variable stars can be identified also using
    {\tt phot\_variable\_flag} from the \texttt{gaia\_source} table, while a classification of the candidate variables by type can be found in the 
    \texttt{vary\_summary} table.
    \item Finally, users may be interested in cleaning the dataset from objects affected by crowding. An assessment of the number of transits that contributed to the generation of the source spectra in \gdr{3} and that were affected by a non-target source within the window (these cases are labelled blended) or by a nearby bright object (contaminated) is provided in the table \texttt{xp\_summary} and in particular in the parameters \texttt{bp}/\texttt{rp\_n\_blended\_transits} and \texttt{bp}/\texttt{rp\_n\_contaminated\_transits}\footnote{Please note that the \texttt{gaia\_source} table contains equivalent counters applicable to the photometric data, i.e. integrated \gbp and \grp.} (including the $\beta$ parameter, used in Sect.~\ref{sec:mps}). It should be mentioned that such assessment is based on the \gdr{2} source catalogue. It is therefore expected that the crowding assessment may not be accurate in very dense regions due to the reduced completeness of the catalogue and in cases of sources with very small angular separation. The fraction of transits flagged as blended or contaminated can be used as an additional criterion to remove data affected by crowding.  
\end{itemize}

Section 6.1 in \cite{DR3-DPACP-118} provides more details on the XP spectral data available in the \gdr{3} archive, instructions on how to download the data, and recommendations regarding the treatment of the data. For the purpose of generating synthetic photometry, we recommend using full, non-truncated XP spectra. Truncation has been introduced to remove spurious features in the spectra due to higher order bases fitting the noise in the observed data, particularly for faint sources or sources with a low number of observations. This is achieved by dropping coefficients that are consistent with being noise. When generating synthetic photometry by effectively integrating the spectrum in a given wavelength range, the precision of the result is not significantly improved by applying truncation. On the contrary, in the case of particularly narrow bands, truncation may introduce some systematic errors. See Section 3.4.3 in \cite{DR3-DPACP-118} for more details.
Appendix \ref{sec:ex_query} shows a few examples of queries to create selections from the GSPC and to extract the corresponding parameters from the main table \texttt{gaia\_source}.

%%%%%%%%%%%%%%%%%%%%%%%%%%%%%%%%%%%%%%%%%%%%%%%% END RECOMMENDATIONS

%%%%%%%%%%%%%%%%%%%%%%%%%%%%%%%%%%%%%%%%%%%%%%%% CONCLUSIONS
%\input{sections/conclusions.tex}
\section{Conclusions and perspectives for the future}\label{sec:conclu}

We present a Gaia-DPAC product made available for the first time with Gaia DR3 that provides the possibility to obtain synthetic photometry in any passband for all the stars with published XP spectra, provided that the passband is entirely included in the XP wavelength range (330 nm - 1050 nm), and that the FWHM of the passband is significantly larger than that of the BP or RP LSF at the considered wavelength ($Rf\ge 1.4$; but see Appendix~\ref{sec:app_Rf} for a thorough discussion).

We show that wide-band photometry is reproducible within a few percent over wide ranges in magnitude and colour. We demonstrate this result for several widely used systems with good internal precision.
The accuracy and precision decrease when considering medium- and narrow-band photometry; however, we show that even with measurements from this kind of passbands, performances are, at least, comparable with state-of-the-art ground-based observations, and fruitful scientific applications are possible. For example, the Gaia C1 system \citep{Jordi2006} was brought into life by \gaia XPSP and we demonstrate its capabilities to deliver the astrophysical information, including stellar temperature, gravity, metallicity, and even $\alpha$ element abundance, an especially challenging task for the very low-resolution XP spectra \citep{Gavei2021}.

The residual shifts and trends affecting XPSP, which are mainly due to known systematic errors in the EC XP spectra, can be corrected down to millimag accuracy in some cases using suitable sets of external photometric standards as a reference, a process that we call standardisation.
We performed the standardisation for the JKC, SDSS, PS1, and Str\"omgren systems, as well as for three wide passbands from two different HST systems. In addition, we demonstrate that XPSP  is suited to calibrating narrow-band photometry for surveys designed to trace emission lines in stars.

We provide a few examples of scientific applications, demonstrating the performance of XPSP to trace multiple populations in globular clusters, classify emission line sources,  and obtain metallicity estimates, and also in the very metal-poor regime. The latter is a realm where the complementarity with the DPAC products directly derived from the analysis of XP spectra \citep[GSP-Phot,][]{DR3-DPACP-156} can be more fruitful. We show that by adopting reliable reddening values from external sources, dedicated photometric indices can give satisfactory performances.

Finally, we provide two publicly available catalogues for general use:(a) the Gaia Synthetic Photometry Catalogue (GSPC), queryable from the Gaia Archive, containing standardised photometry in 13 widely used wide passbands for $\sim 220$ M stars with $G<17.65$ all over the sky 
(table {\tt gaiadr3.synthetic\_photometry\_gspc}), and (b) the Gaia Synthetic Photometry Catalogue for White Dwarfs (GSPC-WD), publicly available as a stand-alone catalogue through CDS, containing synthetic photometry in many bands and DA/non-DA classification for a sample of approximately $100000$ WDs down to $G\simeq 20.0$

We demonstrate that XPSP can provide precise space-based all-sky photometry in any optical band, with performances depending on the passband width and wavelength range. Furthermore, XPSP may significantly impact the photometric calibration of existing observations and the design of planned surveys (see, e.g. Sect.~\ref{sec:Gaia2C1}). For the first time it provides extensive means to refer photometry in different magnitude systems to the same flux scale, for example providing simultaneously homogeneous JKC, SDSS, PS1, and HST photometry for the same set of stars.
In perspective, this should be the essential contribution of the Gaia XPSP: providing an absolute photometric reference for optical photometry, while the astrophysical information of the observed sources  can, in principle, be optimally extracted from the entire XP spectra.

There are sound arguments for believing that the performance we present here can significantly improve in future Gaia data releases \citep[see also][]{EDR3-DPACP-118,EDR3-DPACP-120}. The accumulation of many additional epoch spectra will provide mean XP spectra with higher S/N and consequently more precise XPSP. The release of XP spectra for fainter stars will significantly enhance the photometric depth that can be reached, well beyond the current $G<17.65$ limit.
New, improved releases of the SPSS will provide a more robust basis for a more accurate flux scale of XP ECS and a better calibration of the instrument model, a vital ingredient of the chain leading to XPSP.
There are ideas to improve the calibration of the instrument model by other means; for example by a better calibration of the LSF, of the wavelength scale, and so on, to be implemented in the next cycle of data reduction.
The internal calibration of mean XP spectra will improve in future releases. For example, there is currently a lot of work being done to improve the algorithm for sky subtraction, which could imply substantial mitigation of the hockey-stick effect.
In general, each Gaia data release improves upon the entire process of spectro-photometry, as we gain experience in the instruments and the ways to calibrate for even the smallest of effects, and new pieces of the calibration are activated.

If significant mitigation of residual systematic errors were indeed to be achieved, this would greatly extend the contribution of the \gaia mission to optical photometry.

%%%%%%%%%%%%%%%%%%%%%%%%%%%%%%%%%%%%%%%%%%%%%%%% END CONCLUSIONS

%%%%%%%%%%%%%%%%%%%%%%%%%%%%%%%%%%%%%%%%%%%%%%%% Note added in proof
{\bf \noteaddname} Due to a bug in GaiaXPy  the synthetic photometry for the standardized PS1 y band photometry  published in the GSPC (contained in the fields {\tt y\_ps1\_flux}, {\tt y\_ps1\_flux\_error} and {\tt y\_ps1\_mag}) has been generated without applying the correction for the hockey-stick effect. The Gaia Archive table {\tt gaiadr3.synthetic\_photometry\_gspc} will not be fixed. However, correct synthetic photometry in the standardised PS1 system can be generated using GaiaXPy (with version 1.2.4 or later) on spectra extracted from the archive. Prior to version 1.2.4, the GaiaXPy bug gave the same error for all the PS1 passbands, but y was the only PS1 flux/magnitude included in GSPC.
It has also been discovered that the units of the SDSS and PS1 flux and flux error fields in the GSPC are wrong and should have Hz$^{-1}$ instead of nm$^{-1}$. Only the units are wrong: the data contained in the table is correct (except for the issue described above regarding $y_{PS1}$).
%%%%%%%%%%%%%%%%%%%%%%%%%%%%%%%%%%%%%%%%%%%%%%%% end Note added in proof

%%%%%%%%%%%%%%%%%%%%%%%%%%%%%%%%%%%%%%%%%%%%%%%% ACKNOWLEDGEMENTS
%\input{sections/acknowledgments.tex}
\begin{acknowledgements}
This work presents results from the European Space Agency (ESA) space mission \gaia. \gaia\ data are being processed by the \gaia\ Data Processing and Analysis Consortium (DPAC). Funding for the DPAC is provided by national institutions, in particular the institutions participating in the \gaia\ MultiLateral Agreement (MLA). The \gaia\ mission website is \url{https://www.cosmos.esa.int/gaia}. The \gaia\ archive website is \url{https://archives.esac.esa.int/gaia}. Acknowledgments are given in Appendix \ref{sec:app_ack}.
\end{acknowledgements}

%%%%%%%%%%%%%%%%%%%%%%%%%%%%%%%%%%%%%%%%%%%%%%%% ACKNOWLEDGEMENTS

%--------------------------------------------------------------------

\bibliographystyle{aa} % style aa.bst
\bibliography{refs,dpac} % your references refs.bib

%######################################## APPENDICES ################################################################

\begin{appendix}

%%%%%%%%%%%%%%%%%%%%%%%%%%%%%%%%%%%%%% APPE ACKNO
\section{Acknowledgements}
\label{sec:app_ack}

This paper is dedicated to the memory of Arlo U. Landolt (1935 - 2022) whose contribution to the development of
modern astronomical photometry is invaluable.

We are very grateful to an anonymous Referee for a prompt and constructive report, that improved the quality of the manuscript. We acknowledge also the contribution of the A\&A editor T. Forveille for additional insightful comments and suggestions.
We are very grateful to many colleagues around the world that gave their help to this project by providing data and guidelines to use them, as well as valuable suggestions and advice. We hope that the following list includes all of them: M.S. Bessell, C. Clementini, E.A. Magnier, P.M. Marrese, A. Mucciarelli, S.J. Murphy, D. Nardiello, P. Stetson, M. Scalco. We apologise for those that we inadvertently forgot to mention.
With this work PM, MB, AB and CC acknowledge with gratitude the encouragement of Flavio Fusi Pecci who convinced them to join the adventure of the Gaia mission, seventeen years ago.

The \gaia\ mission and data processing have financially been supported by, in alphabetical order by country:
\begin{itemize}
\item the Algerian Centre de Recherche en Astronomie, Astrophysique et G\'{e}ophysique of Bouzareah Observatory;
\item the Austrian Fonds zur F\"{o}rderung der wissenschaftlichen Forschung (FWF) Hertha Firnberg Programme through grants T359, P20046, and P23737;
\item the BELgian federal Science Policy Office (BELSPO) through various PROgramme de D\'{e}veloppement d'Exp\'{e}riences scientifiques (PRODEX) grants and the Polish Academy of Sciences - Fonds Wetenschappelijk Onderzoek through grant VS.091.16N, and the Fonds de la Recherche Scientifique (FNRS), and the Research Council of Katholieke Universiteit (KU) Leuven through grant C16/18/005 (Pushing AsteRoseismology to the next level with TESS, GaiA, and the Sloan DIgital Sky SurvEy -- PARADISE);  
\item the Brazil-France exchange programmes Funda\c{c}\~{a}o de Amparo \`{a} Pesquisa do Estado de S\~{a}o Paulo (FAPESP) and Coordena\c{c}\~{a}o de Aperfeicoamento de Pessoal de N\'{\i}vel Superior (CAPES) - Comit\'{e} Fran\c{c}ais d'Evaluation de la Coop\'{e}ration Universitaire et Scientifique avec le Br\'{e}sil (COFECUB);
\item the Chilean Agencia Nacional de Investigaci\'{o}n y Desarrollo (ANID) through Fondo Nacional de Desarrollo Cient\'{\i}fico y Tecnol\'{o}gico (FONDECYT) Regular Project 1210992 (L.~Chemin);
\item the National Natural Science Foundation of China (NSFC) through grants 11573054, 11703065, and 12173069, the China Scholarship Council through grant 201806040200, and the Natural Science Foundation of Shanghai through grant 21ZR1474100;  
\item the Tenure Track Pilot Programme of the Croatian Science Foundation and the \'{E}cole Polytechnique F\'{e}d\'{e}rale de Lausanne and the project TTP-2018-07-1171 `Mining the Variable Sky', with the funds of the Croatian-Swiss Research Programme;
\item the Czech-Republic Ministry of Education, Youth, and Sports through grant LG 15010 and INTER-EXCELLENCE grant LTAUSA18093, and the Czech Space Office through ESA PECS contract 98058;
\item the Danish Ministry of Science;
\item the Estonian Ministry of Education and Research through grant IUT40-1;
\item the European Commission?s Sixth Framework Programme through the European Leadership in Space Astrometry (\href{https://www.cosmos.esa.int/web/gaia/elsa-rtn-programme}{ELSA}) Marie Curie Research Training Network (MRTN-CT-2006-033481), through Marie Curie project PIOF-GA-2009-255267 (Space AsteroSeismology \& RR Lyrae stars, SAS-RRL), and through a Marie Curie Transfer-of-Knowledge (ToK) fellowship (MTKD-CT-2004-014188); the European Commission's Seventh Framework Programme through grant FP7-606740 (FP7-SPACE-2013-1) for the \gaia\ European Network for Improved data User Services (\href{https://gaia.ub.edu/twiki/do/view/GENIUS/}{GENIUS}) and through grant 264895 for the \gaia\ Research for European Astronomy Training (\href{https://www.cosmos.esa.int/web/gaia/great-programme}{GREAT-ITN}) network;
\item the European Cooperation in Science and Technology (COST) through COST Action CA18104 `Revealing the Milky Way with \gaia (MW-Gaia)';
\item the European Research Council (ERC) through grants 320360, 647208, and 834148 and through the European Union?s Horizon 2020 research and innovation and excellent science programmes through Marie Sk{\l}odowska-Curie grant 745617 (Our Galaxy at full HD -- Gal-HD) and 895174 (The build-up and fate of self-gravitating systems in the Universe) as well as grants 687378 (Small Bodies: Near and Far), 682115 (Using the Magellanic Clouds to Understand the Interaction of Galaxies), 695099 (A sub-percent distance scale from binaries and Cepheids -- CepBin), 716155 (Structured ACCREtion Disks -- SACCRED), 951549 (Sub-percent calibration of the extragalactic distance scale in the era of big surveys -- UniverScale), and 101004214 (Innovative Scientific Data Exploration and Exploitation Applications for Space Sciences -- EXPLORE);
\item the European Science Foundation (ESF), in the framework of the \gaia\ Research for European Astronomy Training Research Network Programme (\href{https://www.cosmos.esa.int/web/gaia/great-programme}{GREAT-ESF});
\item the European Space Agency (ESA) in the framework of the \gaia\ project, through the Plan for European Cooperating States (PECS) programme through contracts C98090 and 4000106398/12/NL/KML for Hungary, through contract 4000115263/15/NL/IB for Germany, and through PROgramme de D\'{e}veloppement d'Exp\'{e}riences scientifiques (PRODEX) grant 4000127986 for Slovenia;  
\item the Academy of Finland through grants 299543, 307157, 325805, 328654, 336546, and 345115 and the Magnus Ehrnrooth Foundation;
\item the French Centre National d?\'{E}tudes Spatiales (CNES), the Agence Nationale de la Recherche (ANR) through grant ANR-10-IDEX-0001-02 for the `Investissements d'avenir' programme, through grant ANR-15-CE31-0007 for project `Modelling the Milky Way in the \gaia era? (MOD4Gaia), through grant ANR-14-CE33-0014-01 for project `The Milky Way disc formation in the \gaia era? (ARCHEOGAL), through grant ANR-15-CE31-0012-01 for project `Unlocking the potential of Cepheids as primary distance calibrators? (UnlockCepheids), through grant ANR-19-CE31-0017 for project `Secular evolution of galxies' (SEGAL), and through grant ANR-18-CE31-0006 for project `Galactic Dark Matter' (GaDaMa), the Centre National de la Recherche Scientifique (CNRS) and its SNO \gaia of the Institut des Sciences de l?Univers (INSU), its Programmes Nationaux: Cosmologie et Galaxies (PNCG), Gravitation R\'{e}f\'{e}rences Astronomie M\'{e}trologie (PNGRAM), Plan\'{e}tologie (PNP), Physique et Chimie du Milieu Interstellaire (PCMI), and Physique Stellaire (PNPS), the `Action F\'{e}d\'{e}ratrice \gaia' of the Observatoire de Paris, the R\'{e}gion de Franche-Comt\'{e}, the Institut National Polytechnique (INP) and the Institut National de Physique nucl\'{e}aire et de Physique des Particules (IN2P3) co-funded by CNES;
\item the German Aerospace Agency (Deutsches Zentrum f\"{u}r Luft- und Raumfahrt e.V., DLR) through grants 50QG0501, 50QG0601, 50QG0602, 50QG0701, 50QG0901, 50QG1001, 50QG1101, 50\-QG1401, 50QG1402, 50QG1403, 50QG1404, 50QG1904, 50QG2101, 50QG2102, and 50QG2202, and the Centre for Information Services and High Performance Computing (ZIH) at the Technische Universit\"{a}t Dresden for generous allocations of computer time;
\item the Hungarian Academy of Sciences through the Lend\"{u}let Programme grants LP2014-17 and LP2018-7 and the Hungarian National Research, Development, and Innovation Office (NKFIH) through grant KKP-137523 (`SeismoLab');
\item the Science Foundation Ireland (SFI) through a Royal Society - SFI University Research Fellowship (M.~Fraser);
\item the Israel Ministry of Science and Technology through grant 3-18143 and the Tel Aviv University Center for Artificial Intelligence and Data Science (TAD) through a grant;
\item the Agenzia Spaziale Italiana (ASI) through contracts I/037/08/0, I/058/10/0, 2014-025-R.0, 2014-025-R.1.2015, and 2018-24-HH.0 to the Italian Istituto Nazionale di Astrofisica (INAF), contract 2014-049-R.0/1/2 to INAF for the Space Science Data Centre (SSDC, formerly known as the ASI Science Data Center, ASDC), contracts I/008/10/0, 2013/030/I.0, 2013-030-I.0.1-2015, and 2016-17-I.0 to the Aerospace Logistics Technology Engineering Company (ALTEC S.p.A.), INAF, and the Italian Ministry of Education, University, and Research (Ministero dell'Istruzione, dell'Universit\`{a} e della Ricerca) through the Premiale project `MIning The Cosmos Big Data and Innovative Italian Technology for Frontier Astrophysics and Cosmology' (MITiC);
\item the Netherlands Organisation for Scientific Research (NWO) through grant NWO-M-614.061.414, through a VICI grant (A.~Helmi), and through a Spinoza prize (A.~Helmi), and the Netherlands Research School for Astronomy (NOVA);
\item the Polish National Science Centre through HARMONIA grant 2018/30/M/ST9/00311 and DAINA grant 2017/27/L/ST9/03221 and the Ministry of Science and Higher Education (MNiSW) through grant DIR/WK/2018/12;
\item the Portuguese Funda\c{c}\~{a}o para a Ci\^{e}ncia e a Tecnologia (FCT) through national funds, grants SFRH/\-BD/128840/2017 and PTDC/FIS-AST/30389/2017, and work contract DL 57/2016/CP1364/CT0006, the Fundo Europeu de Desenvolvimento Regional (FEDER) through grant POCI-01-0145-FEDER-030389 and its Programa Operacional Competitividade e Internacionaliza\c{c}\~{a}o (COMPETE2020) through grants UIDB/04434/2020 and UIDP/04434/2020, and the Strategic Programme UIDB/\-00099/2020 for the Centro de Astrof\'{\i}sica e Gravita\c{c}\~{a}o (CENTRA);  
\item the Slovenian Research Agency through grant P1-0188;
\item the Spanish Ministry of Economy (MINECO/FEDER, UE), the Spanish Ministry of Science and Innovation (MICIN), the Spanish Ministry of Education, Culture, and Sports, and the Spanish Government through grants BES-2016-078499, BES-2017-083126, BES-C-2017-0085, ESP2016-80079-C2-1-R, ESP2016-80079-C2-2-R, FPU16/03827, PDC2021-121059-C22, RTI2018-095076-B-C22, and TIN2015-65316-P (`Computaci\'{o}n de Altas Prestaciones VII'), the Juan de la Cierva Incorporaci\'{o}n Programme (FJCI-2015-2671 and IJC2019-04862-I for F.~Anders), the Severo Ochoa Centre of Excellence Programme (SEV2015-0493), and MICIN/AEI/10.13039/501100011033 (and the European Union through European Regional Development Fund `A way of making Europe') through grant RTI2018-095076-B-C21, the Institute of Cosmos Sciences University of Barcelona (ICCUB, Unidad de Excelencia `Mar\'{\i}a de Maeztu?) through grant CEX2019-000918-M, the University of Barcelona's official doctoral programme for the development of an R+D+i project through an Ajuts de Personal Investigador en Formaci\'{o} (APIF) grant, the Spanish Virtual Observatory through project AyA2017-84089, the Galician Regional Government, Xunta de Galicia, through grants ED431B-2021/36, ED481A-2019/155, and ED481A-2021/296, the Centro de Investigaci\'{o}n en Tecnolog\'{\i}as de la Informaci\'{o}n y las Comunicaciones (CITIC), funded by the Xunta de Galicia and the European Union (European Regional Development Fund -- Galicia 2014-2020 Programme), through grant ED431G-2019/01, the Red Espa\~{n}ola de Supercomputaci\'{o}n (RES) computer resources at MareNostrum, the Barcelona Supercomputing Centre - Centro Nacional de Supercomputaci\'{o}n (BSC-CNS) through activities AECT-2017-2-0002, AECT-2017-3-0006, AECT-2018-1-0017, AECT-2018-2-0013, AECT-2018-3-0011, AECT-2019-1-0010, AECT-2019-2-0014, AECT-2019-3-0003, AECT-2020-1-0004, and DATA-2020-1-0010, the Departament d'Innovaci\'{o}, Universitats i Empresa de la Generalitat de Catalunya through grant 2014-SGR-1051 for project `Models de Programaci\'{o} i Entorns d'Execuci\'{o} Parallels' (MPEXPAR), and Ramon y Cajal Fellowship RYC2018-025968-I funded by MICIN/AEI/10.13039/501100011033 and the European Science Foundation (`Investing in your future');
\item the Swedish National Space Agency (SNSA/Rymdstyrelsen);
\item the Swiss State Secretariat for Education, Research, and Innovation through the Swiss Activit\'{e}s Nationales Compl\'{e}mentaires and the Swiss National Science Foundation through an Eccellenza Professorial Fellowship (award PCEFP2\_194638 for R.~Anderson);
\item the United Kingdom Particle Physics and Astronomy Research Council (PPARC), the United Kingdom Science and Technology Facilities Council (STFC), and the United Kingdom Space Agency (UKSA) through the following grants to the University of Bristol, the University of Cambridge, the University of Edinburgh, the University of Leicester, the Mullard Space Sciences Laboratory of University College London, and the United Kingdom Rutherford Appleton Laboratory (RAL): PP/D006511/1, PP/D006546/1, PP/D006570/1, ST/I000852/1, ST/J005045/1, ST/K00056X/1, ST/\-K000209/1, ST/K000756/1, ST/L006561/1, ST/N000595/1, ST/N000641/1, ST/N000978/1, ST/\-N001117/1, ST/S000089/1, ST/S000976/1, ST/S000984/1, ST/S001123/1, ST/S001948/1, ST/\-S001980/1, ST/S002103/1, ST/V000969/1, ST/W002469/1, ST/W002493/1, ST/W002671/1, ST/W002809/1, and EP/V520342/1.
\end{itemize}

The \gaia\ project and data processing have made use of:
\begin{itemize}
\item the Set of Identifications, Measurements, and Bibliography for Astronomical Data \citep[SIMBAD,][]{2000AAS..143....9W}, the `Aladin sky atlas' \citep{2000A&AS..143...33B,2014ASPC..485..277B}, and the VizieR catalogue access tool \citep{2000A&AS..143...23O}, all operated at the Centre de Donn\'{e}es astronomiques de Strasbourg (\href{http://cds.u-strasbg.fr/}{CDS});
\item the National Aeronautics and Space Administration (NASA) Astrophysics Data System (\href{http://adsabs.harvard.edu/abstract_service.html}{ADS});
\item the SPace ENVironment Information System (SPENVIS), initiated by the Space Environment and Effects Section (TEC-EES) of ESA and developed by the Belgian Institute for Space Aeronomy (BIRA-IASB) under ESA contract through ESA?s General Support Technologies Programme (GSTP), administered by the BELgian federal Science Policy Office (BELSPO);
\item the Spanish Virtual Observatory (https://svo.cab.inta-csic.es) project funded by MCIN/AEI/10.13039/501100011033/ through grant PID2020-112949GB-I00 \citep{SVO};
\item the software products \href{http://www.starlink.ac.uk/topcat/}{TOPCAT}, \href{http://www.starlink.ac.uk/stil}{STIL}, and \href{http://www.starlink.ac.uk/stilts}{STILTS} \citep{2005ASPC..347...29T,2006ASPC..351..666T};
\item Matplotlib \citep{Hunter:2007};
\item IPython \citep{PER-GRA:2007};  
\item Astropy, a community-developed core Python package for Astronomy \citep{2018AJ....156..123A};
\item  NumPy \citep{harris2020array};
\item pyphot
(http://github.com/mfouesneau/pyphot);
\item R \citep{RManual};
\item the \hip-2 catalogue \citep{2007A&A...474..653V}. The \hip and \tyc catalogues were constructed under the responsibility of large scientific teams collaborating with ESA. The Consortia Leaders were Lennart Lindegren (Lund, Sweden: NDAC) and Jean Kovalevsky (Grasse, France: FAST), together responsible for the \hip Catalogue; Erik H{\o}g (Copenhagen, Denmark: TDAC) responsible for the \tyc Catalogue; and Catherine Turon (Meudon, France: INCA) responsible for the \hip Input Catalogue (HIC);  
\item the \tyctwo catalogue \citep{2000A&A...355L..27H}, the construction of which was supported by the Velux Foundation of 1981 and the Danish Space Board;
\item The Tycho double star catalogue \citep[TDSC,][]{2002A&A...384..180F}, based on observations made with the ESA \hip astrometry satellite, as supported by the Danish Space Board and the United States Naval Observatory through their double-star programme;

\item the first data release of the Pan-STARRS survey 
\citep{panstarrs1,panstarrs1b,panstarrs1c,panstarrs1d,Magnier2020,panstarrs1f}. 
The Pan-STARRS1 Surveys (PS1) and the PS1 public science archive have been made possible through contributions by the Institute for Astronomy, the University of Hawaii, the Pan-STARRS Project Office, the Max-Planck Society and its participating institutes, the Max Planck Institute for Astronomy, Heidelberg and the Max Planck Institute for Extraterrestrial Physics, Garching, The Johns Hopkins University, Durham University, the University of Edinburgh, the Queen's University Belfast, the Harvard-Smithsonian Center for Astrophysics, the Las Cumbres Observatory Global Telescope Network Incorporated, the National Central University of Taiwan, the Space Telescope Science Institute, the National Aeronautics and Space Administration (NASA) through grant NNX08AR22G issued through the Planetary Science Division of the NASA Science Mission Directorate, the National Science Foundation through grant AST-1238877, the University of Maryland, Eotvos Lorand University (ELTE), the Los Alamos National Laboratory, and the Gordon and Betty Moore Foundation;

\item the fifteenth and the sixteenth data releases of the Sloan Digitial Sky Survey, \citep[SDSS DR15,][]{Blanton2017}  and \citep[SDSS DR16,][]{apogee16}. 
Funding for the Sloan Digital Sky Survey IV has been provided by the Alfred P. Sloan Foundation, the U.S. Department of Energy Office of Science, and the Participating Institutions. 
SDSS-IV acknowledges support and resources from the Center for High  Performance Computing  at the University of Utah. The SDSS website is www.sdss.org. SDSS-IV is managed by the 
Astrophysical Research Consortium for the Participating Institutions of the SDSS Collaboration including the Brazilian Participation Group, the Carnegie Institution for Science, Carnegie Mellon University, Center for Astrophysics | Harvard \& Smithsonian, the Chilean Participation 
Group, the French Participation Group, Instituto de Astrof\'isica de Canarias, The Johns Hopkins University, Kavli Institute for the Physics and Mathematics of the Universe (IPMU) / University of Tokyo, the Korean Participation Group, Lawrence Berkeley National Laboratory, Leibniz Institut f\"ur Astrophysik Potsdam (AIP),  Max-Planck-Institut f\"ur Astronomie (MPIA Heidelberg), Max-Planck-Institut f\"ur Astrophysik (MPA Garching), Max-Planck-Institut f\"ur 
Extraterrestrische Physik (MPE), National Astronomical Observatories of China, New Mexico State University, New York University, University of Notre Dame, Observat\'ario Nacional / MCTI, The Ohio State University, Pennsylvania State University, Shanghai Astronomical Observatory, United Kingdom Participation Group, Universidad Nacional Aut\'onoma de M\'exico, University of Arizona, University of Colorado Boulder, University of Oxford, University of Portsmouth, University of Utah, University of Virginia, University of Washington, University of 
Wisconsin, Vanderbilt University, and Yale University.

\item the second release of the SkyMapper catalogue \citep[SkyMapper DR2,][Digital Object Identifier 10.25914/5ce60d31ce759]{2019PASA...36...33O}. The national facility capability for SkyMapper has been funded through grant LE130100104 from the Australian Research Council (ARC) Linkage Infrastructure, Equipment, and Facilities (LIEF) programme, awarded to the University of Sydney, the Australian National University, Swinburne University of Technology, the University of Queensland, the University of Western Australia, the University of Melbourne, Curtin University of Technology, Monash University, and the Australian Astronomical Observatory. SkyMapper is owned and operated by The Australian National University's Research School of Astronomy and Astrophysics. The survey data were processed and provided by the SkyMapper Team at the the Australian National University. The SkyMapper node of the All-Sky Virtual Observatory (ASVO) is hosted at the National Computational Infrastructure (NCI). Development and support the SkyMapper node of the ASVO has been funded in part by Astronomy Australia Limited (AAL) and the Australian Government through the Commonwealth's Education Investment Fund (EIF) and National Collaborative Research Infrastructure Strategy (NCRIS), particularly the National eResearch Collaboration Tools and Resources (NeCTAR) and the Australian National Data Service Projects (ANDS);

\end{itemize}

The GBOT programme  uses observations collected at (i) the European Organisation for Astronomical Research in the Southern Hemisphere (ESO) with the VLT Survey Telescope (VST), under ESO programmes
092.B-0165,
093.B-0236,
094.B-0181,
095.B-0046,
096.B-0162,
097.B-0304,
098.B-0030,
099.B-0034,
0100.B-0131,
0101.B-0156,
0102.B-0174, and
0103.B-0165;
and (ii) the Liverpool Telescope, which is operated on the island of La Palma by Liverpool John Moores University in the Spanish Observatorio del Roque de los Muchachos of the Instituto de Astrof\'{\i}sica de Canarias with financial support from the United Kingdom Science and Technology Facilities Council, and (iii) telescopes of the Las Cumbres Observatory Global Telescope Network.
%%%%%%%%%%%%%%%%%%%%%%%%%%%%%%%%%%%%%% END APPE ACKNO

%%%%%%%%%%%%%%%%%%%%%%%%%%%%%%%%%%%%%% APPE R_F
%\input{sections/appe_Rf}
\section{Minimum width of flux-conserving passbands and suggestions for line photometry}
\label{sec:app_Rf}

In general, the flux through a given passband can be correctly measured from  XPSP only if the characteristic width of the TC is larger than the LSF of the EC XP spectrum in the wavelength range of the passband. To trace the relation between passband width and LSF width we adopt the following simple parameter:

\begin{equation}
    Rf=\frac{FWHM_{passband}(\lambda_0)}{FWHM_{LSF}(\lambda_0)}\label{eq:Rf_definition},
\end{equation}

which is the ratio between the FWHM of the passband  and of the XP LSF at the central/peak wavelength of the passband $\lambda_0$, where the relevant ECS XP LSF width should be taken from \citet[][]{EDR3-DPACP-120}. This simple parameter cannot take into account all the subtleties of the relation we are considering, that is, the FWHM is not fully adequate to describe asymmetric TCs. However, here it is sufficient to address the core of this problem and to provide a simple and general criterion for flux conservation and reproducibility of magnitudes in existing systems. 

In principle, for symmetric passbands and local symmetric and perfectly modelled LSF, $Rf>1$ should guarantee that all the incoming flux through the considered TC can be correctly measured by XPSP in any case. However, the LSF is not symmetric \citep{EDR3-DPACP-120}, the instrument model that is used to transform XP mean spectra into EC is not perfect and, consequently, the mixing between photons of different wavelengths ---which is intrinsic to slit-less spectroscopy--- is not optimally corrected. 

To derive an empirical criterion defining the minimum $Rf$ of a passband whose XPSP correctly measure and/or conserve the flux, we proceed as follows. Consider a spectral feature that is very narrow with respect to the local XP LSF, for example the stellar H$\alpha$ Balmer line, and suppose we attempt to measure the flux in a portion of the spectrum including the line with XPSP using a passband with $Rf<1$. If the source has H$\alpha$ in emission, the XP LSF will move a fraction of photons from the line out of the range covered by the passband, resulting in a loss of H$\alpha$ flux. Photons outside that range would also leak within the passband for the same reason, but the asymmetry between the excess flux in the line and the lower surrounding continuum would end up in a net flux loss. The opposite would happen for H$\alpha$ in absorption: in this case, the asymmetry between the deficit of photons in the line and the flat but higher continuum level outside the passband will lead to the measurement of a spurious excess of flux in the passband, mimicking a lower depth of the line.

The idea is to take a set of stars for which we have EC XP spectra and their external counterparts at much higher spectral resolution (HR spectra\footnote{We note that synthetic photometry from HR spectra, in this context, is fully equivalent to external direct photometry obtained by imaging with photometric filters. The conclusions reached in this section are fully applicable to narrow-line photometry obtained in this way, as, e.g. in IPHAS \citep{drew2005} or the VST Photometric H$\alpha$ Survey \citep[VPHAS+,][]{drew2014}.}; in the specific case, about $R\simeq 1000$, to be compared with $R\simeq 30-80$ of XP ECS) and to compare synthetic photometry from the two source spectra around strong spectral lines using passbands of various width. Dealing with absorption features in the presence of strong lines, the fluxes through an overly narrow passband will be larger when measured from XP than from HR spectra, corresponding to positive magnitude differences ${\rm mag}_{HR}-{\rm mag}_{XP}$. Then, progressively wider passbands can be tested until the magnitude difference becomes null, thus identifying the lower $Rf$ limit allowing correct measurement of the flux in the presence of a strong spectral feature. 

\begin{figure}[!htbp]
    \centering
    \includegraphics[width=(\columnwidth)]{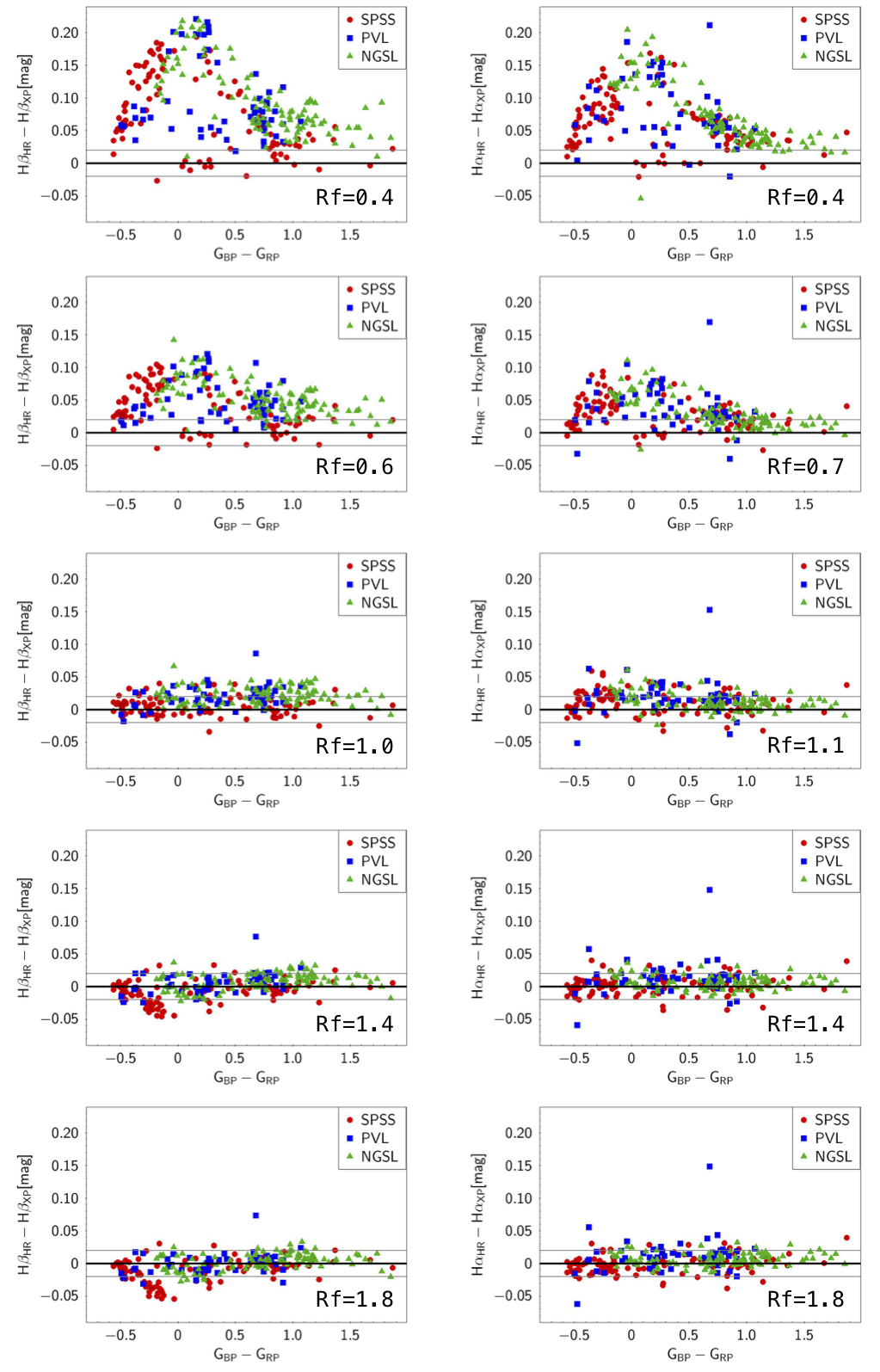}
    \caption{Difference in synthetic magnitudes from HR and XP spectra for SPSS, PVL, and selected NGSL using passbands of increasing FWHM (from top to bottom) to measure the flux around H$\beta$ (left panels) and H$\alpha$ (right panels), as a function of G$_{BP}$-G$_{RP}$ colour. The passbands FWHM adopted in the various panels are, from top to bottom, 5, 8, 13, 18, 23~nm (H$\beta$), and 3, 6, 9, 12, 15~nm (H$\alpha$). The corresponding Rf values are reported in each panel.}
    \label{fig:appe_Rf_delta}
\end{figure}

\begin{figure}[!htbp]
    \centering
    \includegraphics[width=(0.7\columnwidth)]{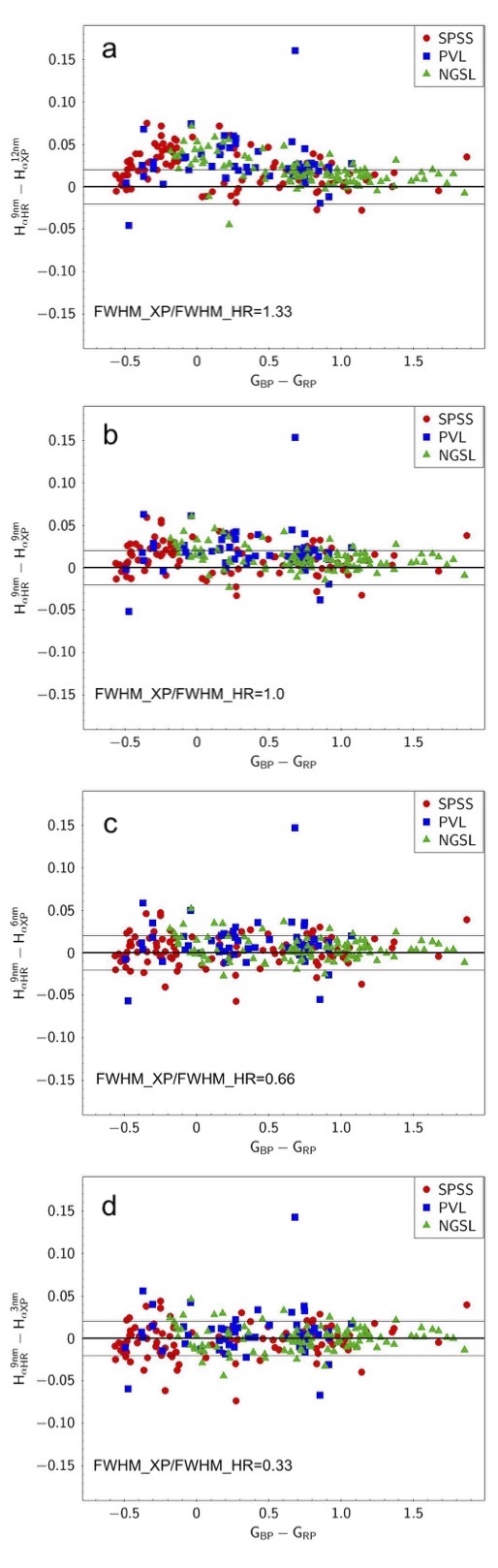}
    \caption{Difference of synthetic H$\alpha$ magnitudes from HR and XP spectra for SPSS, PVL, and selected NGSL as a function of \bprp colour. In this case, the width of the TC adopted for synthetic photometry on HR spectra is kept fixed ($FWHM=9.0$~nm), while the width of the TC used for synthetic photometry on XP spectra is varied from $FWHM=12.0$~nm to  $FWHM=9.0, 6.0, 3.0$~nm, from top to bottom. The  various panels are labelled according to the ratio between the FWHM of the XP and HR TCs. }
    \label{fig:appe_Hadiff}
\end{figure}

Here we perform this test with a set of custom synthetic passbands centred on H$\beta$ ($FWHM_{LSF}= 12.6$~nm) and H$\alpha$ ($FWHM_{LSF}= 8.4$~nm), with FWHM ranging from 1~nm to 25~nm. TCs are centred at the wavelength of the corresponding line and have a strictly symmetric shape, being the junction of two error functions. 

The sample is composed of the calibrating and validating sets of stars including the Gaia SPSS and the PVL \citep{pancino2021} and the selection of NGSL stars \citep{ngsl16} adopted by \citet{EDR3-DPACP-120}. AB magnitudes are considered as they directly trace fluxes.

Figure~\ref{fig:appe_Rf_delta} illustrates the procedure. In the upper pair of panels, passbands significantly narrower than the local LSF are adopted ($Rf=0.4$). Consequently, for the majority of stars, the difference between HR and XP synthetic magnitudes increases from G$_{BP}$-G$_{RP}\simeq -0.5$ to G$_{BP}$-G$_{RP}\simeq 0.0$, reaching its maximum for stars displaying the maximum strength of Balmer absorption lines (A stars; the handful of exceptions are DC and DB WDs with the colour of A stars but lacking strong H lines in their spectra). Then the magnitude difference begins to decrease, reaching a null value for $BP-RP>1.0$, for spectral types later than G. 
As passbands with larger $Rf$ are adopted, the amplitude of the arch of the magnitude difference decreases, until they reach $\simeq 0.0$ over the considered colour range at $Rf\simeq 1.4$, remaining there for larger values of $Rf$. It is important to note that the amplitude of the discrepancy is already as low as $\simeq 0.01-0.02$~mag around $Rf=1.0-1.1$. Still, we prefer to provide a conservative general criterion, possibly accounting also for the approximations involved.

The conclusion of this experiment is that  synthetic fluxes and magnitudes can be accurately measured from EC XP spectra only adopting passbands with $Rf\ge 1.4$. This implies, that { magnitudes from existing systems can be accurately reproduced only if this condition is satisfied, if the TC of the existing system is adopted to obtain the corresponding XPSP} (see below for a different approach that may help to circumvent this rule).
It is reassuring that the same result is consistently found when testing two spectral features that are measured in the different instruments that are used to get mean XP spectra, i.e. BP and (mainly) RP for H$\beta$ and H$\alpha$, respectively. Moreover, in regions of the spectrum lacking strong features, the flux is conserved in XPSP also using passbands with $Rf<1.4$, because, in the absence of any strong flux asymmetry, the losses from inside the passband are compensated by the leaks from outside the passband, leaving the balance near the equilibrium.

The above conclusions refer to the comparison between photometry obtained from different spectra {\em with the same TCs}. However, following up the results shown in Sect.~\ref{sec:narrow_iphas}, now we compare H$\alpha$ magnitudes obtained with a FWHM=9~nm TC ($Rf=1.1$) from the XR spectra with those obtained from XP spectra using TCs of various width, in particular $FWHM=12, 9, 6, 3$~nm. The results of this experiment are presented in Fig.~\ref{fig:appe_Hadiff}. When the passband adopted for the XPSP is wider than that taken as reference for the HR SP (panel a), the distribution is fully analogous to that seen in Fig.~\ref{fig:appe_Rf_delta} for $Rf<1.0$, as, also in this case, the signal from the line is diluted by the continuum. In that case, the dilution was produced by an asymmetric exchange of photons at the thresholds of a passband that is narrower than the local LSF. Here it is due to the inclusion of larger portions of the continuum in the passband adopted for the XP spectra than in the one adopted for the HR spectra. When, as in panel (b), the same passband is adopted in both cases, the performance is determined by the $Rf$, as already established in the previous experiment. However, the comparisons shown in panels (c) and (d) of Fig.~\ref{fig:appe_Hadiff} show that the HR photometry can be satisfactorily matched even for $Rf<1.4$,  with XPSP obtained with a narrower TC than that adopted for the HR spectra. In such a case, the increased sensitivity of the narrower passband offsets the flux lost outside of the passband edges. In the limit of the narrowest synthetic passband, the quantity measured is the height of the line relative to the continuum, after it is convolved with the LSF. A limited set of experiments as well as simple models suggest that with this approach, narrow line photometry in presence of strong spectral features can be reproduced with XPSP down to $Rf\ga 1.0$. In these cases, the best choice of the width of the TC to be adopted for the XPSP should be determined with experiments like those shown in Fig.~\ref{fig:appe_Hadiff}, taking into account the LSF at the wavelegth of the considered line and the TC adopted by the survey that one intend to calibrate with \gaia XPSP. Finally, it is worth noting that while passbands with $Rf<1.0$ cannot conserve or reliably trace the flux of an emission line, they can
still carry useful information on the spectral feature they are targeting. 

%%%%%%%%%%%%%%%%%%%%%%%%%%%%%%%%%%%%%% END APPE R_F

%%%%%%%%%%%%%%%%%%%%%%%%%%%%%%%%%%%%%% APPE ISO
%\input{sections/appe_isochrones}
\section{Comparisons with stellar models}
\label{sec:app_isoc}

In this section, we use stellar models to assess the quality of five standardised photometry systems (Johnson, Pan-STARRS1,SDSS, Str\"omgren , HST ACS/WFC).
We compare the photometry of seven OCs presented in Table  \ref{tab:iso} with the expectations from theoretical isochrones.

\begin{figure}[!htbp]
\center{
\includegraphics[width=\columnwidth]{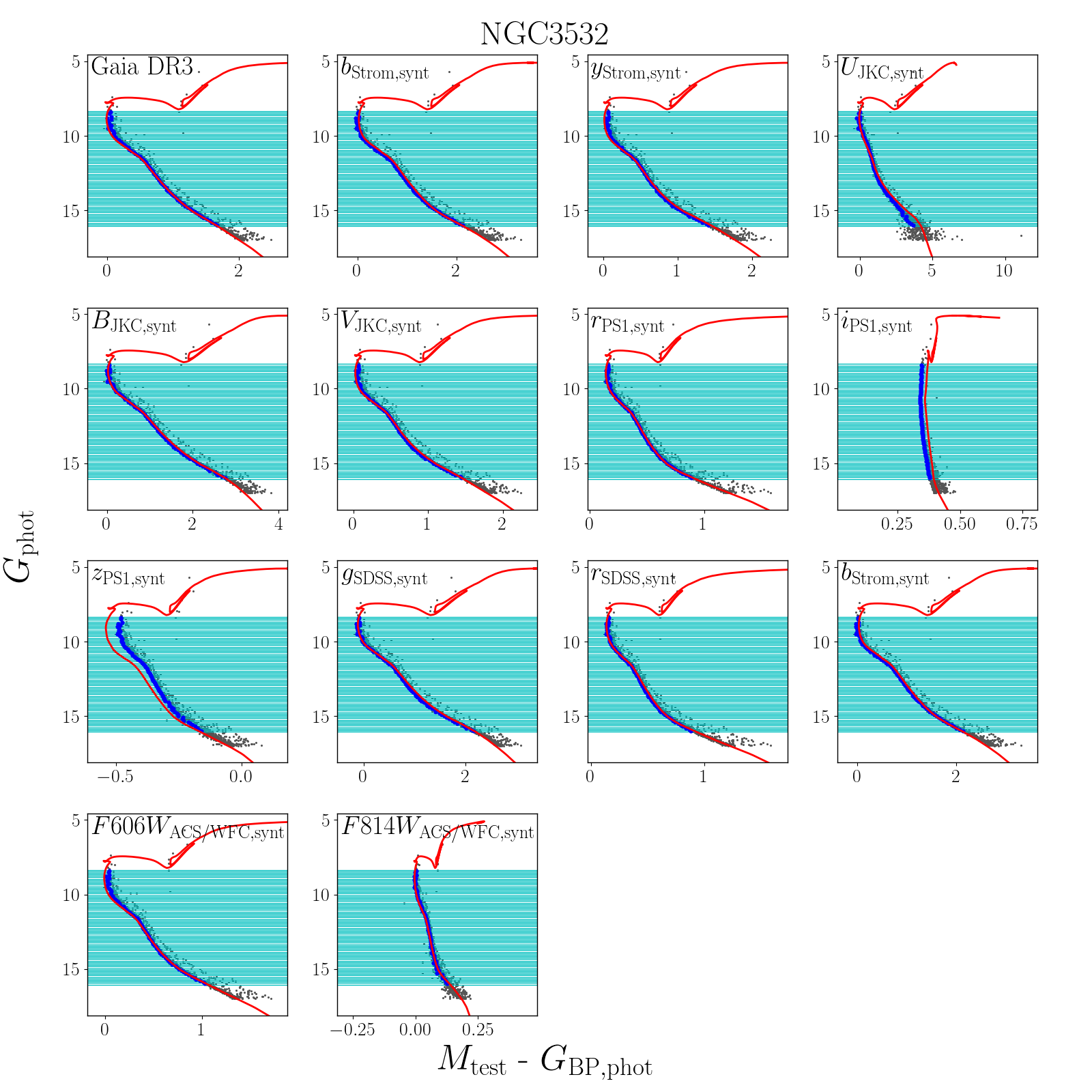}
}
\caption{Example of  isochrone fitting on the cluster NGC~3532.The first panel is  $Gaia$ CMD  while the other CMD contained several synthetic bands from Johnson (B, V), Pan-STARRS1(rp, ip, zp), SLOAN(u, g, r). The red line is the PARSEC isochrone, the cyan lines show the binning definition (see \ref{sec:app_isoc}), and the blue dots are the bluer edge of the colour (x axis) of the stellar distribution.} 
\label{fig:isoCMD}
\end{figure}

\begin{figure}[!htbp]
\center{
\includegraphics[width=\columnwidth]{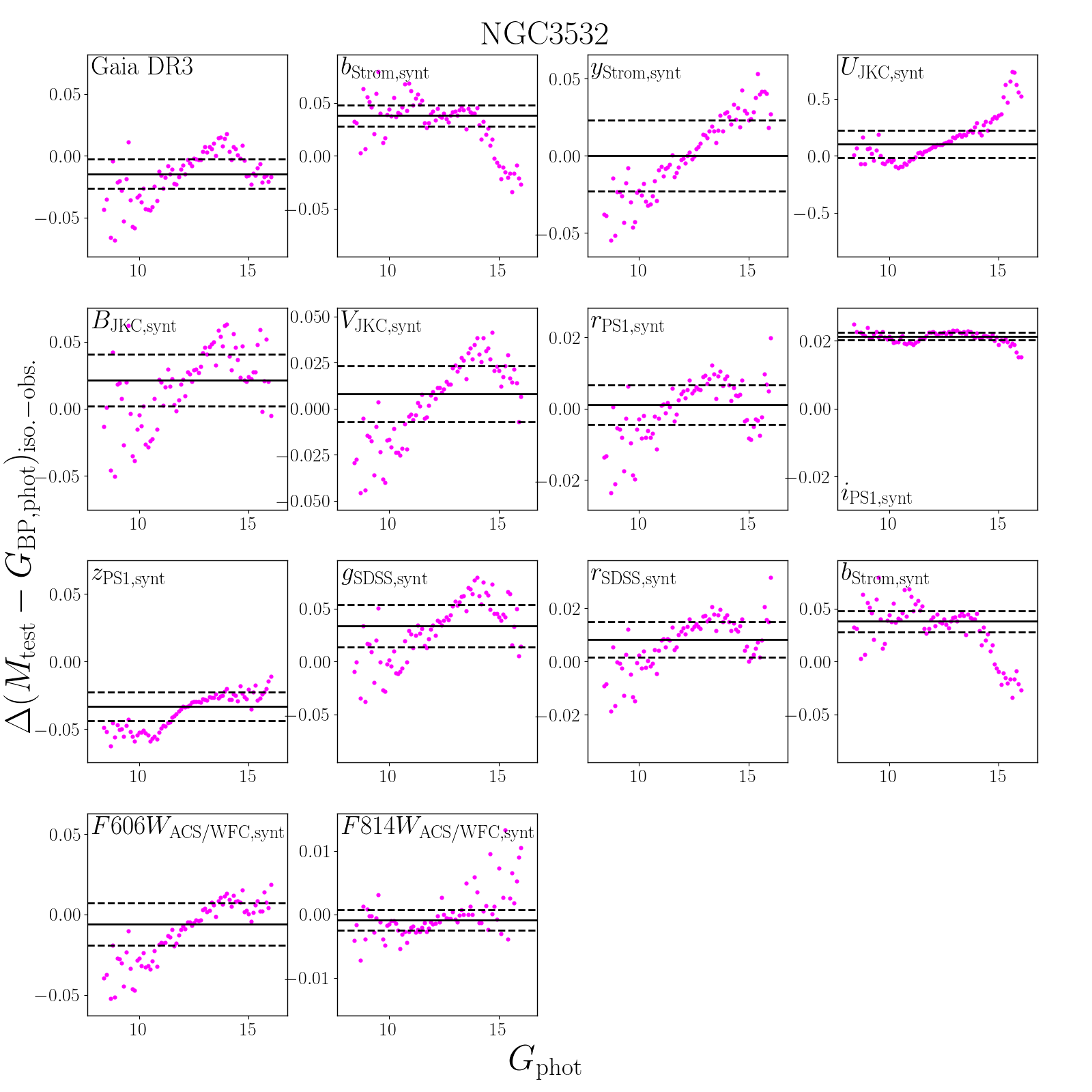}
}
\caption{ Residuals between the isochrone and the bluer edge of the colour (x axis) in the Fig.~\ref{fig:isoCMD} CMDs. The back solid line is the median of the deviations while the dashed line represents the MAD.} 
\label{fig:isoRES}
\end{figure}

\begin{figure}
\center{
\includegraphics[width=\columnwidth]{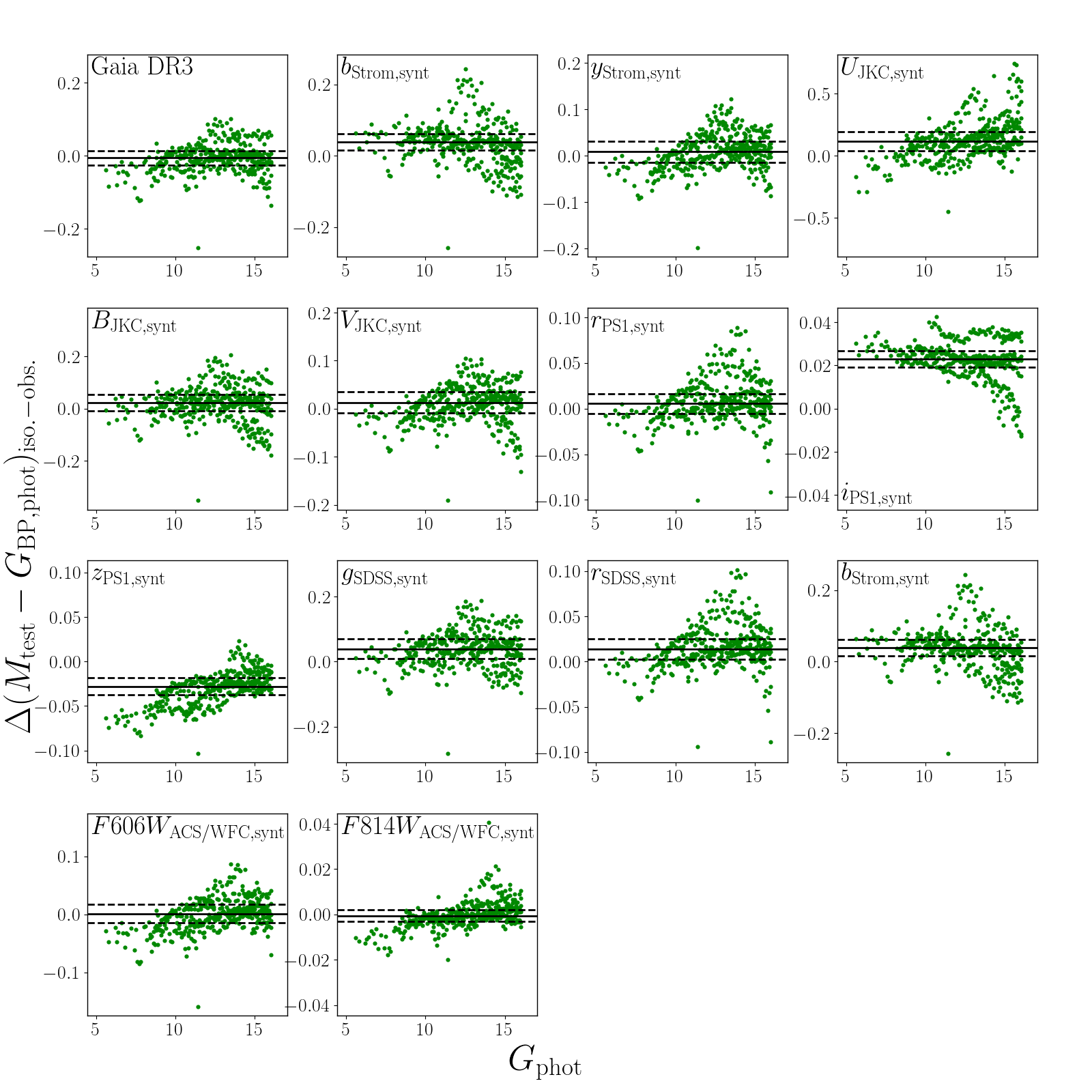}
}
\caption{As in  Fig.~\ref{fig:isoRES}: Residuals between the isochrone and the data for  the selected OCs. The back solid line is the median of the deviations while the dashed line represents the MAD.} 
\label{fig:isoall}
\end{figure}

The set of theoretical isochrones used for the comparison are taken from the PARSEC library\footnote{\url{http://stev.oapd.inaf.it/cgi-bin/cmd}} v2.1 (the \underline{PA}dova and T\underline{R}ieste \underline{S}tellar \underline{E}volution \underline{C}ode; \citealt{Bressan2012}). 
The bolometric correction were calculated using the online tool YBC\footnote{\url{http://stev.oapd.inaf.it/YBC/index.html}} \citep[PARSEC Bolometric Correction;][]{Chen2019}, which interpolates a series of pre-computed bolometric correction tables in $T_\mathrm{eff}$, [Fe/H], $\log{g}$, and $E(B-V)$. For this work, we use exclusively the calculation performed on the Phoenix synthetic spectra \citep{BTSettl}.

Cluster parameters are taken from the literature \citep[e.g. from the catalogue in ][ and summarised in table \ref{tab:iso}]{bossini19}. First, we  verify the agreement of the isochrone on the observational CMD in the passbands $G$ against $G_{\rm BP}$-$G_{\rm RP}$.  Then we compare the isochrones with the standardised photometric bands. 

An example can be seen in figure \ref{fig:isoCMD} for the OC NGC~3532, where we also show our procedure.  
We first divide the main sequence along the $G$ magnitude in bin of 0.1 mag from the turn-off down to $G=16.00$ (blue lines panel). For each bin, we select the blue edge of the main sequence and compare the standardised photometry in JKC, PS1, ACS/WFC, SDSS, and Str\"omgren systems  with isochrone expectations. This procedure allows us to avoid contamination by unresolved binaries and differential reddening that could have blurred the distribution toward the red. In nearby clusters such as Pleiades, we discard  faint main sequence stars, since it is well known that  stellar models do not reproduce the colours of low mass stars. 
The residuals to the isochrones  are reported in figure \ref{fig:isoRES} for two passbands in each photometric systems. 
A similar test is performed on the each of the selected OCs on a total of 4165 stars.
The residuals are shown in figure \ref{fig:isoall}), while  table \ref{tab:isodata} presents the median and median absolute deviation (MAD) of the residual distributions in all the tested  passbands for the whole sample.

The agreement is good. Deviations are of the order of a few hundredths of a magnitude, and reach 0.08 mag for the $U_{\rm JKC,synt}$. The results seems to indicate a small zero-point offset of 0.02 mag in $i_{\rm PS1,synt}$ due either to bolometric corrections or to synthetic photometry.

\begin{table}[!htbp]
    \centering
        \caption{\label{tab:iso}. Adopted parameters for the sample of open clusters.}
{\small
    \begin{tabular}{ccccc
}
cluster &   log($age$) & $m-M$  & E(B-V) & [Fe/H] \\
        &   dex        & mag    &    mag & dex \\
\hline
NGC2168 &   8.60    &   9.33  &   0.15 &  -0.21 \\
M44     &   8.87    &   6.35  &   0.03 &   0.07 \\
M67     &   9.56    &   9.73  &   0.04 &   0.00 \\
NGC2447 &   8.75    &  10.09  &   0.03 &   0.00 \\
NGC3532 &   8.60    &   8.43  &   0.02 &   0.00 \\
NGC6791 &   9.93    &  13.08  &   0.10 &   0.40 \\
NGC6819 &   9.30    &  12.16  &   0.15 &   0.00 \\
Pleiades &  7.94    &   5.67  &   0.05 &   0.00 \\
    \end{tabular}
}
\end{table}

\begin{table}[!htbp]
    \centering
        \caption{\label{tab:isodata}. Median and MAD of the residuals between the isochrone and the bluer edge of the colour (x axis) in the Fig.~\ref{fig:isoCMD} CMDs for all the  tested photometry against $G_{\rm RP,phot}$ in all seven open clusters.}
{\small
    \begin{tabular}{lcc}
Photometric               &   median  & MAD \\
band                      &   mag     & mag \\
\hline
$G_{\rm BP,phot}$         &    -0.005 &     0.020\\
$b_{\rm Strom,synt}$      &     0.039 &     0.023\\
$y_{\rm Strom,synt}$      &     0.009 &     0.023\\
$U_{\rm JKC,synt}$        &     0.116 &     0.076\\
$B_{\rm JKC,synt}$        &     0.023 &     0.032\\
$V_{\rm JKC,synt}$        &     0.013 &     0.022\\
$r_{\rm PS1,synt}$        &     0.005 &     0.011\\
$i_{\rm PS1,synt}$        &     0.023 &     0.004\\
$z_{\rm PS1,synt}$        &    -0.028 &     0.010\\
$g_{\rm SDSS,synt}$       &     0.039 &     0.031\\
$r_{\rm SDSS,synt}$       &     0.014 &     0.011\\
$b_{\rm Strom,synt}$      &     0.039 &     0.023\\
$F606W_{\rm ACS/WFC,synt}$ &     0.001 &     0.016\\
$F814W_{\rm ACS/WFC,synt}$ &    -0.001 &     0.003\\

    \end{tabular}
}
\end{table}

%%%%%%%%%%%%%%%%%%%%%%%%%%%%%%%%%%%%%% END APPE ISO

%\input{sections/appe_standa}
%%%%%%%%%%%%%%%%%%%%%%%%%%%%%%%%%%%%%% APPE RG in SDSS
\section{XPSP of red giants in the SDSS system}
\label{sec:tha21}

The ~\cite{Thanjavur2021} SDSS Stripe 82 standards sample contains relatively few red giants (approximately 1800). In order to explore the behaviour of standardised SDSS XPSP in the regime of red giants 
we selected an additional sample of sources from the SDSS Data Release 17 \texttt{PhotObjAll} by applying the following cuts:

\begin{itemize}
    \item $13<$\texttt{psfmag\_\{u,g,r,i,z\}}$<25$
    \item \texttt{psfmagerr\_\{u,g,r,i,z\}}$<$1
    \item \texttt{type}$==$6
    \item \texttt{psfprob}$>$0
    \item \texttt{ndetect}$==$\texttt{nobserve}
    \item \texttt{clean}$==$1
    \item \texttt{BRIGHT \& EDGE \& BLENDED \& SATURATED \& INTERP\_CENTER \& SATURATED\_CENTER \& PSF\_FLUX\_INTERP} flags set to 0.
\end{itemize}

Additionally, we apply the Gaia filters described above with the addition of:
\begin{itemize}
    \item \texttt{parallax/parallax\_error$>10.0$}
    \item \texttt{in\_dr3$==$True}
    \item $1.72({\rm G}_{BP} - {\rm G}_{RP}) + 0.7  > M_G,$ 
\end{itemize}

where the last filter describes the linear selection of red giants from the colour--absolute magnitude (designated $M_G$) diagram. 
The absolute magnitudes were derived from photo-geometric distances published by \citet{bailer21}.
The final sample contains almost 74 714 candidate red giant stars, half of which within $\simeq 3.2$~kpc of the Sun, and more than 95\% of which within $\simeq 5.0$~kpc. These stars are distributed from the base of the RGB to just above the Red Clump and in the colour range $1.0< G_{BP}-G_{RP}<5.2$, with $\simeq 98$\% of the sources having $G_{BP}-G_{RP}<3.0$. The extension to very red colours is mainly due to relatively large interstellar extinction values. 

\begin{figure}[!htbp]
    \centerline{
    \includegraphics[width=(\columnwidth)]{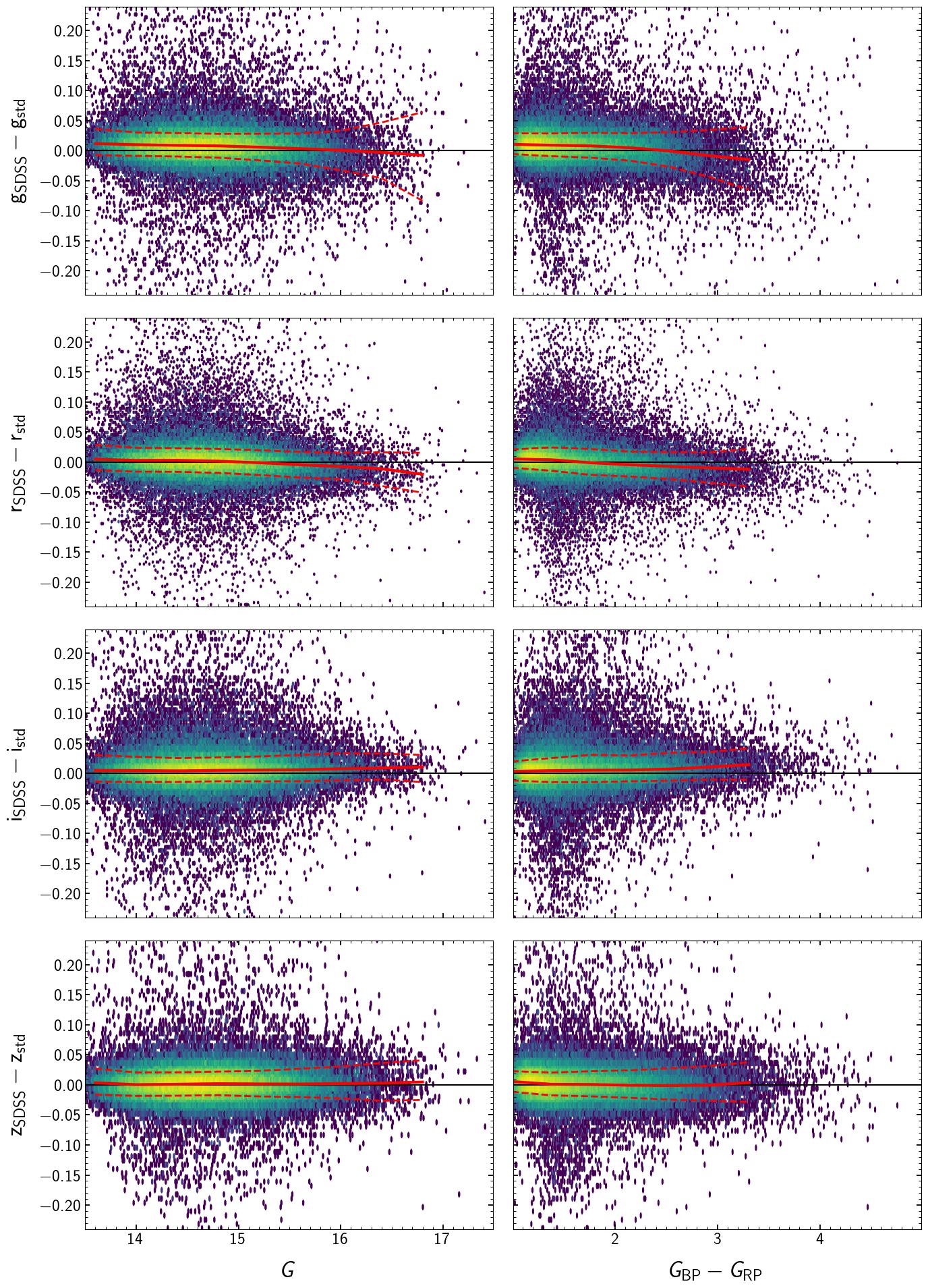}
    }
    \caption{Validation of the performance of the synthetic SDSS $griz$ bands using  the red giant sample sample described in Sect.~\ref{sec:tha21}. The arrangement of the plots and symbols follows the convention used in Fig.~\ref{fig:sdss_delta_app}. The sources in the sample are restricted to those with $G<$17.65 mag and \texttt{in\_dr3} flag set to 1.}
    \label{fig:rg_delta_app}
\end{figure}

The usual plots of $\Delta {\rm mag}$ as a function of $G$ mag and $G_{BP}-G_{RP}$ colour for this sample are presented in Fig.~\ref{fig:rg_delta_app}.
The typical scatter is larger than for the T21 reference sample, $\sigma\simeq 20$~mmag instead of $\sigma\simeq 10$~mmag at $G=15.2$, likely owing to the much higher precision of the T21 Stripe 82 photometry with respect to that available over the entire SDSS area in DR17. However, for $riz$ bands, the median $\Delta {\rm mag}$ is within a few millimag of zero over most of the colour range covered by the sample, and in any case within $\la 10.0$~mmag, while a colour term of amplitude $\simeq 10.0$~mmag in the range $1.0\la G_{BP}-G_{RP}\la2.5$, reaching an  amplitude of $\simeq20.0$~mmag in the range $1.0\la G_{BP}-G_{RP}\la3.5$ is apparent for the more problematic $g$ band (see Sect.~\ref{sec:standa_sdss}). In $u$ band (not shown here), the median $\Delta {\rm mag}$ remains below 10~mmag for $G_{BP}-G_{RP}\le 1.7$, with $\sigma \le 150$~mmag in that range. For $G_{BP}-G_{RP}> 1.7$, the median difference diverges rapidly. However, only very few sources with {\tt flux\_u/flux\_error\_u$>30$} can be found in this red realm.

In summary, the results of this validation experiment suggest that the standardisation of SDSS XPSP we obtained from the dwarf-dominated T21 reference sample should also be valid for red giants, with typical accuracy of better than $0.01$~mag over a large range of colours. However, it is worth noting that this test is mostly limited to K spectral type, and does not probe the coolest M giants.

%%%%%%%%%%%%%%%%%%%%%%%%%%%%%%%%%%%%%% END APPE RG in SDSS

%\input{sections/appe_standa}
%%%%%%%%%%%%%%%%%%%%%%%%%%%%%%%%%%%%%% APPE JKC STETSON

\section{Comparison with Stetson's JKC secondary standard stars}
\label{sec:stet}

\begin{figure}[!htbp]
    \centering
    \includegraphics[width=(\columnwidth)]{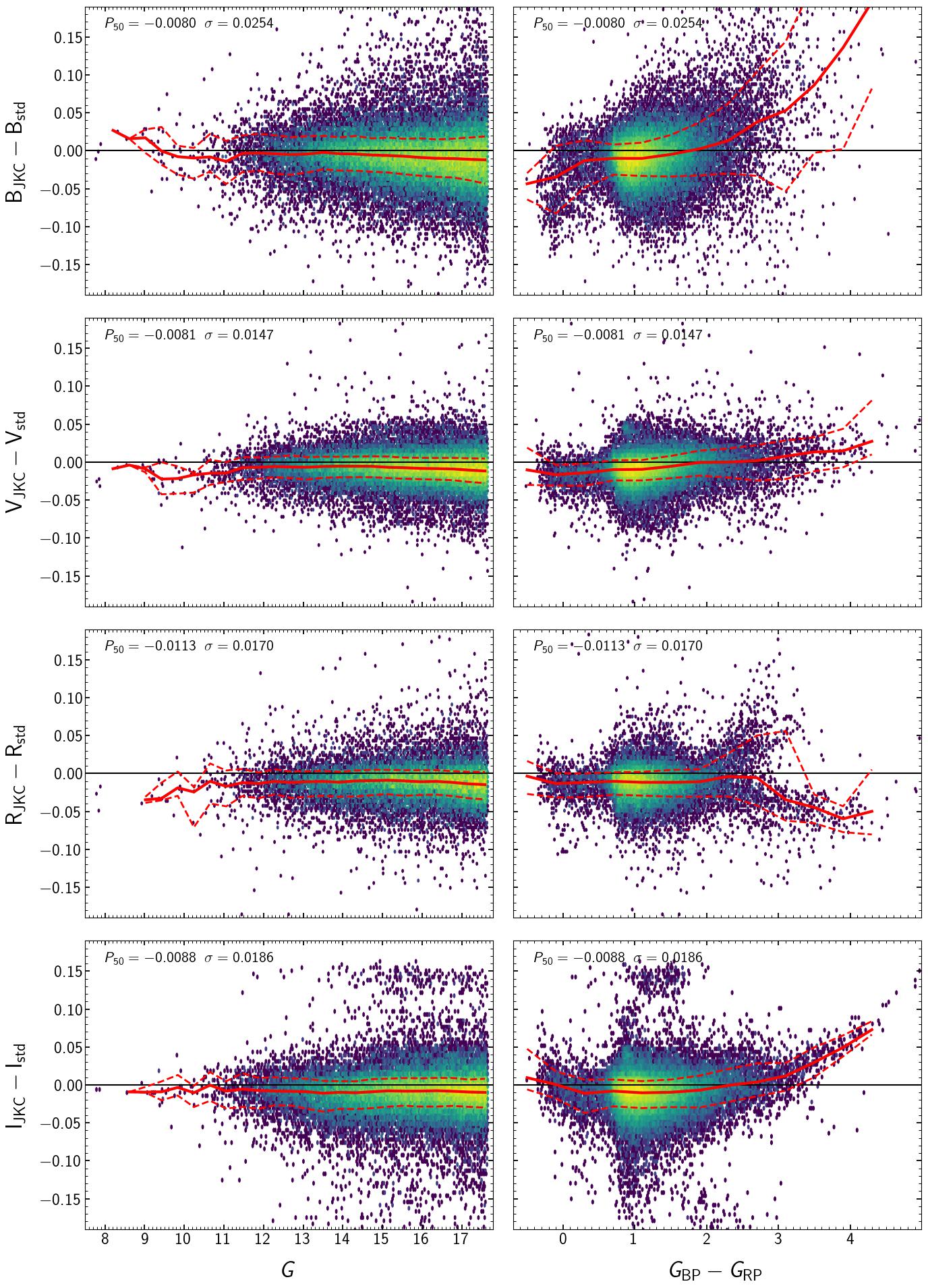}
    \caption{Performances of standardised XPSP in the JKC system ($BVRI$) for the Stetson validating sample. We show $\Delta$mag as a function of $G$ magnitude (left panels) and {\bprp} colour (right panels) for the subsample of reference stars whose XP spectra has been released in {\gdr3} and $G<17.65$ (50468 stars). The arrangement and the meaning of the symbols is the same as in
    \figref{sdss_delta_app}.}
    \label{fig:stet_deltaindr3}
\end{figure}

In Fig.~\ref{fig:stet_deltaindr3} we validate XPSP in the standardised JKC system against the subsample of the \citep{Stetson2019} secondary standard stars described in Sect.~\ref{sec:standa_jkc}, hereafter referred to as Stetson's validating sample. The comparison is limited to the sources with XP spectra released in DR3 and G<17.65. 

For $G_{BP}-G_{RP}<3.0,$ the median of the residuals in V, R, and I magnitudes is within $\simeq$~1\% of zero, with typical $\sigma\simeq 0.02-0.03$~mag. For $G_{BP}-G_{RP}>3.0$ a significant trend with colour is observed in I band, reaching an amplitude of $\simeq 0.1$~mag around  $G_{BP}-G_{RP}\simeq5.0$. 

The bifurcation occurring in $\Delta {\rm R}$ for $G_{BP}-G_{RP}\ga 2.5$ should probably be attributed to the heterogeneity of R TCs used in the observations collected by \citet{Stetson2019}. For the B band, the agreement within 1\% is limited to the range $0.2\la G_{BP}-G_{RP}\la 2.4$, with sizable trends outside, and typical 
$\sigma\simeq 0.04-0.05$~mag, to be attributed to poorer performances in both the photometries in this passband. As we show below, part of the observed scatter may be due to field-to-field inhomogeneities in the the Stetson's sample.

\begin{figure}[!htbp]
    \centering
    \includegraphics[width=(1.0\columnwidth)]{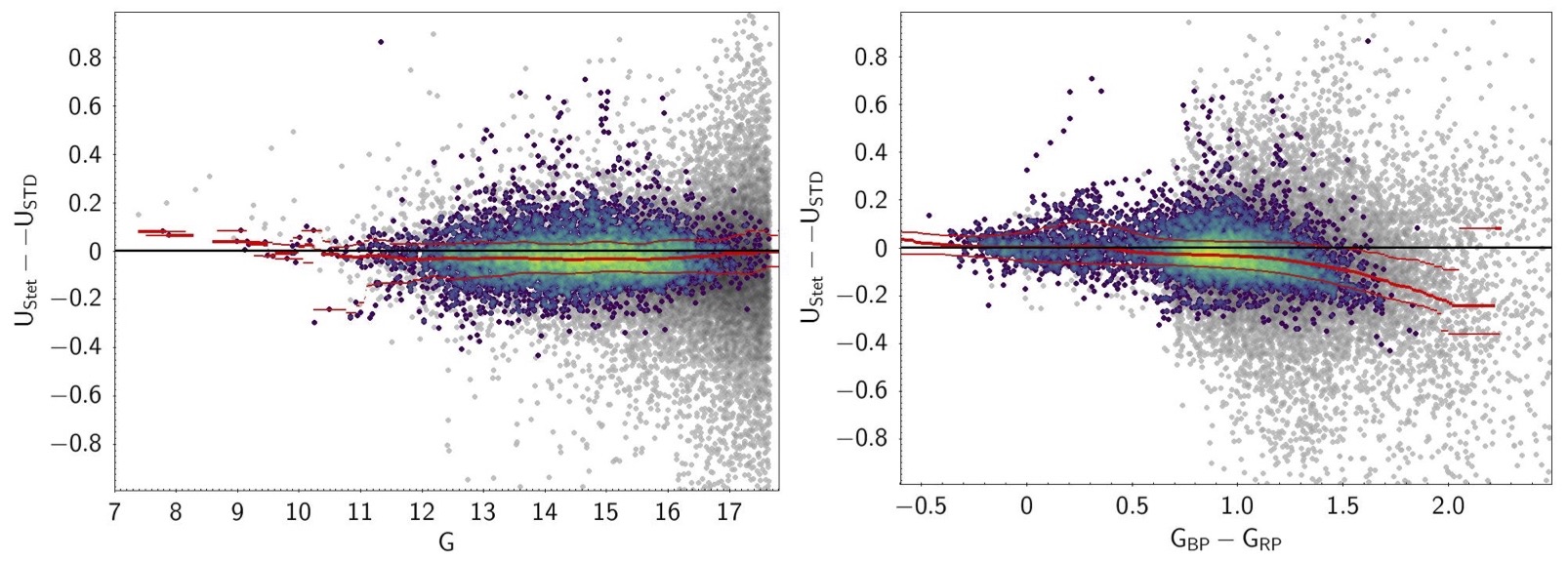}
    \caption{Performances of standardised XPSP in the U band of JKC system for the Stetson validating sample. The meaning of the symbols and the arrangement of the plot are the same as in Fig.~\ref{fig:stet_deltaindr3}, except for the y-axes scale, which is much more expanded here.
    The 29176 stars in the Stetson validating sample having their XP spectra released in \gaia DR3, $G<17.65$, and valid U$_{Stet}$ and U$_{STD}$ magnitudes are represented as grey dots, while those shown as a viridis density maps are the 9157 that also have {\tt flux\_U/flux\_error\_U$>30$}. $P_{50}$, $P_{16}$, and $P_{84}$ lines refer to the latter subsample.}
    \label{fig:stet_deltau}
\end{figure}

Figure~\ref{fig:stet_deltau} focuses on the comparison in the U band. The grey dots shows the entire sample, while those plotted with the viridis density scale have {\tt flux\_U/flux\_error\_U$>30$}. If we limit ourselves to this high-S/N sample and consider the median $\Delta {\rm U}$, we  conclude that the two independent sets of JKC U magnitudes agree within $\simeq 3.0$\% over the range  $-0.4\le G_{BP}-G_{RP}\le 1.0$. For redder colours, the onset of a colour trend is perceivable, reaching $\Delta {\rm U}\le -0.05$~mag for $G_{BP}-G_{RP}\ge 1.3$. The typical scatter is $\sigma\simeq 0.09$, but the distribution of grey points illustrates very clearly how the performances may worsen for S/Ns lower than 30, especially at red colours (here $G_{BP}-G_{RP}\ga 0.8$).
We feel that this plot serves as a further invitation to exercise caution in the use of UV XPSP, even when limited to the set included in the GSPC.

%%%%%%%%%%%%%%%%%%%%%%%%%%%%%%%%%%%%%% END APPE JKC STETSON

%\input{sections/ex_query.tex}
%%%%%%%%%%%%%%%%%%%%%%%%%%%%%%%%%%%%%% APPE EXAMPLES OF QUERIES
\section{Examples of queries}
\label{sec:ex_query}

Thanks to the availability of the GSPC table (\texttt{gaiadr3.synthetic\_photometry\_gspc}) along-side other \gdr{3} tables, it is straightforward to perform various selections and extraction of additional parameters from the archive interface. In this Appendix we provide an example. Users will be able to modify the example query to fulfill their needs.

The query
\begin{lstlisting}
SELECT dr3.source_id, dr3.ra, dr3.dec, dr3.parallax, dr3.parallax_over_error, dr3.ruwe, gspc.g_sdss_mag, gspc.i_sdss_mag FROM gaiadr3.gaia_source AS dr3
INNER JOIN gaiadr3.synthetic_photometry_gspc AS gspc
ON dr3.source_id=gspc.source_id
WHERE ABS(gspc.c_star)<(0.0059898 + 8.817481e-12 * POWER(dr3.phot_g_mean_mag, 7.618399)) 
\end{lstlisting}
joins the GPSC table with the main \texttt{gaia\_source} table and selects a few parameters from each but only for sources that have an absolute corrected BP/RP flux excess factor smaller than the $1-\sigma$ relation suggested in \cite{riello2021}.

The resulting dataset can be reduced in size by using the \texttt{random\_index} available in \texttt{gaia\_source}: For instance, by adding 
\begin{lstlisting}
AND dr3.random_index<1811709
\end{lstlisting}
the query would effectively run on a 0.001 random selection of the \gaia source catalogue.
Similar joins can of course be made with a user-defined input list of source identifier.

The result of this can then be uploaded as a new user-defined table, here called \texttt{gspc\_plus}.
The user could then for instance generate a CMD in SDSS $g-i$ as colour and absolute magnitude $g_{\rm abs}$ (here simply computed using the inverse of the parallax to approximate the distance). The following query shows how to do this:
\begin{lstlisting}
SELECT col_index / 40 AS col, mag_abs_index / 10 AS mag_abs, n FROM (
SELECT 
	floor((g_sdss_mag-i_sdss_mag) * 40) AS col_index,
	floor((g_sdss_mag + 5 * log10(parallax) - 10) * 10) AS mag_abs_index,
	count(*) AS n
FROM user_xxxx.gspc_plus
WHERE parallax_over_error > 5
GROUP BY col_index, mag_abs_index
) AS subquery
\end{lstlisting}

On the other hand, the following query extracts some parameters from {\tt gaiadr3.gaia\_source} and some
from {\tt gaiadr3.synthetic\_photometry\_gspc}, taking all the relevant GSPC quantities
for the selected photometry, for a cone of radius 1.0 deg  centred on the
globular cluster NGC~5139 ($\omega$~Centauri):

\begin{lstlisting}
SELECT dr3.source_id, dr3.ra, dr3.dec, dr3.pmra, dr3.pmra_error, dr3.pmdec, dr3.pmdec_error, dr3.ruwe, 
gspc.c_star, gspc.u_jkc_mag, gspc.u_jkc_flux, 
gspc.u_jkc_flux_error, gspc.u_jkc_flag,
gspc.b_jkc_mag, gspc.b_jkc_flux, 
gspc.b_jkc_flux_error, gspc.b_jkc_flag,
gspc.v_jkc_mag, gspc.v_jkc_flux, 
gspc.v_jkc_flux_error, gspc.v_jkc_flag,
gspc.y_ps1_mag, gspc.y_ps1_flux, 
gspc.y_ps1_flux_error, gspc.y_ps1_flag,
gspc.f606w_acswfc_mag, gspc.f606w_acswfc_flux, 
gspc.f606w_acswfc_flux_error, gspc.f606w_acswfc_flag
FROM gaiadr3.gaia_source AS dr3
JOIN gaiadr3.synthetic_photometry_gspc AS gspc
ON dr3.source_id=gspc.source_id
WHERE 
CONTAINS(
        POINT('ICRS',dr3.ra,dr3.dec),
        CIRCLE(
                'ICRS',
                COORD1(EPOCH_PROP_POS(201.697,-47.479472,.1368,-3.2400,
                -6.7300,234.2800,2000,2016.0)),
                COORD2(EPOCH_PROP_POS(201.697,-47.479472,.1368,-3.2400,
                -6.7300,234.2800,2000,2016.0)),
                1)
)=1

\end{lstlisting}
%
%%%%%%%%%%%%%%%%%%%%%%%%%%%%%%%%%%%%%% END APPE EXAMPLES OF QUERIES

%\input{sections/appe_standa}
%%%%%%%%%%%%%%%%%%%%%%%%%%%%%%%%%%%%%% APPE Delta mag for STANDARDISED SYS

\section{$\Delta$ mag diagrams for standardised systems}
\label{sec:app_stand}

In this Appendix, we show the $\Delta {\rm mag}$ distributions as a function of G magnitude and BP-RP colour, before and after the standardisation process ---as done in Fig.~\ref{fig:sdss_delta_app} and Fig.~\ref{fig:jkc_delta_app} for the SDSS and JKC systems--- for all the remaining standardised systems discussed in Sect.~\ref{sec:stand} and Sect.~\ref{sec:narrow}, including the JKC U and SDSS u bands discussed in Sect.\ref{sec:standa_u}. 
The reference sets of standard stars adopted are described there.
In all the figures, comparisons of $\Delta {\rm mag}$ as a function of G magnitude are performed on the entire reference sample, including $G>17.65$ stars that are required to model the hockey-stick effect, while those of $\Delta {\rm mag}$ as a function of G$_{BP}$-G$_{RP}$ colour are limited to the subsample of stars with XP spectra released in DR3 (see Sect.~\ref{sec:methods}). 

\begin{figure*}[!htbp]
    \centerline{
    \includegraphics[width=(\columnwidth)]{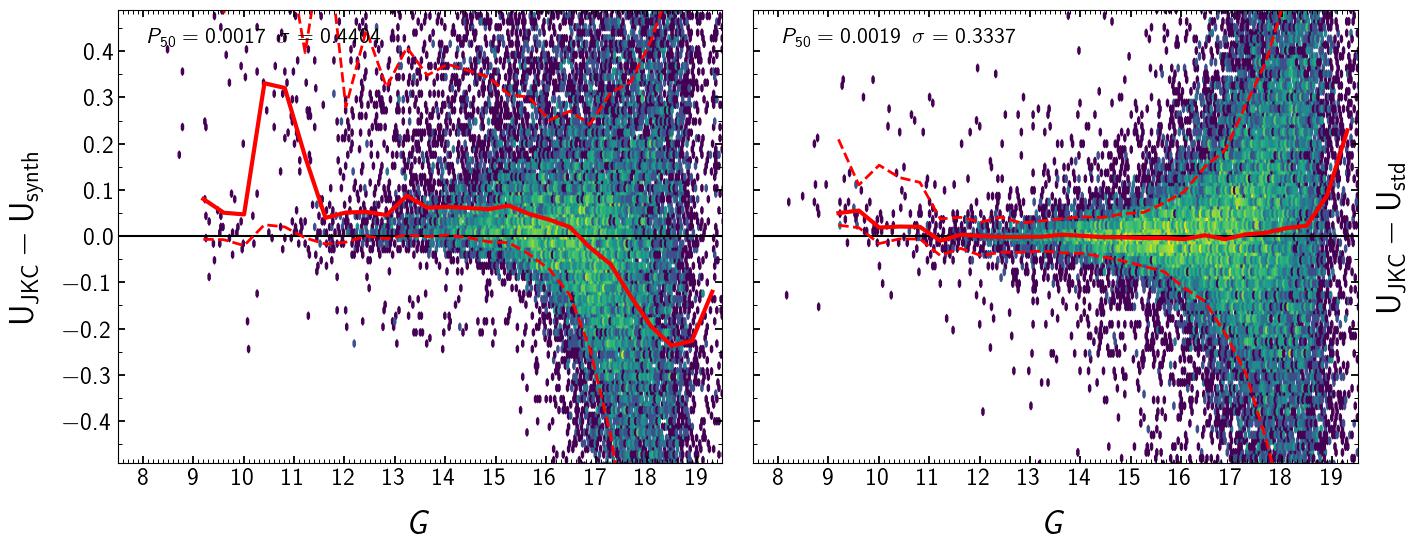}
    \includegraphics[width=(\columnwidth)]{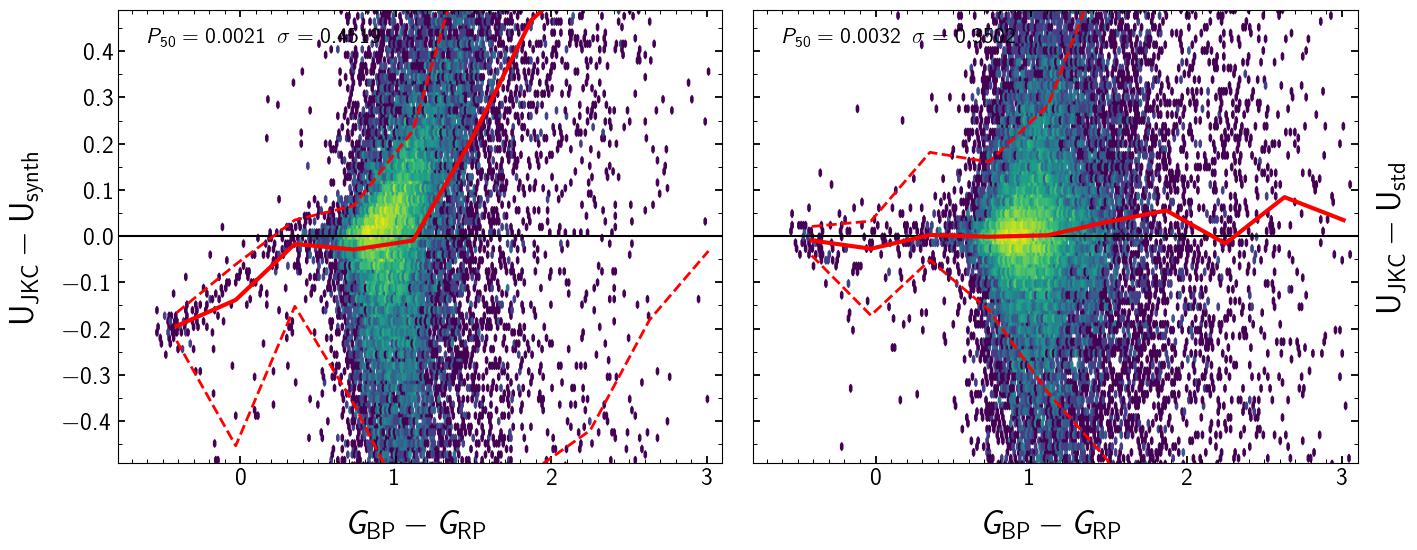}
    }
    \caption{Performance and standardisation of JKC U band XP synthetic magnitudes using the reference sample described in Sect.~\ref{sec:standa_jkc}. Left set of panels: $\Delta$mag as a function of G magnitude for the entire sample using nominal XP synthetic magnitudes (left panel) and standardised XP synthetic magnitudes (right panel). In each panel, the continuous red line connects the median $\Delta$mag computed in 0.2~mag wide bins, the dashed red lines connect the loci of the 15.87\% ($P_{16}$) and the 84.13\% ($P_{84}$) percentile computed in the same bins. The median ($P_{50}$) and the difference between $P_{84}$ and $P_{16}$, here used as a proxy for the standard deviation $\sigma$, for the entire sample are reported in the upper left panel of each panel. Right set of panels: the same for $\Delta$mag as a function of BP-RP colour, limited to the subsample of reference stars having XP spectra released in DR3.}
    \label{fig:JKC_U_delta_app}
\end{figure*}

\begin{figure*}[!htbp]
    \centerline{
    \includegraphics[width=(\columnwidth)]{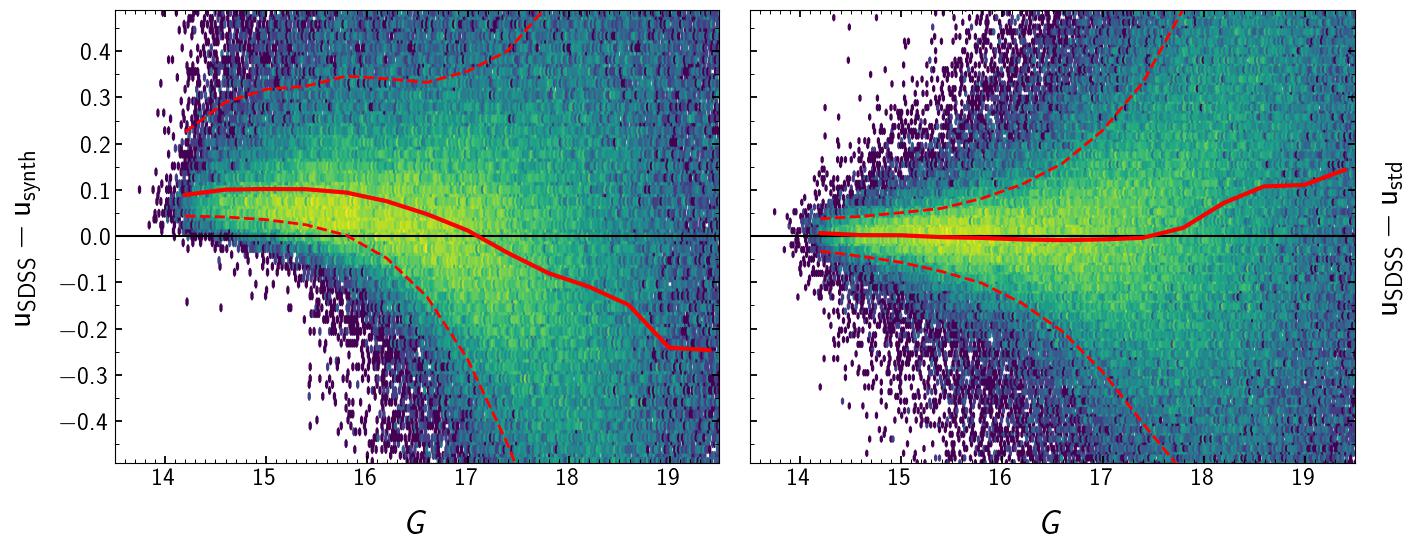}
    \includegraphics[width=(\columnwidth)]{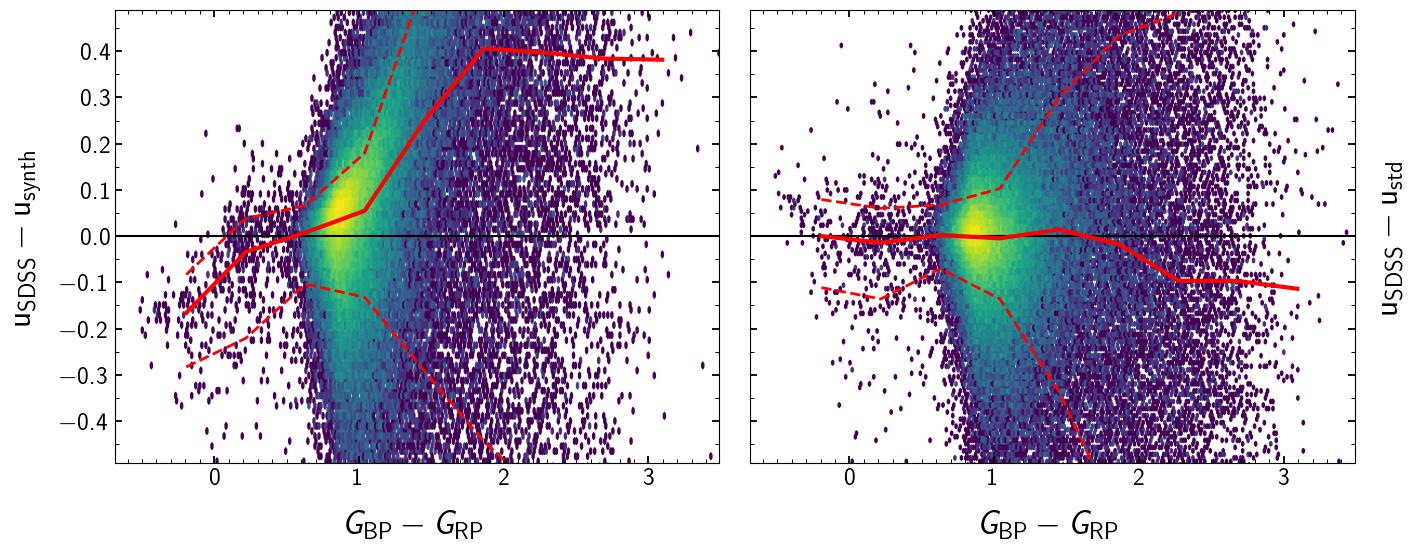}
    }
    \caption{Performance and standardisation of SDSS u band XP synthetic magnitudes using the reference sample described in Sect.~\ref{sec:standa_sdss}. The arrangement of the plots and symbols are the same as Fig.~\ref{fig:JKC_U_delta_app}, above.}
    \label{fig:sdss_u_delta_app}
\end{figure*}

\begin{figure*}[!htbp]
    \centerline{
    \includegraphics[width=(\columnwidth)]{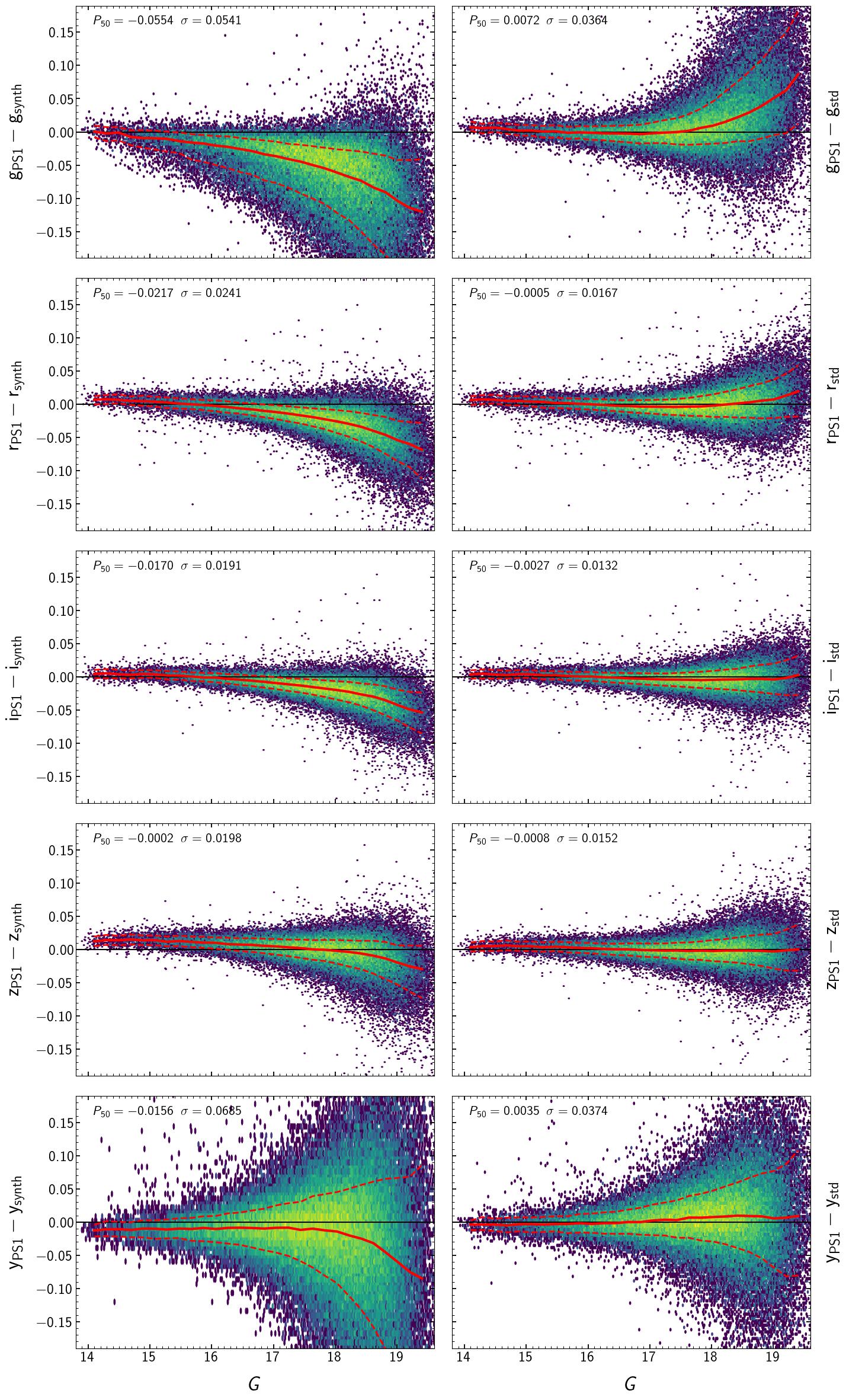}
    \includegraphics[width=(\columnwidth)]{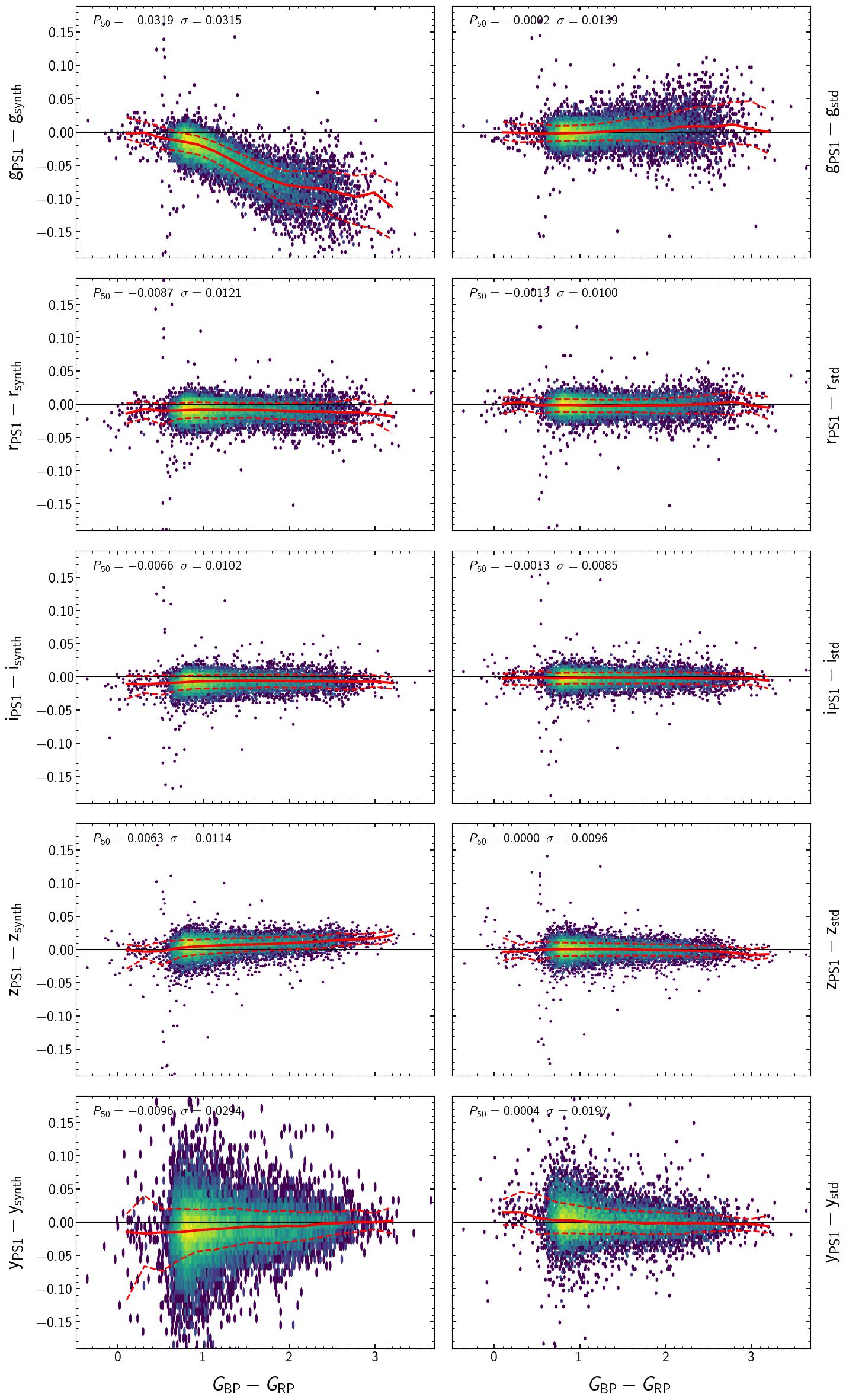}
    }
    \caption{Performance and standardisation of PS1 $griz$ bands XP synthetic magnitudes using the reference sample described in Sect.~\ref{sec:standa_ps1}. The arrangement of the plots and symbols are the same as Fig.~\ref{fig:JKC_U_delta_app}.}
    \label{fig:ps1_delta_app}
\end{figure*}

\begin{figure*}[!htbp]
    \centerline{
    \includegraphics[width=(0.8\columnwidth)]{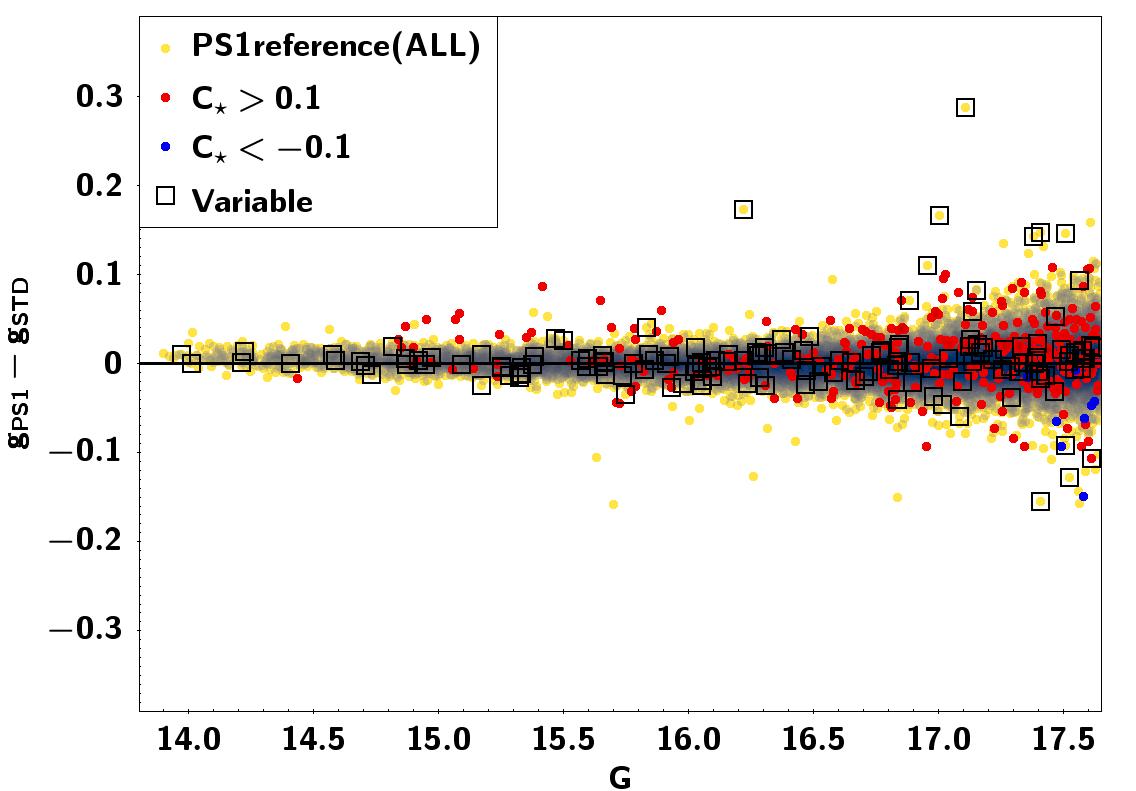}
    \includegraphics[width=(0.8\columnwidth)]{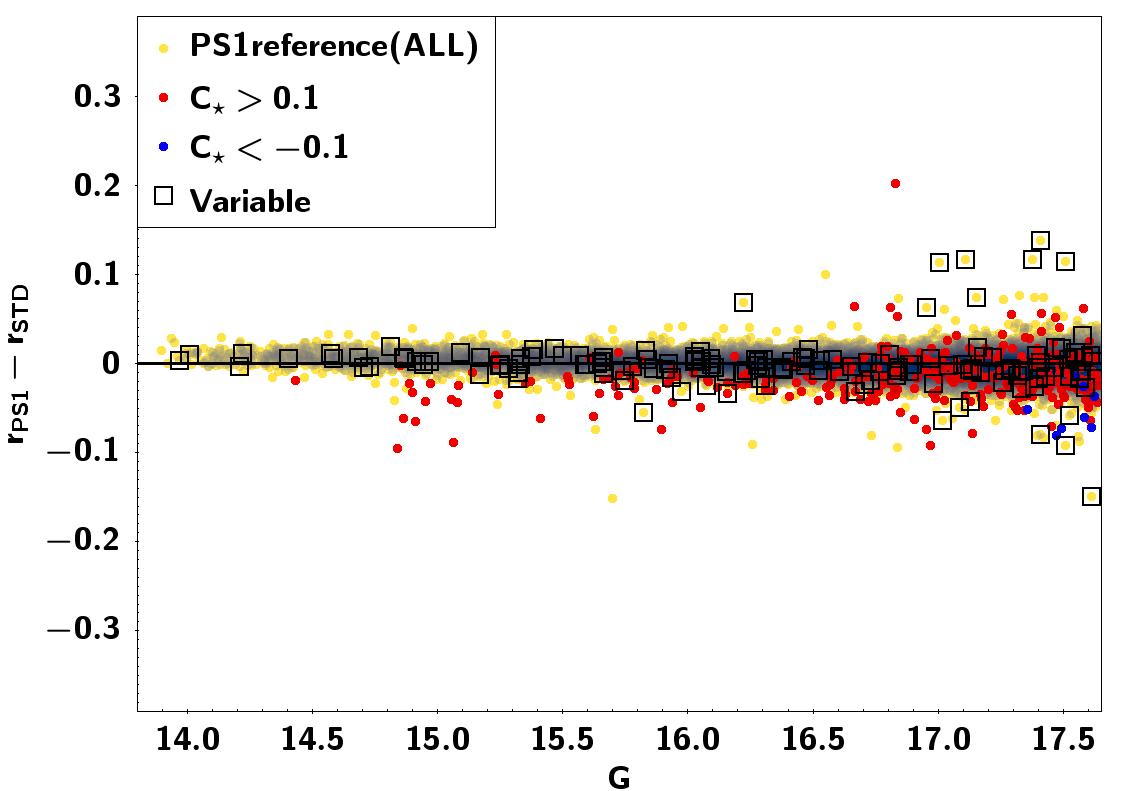}
    }
    \caption{Example of the different distribution of $C_{\star}$ in $\Delta {\rm mag}$ in different passbands using the PS1 reference sample. We note that source with high positive $C_{\star}$ tend to have positive residuals in $g_{PS1}$ and negative residuals in $r_{PS1}$. Also, most of the outliers in both plots are accounted for by sources with (relatively) large absolute $C_{\star}$ values and by sources classified as variable (see also Fig.~\ref{fig:dgcstar}).}
    \label{fig:ps1_cstar_app}
\end{figure*}

\begin{figure*}[!htbp]
    \centerline{
    \includegraphics[width=(\columnwidth)]{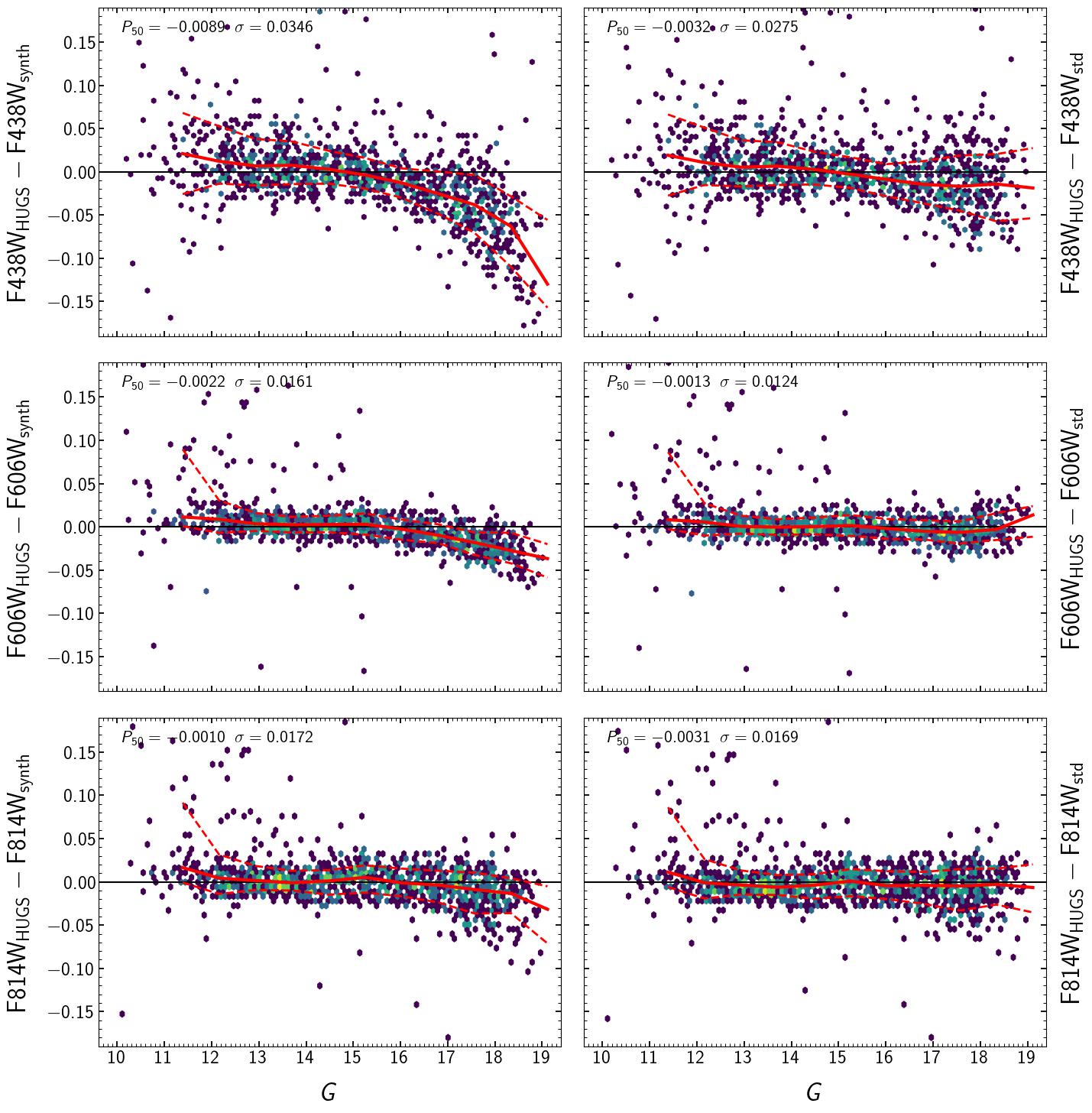}
    \includegraphics[width=(\columnwidth)]{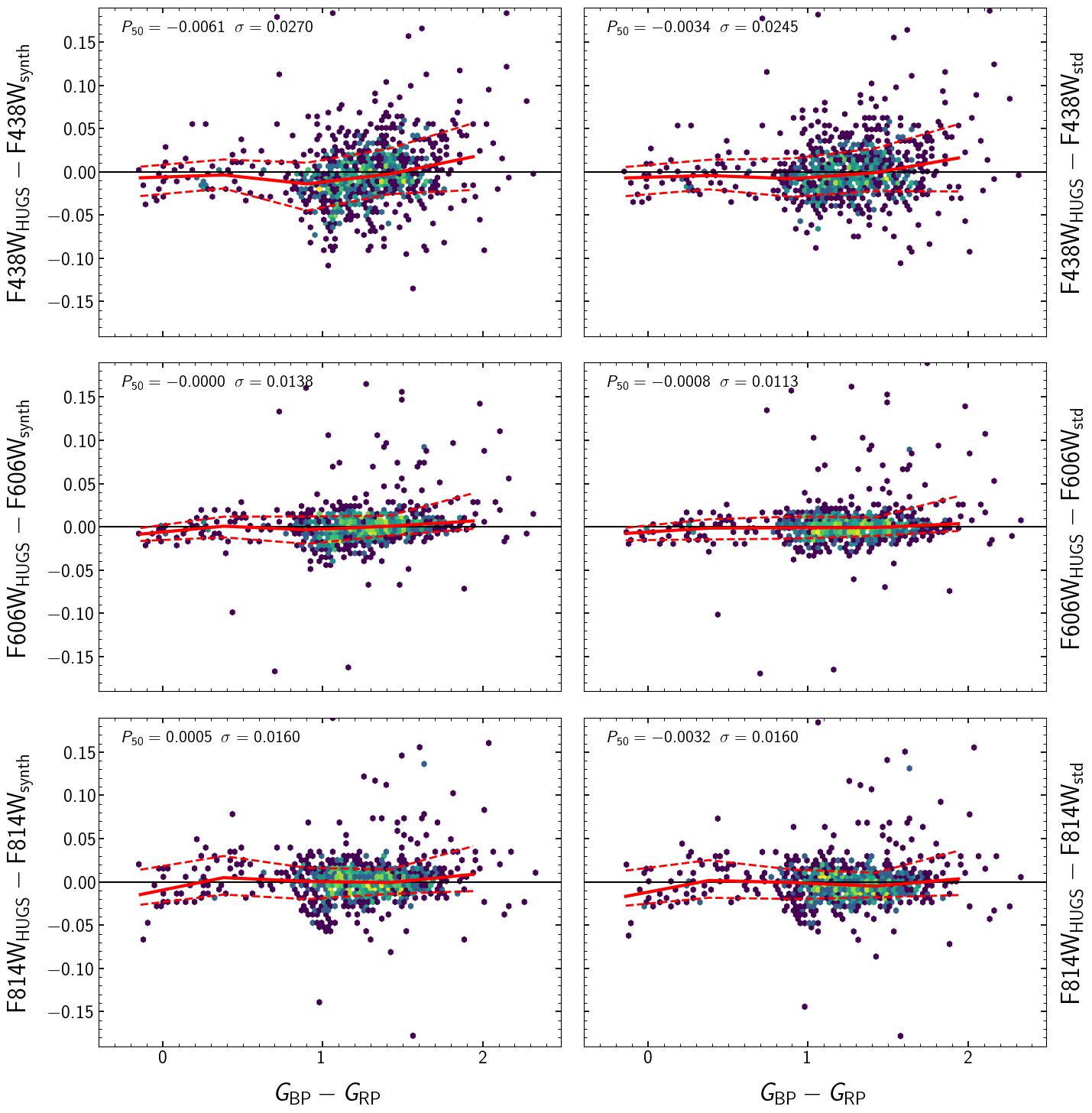}
    }
    \caption{Performance and standardisation of HST F438W$_{WFC3/UVIS}$, F606W$_{ACS/WFC}$, and F814W $_{ACS/WFC}$ bands XP synthetic magnitudes using the reference sample described in Sect.~\ref{sec:standa_hugs}. The arrangement of the plots and symbols are the same as Fig.~\ref{fig:JKC_U_delta_app}, above.}
    \label{fig:hugs_delta_app}
\end{figure*}

\begin{figure*}[!htbp]
    \centerline{
    \includegraphics[width=(\columnwidth)]{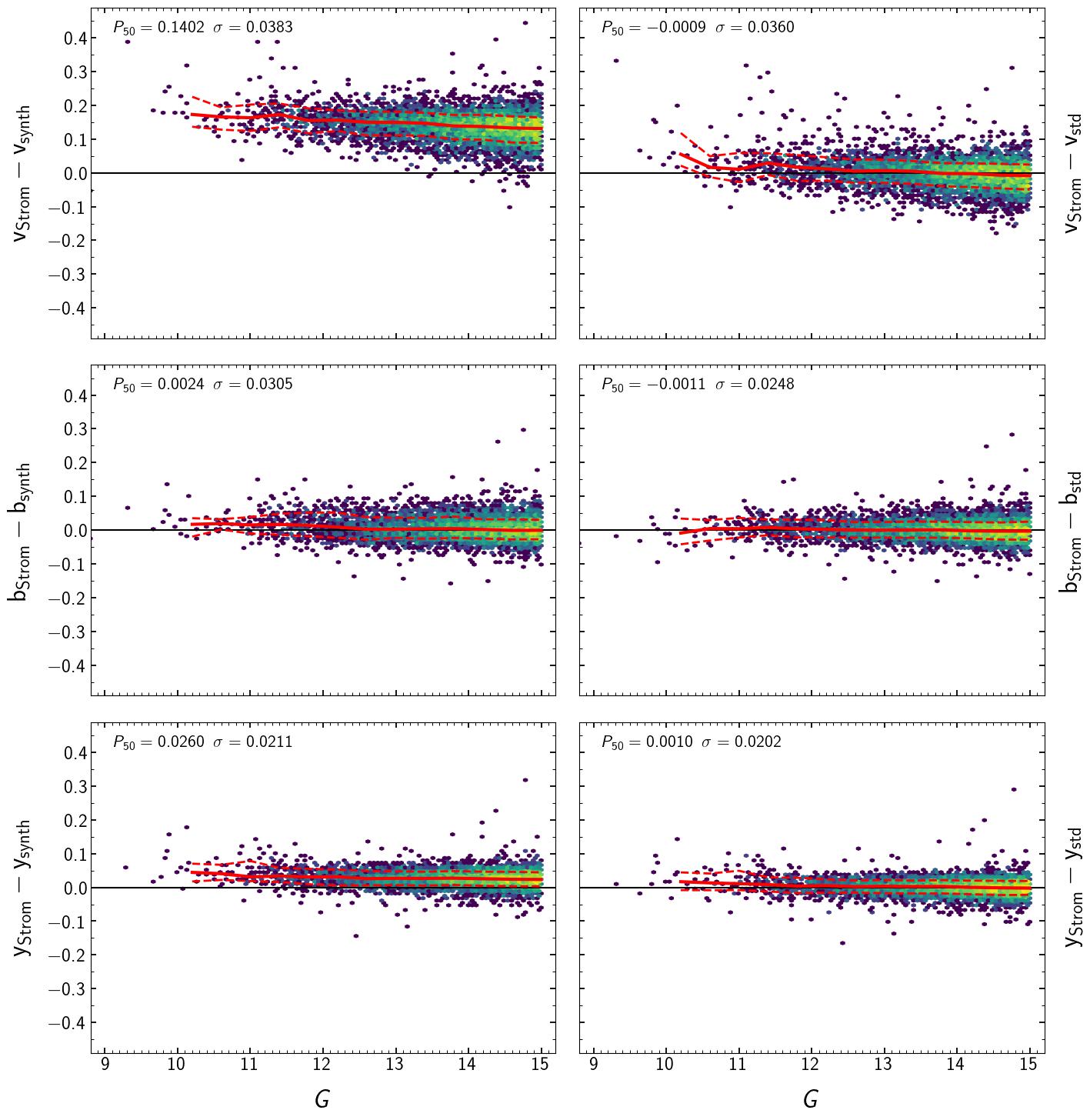}
    \includegraphics[width=(\columnwidth)]{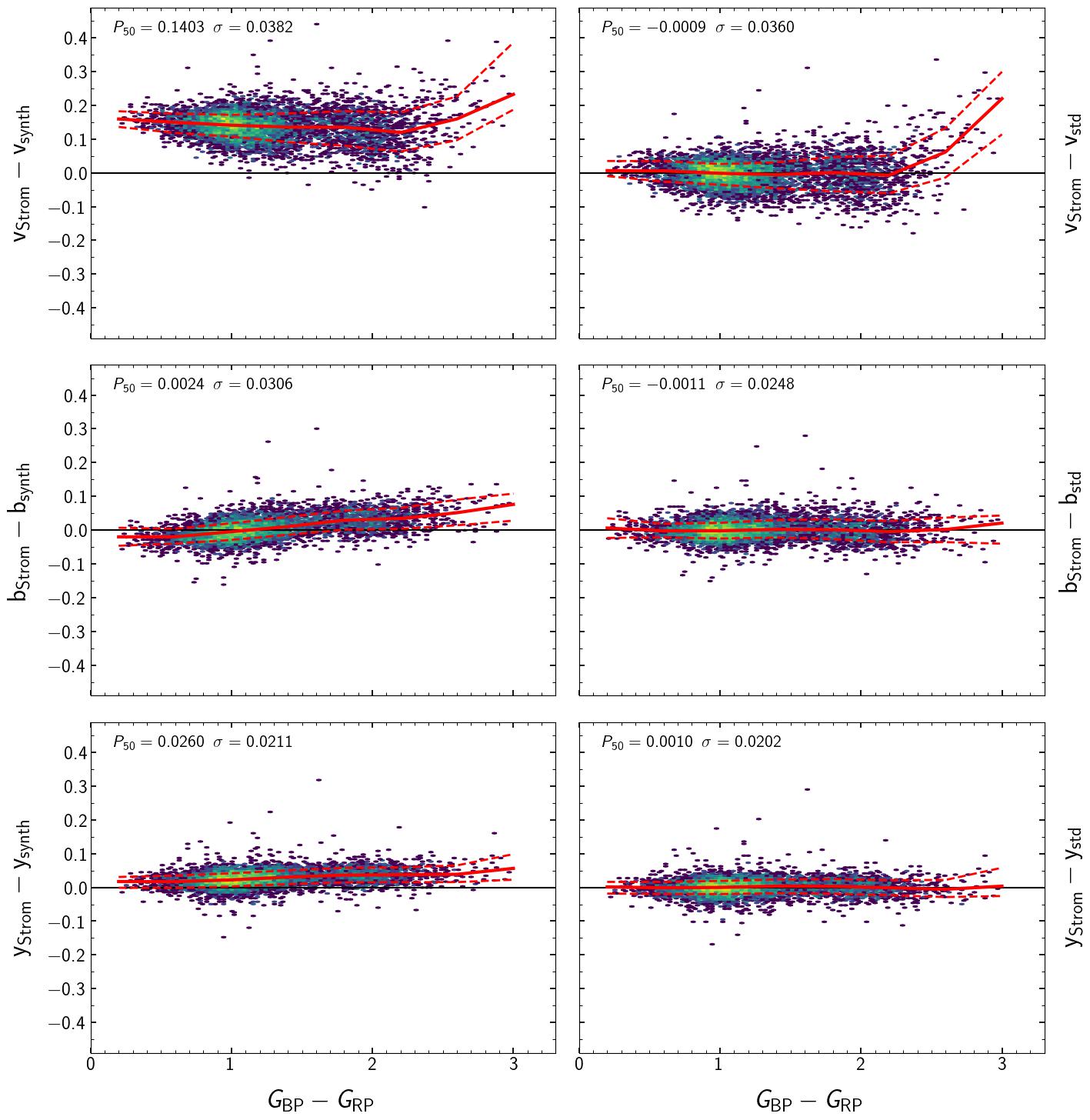}
    }
    \caption{Performance and standardisation of Stromgren vby  XP synthetic magnitudes using the reference sample described in Sect.~\ref{sec:standa_strom}. The arrangement of the plots and symbols are the same as Fig.~\ref{fig:JKC_U_delta_app}, above.}
    \label{fig:strom_delta_app}
\end{figure*}

The $\Delta {\rm mag}$ distributions for the standardised UV magnitudes are presented in Fig.~\ref{fig:JKC_U_delta_app} and Fig.~\ref{fig:sdss_u_delta_app}. Those for the PS1 system are shown in Fig.~\ref{fig:ps1_delta_app}, with a focus on variable stars and high $|C_{\star}|$ stars in Fig.~\ref{fig:ps1_cstar_app}. The cases of the standardised HUGS and Str\"omgren magnitudes are illustrated in Fig.~\ref{fig:hugs_delta_app} and Fig.~\ref{fig:strom_delta_app}, respectively.

The values of $P_{50}$, $P_{16}$, and $P_{84}$ as a function of $G$ magnitude for the $\Delta {\rm mag}$ distributions of stars of the reference samples with XP spectra released in DR3, for the PS1, HUGS, and Str\"omgren standardised magnitudes are listed in Tables~\ref{tab:PS1_median}, \ref{tab:hugs_median}, and ~\ref{tab:strom_median}, respectively, in the same way as done for the SDSS and JKC systems in Tables~\ref{tab:SDSS_median} and ~\ref{tab:JKC_median}, respectively.

%%%%%%%%%%%%%%%%%%%%%%%%%%%%%%%%%%%%%%%%%%%%%%%%%%%%%%%%%%%%%%%%%%%%%%%%%%%%%%%%
\begin{table*}[!htbp]
\setlength{\tabcolsep}{1.2mm}    \centering
        \caption{\label{tab:PS1_median} PS1 system: median ($P_{50)}$) and  15.87\% ($P_{16}$) and 84.13\% ($P_{84}$) percentiles 
of the $\Delta {\rm mag}$ distributions of Fig.~\ref{fig:ps1_deltaindr3}. n$_{\star}$ is the number of sources in the considered bin.}
{\small
    \begin{tabular}{lcccccccccccccccr}
  G  &$P_{50}(\Delta g)$  & P$_{16}$ & P$_{84}$ & $P_{50}(\Delta r)$ & P$_{16}$ & P$_{84}$ & $P_{50}(\Delta i)$ & P$_{16}$ & P$_{84}$ & $P_{50}(\Delta z$) & P$_{16}$ & P$_{84}$ & $P_{50}(\Delta y$) & P$_{16}$ & P$_{84}$ & n$_{\star}$\\
 mag &    mmag &  mmag &  mmag &  mmag &    mmag &  mmag & mmag &    mmag &  mmag & mmag &    mmag &  mmag& mmag &    mmag &  mmag & \\
\hline
14.0  &  8.7 &   1.2 &  14.9 &   6.5 &  -0.1 & 12.0 &   2.7 &  -2.3 & 13.3 &  4.8 &  -3.2 &  9.9 & -4.2 & -10.8 &  2.4 &    40 \\
14.4  &  6.6 &  -0.6 &  13.4 &   8.2 &  -0.0 & 16.1 &   3.8 &  -1.7 & 11.4 &  5.0 &  -2.5 & 12.7 & -3.6 & -11.8 &  7.1 &   138 \\
14.8  &  2.6 &  -4.0 &  11.4 &   5.3 &  -1.0 & 12.8 &   4.1 &  -1.3 &  9.8 &  5.2 &  -1.3 & 13.2 & -3.9 & -14.5 &  6.6 &   308 \\
15.2  &  1.7 &  -7.9 &  10.2 &   4.1 &  -4.1 & 11.8 &   3.3 &  -2.4 &  9.7 &  4.1 &  -2.6 & 11.5 & -3.0 & -14.7 &  7.6 &   584 \\
15.6  & -0.4 & -10.1 &   8.4 &   2.2 &  -5.2 &  9.3 &   2.2 &  -4.4 &  8.1 &  3.5 &  -3.9 & 10.7 & -2.2 & -14.5 & 10.2 &   901 \\
16.0  & -0.7 & -10.8 &   8.9 &   0.7 &  -6.8 &  8.5 &   0.8 &  -5.8 &  7.2 &  1.7 &  -5.5 &  9.6 & -1.8 & -14.9 & 13.4 &  1391 \\
16.5  & -1.9 & -13.5 &   9.5 &  -1.2 &  -9.6 &  6.9 &  -0.7 &  -8.3 &  6.0 &  0.2 &  -8.1 &  8.2 & -1.0 & -16.4 & 14.9 &  2099 \\
16.9  & -2.2 & -15.5 &  11.5 &  -2.9 & -12.3 &  6.3 &  -2.4 & -10.7 &  5.6 & -1.0 & -10.5 &  8.5 &  0.3 & -17.6 & 20.4 &  3178 \\
17.3  &  0.2 & -16.0 &  18.3 &  -3.2 & -14.6 &  7.2 &  -3.7 & -12.9 &  5.2 & -1.4 & -11.8 &  9.1 &  2.9 & -19.7 & 28.1 &  4509 \\
17.7  &  2.6 & -16.6 &  24.0 &  -2.9 & -15.8 &  8.8 &  -3.9 & -14.3 &  6.6 & -2.2 & -14.3 & 10.1 &  6.5 & -21.1 & 35.8 &  2219 \\
\hline
    \end{tabular}
}
\end{table*}
%%%%%%%%%%%%%%%%%%%%%%%%%%%%%%%%%%%%%%%%%%%%%%%%%%%%%%%%%%%%%%%%%%%%%%%%%%%%%%%%

%%%%%%%%%%%%%%%%%%%%%%%%%%%%%%%%%%%%%%%%%%%%%%%%%%%%%%%%%%%%%%%%%%%%%
\begin{table*}[!htbp]
    \centering
        \caption{\label{tab:hugs_median} Standardised HST magnitudes: median ($P_{50)}$),  15.87\% ($P_{16}$) and 84.13\% ($P_{84}$) percentiles 
of the $\Delta {\rm mag}$ distributions of Fig.~\ref{fig:hugs_deltaindr3}. n$_{\star}$ is the number of sources in the considered bin.
F438W is from the WFC3/UVIS passbands set, F606W and F814W from the ACS/WFC set.}
{\small
    \begin{tabular}{lcccccccccr}
  G  &$P_{50}(\Delta F438W)$  & P$_{16}$ & P$_{84}$ & $P_{50}(\Delta F606W)$ & P$_{16}$ & P$_{84}$ & $P_{50}(\Delta F814W)$ & P$_{16}$ & P$_{84}$ & n$_{\star}$\\
 mag &    mmag &  mmag &  mmag &  mmag &    mmag &  mmag & mmag &    mmag &  mmag & \\
\hline
11.4  & 18.2 & -23.4 &  53.8 &	 4.6 &  -2.6 &  91.1 &	10.0 &   -6.1 &  84.0 &    36 \\
12.2  & 10.6 & -22.2 &  50.8 &	 4.7 &  -9.6 &  25.7 &	-1.0 &  -19.3 &  24.0 &    87 \\
13.0  &  3.4 & -17.3 &  36.3 &	 0.4 &  -8.1 &  11.3 &	-3.5 &  -14.9 &  12.2 &   117 \\
13.8  &  2.6 & -18.9 &  30.9 &	-1.0 &  -9.4 &   8.6 &	-6.8 &  -16.0 &   6.9 &   142 \\
14.5  &  2.6 & -14.2 &  23.9 &	 1.7 &  -7.8 &  11.1 &	-2.5 &  -19.0 &  12.6 &   118 \\
15.3  & -4.3 & -20.6 &  14.6 &	 1.4 & -10.4 &  12.9 &	 0.8 &  -16.5 &  14.3 &   132 \\
16.1  & -9.3 & -28.6 &   6.8 &	-2.2 & -13.8 &  11.0 &	-4.1 &  -20.0 &  13.1 &   109 \\
16.9  &-15.4 & -36.8 &  13.1 &	-7.3 & -15.6 &   4.7 &	-4.1 &  -27.1 &  11.7 &   145 \\
17.7  &-21.4 & -43.9 &  14.1 &	-7.7 & -23.8 &   3.5 & -11.2 &  -39.5 &  12.1 &    67 \\
\hline
    \end{tabular}
}
\end{table*}
%%%%%%%%%%%%%%%%%%%%%%%%%%%%%%%%%%%%%%%%%%%%%%%%%%%%%%%%%%%%%%%%%%%%%

%%%%%%%%%%%%%%%%%%%%%%%%%%%%%%%%%%%%%%%%%%%%%%%%%%%%%%%%%%%
\begin{table*}[!htbp]
    \centering
        \caption{\label{tab:strom_median} Standardised Str\"omgren magnitudes: median ($P_{50)}$) and  15.87\% ($P_{16}$) and 84.13\% ($P_{84}$) percentiles 
of the $\Delta {\rm mag}$ distributions of Fig.~\ref{fig:strom_deltaindr3}. n$_{\star}$ is the number of sources in the considered bin.}
{\small
    \begin{tabular}{lcccccccccr}
  G  &$P_{50}(\Delta v)$  & P$_{16}$ & P$_{84}$ & $P_{50}(\Delta b)$ & P$_{16}$ & P$_{84}$ & $P_{50}(\Delta y)$ & P$_{16}$ & P$_{84}$ & n$_{\star}$\\
 mag &    mmag &  mmag &  mmag &  mmag &    mmag &  mmag & mmag &    mmag &  mmag & \\
\hline
11.2  & 31.7 &  -9.1 &  67.3 &	15.3 &  -10.5 &  44.1 &	16.6 &   -5.1 &  40.1 &    67 \\ 
11.6  & 18.6 & -21.6 &  52.9 &	 1.9 &  -23.2 &  27.9 &	 2.7 &  -18.1 &  28.6 &   125 \\ 
12.0  & 12.5 & -22.8 &  43.4 &	 5.4 &  -19.7 &  26.6 &	 5.2 &  -16.6 &  23.3 &   177 \\ 
12.5  &  8.7 & -26.5 &  44.4 &	 0.9 &  -22.1 &  27.6 &	 2.0 &  -16.9 &  21.6 &   289 \\ 
12.9  &  7.3 & -29.5 &  35.7 &	-0.4 &  -21.2 &  23.1 &	 1.6 &  -17.1 &  19.9 &   448 \\ 
13.3  &  7.0 & -29.3 &  39.0 &	 1.1 &  -23.5 &  28.3 &	 3.3 &  -17.8 &  21.2 &   596 \\ 
13.7  & -0.2 & -39.5 &  30.5 &	 1.6 &  -21.5 &  25.7 &	 2.1 &  -18.4 &  20.0 &   882 \\ 
14.1  & -2.9 & -40.5 &  27.3 &	-2.4 &  -24.3 &  22.9 &	-0.3 &  -20.6 &  19.8 &  1111 \\ 
14.6  & -6.1 & -44.7 &  27.0 &	-2.2 &  -26.9 &  22.9 &	-1.8 &  -22.3 &  18.7 &  1419 \\ 
15.0  & -7.4 & -48.6 &  25.7 &	-2.2 &  -28.4 &  26.0 &	-1.9 &  -23.2 &  19.1 &   981 \\ 
\hline
    \end{tabular}
}
\end{table*}

%%%%%%%%%%%%%%%%%%%%%%%%%%%%%%%%%%%%%%%%%%%%%%%%%%%%%%%%%%%
\FloatBarrier

%%%%%%%%%%%%%%%%%%%%%%%%%%%%%%%%%%%%%% END APPE Delta mag for STANDARDISED SYS

%\input{sections/dereddening.tex}
%%%%%%%%%%%%%%%%%%%%%%%%%%%%%%%%%%%%%% APPE DEREDDENING C1
\section{Reddening correction for C1 passbands}
\label{sec:app_dereddening}

In this section, we provide the coefficients of the reddening curve to correct magnitudes in the $C1$ system (\secref{Gaia2C1}) for interstellar extinction. These are obtained by fitting polynomial functions to suitable theoretical simulations. To perform these simulations we used BTSettl library \citep{BTSettl} retrieved from the Spanish Virtual Observatory web server for theoretical spectra\footnote{\webref{http://svo2.cab.inta-csic.es/theory/newov2/}}. 

\begin{table*}[!htbp]
    \centering
        \caption{\label{tab:absorption} Coefficients obtained when fitting \equref{dereddening} to the passbands in the C1 system using the BTSettl SED library \citep{BTSettl}.}
{\small
    \begin{tabular}{cccccccccc
}
$X$&$\alpha$ & $\beta_1$ & $\beta_2$ & $\beta_3$ & $\beta_4$& $\gamma_1$& $\gamma_2$& $\gamma_3$& $\delta$ \\
\hline
$\frac{A_{\rm C1M326}}{A_G}$            
&  1.710 & 0.237 & 0.0131 &-0.00325 & 0.000131 &-0.0631 & 0.000303 &-0.0000476 &-0.000302  
\\
$\frac{A_{\rm C1M379}}{A_G}$            
&  1.533 & 0.257 & 0.00999 &-0.00315 & 0.000134 &-0.0736 &-0.0000418 &-0.0000625 & 0.00168 
\\
$\frac{A_{\rm C1M395}}{A_G}$            
&  1.492 & 0.249 & 0.00892 &-0.00291 & 0.000124 &-0.0716 & 0.000463 &-0.0000892 & 0.00141 
\\
$\frac{A_{\rm C1M410}}{A_G}$            
&  1.442 & 0.241 & 0.00985 &-0.00301 & 0.000127 &-0.0686 &-0.000266 &-0.0000432 & 0.00158 
\\
$\frac{A_{\rm C1M467}}{A_G}$            
&  1.233 & 0.207 & 0.00768 &-0.00251 & 0.000107 &-0.0582 &-0.000239 &-0.0000369 & 0.00147 
\\
$\frac{A_{\rm C1M506}}{A_G}$            
&  1.117 & 0.0187 & 0.00745 &-0.00229 & 0.0000981 &-0.0550 & 0.000577 &-0.0000748 & 0.000623 
\\
$\frac{A_{\rm C1M515}}{A_G}$            
&  1.089 & 0.183 & 0.00658 &-0.00213 & 0.0000905 &-0.0516 & 0.0000669 &-0.0000440 & 0.000948 
\\
$\frac{A_{\rm C1M549}}{A_G}$            
&  1.009 & 0.169 & 0.00671 &-0.00206 & 0.0000868 &-0.0474 &-0.000147 &-0.0000277 & 0.000899 
\\
$\frac{A_{\rm C1M656}}{A_G}$            
&  0.825 & 0.139 & 0.00519 &-0.00167 & 0.0000708 &-0.0412 & 0.000448 &-0.0000675 & 0.000898 
\\
$\frac{A_{\rm C1M716}}{A_G}$            
&  0.733 & 0.122 & 0.00438 &-0.00143 & 0.0000614 &-0.0363 & 0.000402 &-0.0000608 & 0.000821
\\
$\frac{A_{\rm C1M747}}{A_G}$            
&  0.683 & 0.114 & 0.00438 &-0.00137 & 0.0000583 &-0.0336 & 0.000407 &-0.0000535 & 0.000515 
\\
$\frac{A_{\rm C1M825}}{A_G}$            
&  0.567 & 0.0953 & 0.00359 &-0.00116 & 0.0000490 &-0.0266 &-0.000234 &-0.00000838 & 0.000698 
\\
$\frac{A_{\rm C1M861}}{A_G}$            
&  0.523 & 0.0878 & 0.00310 &-0.00105 & 0.0000449 &-0.0264 & 0.000364 &-0.0000513 & 0.000672 
\\
$\frac{A_{\rm C1M965}}{A_G}$            
&  0.433 & 0.0723 & 0.00259 &-0.000885 & 0.0000380 &-0.0218 & 0.000238 &-0.0000400 & 0.000679 
\\
$\frac{A_{\rm C1B431}}{A_G}$            
&  1.367 & 0.183 & 0.0126 &-0.00260 & 0.000101 &-0.0486 & 0.00117 &-0.0000154 &-0.00372 
\\
$\frac{A_{\rm C1B556}}{A_G}$            
&  1.011 & 0.151 & 0.00628 &-0.00167 & 0.0000678 &-0.0423 & 0.00116 &-0.0000564 &-0.00157 
\\
$\frac{A_{\rm C1B655}}{A_G}$            
&  0.828 & 0.138 & 0.00458 &-0.00166 & 0.0000711 &-0.0393 &-0.000192 &-0.0000291 & 0.00138 
\\
$\frac{A_{\rm C1B768}}{A_G}$            
&  0.661 & 0.106 & 0.000155 &-0.00101 & 0.0000480 &-0.0306 &-0.000148 &-0.0000417 & 0.00226 
\\
$\frac{A_{\rm C1B916}}{A_G}$            
&  0.473 & 0.0789 & 0.00239 &-0.000900 & 0.0000390 &-0.0235 & 0.000268 &-0.0000409 & 0.000657 
\\
\hline
    \end{tabular}
}
\end{table*}

\begin{figure*}[!htbp]
\center{
\includegraphics[width=0.20\textwidth]{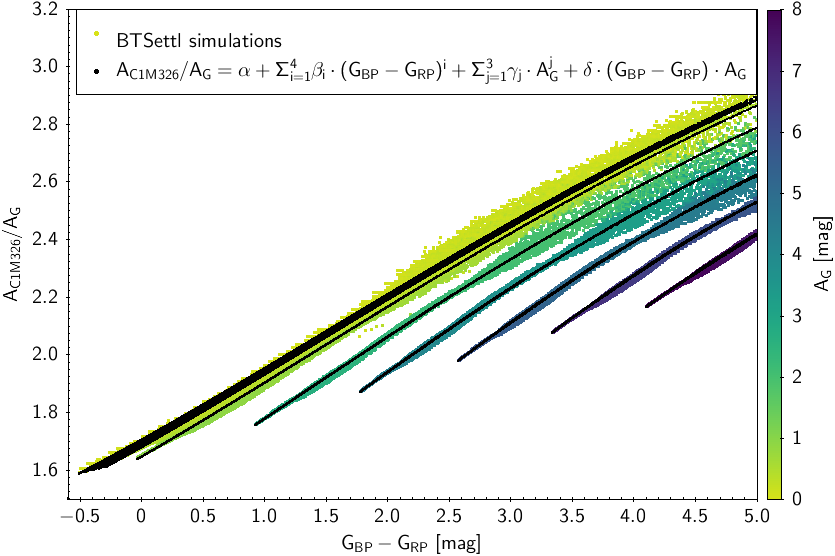}
\includegraphics[width=0.20\textwidth]{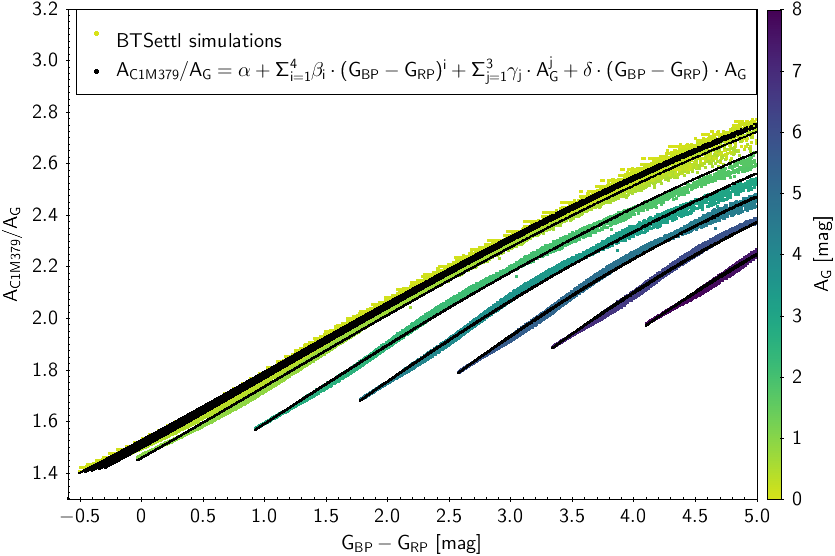}
\includegraphics[width=0.20\textwidth]{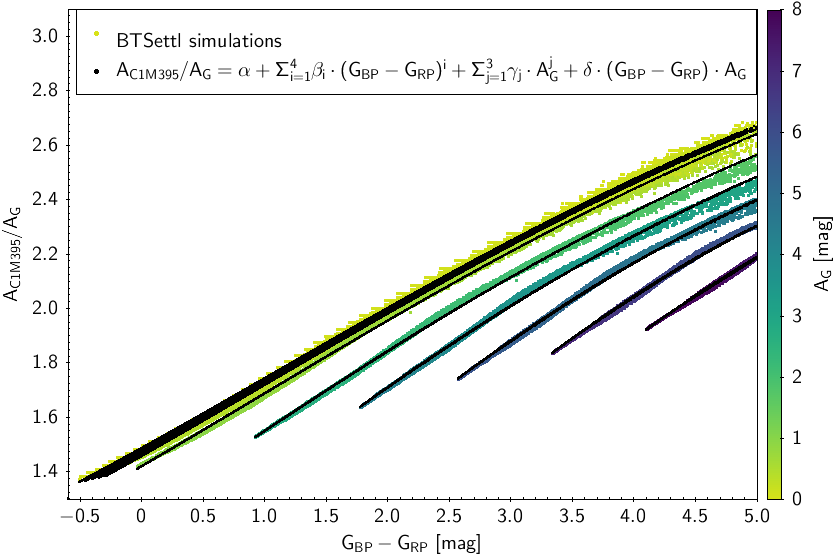}
\includegraphics[width=0.20\textwidth]{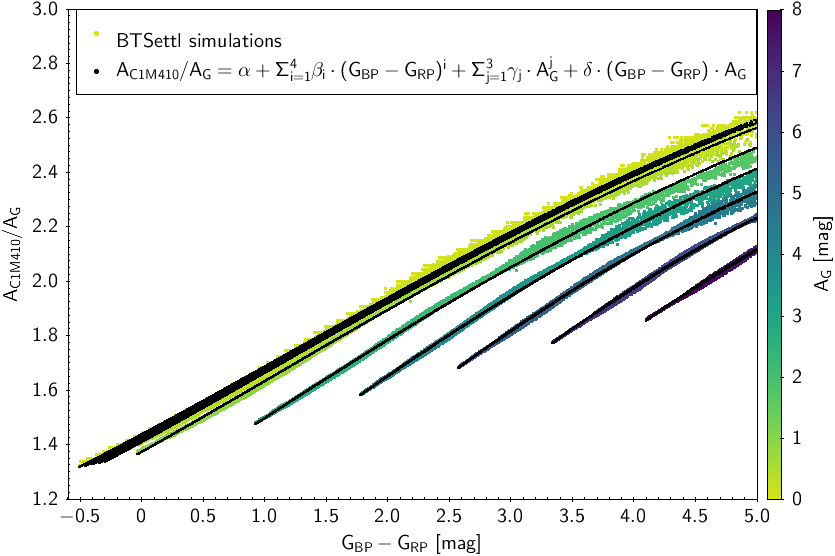}
\includegraphics[width=0.20\textwidth]{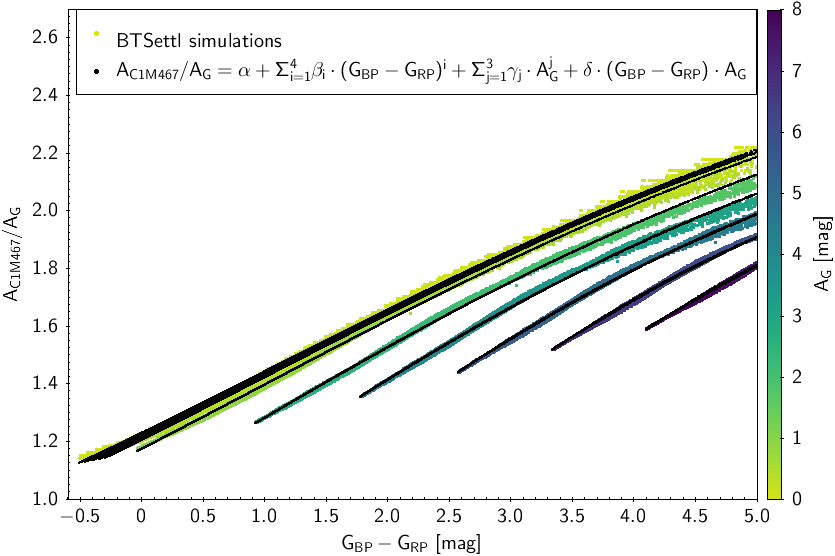}
\includegraphics[width=0.20\textwidth]{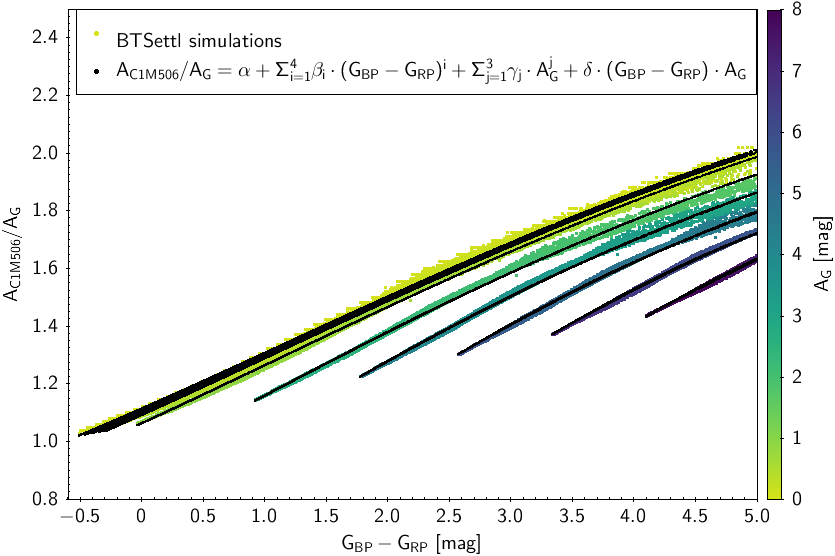}
\includegraphics[width=0.20\textwidth]{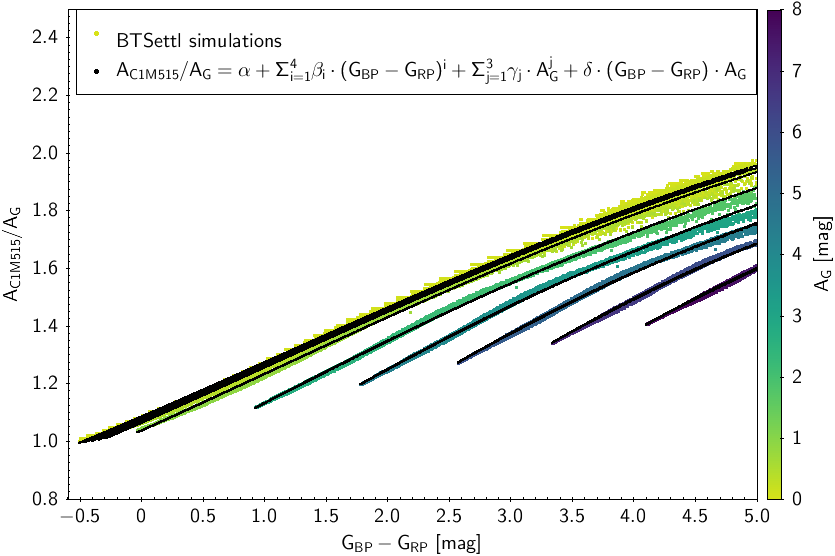}
\includegraphics[width=0.20\textwidth]{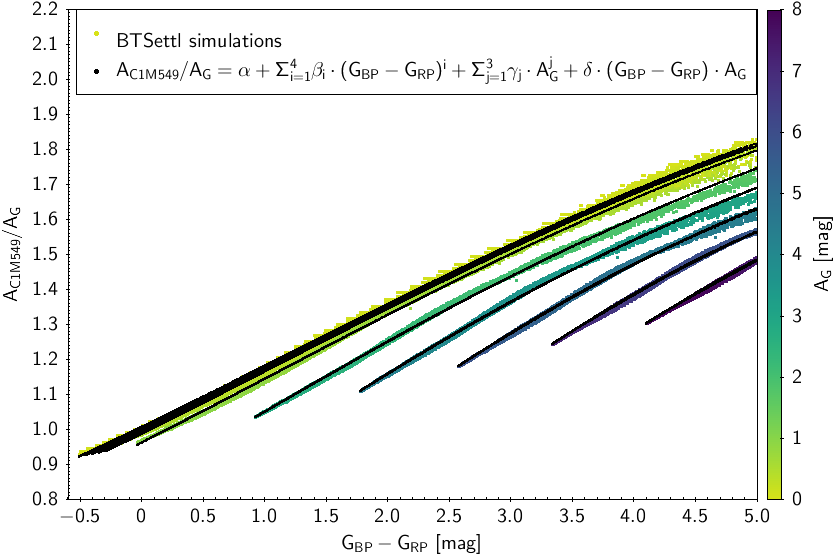}
\includegraphics[width=0.20\textwidth]{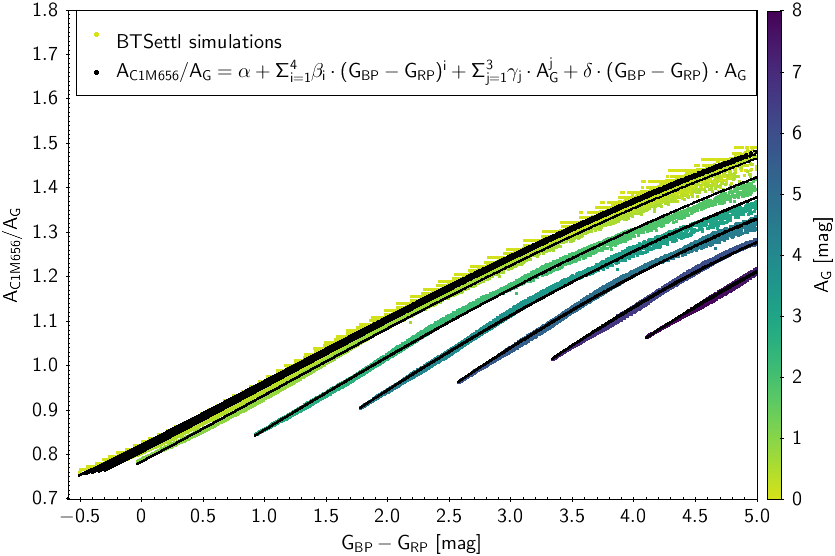}
\includegraphics[width=0.20\textwidth]{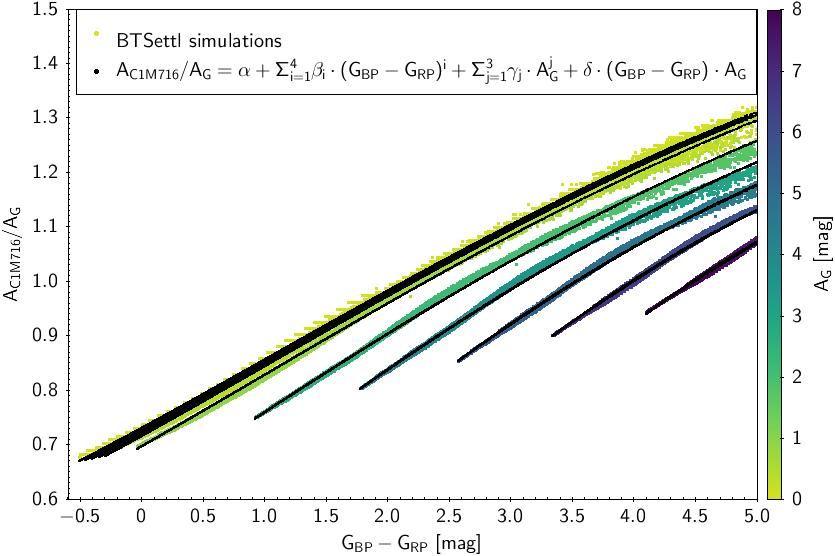}
\includegraphics[width=0.20\textwidth]{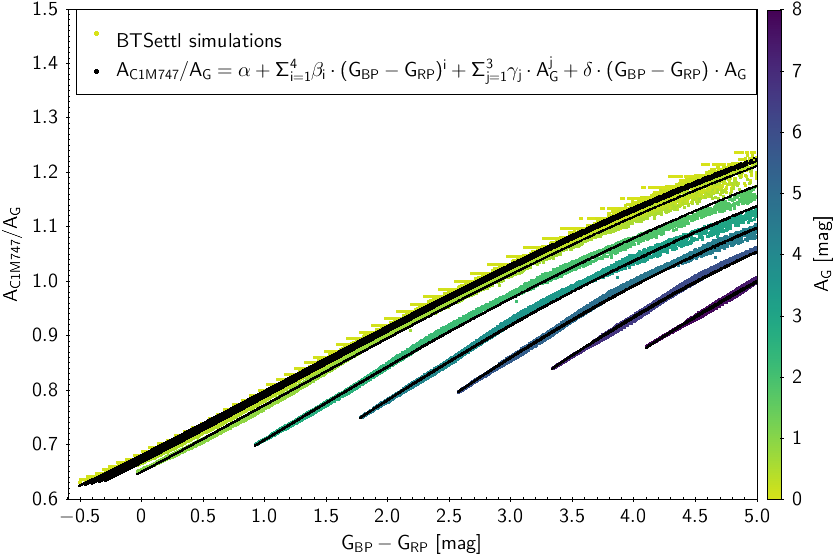}
\includegraphics[width=0.20\textwidth]{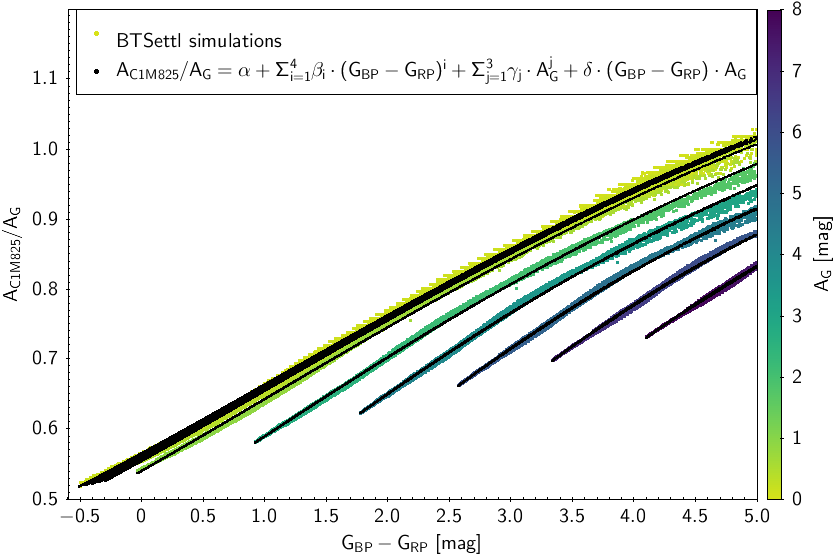}
\includegraphics[width=0.20\textwidth]{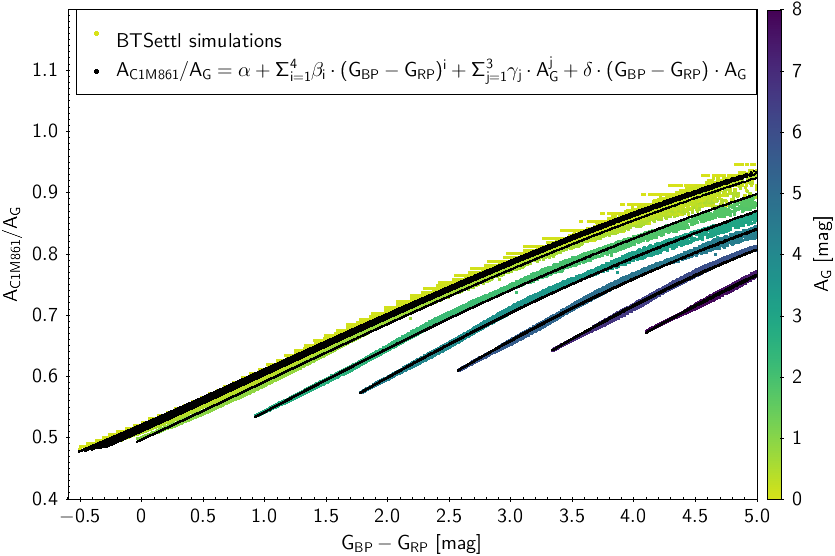}
\includegraphics[width=0.20\textwidth]{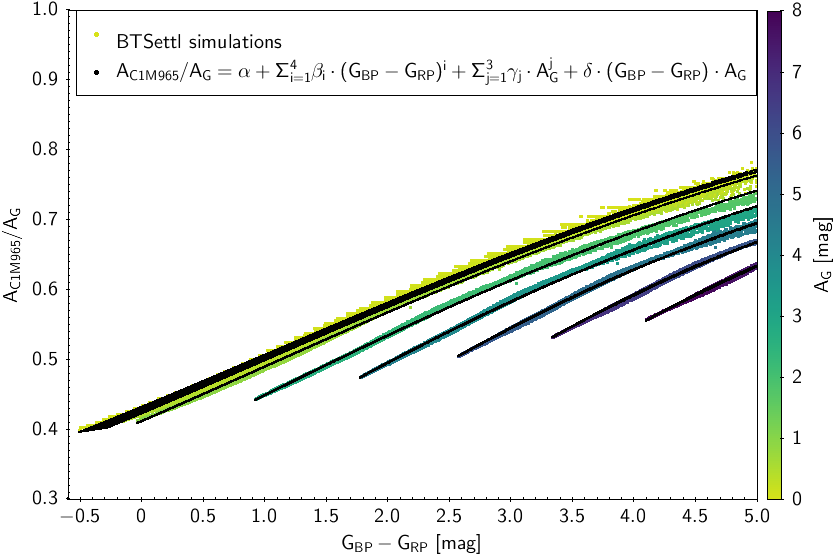}
}
\caption{Fitted relationships (in black) obtained for the simulated C1M photometry using the BTSettl library (coloured points as a function of absorption in Gaia EDR3 $G$ passband as derived by DPAC).
\label{fig:fittingAbsorptionC1M}
} 
\end{figure*}

\begin{figure*}
\center{
\includegraphics[width=0.20\textwidth]{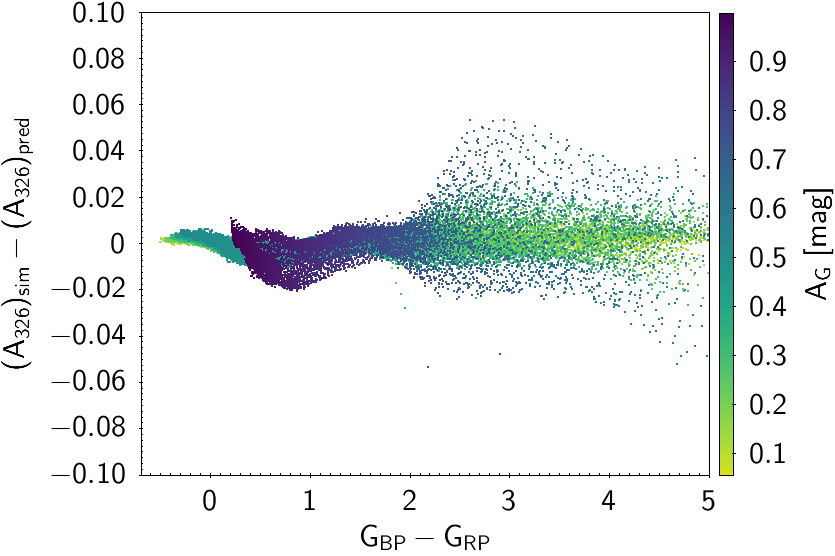}
\includegraphics[width=0.20\textwidth]{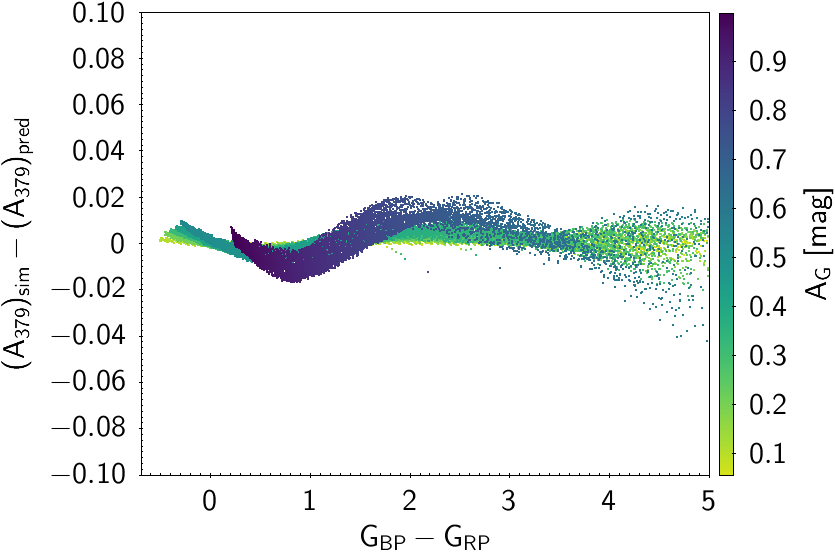}
\includegraphics[width=0.20\textwidth]{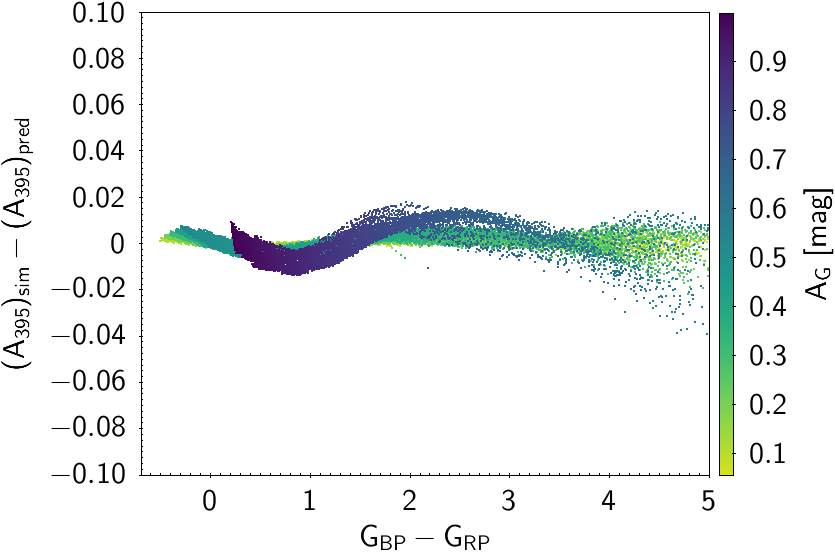}
\includegraphics[width=0.20\textwidth]{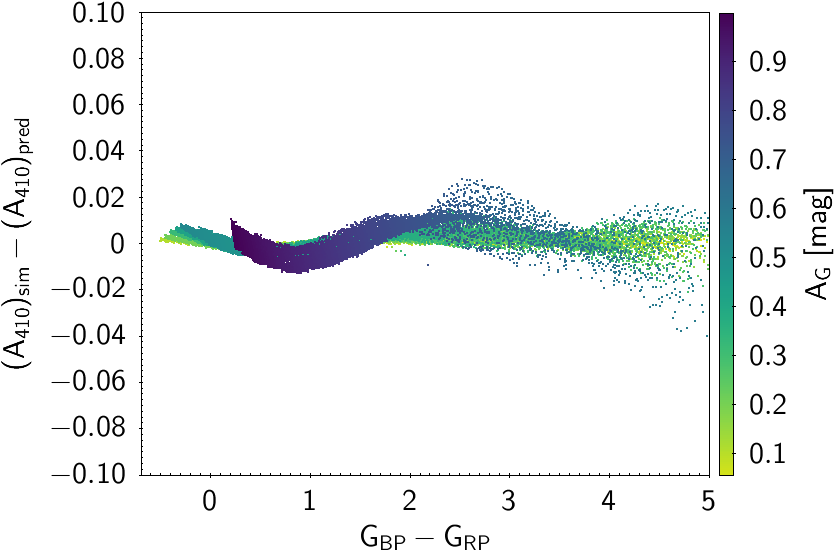}
\includegraphics[width=0.20\textwidth]{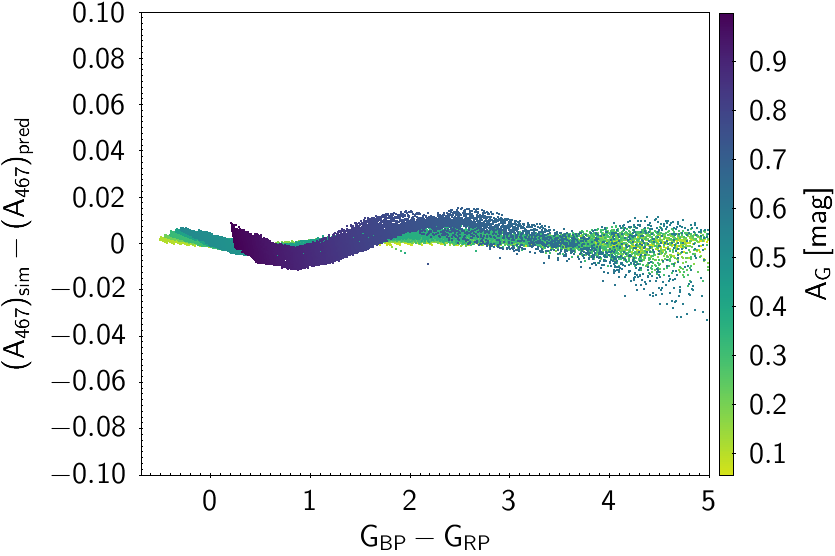}
\includegraphics[width=0.20\textwidth]{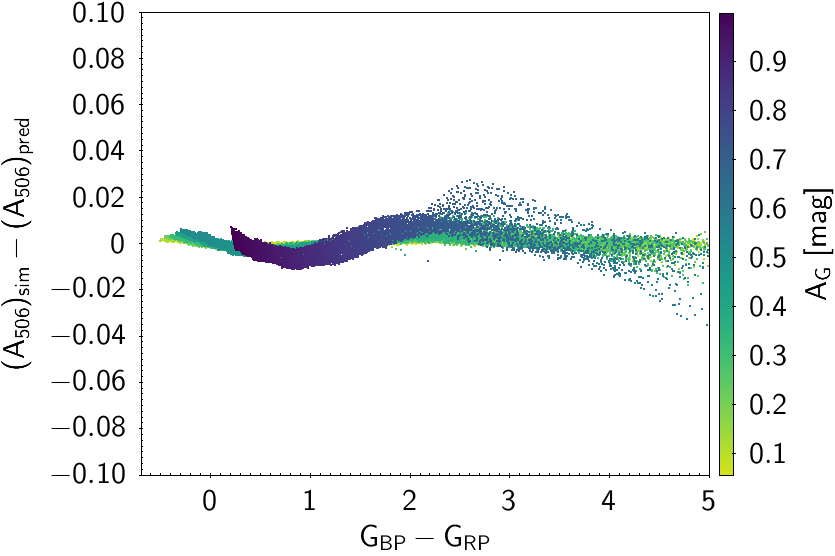}
\includegraphics[width=0.20\textwidth]{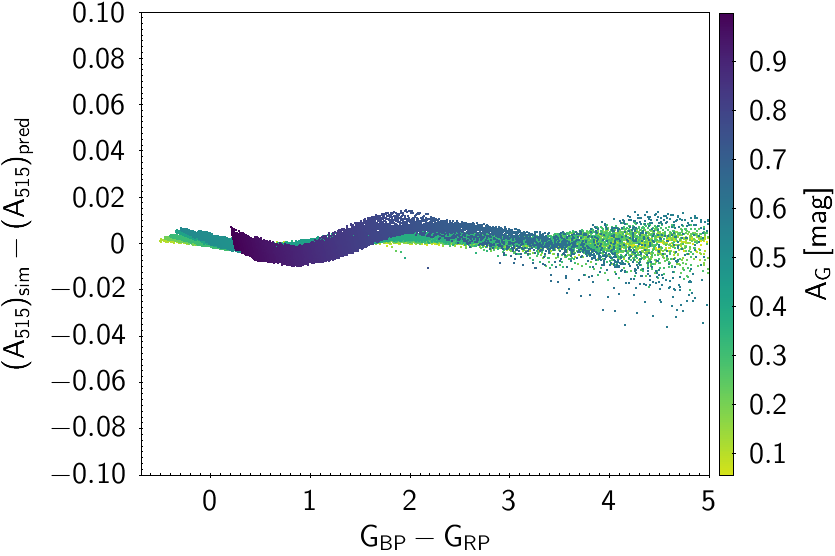}
\includegraphics[width=0.20\textwidth]{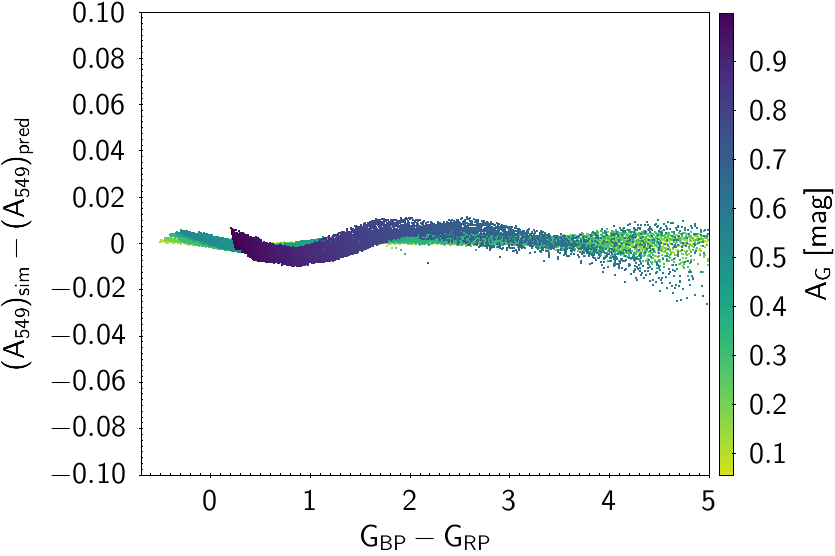}
\includegraphics[width=0.20\textwidth]{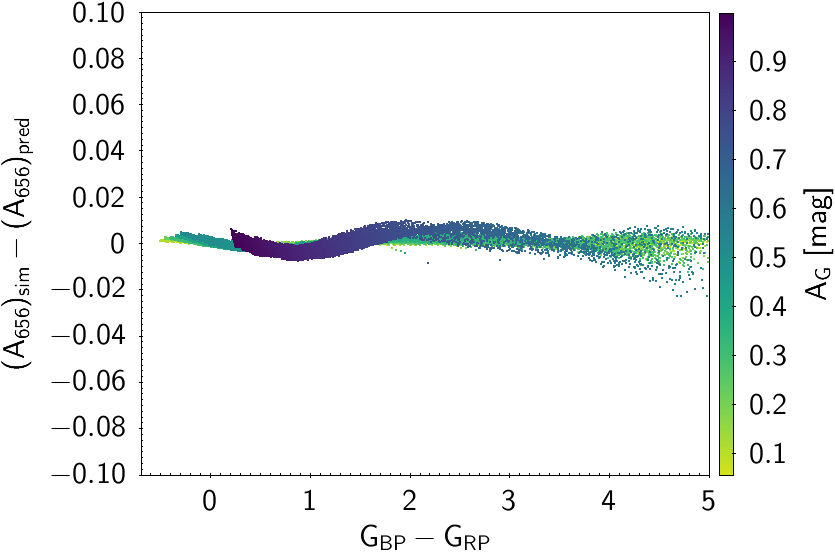}
\includegraphics[width=0.20\textwidth]{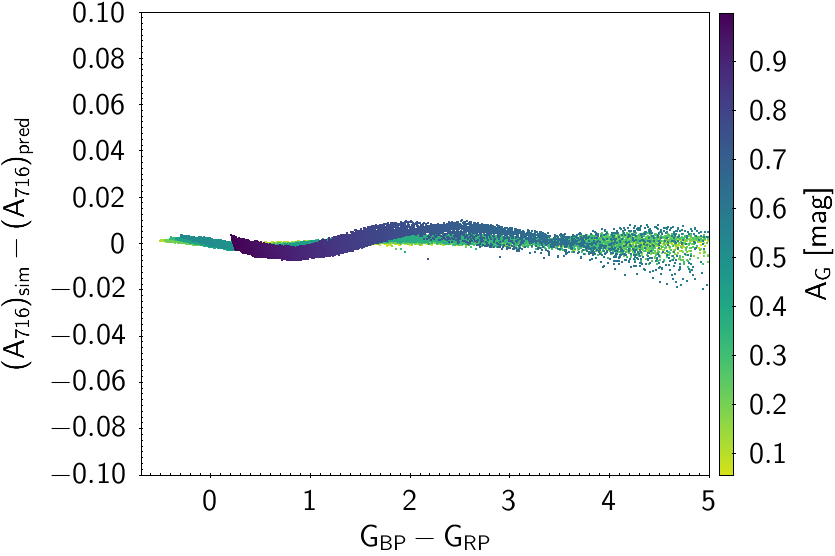}
\includegraphics[width=0.20\textwidth]{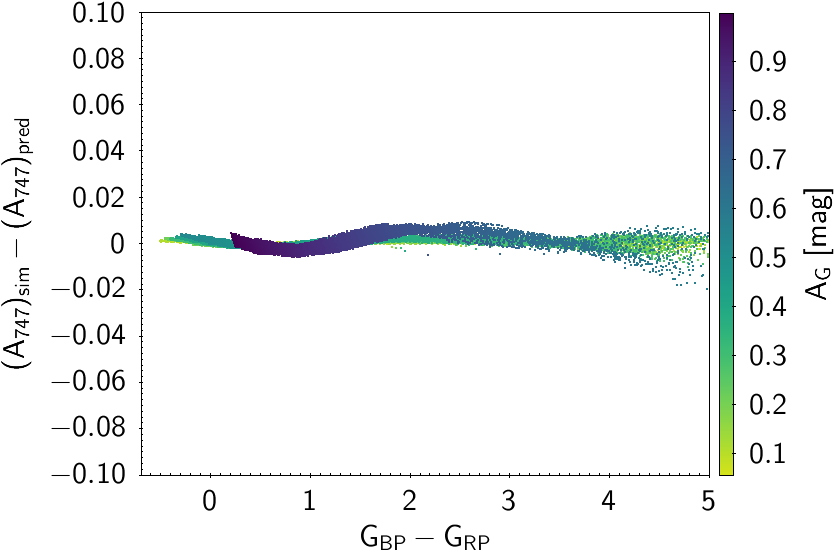}
\includegraphics[width=0.20\textwidth]{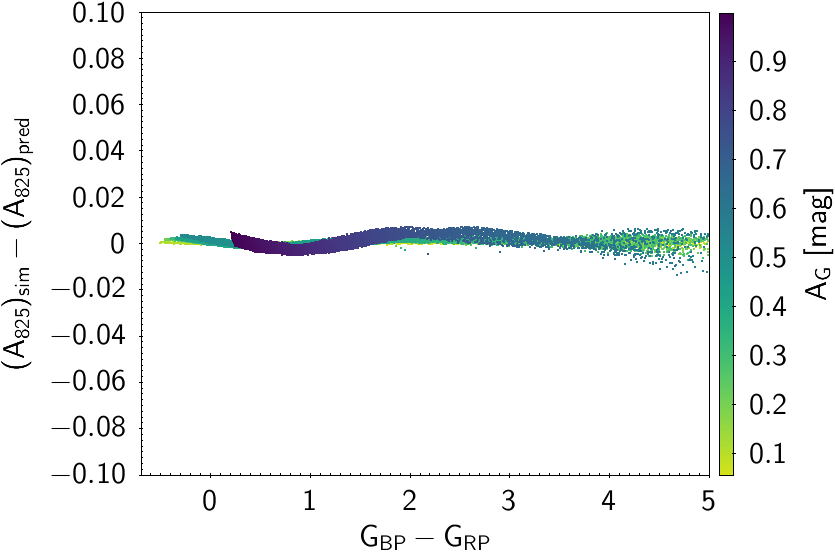}
\includegraphics[width=0.20\textwidth]{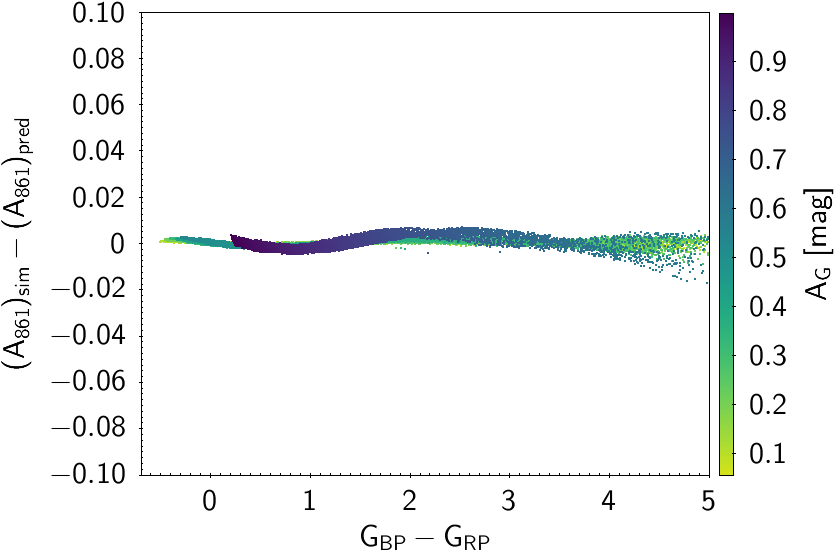}
\includegraphics[width=0.20\textwidth]{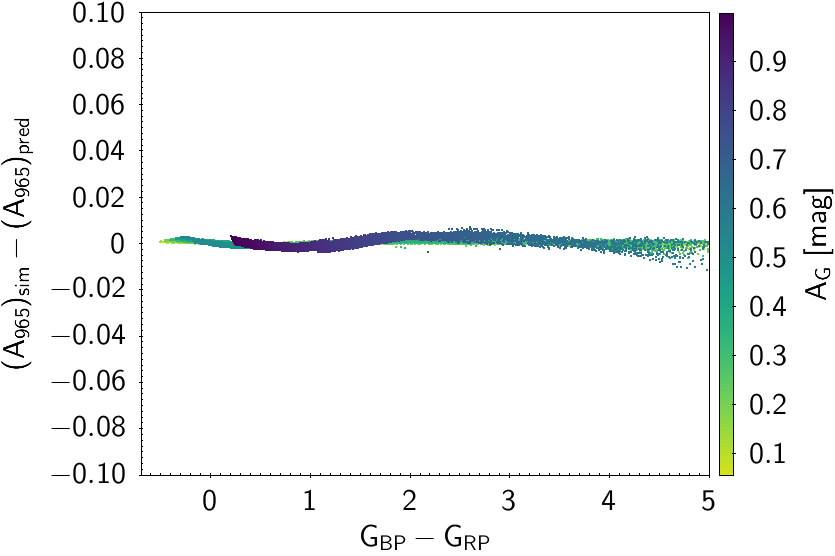}
}
\caption{Residuals obtained for the fitted relationships in Fig.\ref{fig:fittingAbsorptionC1M}.
\label{fig:resabsorptionC1M}
} 
\end{figure*}

Once the C1 photometry is simulated using BTSettl SEDs as input, we fit some polynomial dependencies to derive the absorption in any $X$ band as a function of the global absorption in {\gband}, $A_G$, the {\bprp} colour, and considering also a crossed term between both (see \equref{dereddening}).

\begin{equation}
\label{eq:dereddening}
    \frac{A_X}{A_G}=\alpha+\sum_{i=1}^{4}\beta_i \cdot ({\bprp})^i+\sum_{j=1}^3 \gamma_j \cdot A_G^j + \delta \cdot ({\bprp})\cdot A_G
.\end{equation}

\begin{figure*}
\center{
\includegraphics[width=0.20\textwidth]{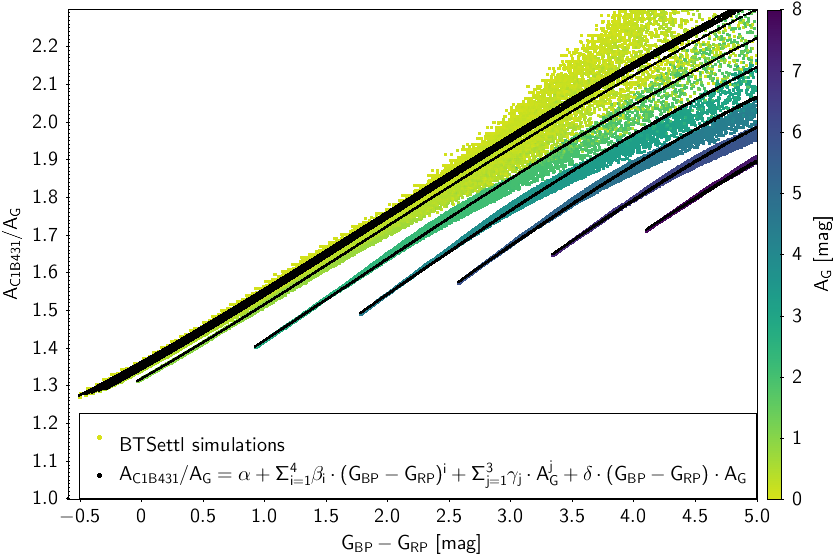}
\includegraphics[width=0.20\textwidth]{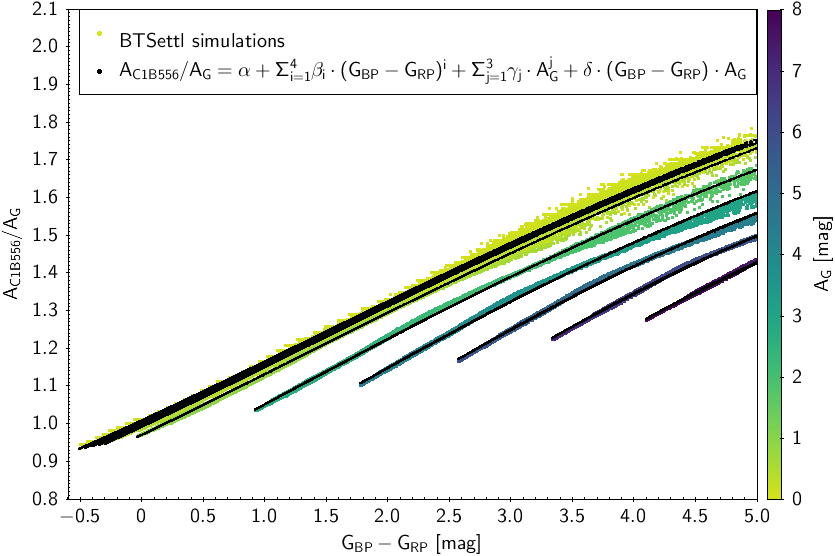}
\includegraphics[width=0.20\textwidth]{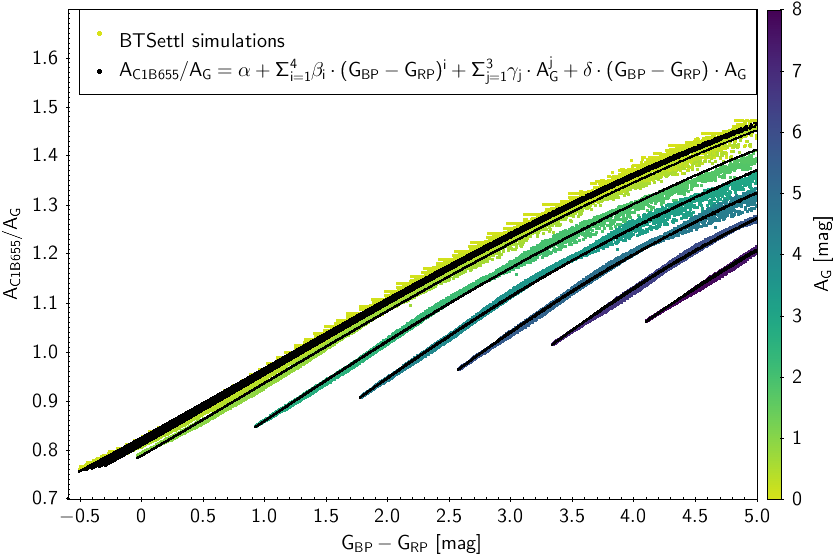}
\includegraphics[width=0.20\textwidth]{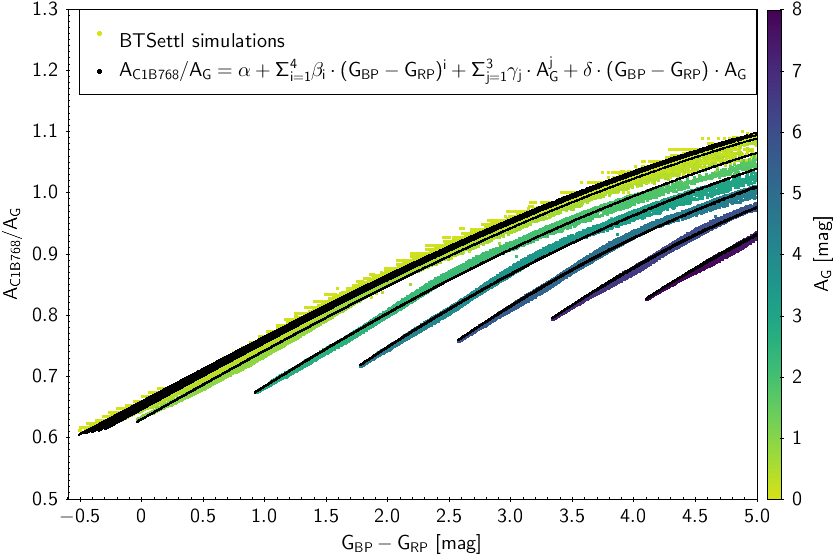}
\includegraphics[width=0.20\textwidth]{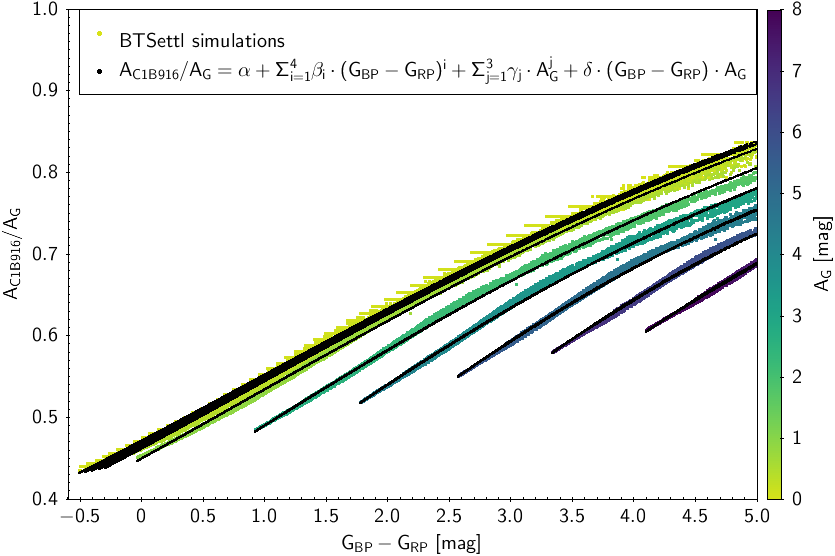}
}
\caption{Same as Fig.~\ref{fig:fittingAbsorptionC1M} but for C1B photometry.
\label{fig:fittingAbsorptionC1B}
} 
\end{figure*}

\begin{figure*}[!htbp]
\center{
\includegraphics[width=0.20\textwidth]{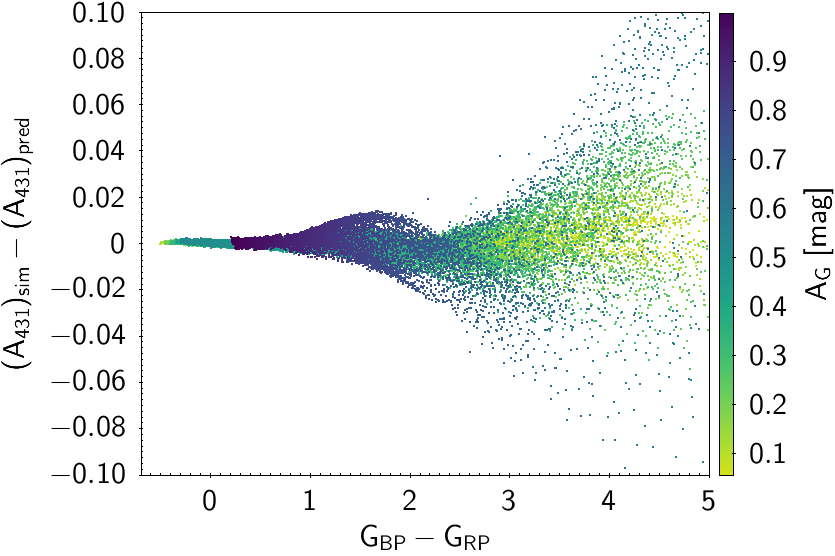}
\includegraphics[width=0.20\textwidth]{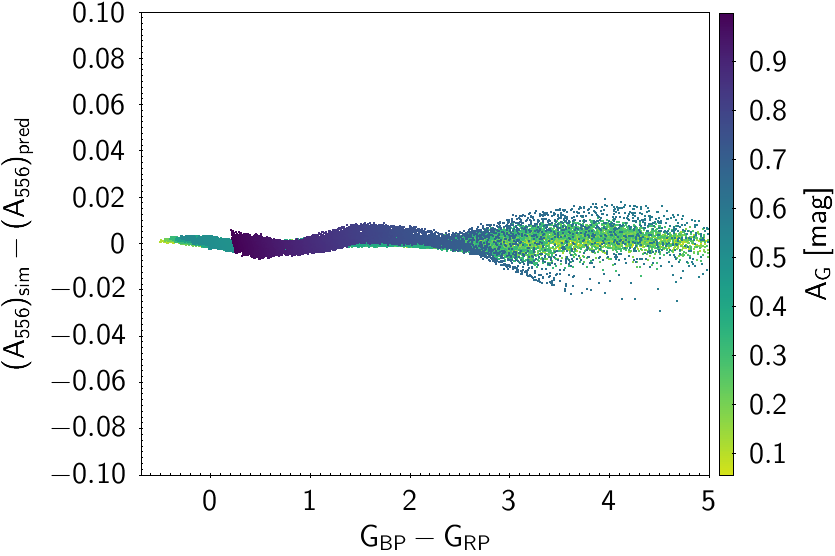}
\includegraphics[width=0.20\textwidth]{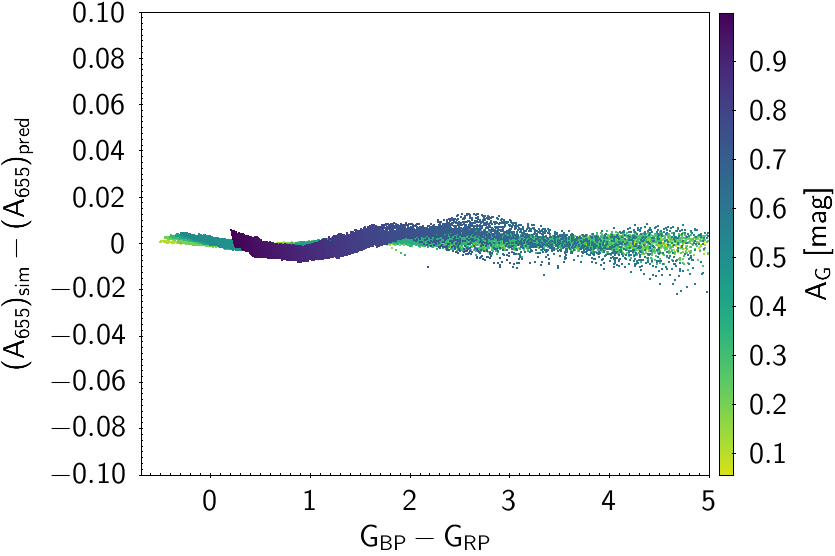}
\includegraphics[width=0.20\textwidth]{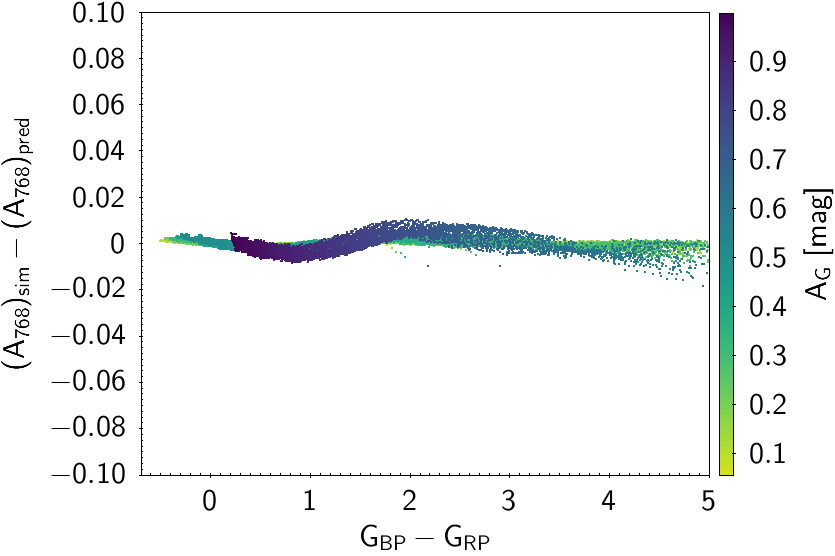}
\includegraphics[width=0.20\textwidth]{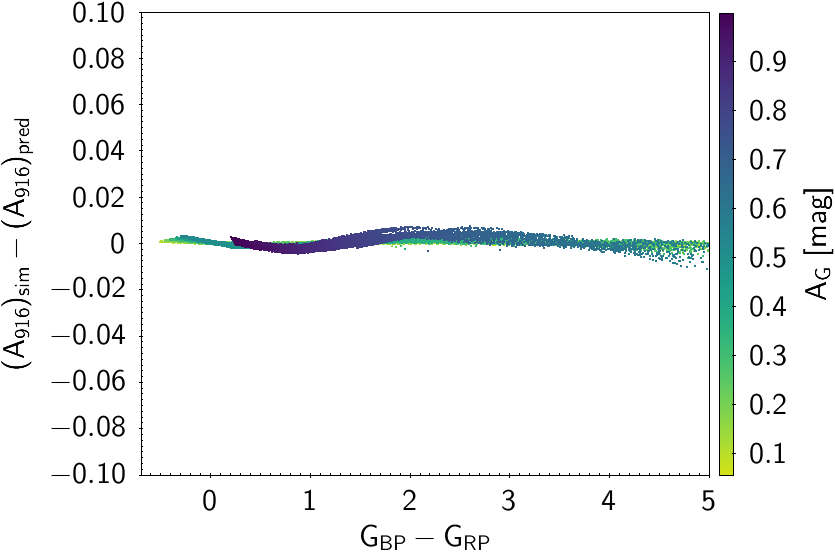}
}
\caption{Same as Fig.~\ref{fig:resabsorptionC1B} but for C1B photometry.
\label{fig:resabsorptionC1B}
} 
\end{figure*}

\afigsref{fittingAbsorptionC1M} and \ref{fig:fittingAbsorptionC1B}
show the obtained fitted laws for every C1M medium and C1B broad passbands, respectively. The coefficients obtained for all C1 passbands are included in \tabref{absorption}. 
Although we produced the fitting using also extremely red sources (brown dwarfs) present in the BTSettl library, we recommend restricting the applicability of these relationships to the intervals plotted in the figures (${\bprp}<5$~mag). The residuals obtained with these polynomials for every passband are plotted in \figsref{resabsorptionC1M} and \ref{fig:resabsorptionC1B} for C1M and C1B, respectively.

\FloatBarrier

%%%%%%%%%%%%%%%%%%%%%%%%%%%%%%%%%%%%%% APPE DEREDDENING C1

%\input{sections/acronyms}
%%%%%%%%%%%%%%%%%%%%%%%%%%%%%%%%%%%%%% APPE ACRONYMS
\section{\gaia-related acronyms}\label{sec:acronyms}

For convenience, we list all the \gaia-related acronyms used in this paper in Table~\ref{tab:acronyms}.

%%%%%%%%%%%%%%%%%%%%%%%%%%%%%%%%%%%%%%%%%%%%%%%%%%%%%%%%%%%%%%
\begin{table*}[!hbp]
    \caption{\gaia-related acronyms used in the paper. Each acronym is also defined at its first occurrence in the paper.}
    \label{tab:acronyms}
    \centering
    \begin{tabular}{l|l|l}
    \hline\hline
Acronym & Description & See \\
\hline
Apsis & Astrophysical parameter inference system & \ref{sec:Gaia2C1}\\
BP & Blue Photometer & \secref{introduction} \\
\cu(s) & Calibration Unit  & \secref{introduction} \\
DPAC & Data Processing and Analysis Consortium & \secref{introduction}\\
ECS & Externally Calibrated (XP) Spectra  & \secref{introduction} \\
ELS & Emission Line Star & \ref{sec:narrow_iphas} \\
ESA & European Space Agency & \secref{introduction} \\
ESP-ELS & DR3 module dealing with ELS & \ref{sec:narrow_iphas} \\
FoV(s) & Field(s) of View & \secref{standa_hugs} \\
G, G$_{BP}$, G$_{BP}$ & Integrated \gaia magnitudes/fluxes &\secref{stand} \\
GCNS & \gaia Nearby Stars Catalogue & \secref{Gaia2C1} \\
GSPC & \gaia Synthetic Photometry Catalogue & \secref{gspc} \\
GSPC-WD & \gaia Synthetic Photometry Catalogue for White Dwarfs& \secref{gwdc} \\
GSP-Phot & DR3 module deriving astrophysical parameters from XP spectra & \secref{perfver}\\
GSP-Spec & DR3 module deriving astrophysical parameters from RVS spectra & \secref{perfver}\\
%  & Instrument Model  & \secref{methods} \\
LSF & Line Spread Function & \secref{introduction} \\
%NGSL & New Generation Spectral Library & \cite{ngsl16} \\
PVL & Passband Validation Library & \secref{app_Rf}\\
RP & Red Photometer & \secref{introduction} \\
RVS & Radial Velocity Spectrometer & \secref{mps} \\ 
SPSS & \gaia Spectro Photometric Standard Stars &  \secref{introduction}\\
XP    & BP and RP (referred to spectra or photometry) & \secref{introduction}\\
XPSP    & synthetic photometry from XP spectra   & \secref{introduction}\\
\end{tabular}
\end{table*}
%%%%%%%%%%%%%%%%%%%%%%%%%%%%%%%%%%%%%%%%%%%%%%%%%%%%%%%%%%%%%%
\FloatBarrier

%%%%%%%%%%%%%%%%%%%%%%%%%%%%%%%%%%%%%% END APPE ACRONYMS

\end{appendix}

\end{document}